\documentclass[letterpaper, 12pt]{article}
\usepackage{graphicx}
\usepackage{epsfig}
\usepackage{multirow}
\usepackage{slashbox}
\usepackage[TABBOTCAP]{subfigure}

\usepackage[ruled]{algorithm2e}
\usepackage{amsbsy}
\usepackage{amssymb}
\usepackage{amsthm}
\usepackage{amsmath}
\usepackage{array}
\usepackage{bm}
\usepackage{rotating}
\usepackage{stackrel}
\usepackage{subfigure}
\usepackage{url}
\usepackage{color}
\usepackage{xspace}
\usepackage{ifpdf}
\ifpdf
  \usepackage{epstopdf} 
\fi
\usepackage[T1]{fontenc}

\usepackage[colorlinks=true, citecolor=blue, filecolor=black, linkcolor=red, urlcolor=blue]{hyperref}
\renewcommand{\[}{\left[}
\renewcommand{\]}{\right]}
\renewcommand{\(}{\left(}
\renewcommand{\)}{\right)}

\newcommand{\pp}[2]{\frac{\partial #1}{\partial #2}}

\newcommand{\norm}[2]{\|\, #1 \,\|_{#2}}
\newcommand{\vvvert}{|\kern-1pt|\kern-1pt|}

\newcommand{\hU}{\hat{U}}

\newcommand{\hg}{\hat{g}}
\newcommand{\hh}{\hat{h}}

\newcommand{\hx}{\hat{x}}

\newcommand{\barh}{\bar{h}}
\newcommand{\barU}{\bar{U}}

\newcommand{\mb}[1]{\mathbf{#1}}

\newcommand{\bC}{\mathbf{C}}

\newcommand{\bE}{\mathbf{E}}

\newcommand{\bG}{\mathbf{G}}
\newcommand{\bI}{\mathbf{I}}

\newcommand{\bY}{\mathbf{Y}}
\newcommand{\bZ}{\mathbf{Z}}
\newcommand{\bb}{\mathbf{b}}
\newcommand{\bd}{\mathbf{d}}

\newcommand{\bii}{\mathbf{i}}

\newcommand{\bn}{\mathbf{n}}

\newcommand{\bx}{\mathbf{x}}
\newcommand{\by}{\mathbf{y}}
\newcommand{\bz}{\mathbf{z}}

\newcommand{\btheta}{\boldsymbol{\theta}}
\newcommand{\bttheta}{\tilde{\boldsymbol{\theta}}}

\newcommand{\bxi}{\boldsymbol{\xi}}
\newcommand{\bTheta}{\boldsymbol{\Theta}}
\newcommand{\bXi}{\boldsymbol{\Xi}}

\newcommand{\EE}{\mathbb{E}}

\newcommand{\NN}{\mathbb{N}}

\newcommand{\RR}{\mathbb{R}}

\newcommand{\CD}{\mathcal{D}}

\newcommand{\CF}{\mathcal{F}}
\newcommand{\CH}{\mathcal{H}}

\newcommand{\CX}{\mathcal{X}}
\newcommand{\CY}{\mathcal{Y}}
\newcommand{\CZ}{\mathcal{Z}}

\newcommand{\Var}{\textrm{Var}}

\def\sgn{\mathop{\rm sign}}

\providecommand{\SetAlgoLined}{\SetLine}
\providecommand{\DontPrintSemicolon}{\dontprintsemicolon}
\SetKwHangingKw{KwSolve}{Solve}

\begin{document}

\title{Gradient-based stochastic optimization methods in \\ Bayesian experimental design}

\author{Xun Huan, Youssef M.~Marzouk\footnote{Corresponding author:
    \href{mailto:ymarz@mit.edu}{ymarz@mit.edu},
    \url{http://web.mit.edu/aeroastro/labs/uqlab/index.html}, 77
    Massachusetts Avenue, Massachusetts Institute of Technology,
    Cambridge, MA 02139, USA.}}

\maketitle

\abstract{Optimal experimental design (OED) seeks experiments expected
  to yield the most useful data for some purpose. In practical
  circumstances where experiments are time-consuming or
  resource-intensive, OED can yield enormous savings. We pursue OED
  for nonlinear systems from a Bayesian perspective, with the goal of
  choosing experiments that are optimal for parameter inference. Our
  objective in this context is the \textit{expected information gain}
  in model parameters, which in general can only be estimated using
  Monte Carlo methods. Maximizing this objective thus becomes a
  stochastic optimization problem. \\
  
  This paper develops gradient-based stochastic optimization methods
  for the design of experiments on a continuous parameter space. Given
  a Monte Carlo estimator of expected information gain, we use
  infinitesimal perturbation analysis to derive gradients of this
  estimator. We are then able to formulate two gradient-based
  stochastic optimization approaches: (i) Robbins-Monro stochastic
  approximation, and (ii) sample average approximation combined with a
  deterministic quasi-Newton method. A polynomial chaos approximation
  of the forward model accelerates objective and gradient evaluations
  in both cases. We discuss the implementation of these optimization
  methods, then conduct an empirical comparison of their
  performance. To demonstrate design in a nonlinear setting with
  partial differential equation forward models, we use the problem of
  sensor placement for source inversion. Numerical results yield
  useful guidelines on the choice of algorithm and sample sizes,
  assess the impact of estimator bias, and quantify tradeoffs of
  computational cost versus solution quality and robustness.}

% Up to seven key words are required. Please provide a list consulting
% the keywords given in the document IJ4UQ-KeyWords.pdf if different
% key words as needed.

%\keywords{stochastic approximation, sample average approximation,
%  polynomial chaos, infinitesimal perturbation analysis, optimal
%  experimental design, mutual information, Bayesian inference}

\section{Introduction}

Experimental data play a crucial role in the development of
models---and the advancement of scientific understanding---across a
host of disciplines. Some experiments are more useful than others,
however, and a careful choice of experiments can translate to enormous
savings of time and financial resources.
Traditional experimental design methods, such as factorial and
composite designs, are largely used as heuristics for exploring the
relationship between input factors and response
variables. \textit{Optimal} experimental design, on the other hand,
uses a model to guide the choice of experiments for a particular
purpose, such as parameter inference, prediction, or model
discrimination. Optimal design has seen extensive development for
linear models
% (where the measured quantities depend linearly on the model
% parameters)
endowed with Gaussian distributions~\cite{atkinson:1992:oed}.
Extensions to nonlinear models are often based on linearization and
Gaussian approximations~\cite{box:1959:doe, ford:1989:rai,
  chaloner:1995:bed}, as analytical results are otherwise impractical
or impossible to obtain. With advances in computational power,
however, optimal experimental design for nonlinear systems can now be
tackled directly using numerical simulation~\cite{loredo:2003:bae,
  ryan:2003:eei, vanDenBerg:2003:onb, loredo:2010:rsa,
  solonen:2012:sbo, huan:2010:abe, huan:2013:sbo}.

This paper pursues nonlinear experimental design from a Bayesian
perspective (e.g.,~\cite{muller:1998:sbo}). The Bayesian statistical
approach~\cite{kennedy:2001:bco, sivia:2006:daa} provides a rigorous
foundation for inference from noisy, indirect, and incomplete data
and a natural mechanism for incorporating physical constraints and
heterogeneous sources of information. We focus on experiments
described by a {continuous} design space, with the goal of choosing
experiments that are optimal for Bayesian {parameter inference}. A
useful objective function for this purpose is the \textit{expected
  information gain} in model parameters~\cite{lindley:1956:oam,
  lindley:1972:bsa}---or equivalently, the \textit{mutual information}
between parameters and observables, conditioned on the design
variables. This objective can be derived in a decision theoretic
framework, using the Kullback-Leibler divergence from posterior to
prior as a utility function~\cite{chaloner:1995:bed}. From the
numerical perspective, however, it is a complicated quantity. In
general, it must be approximated using a Monte Carlo
method~\cite{ryan:2003:eei, terejanu:2011:bed}. Consequently, only
noisy estimates of the objective function are available and the
optimal design problem becomes a stochastic optimization problem.

There are many approaches for solving continuous optimization problems
with stochastic objectives. While some do not require the direct
evaluation of gradients (e.g., Nelder-Mead~\cite{nelder:1965:asm},
Kiefer-Wolfowitz~\cite{kiefer:1952:seo}, and simultaneous perturbation
stochastic approximation~\cite{spall:1998:aoo}), other algorithms can
use gradient evaluations to great advantage. Broadly, these algorithms
involve either stochastic approximation (SA)~\cite{kushner:2003:saa}
or sample average approximation (SAA)~\cite{shapiro:1991:aao}, where
the latter approach must also invoke a gradient-based deterministic
optimization algorithm. Hybrids of the two approaches are possible as
well. In either case, for model-based experimental design, one must
employ gradients of the information gain objective described
above. This objective function itself involves nested integrations
over possible model outputs and over the input parameter space, where
the model output may be a functional of the solution of a partial
differential equation. In many practical cases, the model may be
essentially a black box; while in other cases, even if gradients can
be evaluated with adjoint methods, using the full model to evaluate
the expected information gain or its gradient is computationally
prohibitive. Previous work \cite{huan:2013:sbo} has addressed these
difficulties by constructing polynomial surrogates for the the model
output, i.e., polynomial chaos expansions~\cite{wiener:1938:thc,
  ghanem:1991:sfe, xiu:2002:twa, debusschere:2004:nci, najm:2009:uqa,
  xiu:2009:fnm, lemaitre:2010:smf} that capture dependence on both
uncertain parameters and design variables.

The main contributions of this paper are as follows. First, we show
how to use infinitesimal perturbation analysis to derive gradients of
a Monte Carlo estimator of the expected information gain. 
% This task is non-trivial because the estimator is not a simple Monte
% Carlo sum; rather it contains nested Monte Carlo estimates.
When the estimator incorporates a polynomial surrogate, we show how
this surrogate can be readily extended to provide analytical gradient
estimates. We then conduct a systematic empirical comparison of two
gradient-based stochastic optimization approaches for nonlinear
experimental design: (1) Robbins-Monro (RM) stochastic approximation,
and (2) sample average approximation combined with a deterministic
quasi-Newton method. The comparison is performed in the context of a
physics-based sensor placement application, where the forward model is
given by a partial differential equation. From the numerical results,
we are able to assess the impact of estimator bias, extract useful
guidelines on the choice of algorithm and sample sizes, and quantify
tradeoffs of computational cost versus solution quality and
robustness.

The RM algorithm~\cite{robbins:1951:asa} is the original and perhaps
most widely used stochastic approximation method, and has become a
prototype for many subsequent algorithms. It involves an iterative
update that resembles steepest descent, except that it uses stochastic
gradient information.  Sample average approximation (SAA) (also known
as the retrospective method~\cite{healy:1991:rsr} or the sample-path
method~\cite{gurkan:1994:spo}) is a more recent approach, with
theoretical analysis initially appearing in the
1990s~\cite{shapiro:1991:aao, gurkan:1994:spo,
  kleywegt:2001:tsa}. Convergence rates and stochastic bounds,
although useful, do not necessarily reflect empirical performance
under finite computational resources and with imperfect numerical
optimization schemes. To the best of our knowledge, extensive
numerical testing of SAA has focused on stochastic programming
problems with special structure (e.g., linear programs with discrete
design variables)~\cite{ahmed:2002:tsa, verweij:2003:tsa,
  benisch:2004:asp, greenwald:2006:sut, schutz:2009:scd}.  While
numerical improvements to SAA have seen continual development (e.g.,
estimators of optimality gap~\cite{norkin:1998:aba, mak:1999:mcb} and
sample size adaptation~\cite{chen:1994:raa, chen:2001:srf}), the
practical behavior of SAA in more general optimization settings is
largely unexplored. The numerical assessment of SAA conducted here, in
a nonlinear and continuous variable design setting, is thus expected
to be of practical interest.

SAA is frequently compared to stochastic approximation methods such as
RM. For example, \cite{shapiro:2003:spb} suggests that SAA is more
robust than SA because of sensitivity to step size choice in the
latter. On the other hand, variants of SA have been developed that,
for certain classes of problems (e.g., \cite{nemirovski:2009:rsa}),
reach solution quality comparable to that of SAA in substantially less
time. The comparison of SA and SAA presented here focuses on their
performance in the Bayesian experimental design problem. We do not aim
to identify one approach as superior to the other; instead, we will
simply illustrate the differences between the two algorithms in this
context and provide some selection guidelines based on their
properties.

% {\color{red} The key contributions of this paper are as
%   follows. First, we introduce the use of gradient-based stochastic
%   optimization methods in the context of Bayesian experimental design
%   for nonlinear and PDE-based models, while maintaining computational
%   feasibility (but this last bit makes it sound by doing that we made
%   it more expensive but still feasible... rephrase or
%   remove). Specifically, access to gradient information is now made
%   possible through the use of polynomial chaos surrogates for the
%   forward model, and we provide detailed analytical derivation of the
%   gradient formula. Second, we demonstrate the algorithm for a
%   non-trivial, physics-based sensor placement application. Through the
%   numerical results, we provide guidance on how to select the
%   algorithm sample sizes, and highlight the advantages and
%   disadvantages of the different stochastic optimization methods. }

This paper is organized as follows. Section~\ref{s:experimentalDesign}
introduces optimal Bayesian experimental design (\S\ref{s:oedform})
and extracts the underlying stochastic optimization problem
(\S\ref{s:stochasticOptimization}), then presents the RM
(\S\ref{s:SARM}) and SAA-BFGS (\S\ref{s:SAA}) algorithms. The challenge
of evaluating gradient information appropriate to each of these
algorithms is described in Section~\ref{s:problem}.
Section~\ref{s:pce} and Section~\ref{s:ipa} describe how to obtain
gradients (or gradient estimators) for the experimental design
objective using polynomial chaos expansions and infinitesimal
perturbation analysis. Section~\ref{s:diffusion} then
analyzes the numerical performance of RM and SAA-BFGS on an optimal
sensor placement problem involving contaminant diffusion.
Conclusions on the algorithms and the relative strengths of SA and SAA
for optimal experimental design are provided in
Section~\ref{s:conclusions}.

\section{Optimal Bayesian Experimental Design}
\label{s:experimentalDesign}

\subsection{Background}
\label{s:oedform}

We are interested in choosing the ``best'' experiments\footnote{These
  design choices will be made all-at-once; this setup corresponds to
  batch or \textit{open-loop} design. In contrast, sequential or
  \textit{closed-loop} design allows the results of one set of
  experiments to guide the next set. Rigorous approaches to optimal
  closed-loop design are more challenging, and will not be tackled in
  this paper.} from a continuously parameterized design space, for the
purpose of inferring model parameters from noisy and indirect
observations. In other words, we seek experiments that are optimal for
parameter inference (in a sense to be precisely defined below), with
inference performed in a Bayesian setting. In the problems considered
here, the mean observations are nonlinear functions of the model
parameters, and the observations and model parameters are continuous
random variables.

% ... YMM: this footnote is interesting and well written, but at this
% stage it seems like a digression. Either we should move it elsewhere
% (a side note to the discusssion) or omit it from this paper.
%
% \footnote{The design parameters in this theoretical formulation (i.e.,
% in the absence of numerical algorithms or approximations that
% utilize random numbers) are not random variables, as there are no
% inherent underlying uncertainties associated with them. However,
% they do become random once numerical algorithms and approximations
% are prescribed: for example, if random initial positions are imposed
% for an optimization process or if Monte Carlo sampling is used to
% approximate the objective function. Consequently, the design points
% encountered in the overall process will be stochastic, with a
% complicated underlying distribution affected by the algorithms and
% sample realizations.
  
% The random variable view of the design parameters is especially
% important in the numerical treatment of closed-loop design
% processes, but it is not crucial to our open-loop design study. We
% elect not to treat them as random variables in this study, but for
% ease of notation, we may still condition on particular values of the
% design parameters.}

Bayes' rule describes the parameter update process:
\begin{eqnarray}
  f_{\bTheta|\bY,\bd}(\btheta|\by,\bd) =
  \frac{f_{\bY|\bTheta,\bd}(\by|\btheta,\bd) f_{\bTheta|\bd}(\btheta|\bd)}
  {f_{\bY|\bd}(\by|\bd)}. \label{e:Bayes}
\end{eqnarray}
Here $\bTheta$ represents the uncertain parameters of interest, $\bY$
the observations, and $\bd$ the design variables. Like the
observations and parameters, the design parameters are
\textit{continuous}. Also $f_{\bTheta|\bd}$ is the prior
density, $f_{\bY|\bTheta,\bd}$ is the likelihood
function, $f_{\bTheta|\bY,\bd}$ is the posterior
density, and $f_{\bY|\bd}$ is the evidence. It is reasonable
to assume that prior knowledge on $\bTheta$ does not vary with the
design choice, leading to the simplification
$f_{\bTheta|\bd}(\btheta|\bd) = f_{\bTheta}(\btheta)$.

% \footnote{The design variables are not treated as a random variables
%   in this study, thus there is no corresponding ``$\bD$'' but only
%   realizations ``$\bd$''. However, for the ease of notation, we may
%   still condition on $\bd$: $f_{\bTheta|\bd}(\btheta|\bd)$,
%   $f_{\bY|\bTheta,\bd}(\by|\btheta,\bd)$,
%   $f_{\bTheta|\bY,\bd}(\btheta|\by,\bd)$, and $f_{\bY|\bd}(\by|\bd)$,
%   where it is understood that $\bd$ is a distribution parameter, and
%   it would not make sense to apply Bayes' theorem to obtain
%   distributions on $\bd$.}

% \footnote{Since
%   $\bd$ is not treated as a random variable in this study, we 
%   view it  as a distribution parameter and adopt the notation of
%   $f_{\bTheta}(\btheta;\bd)$, $f_{\bY|\bTheta}(\by|\btheta;\bd)$,
%   $f_{\bTheta|\bY}(\btheta|\by;\bd)$, and $f_{\bY}(\by;\bd)$, instead
%   of the random variable approach of $f_{\bTheta|\bD}(\btheta|\bd)$,
%   $f_{\bY|\bTheta,\bD}(\by|\btheta,\bd)$,
%   $f_{\bTheta|\bY,\bD}(\btheta|\by,\bd)$, and
%   $f_{\bY|\bD}(\by|\bd)$.}

Taking the decision theoretic approach proposed by
Lindley~\cite{lindley:1956:oam, lindley:1972:bsa}, we use the
Kullback-Leibler (KL) divergence \cite{cover:2006:eoi, mackay:2006:eoi}
from the posterior to the prior as a utility function, and take its
expectation under the prior predictive distribution of the data to
obtain an \textit{expected utility} $U(\bd)$:
% \footnote{The KL divergence reflects the difference in
%   information carried by two distributions in units of nats
%   \cite{cover:2006:eoi, mackay:2006:eoi}.} 
\begin{eqnarray}
  U(\bd) & = & \int_{\CY} \int_{\CH}
  f_{\bTheta|\bY,\bd}(\btheta|\by,\bd)
  \ln\left[\frac{f_{\bTheta|\bY,\bd}(\btheta|\by,\bd)}{f_{\bTheta}(\btheta)}\right]
  \,d\btheta \,f_{\bY|\bd}(\by|\bd) \,d\by \label{e:expectedUtility}
  \\[8pt] & = &
  \EE_{\bY|\bd} \left [ D_{\mathrm{KL}} \left ( f_{\bTheta|\bY,\bd}
      ( \cdot |\bY, \bd) || f_{\bTheta} ( \cdot ) \right )
  \right ] . \nonumber
\end{eqnarray}
Here $\CH$ is the support of $f_{\bTheta}(\btheta)$ and $\CY$ is the
support of $f_{\bY|\bd}(\by|\bd)$. Because the observation $\bY$
cannot be known before the experiment is performed, taking the
expectation over the prior predictive $f_{\bY|\bd}$ lets the resulting
utility function reflect the information gain \textit{on average},
over all anticipated outcomes of the experiment. The KL divergence may
be understood as information gain: larger KL divergence from
posterior to prior implies that the data $\bY$ decrease entropy in
$\bTheta$ by a larger amount, and hence are more informative for
parameter inference. The expected utility $U(\bd)$ is thus the
\textit{expected information gain} due to an experiment performed at
conditions $\bd$, which is equivalent to the \textit{mutual
  information} between the parameters $\btheta$ and the observables
$\by$ conditioned on $\bd$. A more detailed derivation and discussion
can be found in~\cite{huan:2013:sbo}.

Typically, the expected utility in (\ref{e:expectedUtility}) has no
closed form (even if the predictive mean of the data is, for example,
a polynomial function of $\btheta$). Instead, it must be approximated
numerically. By applying Bayes' rule to the quantities inside and
outside the logarithm in (\ref{e:expectedUtility}), and then
introducing Monte Carlo approximations for the resulting integrals, we
obtain the nested Monte Carlo estimator proposed by
Ryan~\cite{ryan:2003:eei}:
\begin{eqnarray}
  U(\bd) \approx \hU_{N,M}(\bd,\btheta_s,\by_s) \equiv
  \frac{1}{N} \sum_{i=1}^{N} 
  \left\{\ln\[f_{\bY|\bTheta,\bd}(\by^{(i)}|\btheta^{(i)},\bd)\] -
    \ln\[\frac{1}{M} \sum_{j=1}^{M}
    f_{\bY|\bTheta,\bd}(\by^{(i)}|\bttheta^{(i,j)},\bd) \]
  \right\}, \label{e:EUEstimator1}
\end{eqnarray}
where $\btheta_s \equiv \left\{ \btheta^{(i)} \right \} \cup \left
\{\bttheta^{(i,j)}\right\}$, $i=1\ldots N$, $j=1\ldots M$, are
i.i.d.\ samples from the prior $f_{\bTheta}$; and $\by_s \equiv
\left\{\by^{(i)}\right\}$, $i=1 \ldots N$, are independent samples
from the likelihoods $f_{\bY|\bTheta,\bd}(\cdot | \btheta^{(i)}, \bd
)$.  The variance of this estimator is approximately ${A(\bd)}/{N}+{B(\bd)}/{NM}$
and its bias is (to leading order) ${C(\bd)}/{M}$~\cite{ryan:2003:eei}, where $A$, $B$,
and $C$ are terms that depend only on the distributions at hand.
While the estimator $\hU_{N,M}$ is biased for finite $M$,
it is asymptotically unbiased.

Finally, the expected utility must be maximized over the design space
$\CD$ to find the optimal experiment(s):
\begin{eqnarray}
  \bd^\ast = \mathrm{arg } \max_{\bd\in\CD}\, U(\bd).
  \label{e:optimization}
\end{eqnarray}
Since $U$ can only be approximated by Monte Carlo estimators such as
$\hU_{N,M}$, optimization methods for stochastic objective functions
are needed.

\subsection{Stochastic optimization}
\label{s:stochasticOptimization}

In this section we describe two gradient-based stochastic optimization
approaches: Robbins-Monro stochastic approximation, and sample average
approximation with the Broyden-Fletcher-Goldfarb-Shanno
method. Both approaches require some flavor of gradient information,
but they do not use the exact gradient of $U(\bd)$. Calculating the
latter is generally not possible, given that we only have a Monte
Carlo estimator (\ref{e:EUEstimator1}) of $U(\bd)$.

% As seen later on in Sections~\ref{s:pce} and~\ref{s:ipa}, the use of
% of polynomial chaos surrogates and infinitesimal perturbation analysis
% will make this information available.

For simplicity, in this section only (\S
\ref{s:stochasticOptimization}), we will use a more generic notation
to describe the stochastic optimization problem at hand. This will
allow the essential ideas to be presented before tackling the
additional complexities of the expected information gain estimator
above. The problem to be solved is of the form
\begin{eqnarray}
  x^\ast = \mathrm{arg } \min_{x\in\CX}\, \left\{ h(x) =
    \EE_{W} \[\hh(x, W)\]\right\},
  \label{e:true}
\end{eqnarray}
where $x$ is the design variable, $W$ is the (generally
design-dependent) ``noise'' random variable, and $\hh(x,w)$ is an
unbiased estimator of the unavailable objective function $h(x)$.

\subsubsection{Robbins-Monro (RM) stochastic approximation}
\label{s:SARM}

The iterative update of the Robbins-Monro method is
\begin{eqnarray}
  x_{k+1} = x_k - a_k \hg(x_k, w^\prime),
\label{e:rm}
\end{eqnarray}
% \hg can actually be \hg_k (dep on k), and perhaps w^\prime_k
% too, but may open up a new can of worms.
where $k$ is the iteration index and $\hg(x_k,w^\prime)$ is an unbiased
estimator of the gradient (with respect to $x$) of $h(x)$ evaluated at
$x_k$. In other words,
$\EE_{W^\prime} \[\hg(x, W^\prime)\]  = \nabla_x
h(x)$, but $\hg$ is not necessarily equal to
$\nabla \hh$. Also, $W^\prime$ and $W$ may, but need not, be related. The
gain sequence $a_k$ should satisfy the following properties:
\begin{eqnarray}
  \sum_{k=0}^{\infty} a_k = \infty & \mathrm{and} &
  \sum_{k=0}^{\infty} a_k^2 < \infty.
\end{eqnarray}
One natural choice, used in this study, is the harmonic step size
sequence $a_k={\beta}/{k}$, where $\beta$ is some appropriate scaling
constant. For example, in the diffusion problem of
Section~\ref{s:diffusion}, $\beta$ is chosen to be 1.0 since the
design space is $[0,1]^2$.
With various technical assumptions on $\hg$ and $g$, it can be shown
that RM converges to the exact solution of (\ref{e:true}) almost
surely~\cite{kushner:2003:saa}.
% ... XH: By convergence in L2, I mean in the 2nd moment or mean
% square (lim E[(X_n-X*)^2] = 0. This implies convergence in
% probability.
% 
% By uniform bounded, I actually meant for the noisy gradient to have compact
% support. This is violated in our examples since we use additive
% Gaussian. Robbins and Monro also requires the function to be convex
% for its proof.  Look into looser convergence
% conditions developed afterwards.
%
% YMM: FWIW, I think that Gaussian noise is actually okay for showing
% a.s. convergence of the RM algorithm. Kushner seems to require that
% the observational noise (additively corrupting the gradient) is a
% "martingale difference sequence" with uniformly bounded
% *variances*. 

% From Kushner and Yin:  See for example, remark on p 127. Under
% certain conditions, 

Choosing the sequence $a_k$ is often viewed as the Achilles' heel of
RM, as the algorithm's performance can be very sensitive to step
size. We acknowledge this fact and do not downplay the difficulty of
choosing an appropriate gain sequence, but we will try to show that
there exist logical approaches to selecting $a_k$ that yield
reasonable performance. More sophisticated strategies, such as
search-then-converge learning rate schedules~\cite{darken:1990:nol},
adaptive stochastic step size rules~\cite{benveniste:1990:aaa}, and
iterate averaging methods~\cite{kushner:2003:saa, polyak:1992:aos},
have been developed and successfully demonstrated in applications. For
simplicity, however, we will use only the harmonic step size sequence
in this paper.

We will also use relatively simple stopping criteria for the RM
iterations: the algorithm will be terminated when changes in $x_k$
stall (e.g., $\norm{x_{k}-x_{k-1}}{}$ falls below some designated
tolerance for 5 successive iterations) or when a maximum number of
iterations has been reached (e.g., 50 iterations in the numerical
experiments of Section~\ref{s:stochopt-results}.)

\subsubsection{Sample average approximation (SAA)}
\label{s:SAA}

\paragraph{Transformation to design-independent noise.}
The central idea of sample average approximation is to reduce the
stochastic optimization problem to a deterministic problem, by fixing the
noise throughout the entire optimization process. In practice, if the
noise $W$ is design-dependent, it is first transformed to a
design-independent random variable by moving all the design dependence
into the function $\hh$. (An example of this transformation is given
in Section~\ref{s:ipa}.) The noise variables at different $x$ then
share a common distribution, and a common set of realizations is
employed at all values of $x$.

Such a transformation is always possible in practice, since the random
numbers in any computation are fundamentally generated from uniform
random (or really pseudorandom) numbers. Thus one can always
transform $W$ back into these uniform random variables, which are of
course independent of $x$.\footnote{One does not need to go all the
  way to the uniform random variables; any higher-level
  ``transformed'' random variable, as long as it remains independent
  of $x$, suffices.} For the remainder of this section (\S\ref{s:SAA})
we shall, without loss of generality, assume that $W$ is independent
of $x$.

\paragraph{Reduction to a deterministic problem.}
SAA approximates the true optimization problem in
(\ref{e:true}) with
\begin{eqnarray}
  \hx_s = \mathrm{arg} \min_{x\in \CX} \left\{\hh_{N}(x, w_s) \equiv \frac{1}{N}
    \sum_{i=1}^N \hh(x,w_i)\right\}, \label{e:approx}
\end{eqnarray}
where $\hx_s$ and $\hh_{N}(\hx_s, w_s)$ are the optimal design and
objective values under a particular set of $N$ realizations of the
random variable $W$, $w_s \equiv \{w_i\}_{i=1}^N$. The \textit{same} set
of realizations is used for different values of $x$ during the
optimization process, thus making the minimization problem in
(\ref{e:approx}) deterministic. (One can view this approach as
an application of common random numbers.)  A deterministic
optimization algorithm can then be chosen to find $\hx_s$ as an
approximation to $x^\ast$. 

Estimates of $h(\hx_s)$ can be improved by using
$\hh_{N^\prime}(\hx_s, w_{s^\prime})$ instead of $\hh_{N}(\hx_s,
w_s)$, where $\hh_{N^\prime}(\hx_s, w_{s^\prime})$ is computed from a
larger set of realizations $w_{s^\prime} \equiv
\{w_j\}_{j=1}^{N^\prime}$ with $N^\prime>N$, in order to attain a
lower variance.  Finally, multiple (say $T$) optimization runs are
often performed to obtain a sampling distribution for the optimal
design values and the optimal objective values, i.e., $\hx_s^t$ and
% $\hh_{N^\prime}(\hx_s^t, w_{s^\prime}^t)$, for $t=1 \ldots T$. 
$\hh_{N}(\hx_s^t, w_s^t)$, for $t=1 \ldots T$. 
The sets $w_s^t$ % and $w_{s^\prime}^t$
are independently chosen for each optimization run, but remain fixed within
each run.  Under certain assumptions on the objective function and the
design space,
%\footnote{Convergence proofs assume that the true global optimum
%  of each deterministic problem is obtained, which is almost never the
%  case in practice if finite-precision numerical methods are used and
%  if the problem is, for example, nonconvex.} 
%
% ... XH: my main concern here is that exact arg-mins are never
% attained in practice with numerical methods. I now think it is not
% worth mentioning here. There are possibly many assumptions from the
% convergence proofs that are violated. This is just one obvious and
% all-applicable one.
%
the optimal design and objective estimates in SAA generally converge
to their respective true values \textit{in distribution} at a rate of
$1/\sqrt{N}$~\cite{shapiro:1991:aao, kleywegt:2001:tsa}.\footnote{More
  precise properties of these asymptotic distributions depend on
  properties of the objective and the set of optimal solutions to the
  true problem. For instance, in the case of a singleton optimum
  $x^\ast$, the SAA estimates $\hh_N(\hx_s,\cdot)$ converge to a
  Gaussian with variance $\Var_W [\hh(x^\ast, W)] / N$. Faster
  convergence to the optimal objective value may be obtained when the
  objective satisfies stronger regularity conditions. The SAA
  solutions $\hx_s$ are not in general asympotically normal,
  however. Furthermore, discrete probability distributions lead to
  entirely different asymptotics of the optimal solutions.}

For the solution of a particular deterministic problem $\hx^t_s$,
stochastic bounds on the true optimal value can be constructed by
estimating the optimality gap
$h(\hx^t_s)-h(x^{\ast})$~\cite{norkin:1998:aba, mak:1999:mcb}. The
first term can simply be approximated using the unbiased estimator
$\hh_{N^\prime}(\hx_s^t, w_{s^\prime}^t)$ since
$\EE_{W_{s^\prime}}\[\hh_{N^\prime}(\hx_s^t, W_{s^\prime})\]=
h(\hx_s^t)$.  The second term may be estimated using the average of
the approximate optimal objective values across the $T$ replicate
optimization runs (based on $w_s^t$, rather than $w_{s^\prime}^t$):
\begin{eqnarray}
  \barh_{N} = \frac{1}{T}\sum_{t=1}^T \hh_{N}(\hx_s^t, w_{s}^t) .
  \label{e:lowerBound}
\end{eqnarray}
This is a negatively biased estimator and hence a stochastic lower
bound on $h(x^{\ast})$~\cite{norkin:1998:aba, mak:1999:mcb,
  shapiro:2007:ato}.\footnote{Short proof
  from~\cite{shapiro:2007:ato}: For any $x\in\CX$, we have that
  $\EE_{W_s}\[\hh_{N}(x, W_{s})\] = h(x)$, and that $\hh_N(x,w_s^t)
  \geq \min_{x^\prime\in\CX}\hh_N(x^\prime,w_s^t)$. Then
  $h(x)=\EE_{W_s}\[\hh_N(x,W_s)\] \geq
  \EE_{W_s}\[\min_{x^\prime\in\CX}\hh_N(x^\prime,W_s)\]=\EE_{W_s}\[\hh_N(\hx_s^t,W_s)\]
  = \EE_{W_s}\[\barh_N\]$.}\textsuperscript{,}\footnote{The bias
  decreases monotonically with $N$~\cite{norkin:1998:aba}.}  The
difference $\hh_{N^\prime}(\hx_s^t, w_{s^\prime}^t) - \barh_{N}$ is
thus a stochastic upper bound on the true optimality gap $h(\hx^t_s) -
h(x^{\ast})$.
% The optimality gap for each $\hx_s^t$ thus can be approximated by
% \begin{eqnarray}
%   \hh_{N^\prime}(\hx_s^t, w_{s^\prime}^t) - \barh_{N} \approx
%   h(\hx^t_s) - h(x^{\ast}) .
%   \label{e:optGap}
% \end{eqnarray}
The variance of this optimality gap estimator can be derived from the
Monte Carlo standard error formula~\cite{ahmed:2002:tsa}. One could
then use the optimality gap estimator and its variance to decide
whether more runs are required, or which approximate optimal designs
are most trustworthy.

Pseudocode for the SAA method is presented in Algorithm~\ref{m:SAA}. At
this point, we have reduced the stochastic optimization problem to a
series of deterministic optimization problems; a suitable
deterministic optimization algorithm is still needed to solve them.

\begin{algorithm}
  \SetAlgoLined
  Set optimality gap tolerance $\eta$ and number of replicate optimization runs
  $T$\;
  $t=1$\;
  \While{optimality gap estimate $> \eta$ and $t \leq T$}
  {
    Sample the set $w_s^t=\{w_i^t\}_{i=1}^N$\;
    Perform a deterministic optimization run and find $\hx_s^t$ (see
    Algorithm~\ref{a:BFGS})\;
    Sample the larger set $w_{s^\prime}^t=\{w_j^t\}_{j=1}^{N^\prime}$ where $N^\prime>N$\;
    Compute $\hh_{N^\prime}(\hx_s^t, w_{s^\prime}^t)=\frac{1}{N^\prime}\sum_{j=1}^{N^\prime}
    \hh\(\hx_s^t,w_j^t\)$\;
    Estimate the optimality gap and its variance\;
    $t=t+1$\;
  }
  Output the sets $\{\hx_s^t\}_{t=1}^T$ and $\{\hh_{N^\prime}(\hx_s^t, w_{s^\prime}^t)\}_{t=1}^T$ for
  post-processing\;
  \caption{SAA method in pseudocode.}
  \label{m:SAA}
\end{algorithm}

\paragraph{BFGS method.}
\label{s:SAA_BFGS}

The Broyden-Fletcher-Goldfarb-Shanno (BFGS)
method~\cite{nocedal:2006:no} is a gradient-based method for solving
deterministic nonlinear optimization problems, widely used for its
robustness, ease of implementation, and efficiency. It is a
quasi-Newton method, iteratively updating an approximation to the
(inverse) Hessian matrix from objective and gradient evaluations at
each stage. Pseudocode for the BFGS method is given in
Algorithm~\ref{a:BFGS}. In the present implementation, a simple
backtracking line search is used to find a stepsize that satisfies the
first (Armijo) Wolfe condition only. The algorithm can
  be terminated according to many commonly used criteria: for example,
  when the gradient stalls, the line search stepsize falls below a
  prescribed tolerance, the design variable or function value stalls,
  or a maximum allowable number of iterations or objective evaluations
  is reached. BFGS is shown to converge super-linearly to a local
minimum if a quadratic Taylor expansion exists near that
minimum~\cite{nocedal:2006:no}.

The limited memory BFGS (L-BFGS)~\cite{nocedal:2006:no} method can
also be used when the design dimension becomes very large (e.g., more
than $10^4$), such that the dense inverse Hessian cannot be stored
explicitly.

\begin{algorithm}
  \SetAlgoLined
  Initialize starting point $x_0$, inverse Hessian
  approximation $H_0$, gradient termination tolerance $\varepsilon$\;
  Initialize any other termination conditions and parameters\;
  $k=0$\;
  \While{$\norm{\nabla \hh_{N}(x_k, w_s^t)}{} > \varepsilon$ and other
    termination conditions are not met}
  {
    Compute search direction $p_k = -H_k\nabla
    \hh_{N}(x_k, w_s^t)$\;
    Find acceptable stepsize $\alpha_k$ via line search\;
    Update position $x_{k+1} = x_k + \alpha_k p_k$\;
    Define vectors $s_k=x_{k+1}-x_k$ and
    $u_k = \nabla \hh_{N}(x_{k+1}, w_s^t) - \nabla \hh_{N}(x_k, w_s^t)$ \;
    Update inverse Hessian approximation $H_{k+1} = \(I - \frac{s_k
      u_k^{T}}{s_k^{T} u_k}\) H_k \(I - \frac{u_k
      s_k^{T}}{u_k^{T} s_k}\) + 
    \frac{s_k s_k^{T}}{s_k^{T}u_k}$\;
    $k=k+1$\;
  }
  Output $\hx_s^t = x_k$\;
  \caption{BFGS algorithm in pseudocode. In this context, $\hh_{N}(x,
    w_s^t)$ is the deterministic objective function we want to
    minimize (as a function of $x$).}
  \label{a:BFGS}
\end{algorithm}

\subsection{Application to optimal design}
\label{s:problem}

The main challenge in applying the aforementioned stochastic
optimization algorithms to optimal Bayesian experimental design is the
lack of readily-available gradient information. For RM, we need an
\textit{unbiased estimator of the gradient} of the expected utility,
i.e., $\hg$ in (\ref{e:rm}). For SAA-BFGS, we need the
\textit{gradient of the finite-sample Monte Carlo approximation} of
the expected utility, i.e., $\nabla \hh_{N}(\cdot, w_s^t)$.

We address these needs by introducing two concepts:
\begin{enumerate}
\item A simple surrogate model, based on \textit{polynomial chaos}
  expansions (see Section~\ref{s:pce}), replaces the often
  computationally-intensive forward model. The purpose of the
  surrogate is twofold. First, it allows the nested Monte Carlo
  estimator (\ref{e:EUEstimator1}) to be evaluated in a
  computationally tractable manner. Second, its polynomial form allows
  the gradient of (\ref{e:EUEstimator1}), $\nabla \hh_{N}(\cdot,
  w_s^t)$, to be derived analytically. These gains come at the expense
  of introducing additional error via the polynomial
  approximation of the original forward model, however. In other words,
  \textit{given} a surrogate for the forward model and the resulting
  expected information gain, we can derive exact gradients of a Monte
  Carlo approximation of this expected information gain, and use these
  gradients in SAA.
  % YMM: Is it clear from the above language that our SAA approach
  % maximizes an expected information gain approximation that uses
  % a surrogate model, rather than the maximizing the exact expected
  % information gain?
  %
  % XH: see revised.
  % YMM: I revised it a little more.

\item \textit{Infinitesimal perturbation analysis} (see
  Section~\ref{s:ipa}) applied to (\ref{e:expectedUtility}), along
  with the estimator in (\ref{e:EUEstimator1}) and the polynomial
  surrogate model, allows the analytical derivation of an unbiased
  gradient estimator $\hg$, as required for the RM
  approach. 
  % \footnote{As we will see in Section~\ref{s:ipa}, the estimator is
  % actually biased with respect to the \textit{expected
  % utility}. However, the bias decreases with $M$, thus asymptotic
  % convergence of the optimized result is expected.}
  %... YMM: I feel like this point is much better explained in the IPA
  % section. Let the reader find it out there? Now we say "appropriate"
  % and thus keep it vague.
  % 
  % XH: agreed, it would be better there. Where did you use ``appropriate''?
  % YMM: I must have deleted it after writing comment. Anyway, all is
  % fine here.
\end{enumerate}

\section{Polynomial chaos surrogates}
\label{s:pce}

\subsection{Background}

This section introduces the first of two computational tools used to
address the challenges described in
Section~\ref{s:problem}. Polynomial expansions will be used to
mitigate the cost of repeated forward model evaluations. Later (see
Section~\ref{s:ipa}) they will also be used to help evaluate
appropriate gradient information for stochastic optimization
methods.

Mathematical models of the experiment enter the inference and design
formulation through the likelihood function
$f_{\bY|\bTheta,\bd}$. For example, a simple
likelihood function might allow for an additive discrepancy $\bE$
between experimental observations and model predictions
\begin{eqnarray}
  \bY = \bG( \bTheta, \bd ) + \bE.
\end{eqnarray}
Here $\bG(\btheta,\bd)$ is the ``forward model'' describing the
experiment; it is a function that maps both the design variables and
the parameters into the data space. The discrepancy $\bE$ is often
taken to be a Gaussian random variable, but is by no means limited to
this; we will use $f_{\bE}$ to denote its probability
density. Computationally intensive forward models can render Monte
Carlo estimation of the expected information gain impractical. In
particular, drawing a sample from
$f_{\bY|\bTheta,\bd}(\by|\btheta,\bd)$ requires evaluating $\bG$
at a particular $(\btheta, \bd)$. Evaluating the density
$f_{\bY|\bTheta,\bd}(\by|\btheta,\bd) = f_{\bE} (\by - \bG ( \btheta,
\bd ) )$ again requires evaluating $\bG$.

% YMM: too nitpicky! :)
% \footnote{Although the output of $\bG(\bTheta,\bd)$ is indeed a random
%   variable, the upper case $\bG$ does not denote that random variable
%   in this case, but rather, the function.}
%

To make these calculations tractable, one would like to replace $\bG$
with a cheaper ``surrogate'' model that is accurate over the entire
prior support $\CH$ and the entire design space $\CD$. Many options
exist, ranging from projection-based model
reduction~\cite{buithanh:2007:mrl,frangos:2010:srm} to spectral
methods based on polynomial chaos (PC)
expansions~\cite{wiener:1938:thc, ghanem:1991:sfe, xiu:2002:twa,
  debusschere:2004:nci, najm:2009:uqa, xiu:2009:fnm,
  lemaitre:2010:smf,conrad:2012:asp}. The latter approaches do not
reduce the internal physics of the deterministic model; rather, they
exploit regularity in the dependence of model outputs on uncertain
input parameters and design variables.

Polynomial chaos has seen extensive use in a range of engineering
applications (e.g.,~\cite{hosder:2006:ani, reagan:2003:uqi,
  walters:2003:tsf, xiu:2003:ans}) including parameter estimation and
inverse problems
(e.g.,~\cite{marzouk:2007:ssm,marzouk:2009:asc,marzouk:2009:dra}).
More recently, it has also been used in open-loop optimal Bayesian
experimental design~\cite{huan:2010:abe, huan:2013:sbo}, with
excellent accuracy and multiple order-of-magnitude speedups over
direct evaluations of forward model. Unlike the present work, however,
our earlier study \cite{huan:2013:sbo} used only gradient-free
stochastic optimization methods (Nelder-Mead and simultaneous
perturbation stochastic approximation).

\subsection{Formulation}

Any random variable $Z$ with finite variance can be
represented by an infinite series
\begin{eqnarray}
  Z = 
  \sum_{|\bii|=0}^{\infty} a_{\bii}
  \Psi_{\bii}(\Xi_1,\Xi_2,\ldots),
  \label{e:PCEForm2}
\end{eqnarray}
where $\bii=\(i_1,i_2,\ldots\),\,i_j\in\NN_0$, is an
infinite-dimensional multi-index; $|\bii|=i_1+i_2+\ldots$ is the $l_1$
norm; $a_{\bii} \in \RR$ are the expansion coefficients; $\Xi_i$ are
independent random variables; and
\begin{eqnarray}
  \Psi_{\bii}(\Xi_1,\Xi_2,\ldots) = \prod_{j=1}^{\infty}
  \psi_{i_j}(\Xi_j)
\end{eqnarray}
are multivariate polynomial basis functions~\cite{xiu:2002:twa}. Here
$\psi_{i_j}$ is an orthogonal polynomial of order $i_j$ in the
variable $\Xi_j$, where orthogonality is with respect to the density
of $\Xi_j$,
\begin{eqnarray}
  \EE_{\Xi}\[\psi_m(\Xi)\psi_n(\Xi)\] = \int_{\CF}
  \psi_m(\xi)\psi_n(\xi)f_{\Xi}(\xi)\,d\xi =
  \delta_{m,n}\EE_{\Xi}\[\psi_m^2(\Xi)\],
  \label{e:orthogonality}
\end{eqnarray}
and $\CF$ is the support of $f_{\Xi}(\xi)$. The
expansion~(\ref{e:PCEForm2}) is convergent in the mean-square
sense~\cite{cameron:1947:tod}. For computational purposes, the
infinite sum in (\ref{e:PCEForm2}) must be truncated to some finite
stochastic dimension $n_s$ and a finite number of polynomial terms. A
common choice is the ``total-order'' truncation $|\bii|\leq p$, but
other truncations that retain fewer cross terms, a larger number of
cross terms, or anisotropy among the dimensions are certainly
possible~\cite{conrad:2012:asp}.

% \begin{eqnarray}
%   Z &\approx& 
%   \sum_{|\bii|\leq p} a_{\mb{i}}
%  \Psi_{\bii}(\Xi_1,\Xi_2,\ldots,\Xi_{n_s}) 
%   \label{e:truncatedPC} \\
%   \Psi_{\bii}(\Xi_1,\Xi_2,\ldots,\Xi_{n_s}) &=& \prod_{j=1}^{n_s}
%   \psi_{i_j}(\Xi_j).
%   \label{e:truncatedPsi}
% \end{eqnarray}
% The total number of terms in this expansion is
% \begin{eqnarray}
%   n_{\mathrm{PC}} = \(\begin{array}{c}n_s+p \\ p \end{array} \) =
%   \frac{(n_s+p)!}{n_s!p!}.
%   \label{e:npc}
% \end{eqnarray}
% The choice of $p$ is influenced by the degree of nonlinearity in the
% relationship between $Z$ and $\Xi_j$, and the choice of $n_s$ reflects
% the degrees of freedom needed to capture the stochasticity of the
% system. These choices might also be constrained by the availability of
% computational resources, as $n_{\mathrm{PC}}$ grows quickly when these
% numerical parameters are increased.

In the optimal Bayesian experimental design context, the model outputs
depend on both the parameters and the design variables. Constructing a
new polynomial expansion at each value of $\bd$ encountered during
optimization is generally impractical. Instead, we can construct a
\textit{single} PC expansion for each component of $\bG$, depending
jointly on $\bTheta$ and $\bd$~\cite{huan:2013:sbo}. To proceed, we
assign one stochastic dimension to each component of $\bTheta$ and one
to each component of $\bd$. Further, we assume an affine
transformation between each component of $\bd$ and the corresponding
$\Xi_i$; any realization of $\bd$ can thus be uniquely associated with
a vector of realizations $\xi_i$. Since the design variables will
usually be supported on a bounded domain (e.g., inside some
hyper-rectangle), the corresponding $\Xi_i$ are endowed with uniform
distributions. The associated univariate $\psi_i$ are thus Legendre
polynomials.  These distributions effectively define a uniform weight
function over the design space $\CD$ that governs where the
$L^2$-convergent PC expansions should be most accurate.\footnote{In
  the present context, it is appropriate to view $\bd$ as a
  deterministic design variable. Since the stochastic optimization
  algorithms used later all involve some level of randomness, however,
  the $\bd$ values encountered during optimization may also be viewed
  as realizations from some probability distribution. This
  distribution, if known, could replace the uniform distribution and
  define a more efficient weighted $L^2$ norm; however, it is almost
  always too complex to extract in practice.}

% Denoting
% $\left\{\xi_{j}\right\}_{j=n_s+1}^{n_s+n_d}=\left\{d_k\right\}_{k=1}^{n_d}$,
% the new stochastic dimension becomes $n=n_s+n_d$. Uniform ``priors'',
% for example, can be assigned to $\bd$ within the regions where the
% experiments are physically available; however, they are not truly
% priors in the inference sense, because the design variables are not
% uncertain; the ``priors'' may be interpreted as weight functions in
% emphasizing where in the design space the PC expansion should be made
% more accurate. Moreover, one would never sample
% $\left\{\xi_{j}\right\}_{j=n_s+1}^{n_s+n_d}$, because there is no
% associated randomness in $\bd$; instead, when a particular value of
% $\bd$ is desired, the corresponding
% $\left\{\xi_{j}\right\}_{j=n_s+1}^{n_s+n_d}$ are simply computed by
% inverting the PC expansion of $\bd$.\footnote{Inversion can be done
%   for only affine functions in the design-variable dimensions of the
%   expansion; but expansions in the uncertain-parameter dimensions may
%   still be any order.}
% % XH: added new footnote above, see if its necessary.

Constructing the PC expansion involves computing the coefficients
$a_{\bii}$. This computation generally can proceed via two alternative
approaches, intrusive and nonintrusive. The intrusive approach
results in a new system of equations that is larger than the original
deterministic system, but it needs be solved only once. The difficulty
of this latter step depends strongly on the character of the original
equations, however, and may be prohibitive for arbitrary nonlinear
systems. The nonintrusive approach computes the expansion
coefficients by directly projecting the quantity of interest (e.g.,
the model outputs) onto the basis functions $\Psi_{\bii}$. One
advantage of this method is that the deterministic solver can be
reused and treated as a black box. The deterministic problem then
needs to be solved many times, but typically at carefully chosen
parameter and design values. The nonintrusive approach also offers
flexibility in choosing arbitrary functionals of the state trajectory
as observables; these functionals may depend smoothly on $\bXi$ even
when the state itself has a less regular dependence. Here, we will
employ a nonintrusive approach.

Applying orthogonality, the PC coefficients are simply
\begin{eqnarray}
  G_{c,\bii} = \frac{\EE_{\bXi}\[G_c(\bTheta(\bXi),\bd(\bXi))
    \Psi_{\bii}(\bXi)\]}{\EE_{\bXi}\[\Psi_{\bii}^2(\bXi) \]} =
  \frac{\int_{\CF} G_c(\btheta(\bxi),\bd(\bxi)) \Psi_{\bii}(\bxi)
    f_{\bXi}(\bxi) \,d\bxi}{\int_{\CF} \Psi_{\bii}^2(\bxi)
    f_{\bXi}(\bxi) \,d\bxi},
  \label{e:NISP}
\end{eqnarray}
where $G_{c,\bii}$ is the coefficient of $\Psi_{\bii}$ for the $c$th
component of the model outputs. Analytical expressions are available
for the denominators $\EE_{\bXi}\[\Psi_{\bii}^2(\bXi)\]$, but the
numerators must be evaluated numerically.  
% The resulting approach is termed pseudospectral approximation or
% nonintrusive spectral projection.
When the evaluations of the integrand (and hence the forward model)
are expensive and $n_s$ is large, an efficient method for numerical
integration in high dimensions is essential.

To evaluate the numerators in (\ref{e:NISP}), we employ Smolyak sparse
quadrature based on one-dimensional Clenshaw-Curtis quadrature
rules~\cite{clenshaw:1960:amf}. Care must be taken to avoid
significant aliasing errors when using sparse quadrature to construct
polynomial approximations, however. Indeed, it is advantageous to
recast the approximation as a Smolyak sum of constituent full-tensor
polynomial approximations, each associated with a tensor-product
quadrature rule that is appropriate to its
polynomials~\cite{conrad:2012:asp, constantine:2012:spa}. This type of
approximation may be constructed \textit{adaptively}, thus taking
advantage of weak coupling and anisotropy in the dependence of $\bG$
on $\bTheta$ and $\bd$. More details can be found
in~\cite{conrad:2012:asp}.

% dimension-adaptive sparse quadrature (DASQ) algorithm of Gerstner and
% Griebel~\cite{gerstner:2003:dat} in combination with Clenshaw-Curtis
% quadrature~\cite{clenshaw:1960:amf}. DASQ is an efficient extension of
% Smolyak sparse quadrature that \textit{adaptively} tensorizes
% quadrature rules in each coordinate direction, thus taking advantage
% of regularity and anisotropy in the dependence of $\bG$ on $\bTheta$
% and $\bd$. It also has a weak dependence on dimension.  More details
% can be found in~\cite{huan:2010:abe}, which also proposes
% modifications to reuse values from nested quadrature points, and to
% prevent repeated forward model evaluation by guiding the integrations
% of all components via a single adaptation route.

At this point, we may substitute the polynomial approximation of $\bG$
into the likelihood function $f_{\bY|\bTheta,\bd}$, which in turn
enters the expected information gain estimator
(\ref{e:EUEstimator1}). This enables fast evaluation of the expected
information gain. 
%
% Given a fixed set of samples, we may also differentiate this
% expression analytically with respect to $\bd$, in order to obtain
% $\nabla_{\bd} \hh_{N}(\cdot, w_s^t)$
%
The computation of appropriate gradient information is discussed next.

% $\nabla \hh_{N}(\cdot, w_s^t)$ analytically by substituting the
% polynomial approximation of $\bG$ into Equation~\ref{e:EUEstimator1}
% and differentiating. However, we defer the details until the end of
% next section, where the same derivation mechanics will be used.
%
% YMM: I wanted to defer this statement, since we are really
% computing derivatives of the MC estimator that uses the approximate
% approximate likelihood, not of the MC estimator involving the full
% forward model!

\section{Infinitesimal Perturbation Analysis}
\label{s:ipa}

This section applies the method of infinitesimal perturbation analysis
(IPA)~\cite{ho:1983:paa, glasserman:1990:gev, asmussen:2007:ssa} to
construct an unbiased estimator $\hg$ of the gradient of the expected
information gain, for use in RM. The same procedure yields the
gradient $\nabla \hh_{N,M}(\cdot, w_s^t)$ of a finite-sample Monte Carlo
approximation of the expected information gain, for use in SAA. The
central idea of IPA is that under certain conditions, an unbiased
estimator of the gradient of a function can be obtained by simply
taking the gradient of an unbiased estimator of the function. We apply
this idea in the context of optimal Bayesian experimental design.

The first requirement of IPA is the availability of an unbiased
estimator of the function. Unfortunately, as described in
Section~\ref{s:oedform}, $\hU_{N,M}$ in (\ref{e:EUEstimator1}) is a biased
estimator of $U$ for finite $M$~\cite{ryan:2003:eei}. To circumvent
this technicality, let us optimize the following objective function
instead of $U$:
\begin{eqnarray}
  \barU_M(\bd) &\equiv&
  \EE_{\bTheta_s,\bY_s|\bd}\[\hU_{N,M}(\bd,\bTheta_s,\bY_s)\] \nonumber\\
  &=&
  \int_{\CY_s} \int_{\CH_s} 
  \hU_{N,M}(\bd, \btheta_s,\by_s) f_{\bTheta_s,\bY_s|\bd}(\btheta_s,\by_s|\bd) \,d\btheta_s\,d\by_s
  \nonumber\\
  &= &
  \int_{\CY_s} \int_{\CH_s} 
  \hU_{N,M}(\bd,\btheta_s,\by_s)
  \prod_{(i,j)=(1,1)}^{(N,M)}
  f_{\bY|\bTheta,\bd}(\by^{(i)}|\btheta^{(i)}, \bd)
  f_{\bTheta}(\btheta^{(i)}) 
  f_{\bTheta}(\bttheta^{(i,j)}) \,d\btheta_s\,d\by_s,
\end{eqnarray}
where $\CH_s \times \CY_s$ is the support of the joint density
$f_{\bTheta_s,\bY_s|\bd}(\btheta_s,\by_s|\bd)$. Our original estimator
$\hU_{N,M}$ is now unbiased for the new objective $\barU_M$ by
construction! The tradeoff, of course, is that the function being
optimized is no longer the true $U$. But it is consistent in that
$\barU_M(\bd) \to U(\bd)$ as $M \to \infty$, for any $N > 0$. (To
illustrate this convergence, realizations of $\hU_{N,M}$, i.e., Monte
Carlo approximations of $\barU_M$, are plotted in
Figure~\ref{f:EUMC3D} for varying $M$.)

The second requirement of IPA comprises conditions allowing an
unbiased gradient estimator to be constructed by taking the gradient
of the unbiased function estimator. Standard conditions (see, for
example,~\cite{asmussen:2007:ssa}) require that the random quantity
(e.g., $\hU_{N,M}$) be almost surely continuous and differentiable. Here,
because $\hU_{N,M}$ is parameterized by continuous random variables that
have densities with respect to Lebesgue measure, we can take a
perspective that relies on Leibniz's rule with the following
conditions:
% YMM: How are our conditions different from the formal requirements
% in asmussen? Are our conditions less complete or less sufficient in
% any way?
%
% XH: the formal conditions are for the objective function to be
% a.s. differentiable at the design value of interest (d_0), and that the
% objective function a.s. satisfies the Lipschitz condition for two
% points in a nonrandom neighborhood of d_0, where the expected
% value of the Lipschitz constant is finite. Then the expectation and
% differentiation are interchangeable at d_0. (Asmussen p216, proposition 2.3).
%
% After some more thought, I think the Lipschitz condition is for
% random-sample-wise (i.e., for a particular omega from the sample
% space across all d's). I would then expect all
% ``common-random-number'' realizations of these surfaces to be
% differentiable and Lipschitz continuous, but it would still be
% difficult to prove rigorously.  

\begin{enumerate}

\item $\hU_{N,M}$ and $\nabla_{\bd} \( \hU_{N,M} \)$ are continuous over the
  product space of design variables and random variables, $\CD \times
  \CH_s \times \CY_s$;

\item the density of the ``noise'' random variable is independent of
  $\bd$.
\end{enumerate}

The first condition supports the interchange of differentiation and
integration according to Leibniz's rule. This condition might be difficult
to verify in arbitrary cases, but the use of finite-order
polynomial forward models and continuous distributions for
the prior and observational noise ensures that we meet the
requirement.

The second condition is needed to preserve the form of the
expectation. If it is violated, differentiation with respect to $\bd$
must be performed on the
$f_{\bTheta_s,\bY_s|\bd}(\btheta_s,\by_s|\bd)$ term as well via the
product rule, in which case the additional term $ \int_{\CY_s}
\int_{\CH_s} \hU_{N,M}(\bd,\btheta_s,\by_s) \, \nabla
\[f_{\bTheta_s,\bY_s|\bd}(\btheta_s,\by_s|\bd)\] \,d\btheta_s\,d\by_s$
would no longer be an expectation with respect to the original
density. The likelihood-ratio method may be used to restore the
expectation~\cite{glynn:1990:lrg, asmussen:2007:ssa}, but it is not
pursued here. Instead, it is simpler to transform the noise to a design-independent
random variable as described in Section~\ref{s:SAA}.

In the context of optimal Bayesian experimental design, the outcome of
the experiment $\bY$ is a stochastic quantity that depends on the
design $\bd$. From the stochastic optimization perspective, $\bY$ is thus
a noise variable. To demonstrate the transformation to
design-independent noise, we assume a likelihood where the data result
from an additive Gaussian perturbation to the forward model:
\begin{eqnarray}
  \bY &=& \bG(\bTheta,\bd)+\bE \nonumber\\
  &=& \bG(\bTheta,\bd)+\bC(\bTheta,\bd)\bZ . \label{e:likelihood}
\end{eqnarray}
Here $\bC$ is a diagonal matrix with non-zero entries reflecting the
dependence of the noise standard deviation on other quantities, and
$\bZ$ is a vector of i.i.d.\ standard normal random variables. For
example, ``10\% Gaussian noise on the $c$th component'' would
translate to $C_{c,i} = \delta_{ci}0.1 |G_c(\bTheta,\bd)|$, where
$\delta_{ci}$ is the Kronecker delta function. For other forms of the
likelihood, the right-hand side of (\ref{e:likelihood}) is
simply replaced by a generic function of $\bTheta$, $\bd$, and some
random variable $\bZ$. Here, however, we will focus on the additive
Gaussian form in order to derive illustrative expressions.

By extracting a design-independent random variable $\bZ$ from the
noise term $\bE\equiv \bC(\bTheta,\bd)\bZ$, we will satisfy the second
condition above. The design-dependence of $\bY$ is incorporated into
$\hU_{N,M}$ by substituting (\ref{e:likelihood}) into
(\ref{e:EUEstimator1}):
\begin{eqnarray}
  \hU_{N,M}(\bd,\btheta_s,\bz_s) &=& \frac{1}{N} \sum_{i=1}^{N}
  \left\{\ln\[f_{\bY|\bTheta,\bd}\(\left. \bG(\btheta^{(i)},\bd)+\bC(\btheta^{(i)},\bd)\bz^{(i)}\right|\btheta^{(i)},\bd\)\]\right. \nonumber\\
  && - \left.
    \ln\[\frac{1}{M} \sum_{j=1}^{M}
    f_{\bY|\bTheta,\bd}\(\left.\bG(\btheta^{(i)},\bd)+\bC(\btheta^{(i)},\bd)\bz^{(i)}\right|\btheta^{(i,j)},\bd\) \]
  \right\}, \label{e:EUEstimator2}
\end{eqnarray}
where $\bz_s = \left\{\bz^{(i)}\right\}$. The new noise
variables are now
independent of $\bd$. The samples of $\by^{(i)}$ drawn from the
likelihood are instead realized by drawing $\bz^{(i)}$ from
$N(\mb{0},\bI)$, then multiplying these samples by $\bC$ and adding
them to the model output.

With all conditions for IPA satisfied, an unbiased estimator of the
gradient of $\barU_M$, corresponding to $\hg$ in (\ref{e:rm}), is simply
$\nabla_{\bd}\hU_{N,M}(\bd,\btheta_s,\bz_s)$ since
\begin{eqnarray}
  \EE_{\bTheta_s,\bZ_s}\[\nabla_{\bd}\hU_{N,M}(\bd,\bTheta_s,\bZ_s)\] &=& 
  \int_{\CZ_s} \int_{\bTheta_s} \nabla_{\bd}\hU_{N,M}(\bd,\btheta_s,\bz_s)
  f_{\bTheta_s,\bZ_s}(\btheta_s,\bz_s) \,d\btheta_s\,d\bz_s
  \nonumber\\  &=& \nabla_{\bd}
  \int_{\CZ_s} \int_{\bTheta_s} \hU_{N,M}(\bd,\btheta_s,\bz_s)
  f_{\bTheta_s,\bZ_s}(\btheta_s,\bz_s)
  \,d\btheta_s\,d\bz_s
  \nonumber\\ &=& \nabla_{\bd} \EE_{\bTheta_s,\bZ_s}\[\hU_{N,M}(\bd,\bTheta_s,\bZ_s)\]
  \nonumber\\ &=& \nabla_{\bd}\barU_M(\bd),
\end{eqnarray}
where $\CZ_s$ is the support of $f_{\bZ_s}(\bz_s)$. This gradient
estimator is therefore suitable for use in RM.

The gradient of the finite-sample Monte Carlo approximation of
$U(\bd)$, i.e., $\nabla \hh_{N,M}(\cdot, w_s^t)$ used in SAA, takes
exactly the same form. The only difference between the two is that
$\hg$ lets $\bTheta_s$ and $\bZ_s$ be random at every iteration of the
optimization process. When used as $\nabla \hh_{N,M}(\cdot, w_s^t)$,
$\bTheta_s$ and $\bZ_s$ are frozen at some realization throughout the
optimization process. In either case, these gradient expressions
contain derivatives of the likelihood function and thus derivatives
$\nabla_{\bd} \bG (\btheta, \bd)$. When $\bG$ is replaced with a
polynomial expansion, these derivatives can be computed
inexpensively. Detailed derivations of the gradient estimator using
orthogonal polynomial expansions can be found in the Appendix.

% In general, given an unbiased estimator $\hg$, we have, by
% construction,
% \begin{eqnarray}
%   g(\bd) = \EE_{\bW}\[\hg\(\bW,\bd\)\] = \int_{\CW} 
%   \hg(\bw,\bd) f_{\bW}(\bw) \,d\bw,
% \end{eqnarray}
% where $\bW$ is some random variable with PDF $f_{\bW}(\bw)$ and
% support $\CW$. An unbiased gradient estimator is simply
% $\nabla\hg$, since
% \begin{eqnarray}
%   \EE_{\bW}\[\nabla\hg(\bW,\bd)\] = 
%   \int_{\CW} \nabla\hg(\bw,\bd)
%   f_{\bW}(\bw) \,d\bw = \nabla
%   \int_{\CW} \hg(\bw,\bd) f_{\bW}(\bw)
%   \,d\bw
%   = \nabla \EE_{\bW}\[\hg(\bW,\bd)\] = \nabla g(\bd),
% \end{eqnarray}

% An unbiased gradient estimator is simply $\nabla\hg$, since
% \begin{eqnarray}
%   \EE_{\bW}\[\nabla\hg(\bW,\bd)\] = 
%   \int_{\CW} \nabla\hg(\bw,\bd)
%   f_{\bW}(\bw) \,d\bw = \nabla
%   \int_{\CW} \hg(\bw,\bd) f_{\bW}(\bw)
%   \,d\bw
%   = \nabla \EE_{\bW}\[\hg(\bW,\bd)\] = \nabla g(\bd),
% \end{eqnarray}

\section{Source Inversion Problem}
\label{s:diffusion}

\subsection{Governing equations}
\label{s:diffusion-setup}

\begin{figure}[htb]
  \centering 
  \includegraphics[width=0.60\textwidth]{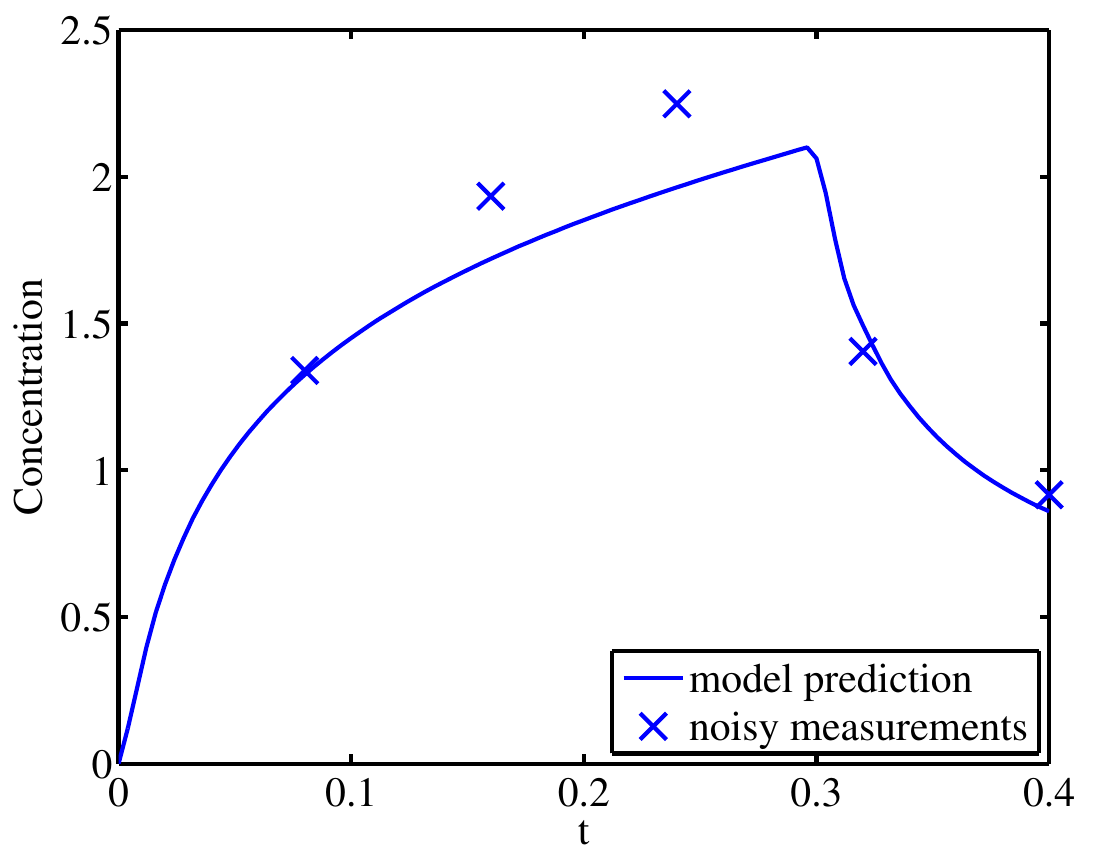}
  \caption{Example forward model solution and realizations from the
    likelihood. In particular, the solid line represents the
    time-dependent contaminant concentration $w(\bx, t;
    \bx_{\mathrm{src}})$ at $\bx = \bx_{\mathrm{sensor}}=(0.0, 0.0)$,
    given a source centered at $\bx_{\mathrm{src}}=(0.1, 0.1)$, source
    strength $s=2.0$, width $h=0.05$, and shutoff time
    $\tau=0.3$. Parameters are defined in the diffusion equation
    (\ref{e:diffusion}). The five crosses represent noisy measurements
    at five designated measurement times.}
  \label{f:uHistory}
\end{figure}

We demonstrate the optimal Bayesian experimental design formulation
and our stochastic optimization tools on a two-dimensional contaminant
identification problem. The goal is to place a single sensor that
yields maximum information about the location of the contaminant
source. Contaminant transport is governed by a scalar diffusion
equation on a square domain:
\begin{eqnarray}
  \pp{w}{t} = \nabla^2 w + S\(\bx_{\mathrm{src}},\bx,t\),\qquad \bx\in
  \CX=\[0,1\]^2,
\label{e:diffusion}
\end{eqnarray}
where $w(\bx,t; \bx_{\mathrm{src}})$ is the space-time concentration
field parameterized by the coordinate of the source center
$\bx_{\mathrm{src}}$. We impose homogeneous Neumann
boundary conditions
\begin{eqnarray}
  \nabla w \cdot \bn = 0 \qquad \textrm{ on } \partial \CX,
\end{eqnarray}
along with a zero initial condition
\begin{eqnarray}
  w(\bx, 0; \bx_{\mathrm{src}} ) = 0.
\end{eqnarray}
The source function has a Gaussian spatial profile
\begin{eqnarray}
  S\(\bx_{\mathrm{src}},\bx,t\) =
  \left\{\begin{array}{cc}\frac{s}{2\pi h^2}
  \exp\(-\frac{\|\bx_{\mathrm{src}} - \bx\|^2}{2h^2}\), & 0 \leq t <
  \tau \\ 0, & t \geq \tau \end{array} \right.
\end{eqnarray}
where $s$, $h$, and $\tau$ are \textit{known} (prescribed) source
intensity, width, and shutoff time parameters, respectively, and
$\bx_{\mathrm{src}} \equiv \(\Theta_x,\Theta_y\)$ is the unknown
source location that we would ultimately like to infer. The design
variable is the location of a single sensor, $\bx_{\mathrm{sensor}}
\equiv \(d_x,d_y\)$, and the measurement data $\{Y_i\}_{i=1}^5$
comprise five noisy point observations of $w$ at the sensor location
and at five equally-spaced sample times. For this study, we choose
$s=2.0$, $h=0.05$, $\tau=0.3$; a uniform prior $\Theta_x, \Theta_y
\sim \mathcal{U} \(0,1\)$; and an additive error model $Y_i =
w\(\bx_{\mathrm{sensor}},t_i, ; \bx_{\mathrm{src}} \) + E_i, i = 1
\ldots 5$, such that the $E_i$ are zero-mean Gaussian random
variables, mutually independent given $\bx_{\mathrm{sensor}}$ and
$\bx_{\mathrm{src}}$, each with standard deviation $\sigma_i = 0.1 +
0.1 \left | w \( \bx_{\mathrm{sensor}}, t_i; \bx_{\mathrm{src}} \)
\right |$. In other words, the error associated with the data has a
``floor'' value of 0.1 plus an additional contribution that is 10\% of
the signal. The sensor may be placed anywhere in the square domain,
such that the design space is $\(d_x,d_y\) \in[0,1]^2$.
Figure~\ref{f:uHistory} shows an example concentration profile and
measurements.

Evaluating the forward model thus requires solving the partial
differential equation (\ref{e:diffusion}) at fixed realizations of
$\btheta = \bx_{\mathrm{src}}$ and extracting the solution field at
the design location $\bd = \bx_{\mathrm{sensor}}$. We discretize
(\ref{e:diffusion}) using 2nd-order centered differences on a 25
$\times$ 25 spatial grid and a 4th-order backward differentiation
formula for time integration. As described in Section~\ref{s:pce}, we
replace the full forward model with a polynomial chaos surrogate, for
computational efficiency. To this end, we construct a Legendre
polynomial approximation of the forward model output over the
4-dimensional joint parameter and design space, using a total-order
polynomial truncation of degree $12$ and $10^6$ forward
  model evaluations. This high polynomial degree and rather large
  number of forward model evaluations were deliberately selected in
  order to render truncation and aliasing error insignificant in our
  study. Optimal experimental design results of similar quality may be
  obtained for this problem with surrogates of lower order and
  with far fewer quadrature points (e.g., degree 4 with $10^4$ forward
  model evaluations) but for brevity they are not included here.  The
relative $L^2$ errors of the current surrogate range from $6 \times
10^{-3}$ to $10^{-6}$.
%
% XH: 10^-2 may seem large, but it is quite small compared to the
% likelihood noise. Also, this is the worst component of the 5
% measurements; the best ones are 1e-5. Perhaps instead of trying to
% explain, let's just not put these numbers in. 

% Previous investigations from the authors have shown the validity of
% the polynomial chaos surrogates via different
% problems~\cite{huan:2010:abe, huan:2013:sbo}.

The optimal Bayesian experimental design formulation now seeks the
sensor location $\bx^*_{\mathrm{sensor}}$ such that when the
experiment is performed, \textit{on average}---i.e., averaged over all
possible source locations according to the prior, and over all
possible resulting concentration measurements according to the
likelihood---the five concentration readings
$\left\{Y_i\right\}_{i=1}^5$ yield the greatest information gain from
prior to posterior.

\subsection{Results}

\subsubsection{Objective function}

Before we present the results of numerical optimization, we first
explore the properties of the expected information gain
objective. Numerical realizations of $\hU_{N,M}$ for $N=1001$ and
$M=2$, 11, 101, and 1001 are shown in Figure~\ref{f:EUMC3D}. These
plots can be interpreted as 1-sample Monte Carlo approximations of
$\barU_M = \mathbb{E} [ \hU_{N,M} ]$, or equivalently, as $l$-sample Monte
Carlo approximations of $\barU_M = \mathbb{E} [ \hU_{(N/l),M} ]$.
% YMM: nice, but we probably can't get away with that remark :-)
% XH: awww :p
% \footnote{This is due to the additive nature of Monte Carlo. While
%   technically true, this may be confusing. But all good papers confuse
%   their readers at some point. }
%
%As $N$ and $M$ grow, $\hU_{N,M}$ becomes a better approximation to
%$\barU_M$. 
%
As $N$ grows, $\hU_{N,M}$ becomes a better approximation to $\barU_M$
and as $M$ grows, $\barU_M$ becomes a better approximation to $U$.
The figures show that values of $\hU_{N,M}$ increase when $M$
increases (for fixed $N$), suggesting a negative bias at finite
$M$. At the same time, the objective becomes less flat in $\bd$; since
$U$ is certainly closer to the $M=1001$ surface than the $M=2$
surface, these results suggest that $U$ is not particularly flat in
$\bd$.  This feature of the current design problem is encouraging,
since stochastic optimization problems with higher curvature can be
more easily solved; in the context of SA, for example, they
effectively have a higher signal-to-noise ratio.
%\btodo Is it gradient
%or curvature that is important here?  \etodo
% XH: I think they are closely related here. In the end, i think it's
% the curvature. 

The expected information gain objective inherits symmetries from the
square, as expected from the physical nature of the problem. The plots
also suggest a smooth albeit nonconvex underlying objective $U$, with
inflection points lying on an interior circle and four local maxima
symmetrically located at the corners of the design space.  The best
placement for a single sensor is therefore at the corners of the
design space, while the worst placement is at the center. The reason
for this perhaps counterintuitive result is that the diffusion
process is isotropic: a series of concentration measurements can only
determine the distance of the source from the sensor, not its
orientation. The posterior distribution thus resembles an annulus of
constant radius surrounding the sensor. A sensor placement that
minimizes the area of these annuli, averaged over all possible source
locations according to the prior, tends to be optimal. In this
problem, because of the domain geometry and the magnitude of the
observational noise, these optimal locations happen to be the furthest
points from the domain center, i.e., the corners.

\begin{figure}[htb]
  \centering 
  \mbox{\subfigure[$\bx_{\mathrm{sensor}}=(0.0,0.0)$]
    {\includegraphics[width=0.45\textwidth]{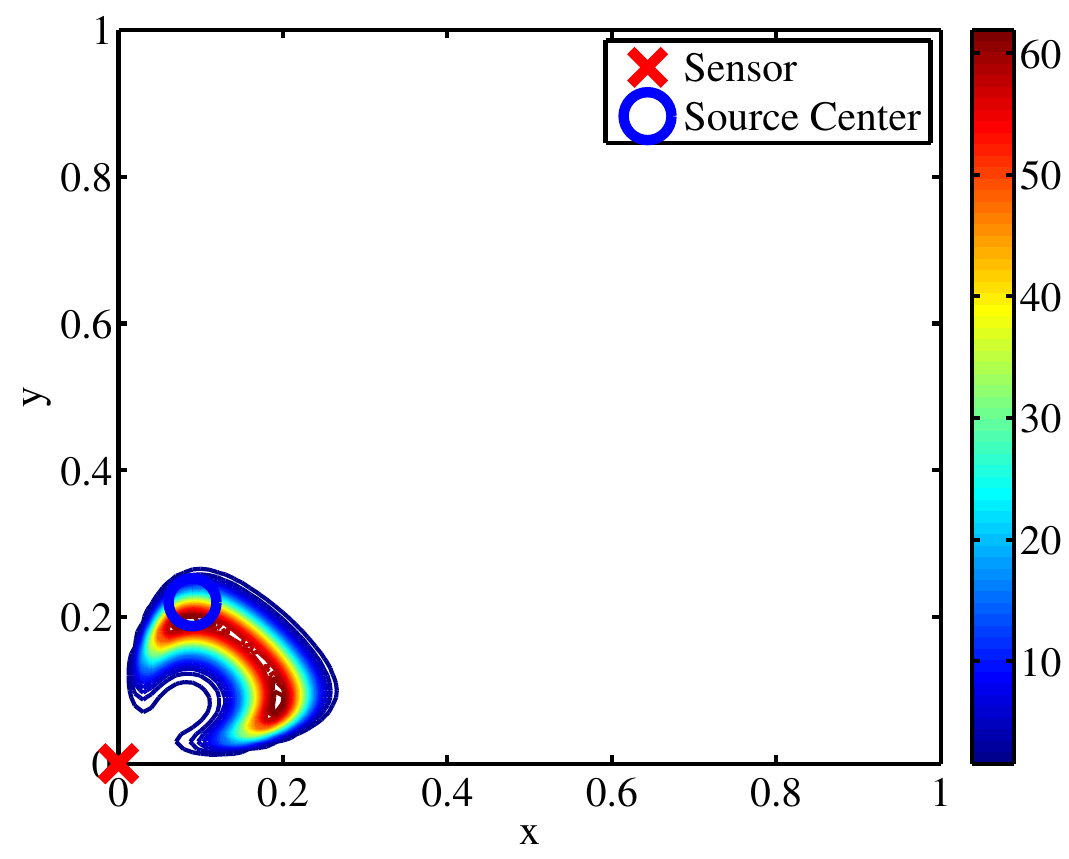}}} 
  \mbox{\subfigure[$\bx_{\mathrm{sensor}}=(0.0,1.0)$]
    {\includegraphics[width=0.45\textwidth]{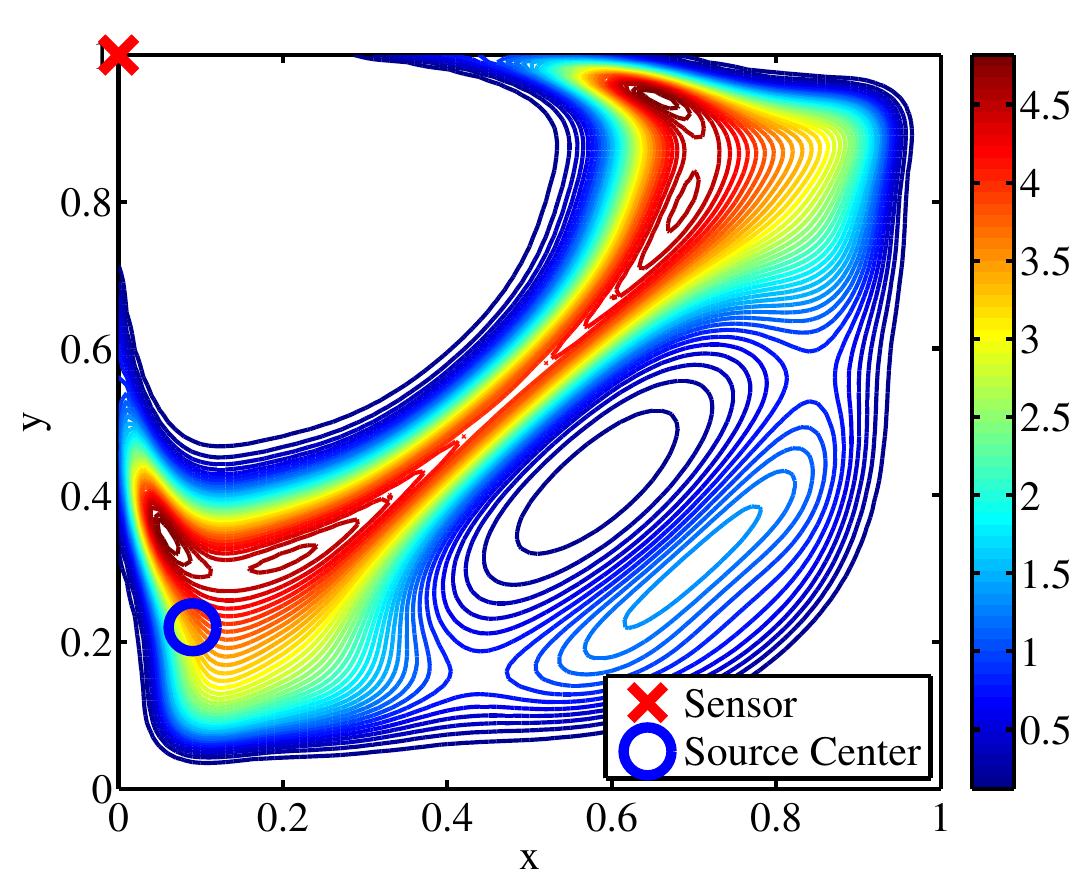}}} 
  \mbox{\subfigure[$\bx_{\mathrm{sensor}}=(1.0,0.0)$]
    {\includegraphics[width=0.45\textwidth]{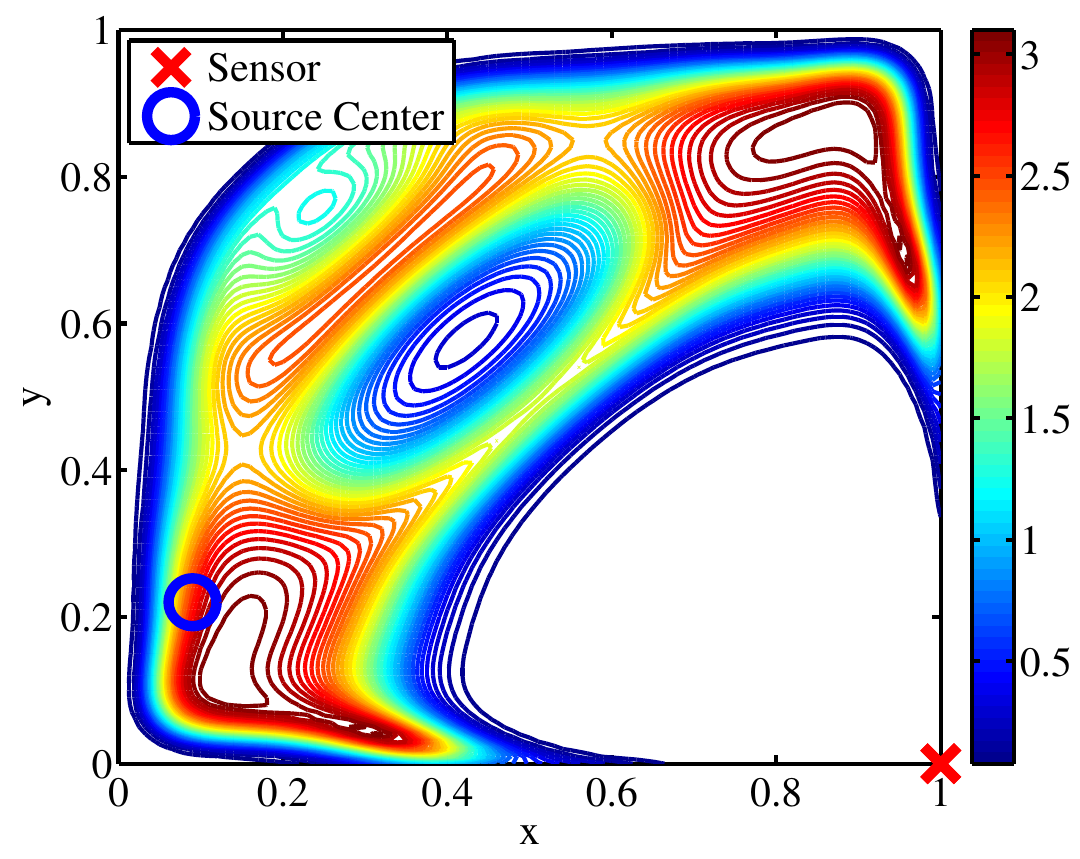}}} 
  \mbox{\subfigure[$\bx_{\mathrm{sensor}}=(1.0,1.0)$]
    {\includegraphics[width=0.45\textwidth]{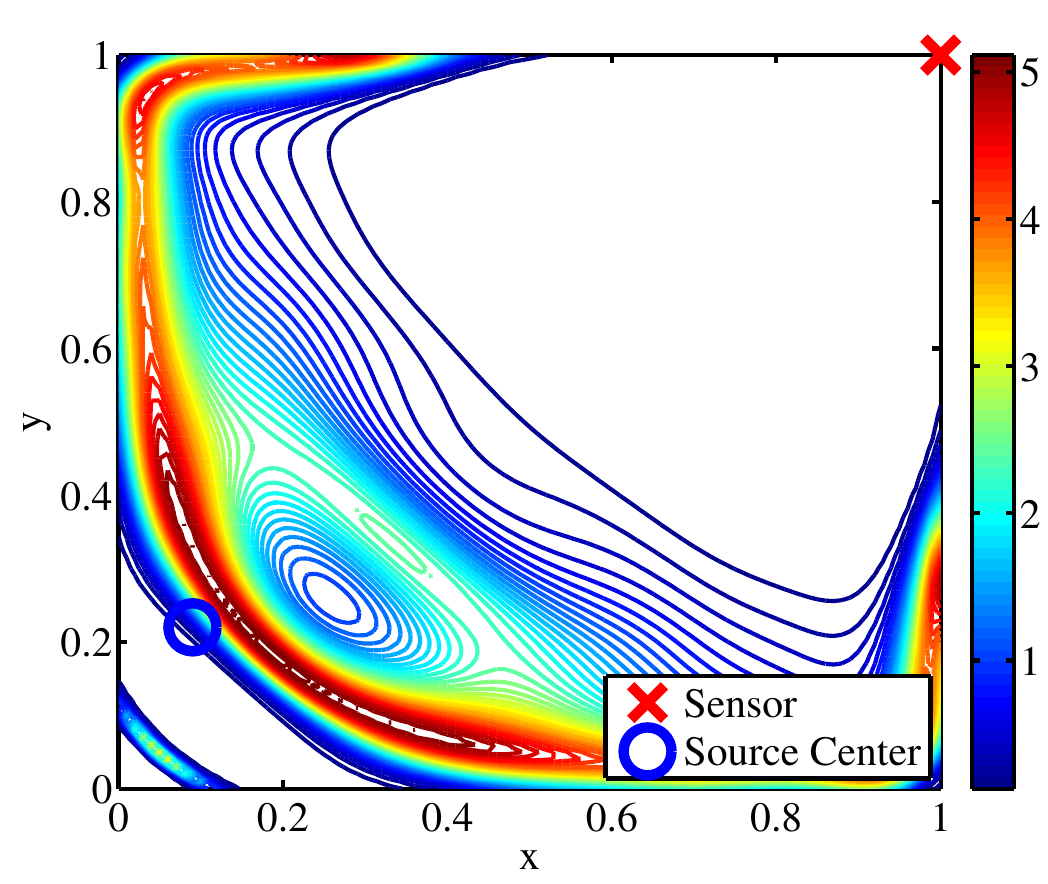}}} 
  \mbox{\subfigure[$\bx_{\mathrm{sensor}}=(0.5,0.5)$]
    {\includegraphics[width=0.45\textwidth]{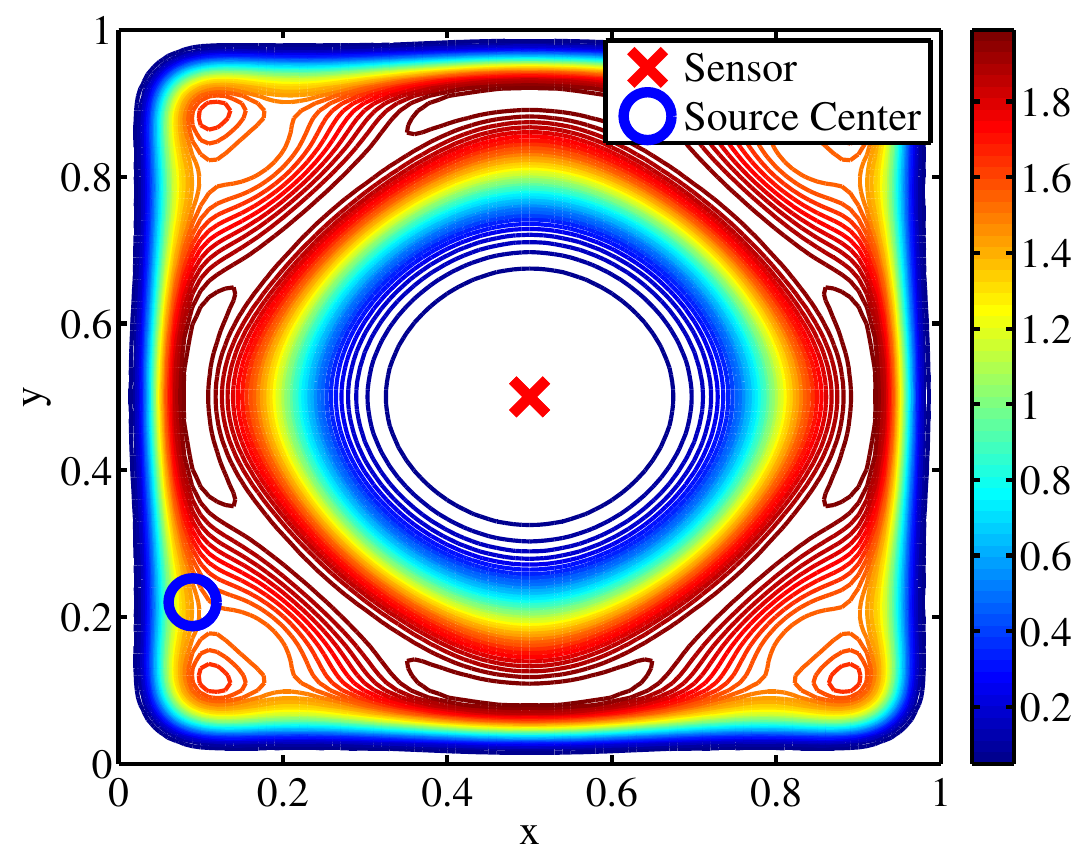}}} 
  \caption{Contours of posterior probability density for the source
    location, given different sensor placements. The true source
    location, marked with a blue circle, is
    $\bx_{\mathrm{src}}=(0.09,0.22)$.}
  \label{f:posteriorsSourcex0_09y0_22}
\end{figure}

Figure~\ref{f:posteriorsSourcex0_09y0_22} shows posterior probability
densities for the source location, under different sensor placements,
given data generated from a ``true'' source centered at
$\bx_{\mathrm{src}}=(0.09,0.22)$. The posterior densities are
evaluated using the polynomial chaos surrogate, while the data are
generated by directly solving the diffusion equation on a denser (101
$\times$ 101) spatial grid than before and then adding the Gaussian
noise described in Section~\ref{s:diffusion-setup}. Note that the
posteriors are extremely non-Gaussian. Moreover, they generally
include the true source location, but do not center on it. Reasons for
not expecting the posterior mode to match the true source location are
twofold: first, we have only 5 measurements, each perturbed with a
relatively significant random noise; second, there is model error, due
to mismatch between the polynomial chaos approximation constructed
from the coarser spatial discretization of the PDE and the more
finely discretized PDE model used to simulate the
data.\footnote{Indeed, there are two levels of model error: (1)
  between the PC expansion and the PDE model used to construct the PC
  expansion, which has a $\Delta x$ = $\Delta y$ = $1/24$ spatial
  discretization; (2) between this PDE model and the more finely
  discretized ($\Delta x$ = $\Delta y$ = $1/100$) PDE model used to
  simulate the noisy data.}\textsuperscript{,}\footnote{Model error is
  an extremely important aspect of uncertainty
  quantification~\cite{kennedy:2001:bco}, but its treatment is beyond
  the scope of this study. Understanding the impact of model error on
  optimal experimental design is an important direction for future
  work.} For this source configuration, it appears that a sensor
placed at any of the corners yields a ``tighter'' posterior than a
sensor placed at the center. But we must keep in mind that this result
is not guaranteed for \textit{all} source locations and data
realizations; it depends on where the source actually is. [Imagine,
for example, if the source happened to be very close to the center of
the domain; then the sensor at (0.5, 0.5) would yield the tightest
posterior.] What the optimal experimental design method yields is the
optimal sensor placement \textit{averaged} over the prior distribution
of the source location and the predictive distribution of the data.

\subsubsection{Stochastic optimization results}
\label{s:stochopt-results}

We now analyze the optimization results, first assessing the behavior
of the two stochastic optimization methods individually, and then
comparing their performance.

\begin{figure}[htb]
  \centering 
  \mbox{\subfigure[$N=1001$, $M=2$]
    {\includegraphics[width=0.49\textwidth]{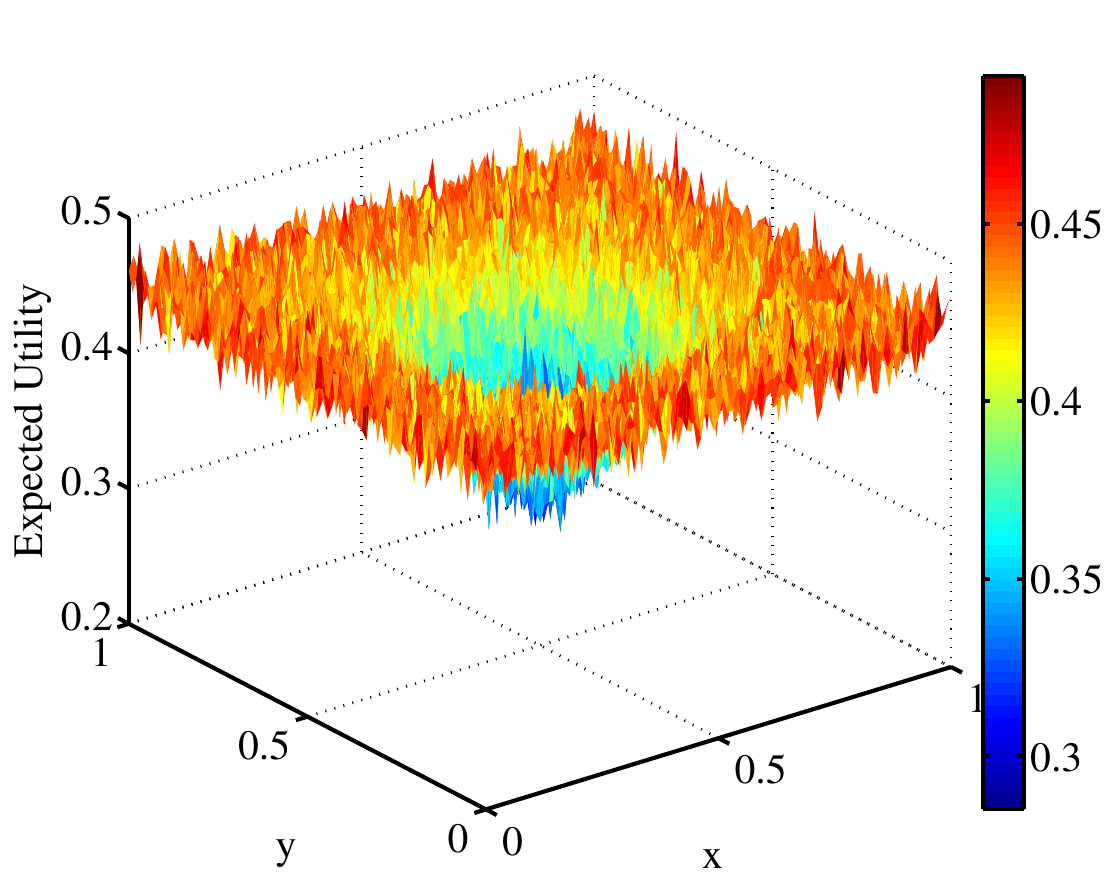}}} 
  \mbox{\subfigure[$N=1001$, $M=11$]
    {\includegraphics[width=0.49\textwidth]{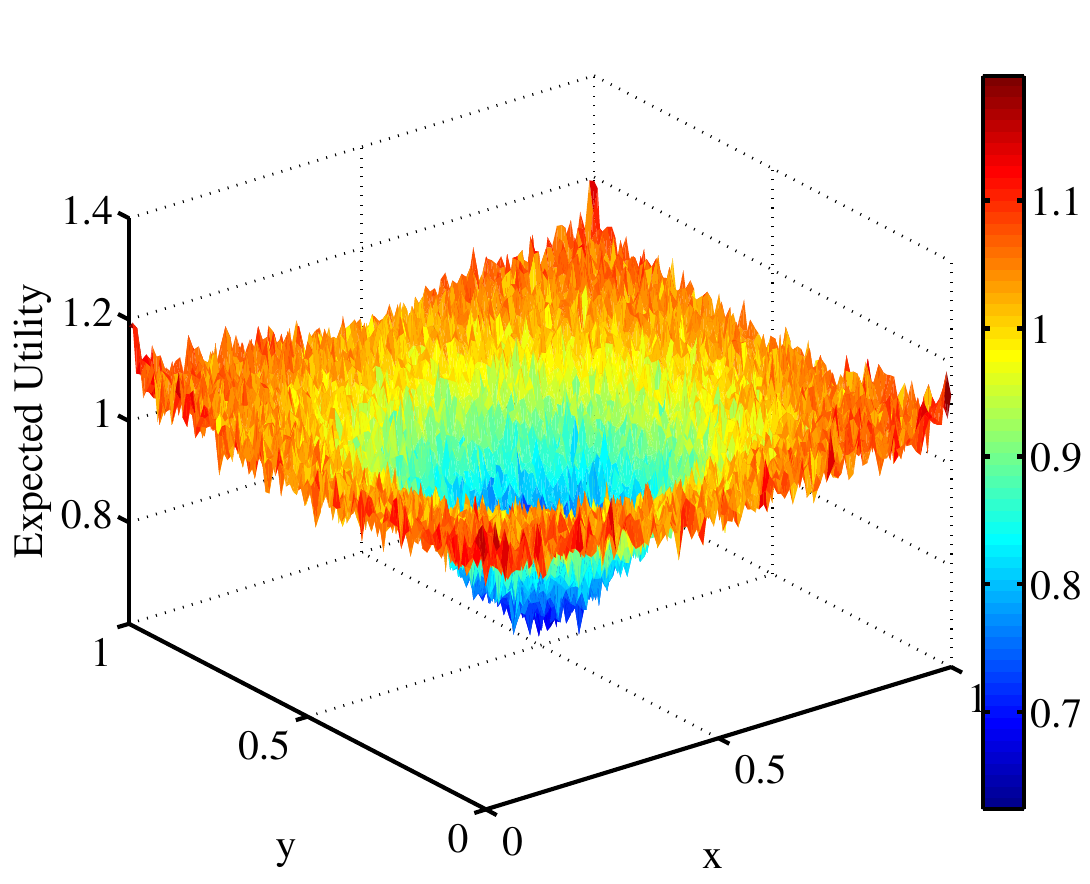}}} 
  \mbox{\subfigure[$N=1001$, $M=101$]
    {\includegraphics[width=0.49\textwidth]{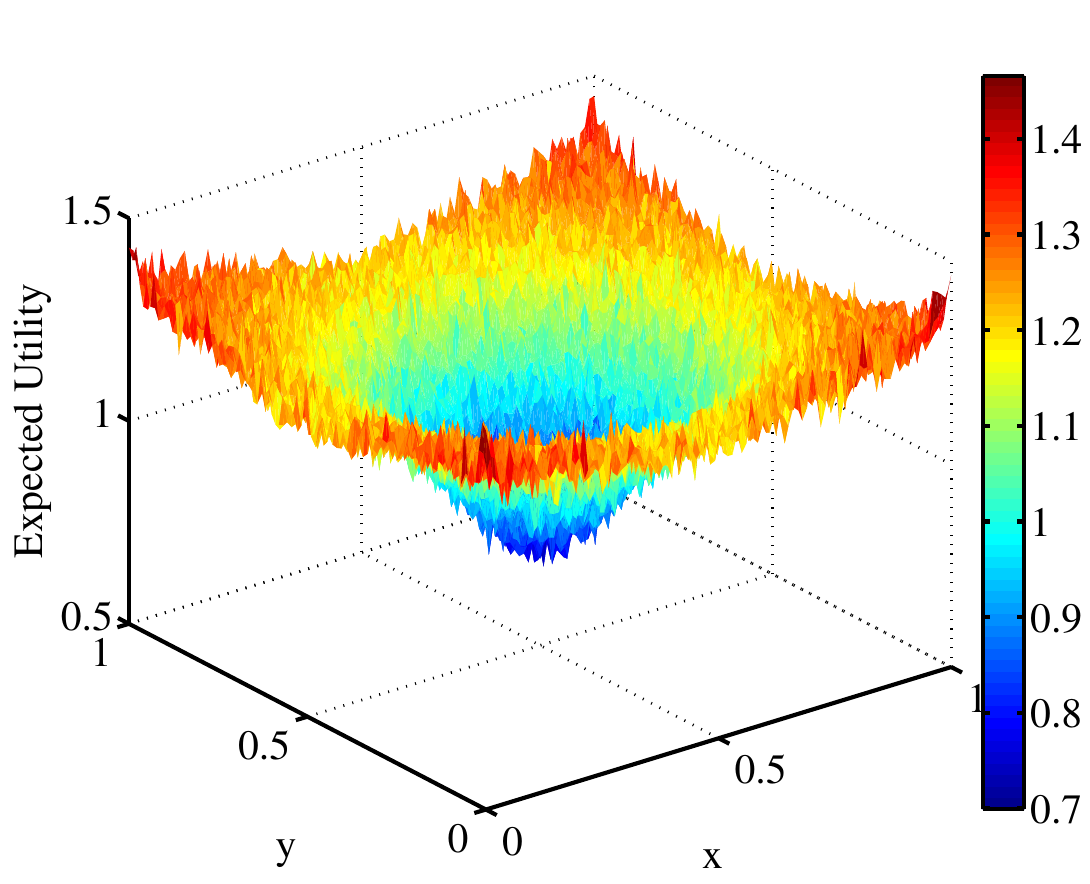}}} 
  \mbox{\subfigure[$N=1001$, $M=1001$]
    {\includegraphics[width=0.49\textwidth]{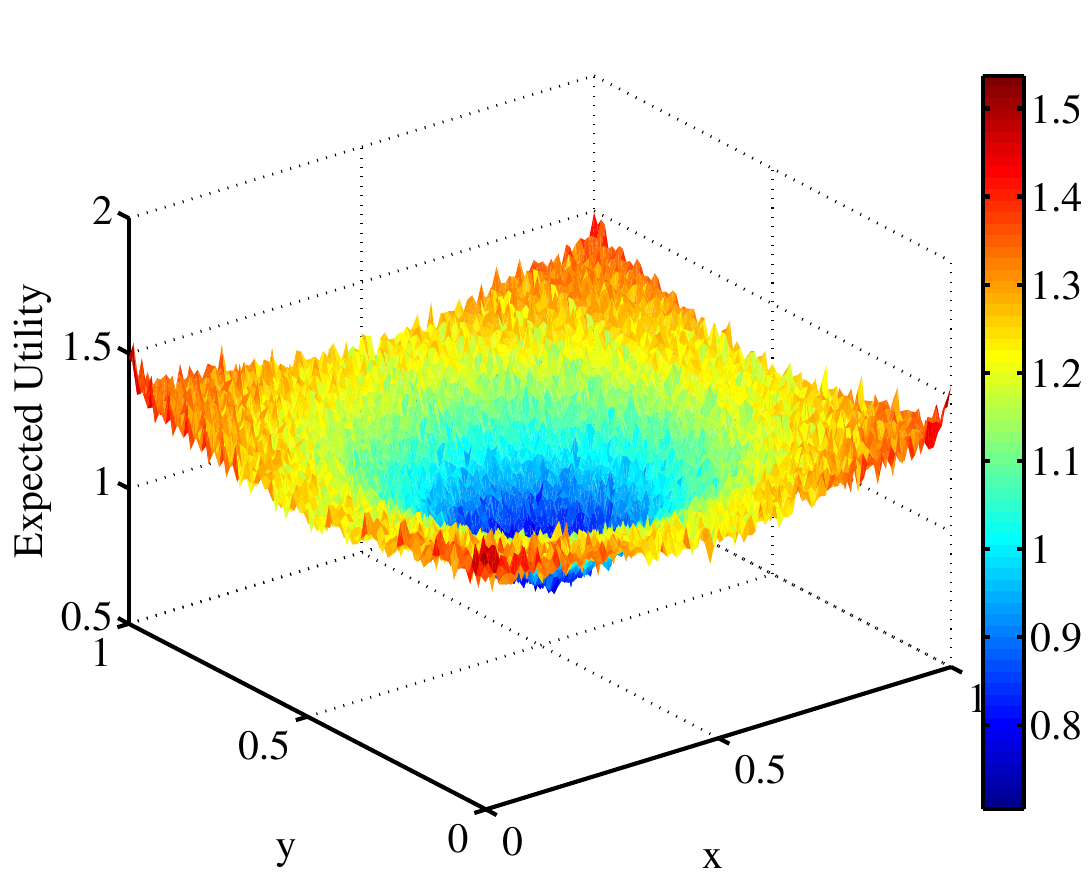}}}   
  \caption{Surface plots of independent $\hU_{N,M}$ realizations,
    evaluated over the entire design space $[0,1]^2 \ni \bd =
    (x,y)$. Note that the vertical axis ranges and color scales vary
    among the subfigures.}
  \label{f:EUMC3D}
\end{figure}

\begin{figure}[htb]
  \centering 
  \mbox{\subfigure[$N=1$, $M=2$]
    {\includegraphics[width=0.48\textwidth]{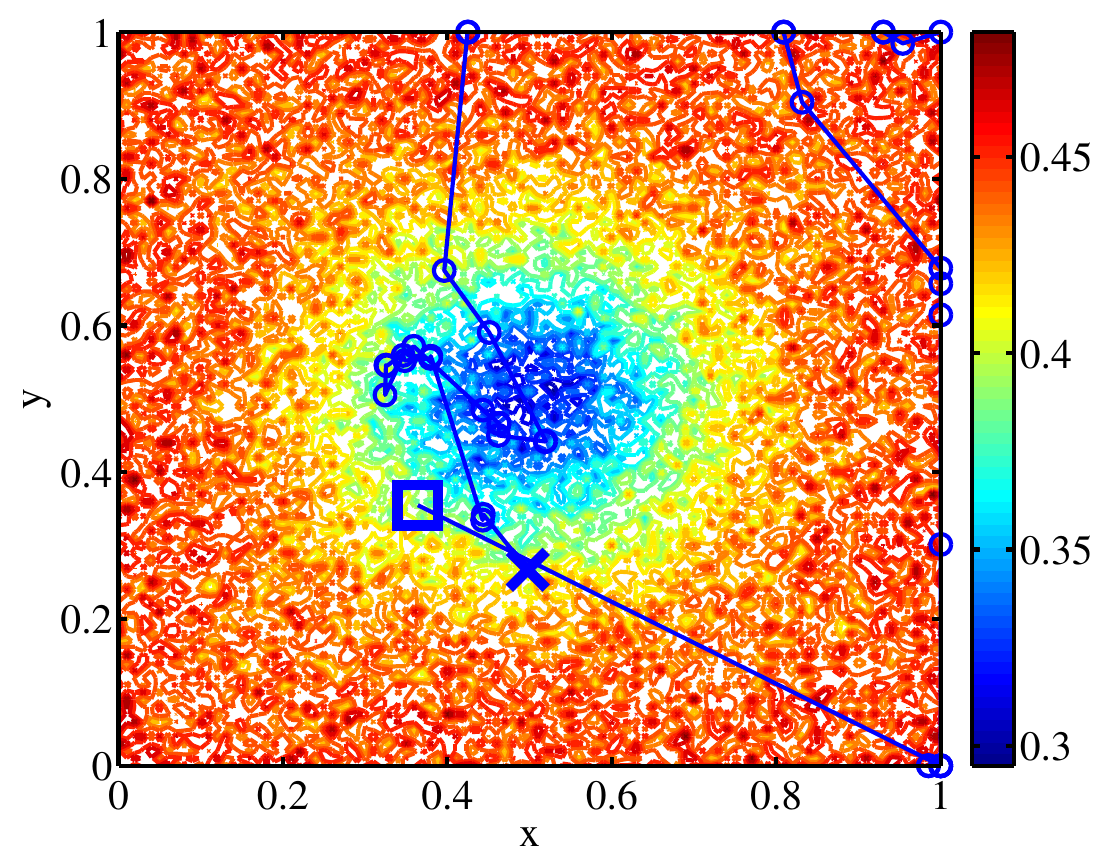}}} 
  \mbox{\subfigure[$N=1$, $M=11$]
    {\includegraphics[width=0.48\textwidth]{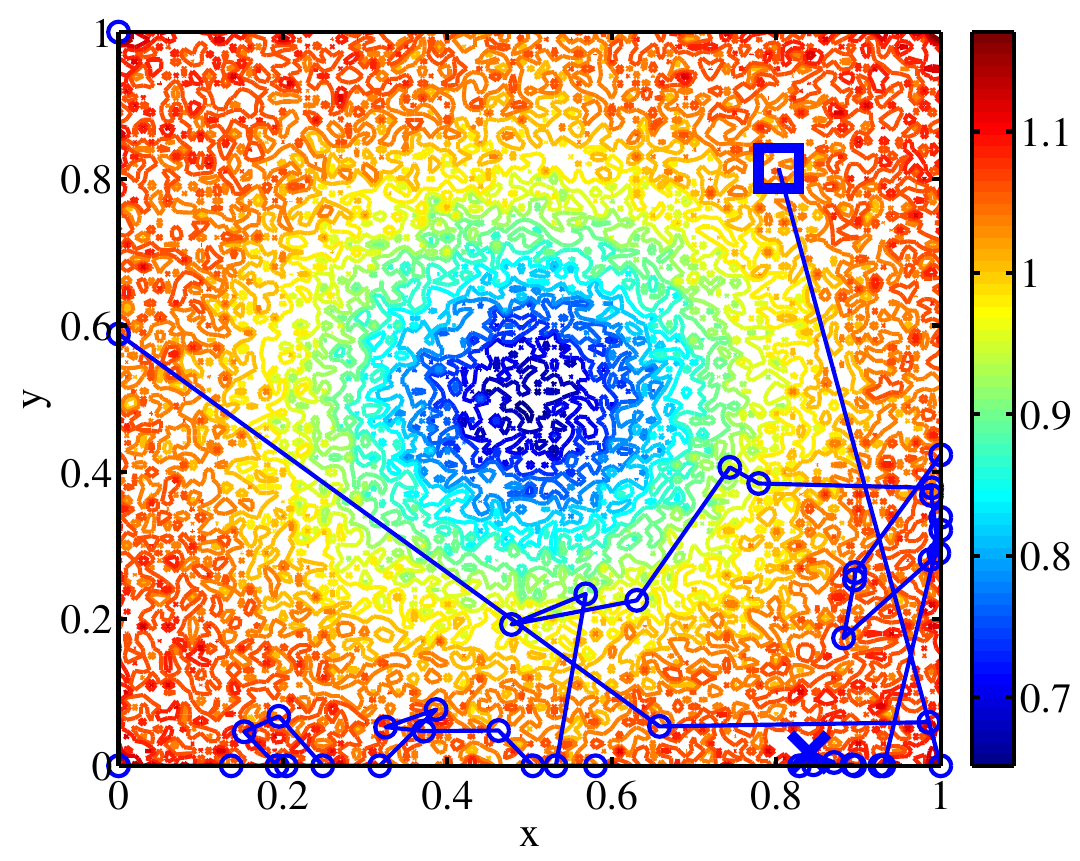}}} 
  \mbox{\subfigure[$N=1$, $M=101$]
    {\includegraphics[width=0.48\textwidth]{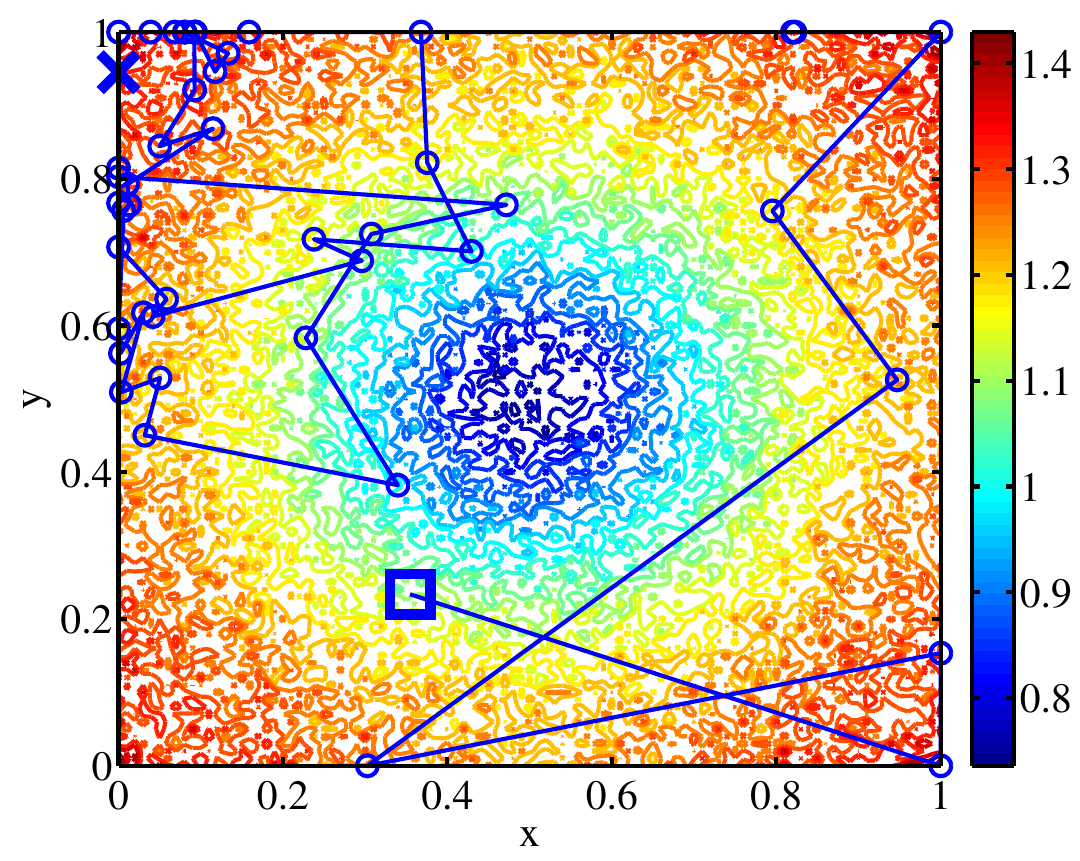}}} 
  \mbox{\subfigure[$N=1$, $M=1001$]
    {\includegraphics[width=0.48\textwidth]{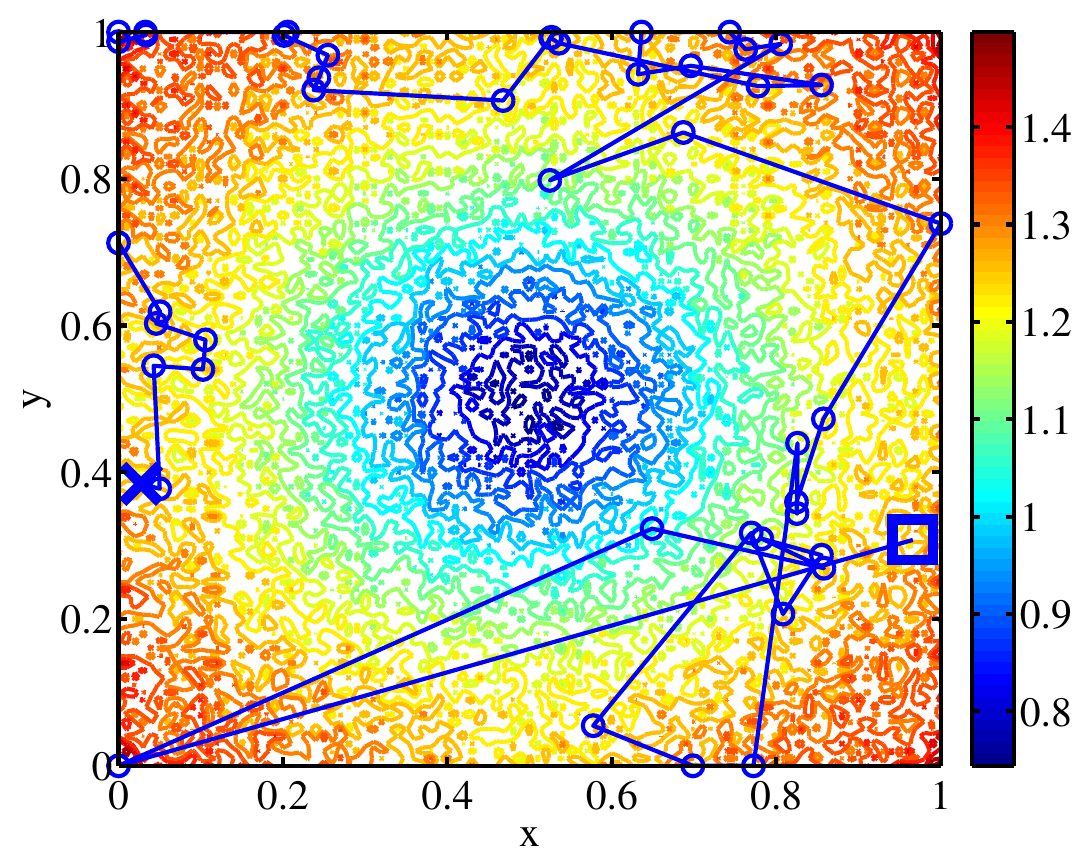}}} 
  \caption{Sample paths of the RM algorithm with $N=1$, overlaid on
    $\hU_{N,M}$ surfaces from Figure~\ref{f:EUMC3D} with the
    corresponding $M$ values. The large $\square$ is the starting
    position and the large $\times$ is the final position.}
  \label{f:SARMRunPath1}
\end{figure}

\begin{figure}[htb]
  \centering 
  \mbox{\subfigure[$N=11$, $M=2$]
    {\includegraphics[width=0.48\textwidth]{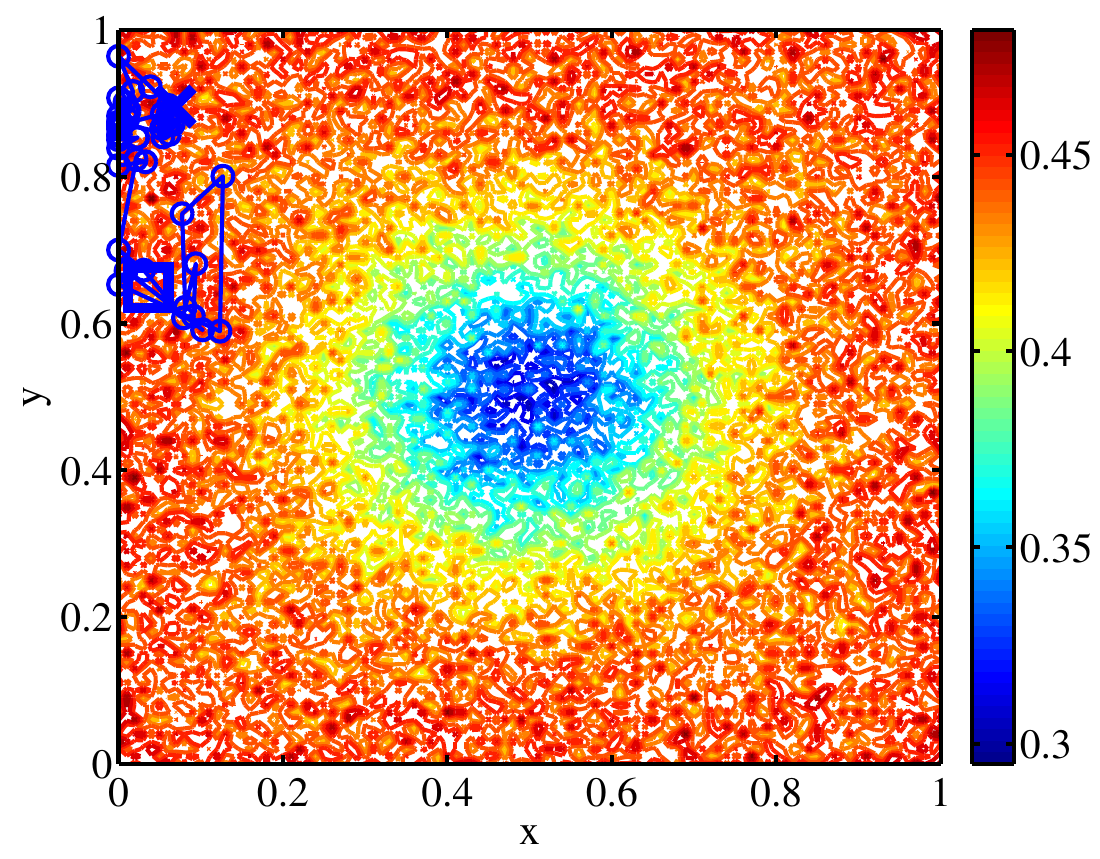}}} 
  \mbox{\subfigure[$N=11$, $M=11$]
    {\includegraphics[width=0.48\textwidth]{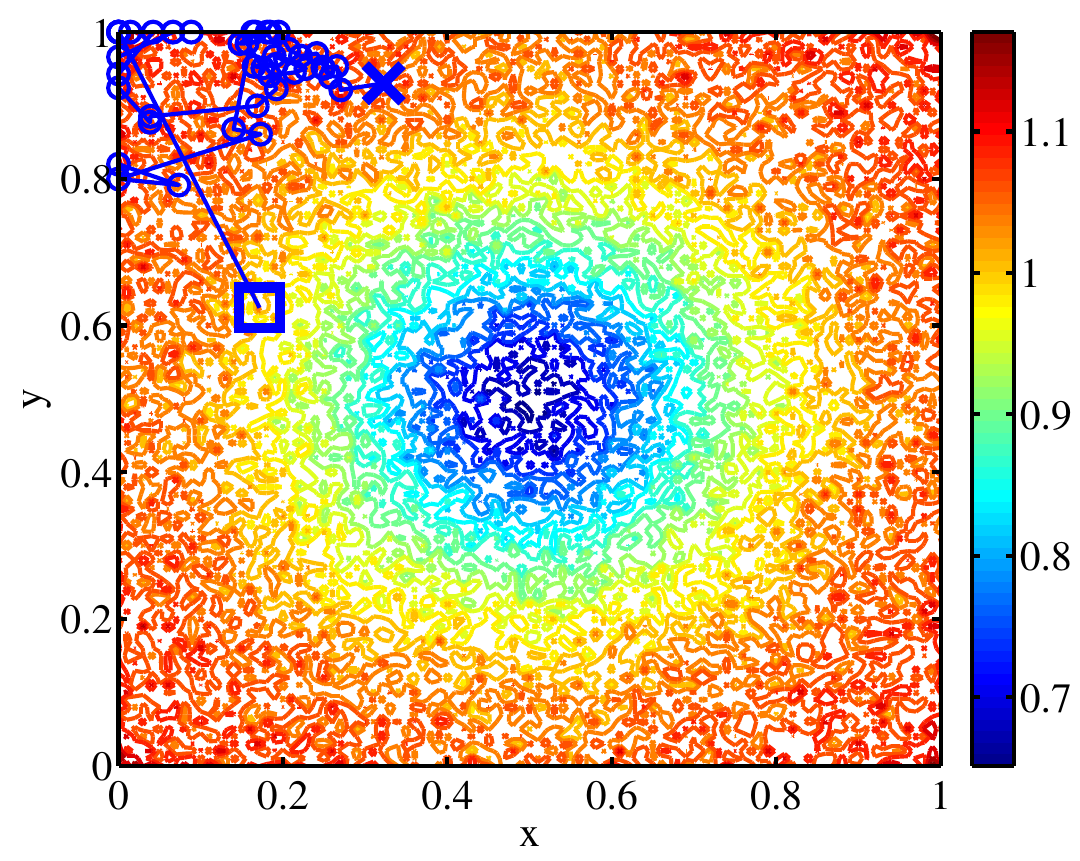}}} 
  \mbox{\subfigure[$N=11$, $M=101$]
    {\includegraphics[width=0.48\textwidth]{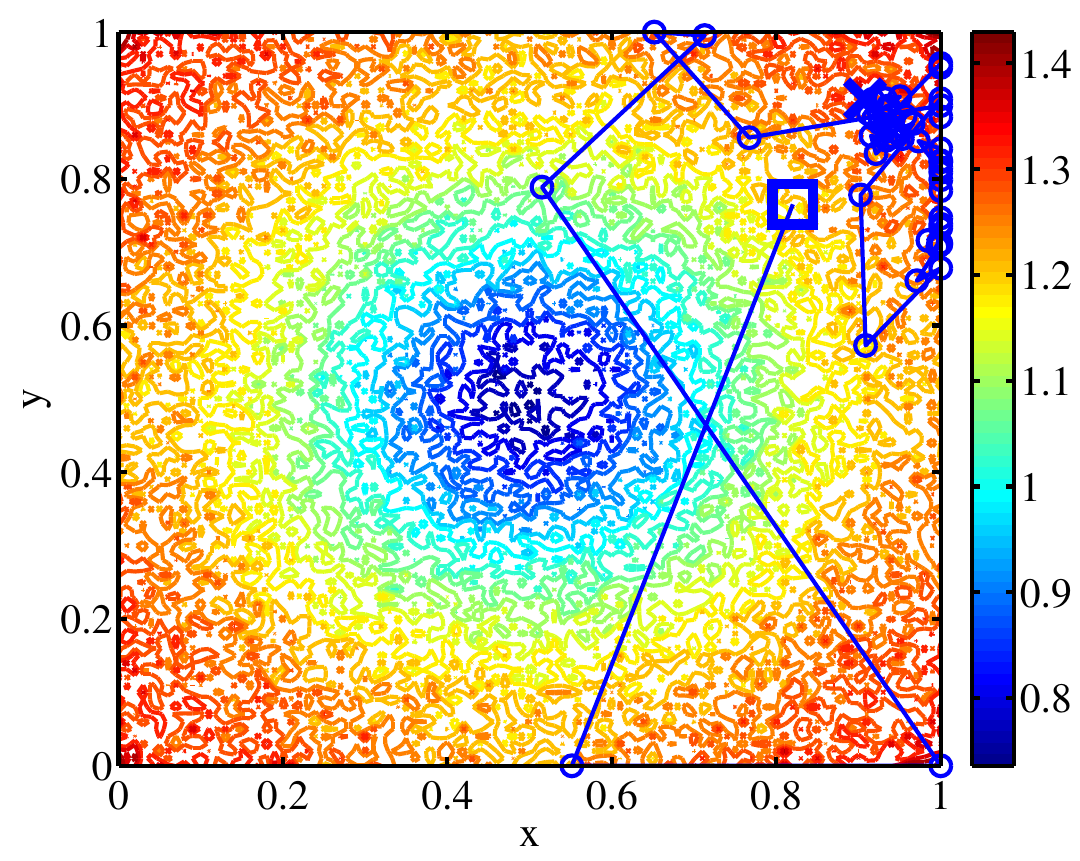}}} 
  \mbox{\subfigure[$N=11$, $M=1001$]
    {\includegraphics[width=0.48\textwidth]{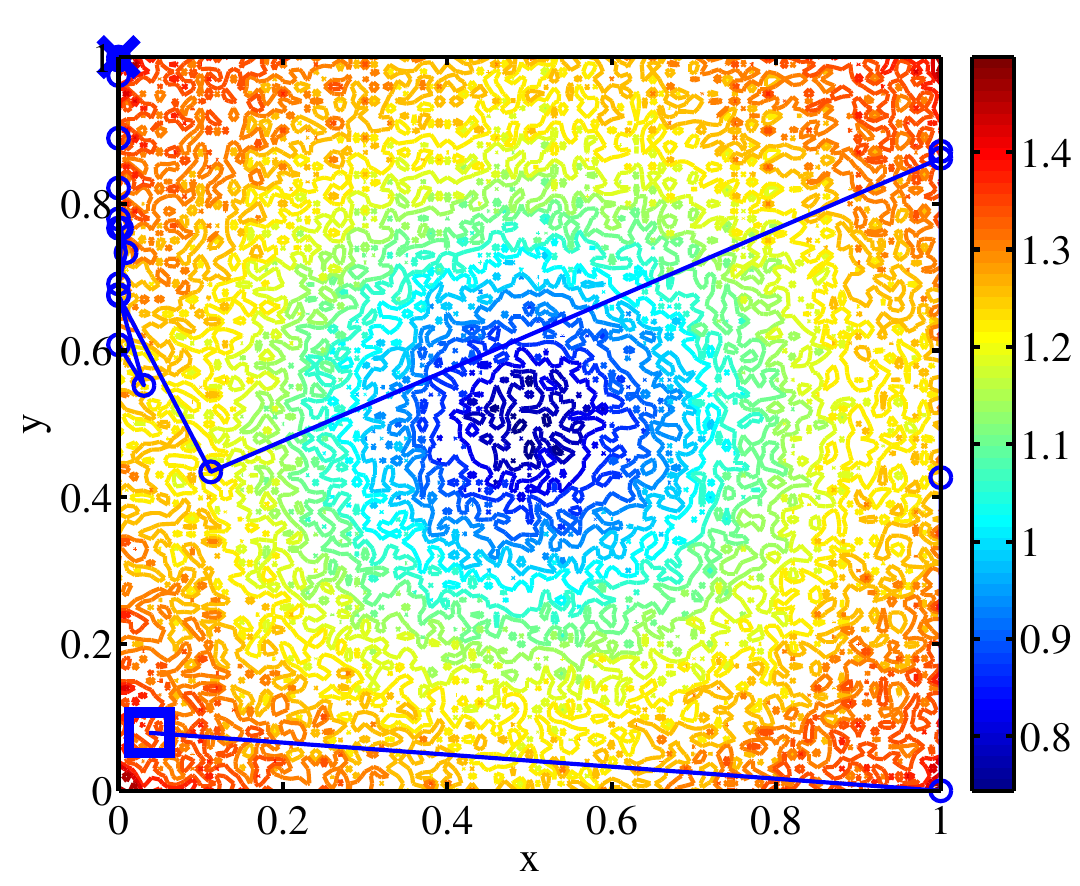}}} 
  \caption{Sample paths of the RM algorithm with $N=11$, overlaid on
    $\hU_{N,M}$ surfaces from Figure~\ref{f:EUMC3D} with the
    corresponding $M$ values. The large $\square$ is the starting
    position and the large $\times$ is the final position.}
  \label{f:SARMRunPath2}
\end{figure}

\begin{figure}[htb]
  \centering 
  \mbox{\subfigure[$N=101$, $M=2$]
    {\includegraphics[width=0.48\textwidth]{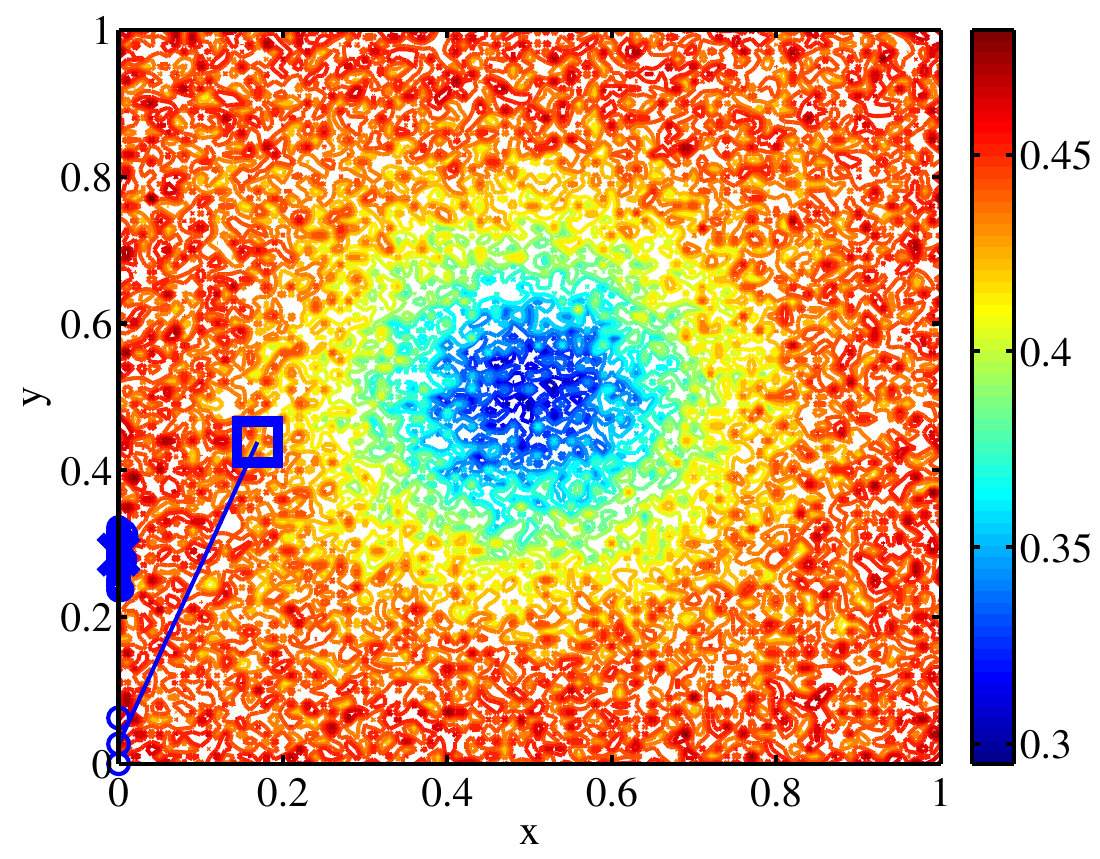}}} 
  \mbox{\subfigure[$N=101$, $M=11$]
    {\includegraphics[width=0.48\textwidth]{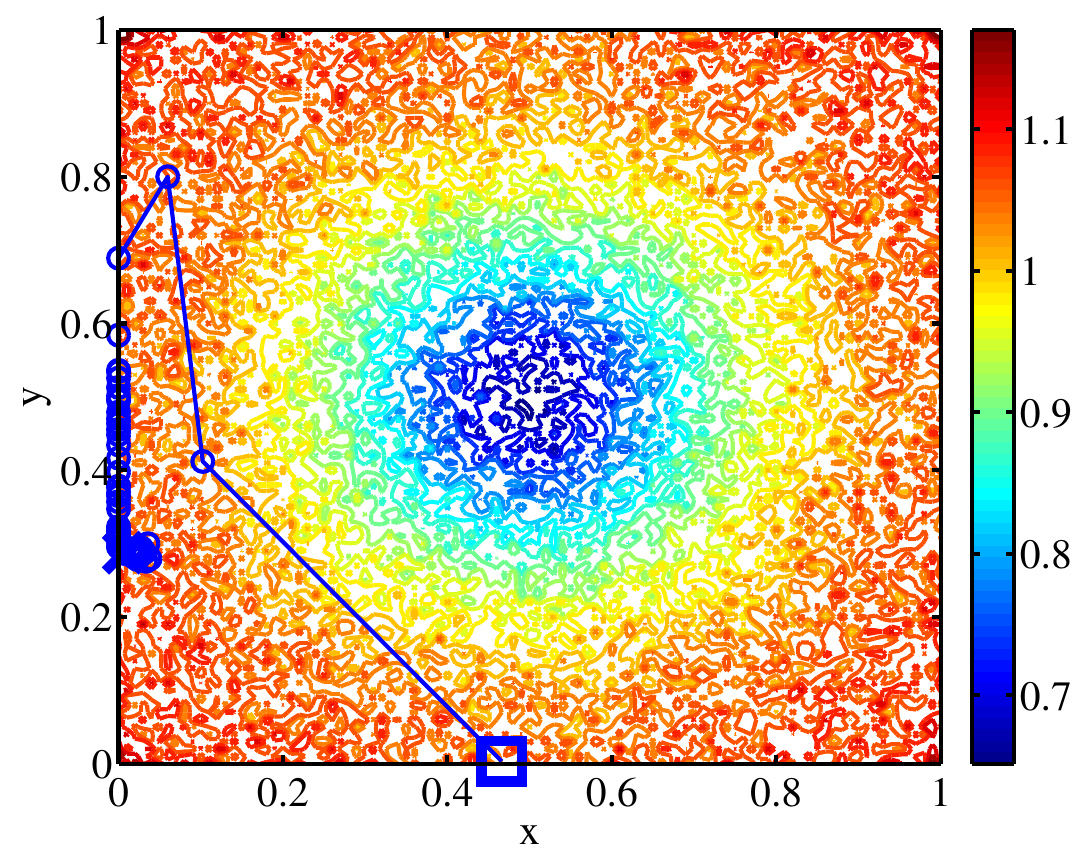}}} 
  \mbox{\subfigure[$N=101$, $M=101$]
    {\includegraphics[width=0.48\textwidth]{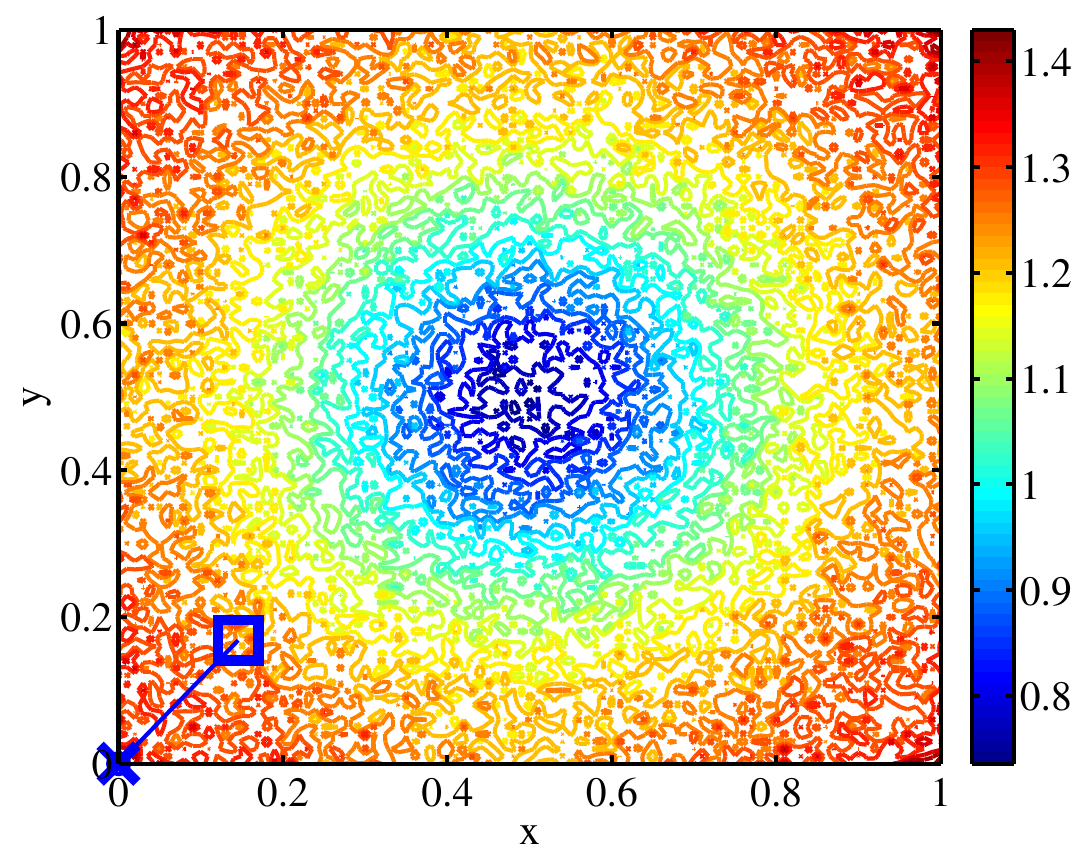}}} 
  \mbox{\subfigure[$N=101$, $M=1001$]
    {\includegraphics[width=0.48\textwidth]{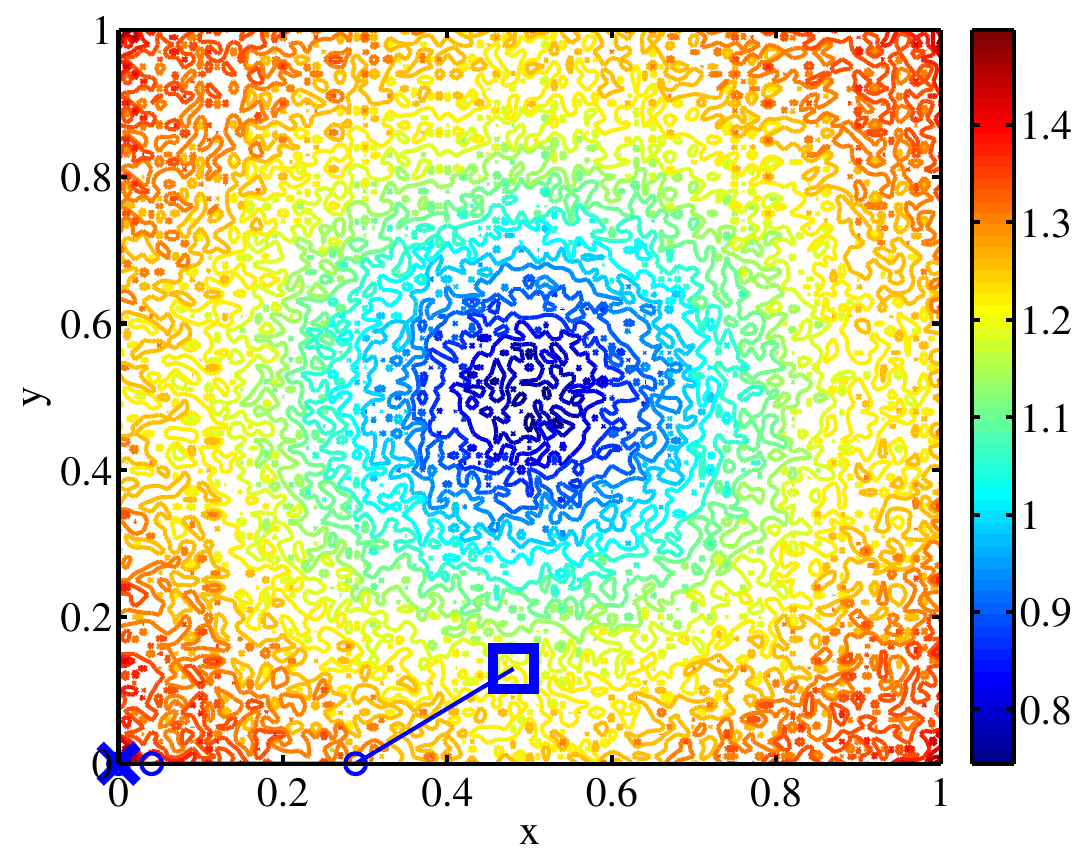}}} 
  \caption{Sample paths of the RM algorithm with $N=101$, overlaid on
    $\hU_{N,M}$ surfaces from Figure~\ref{f:EUMC3D} with the
    corresponding $M$ values. The large $\square$ is the starting
    position and the large $\times$ is the final position.}
  \label{f:SARMRunPath3}
\end{figure}

Recall that the RM algorithm is essentially a steepest-ascent method
(since we are maximizing the objective) with a stochastic gradient
estimate. Figures~\ref{f:SARMRunPath1}--\ref{f:SARMRunPath3} each show
four sample RM optimization paths overlaid on the $\hU_{N,M}$ surfaces
from Figure~\ref{f:EUMC3D}.  The optimization does not always proceed
in an ascent direction, due to the noise in the gradient estimate, but
even a noisy gradient can be useful in eventually guiding the
algorithm to regions of high objective value. Naturally, fewer
iterations are needed and good designs are more likely to be found
when the variance of the gradient estimator is reduced by increasing
$N$ and $M$. Note that one must be cautious not to over-generalize
from these figures, since the paths shown in each plot are not
necessarily representative. Instead, their purpose is to provide
intuition about the optimization mechanics. Data derived from many
runs are more appropriate performance metrics, and will be used later
in this section.

\begin{figure}[htb]
  \centering 
  \mbox{\subfigure[$N=1$, $M=2$]
    {\includegraphics[width=0.48\textwidth]{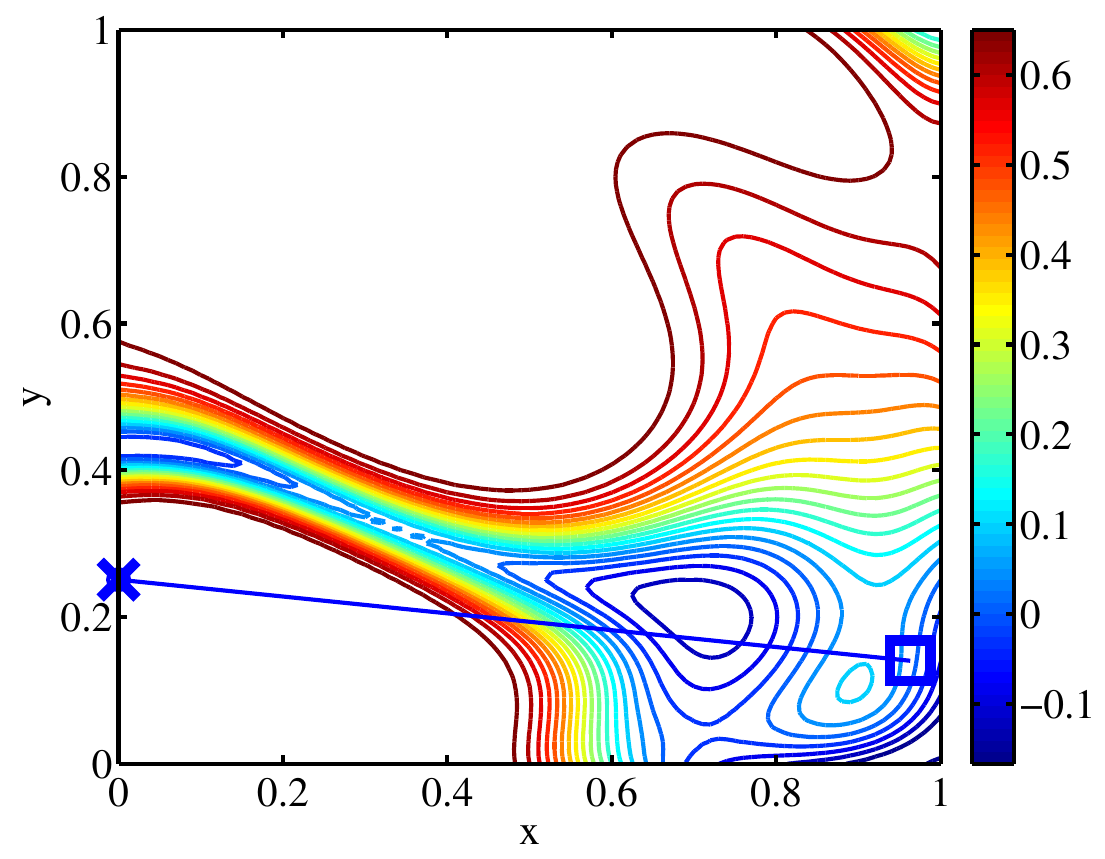}}} 
  \mbox{\subfigure[$N=1$, $M=11$]
    {\includegraphics[width=0.48\textwidth]{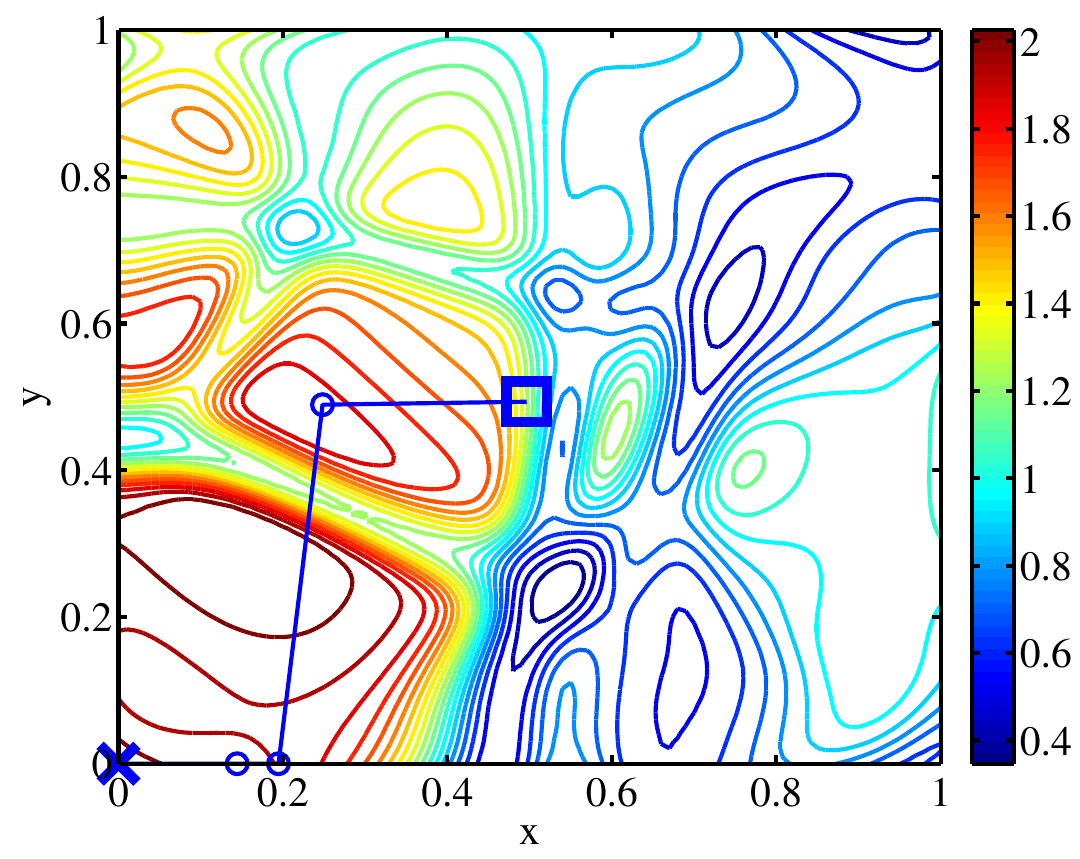}}} 
  \mbox{\subfigure[$N=1$, $M=101$]
    {\includegraphics[width=0.48\textwidth]{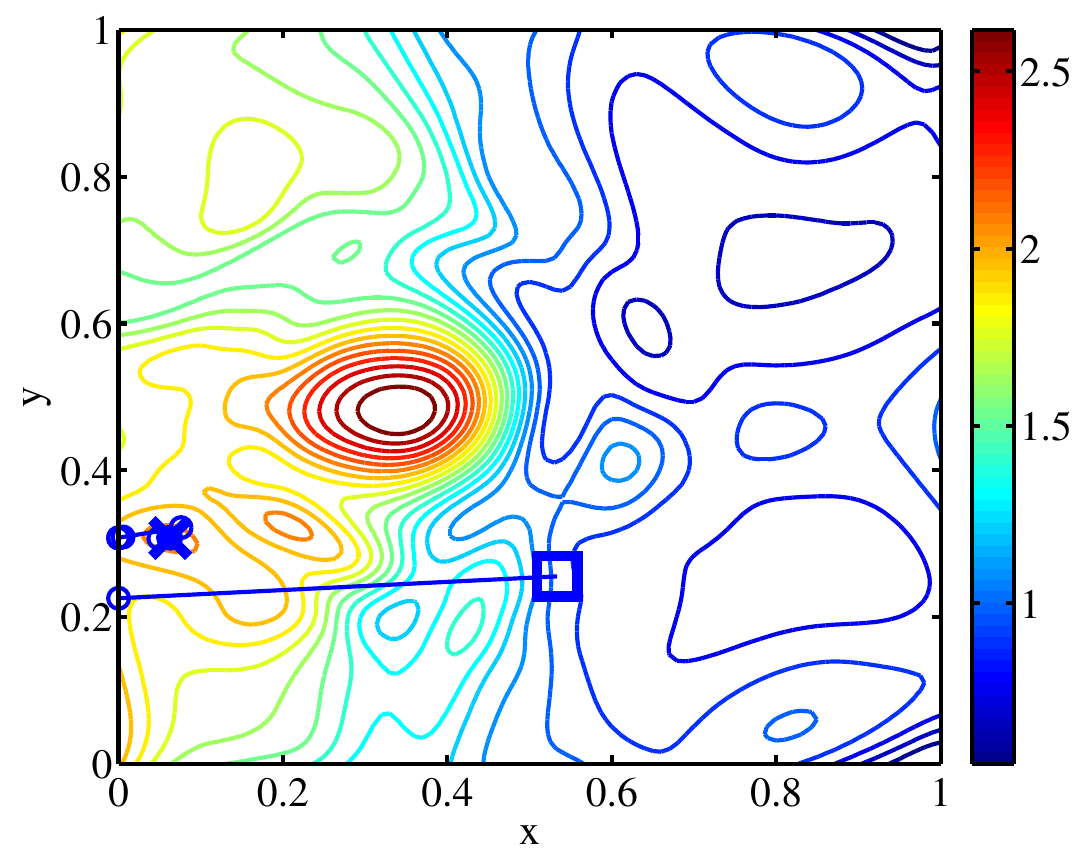}}} 
  \mbox{\subfigure[$N=1$, $M=1001$]
    {\includegraphics[width=0.48\textwidth]{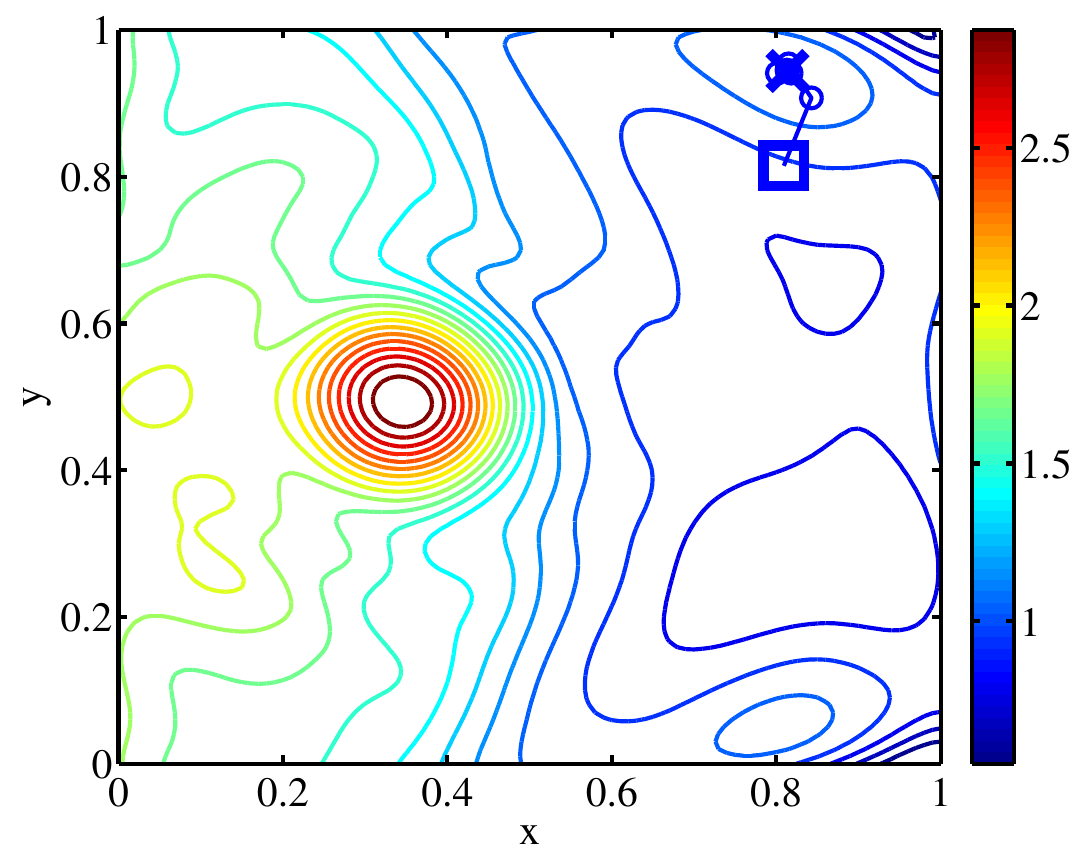}}} 
  \caption{Realizations of the objective function surface using SAA,
    and corresponding steps of BFGS, with $N=1$. The large
    $\square$ is the starting position and the large $\times$ is the
    final position.}
  \label{f:SAAInstance11}
\end{figure}

\begin{figure}[htb]
  \centering 
  \mbox{\subfigure[$N=11$, $M=2$]
    {\includegraphics[width=0.48\textwidth]{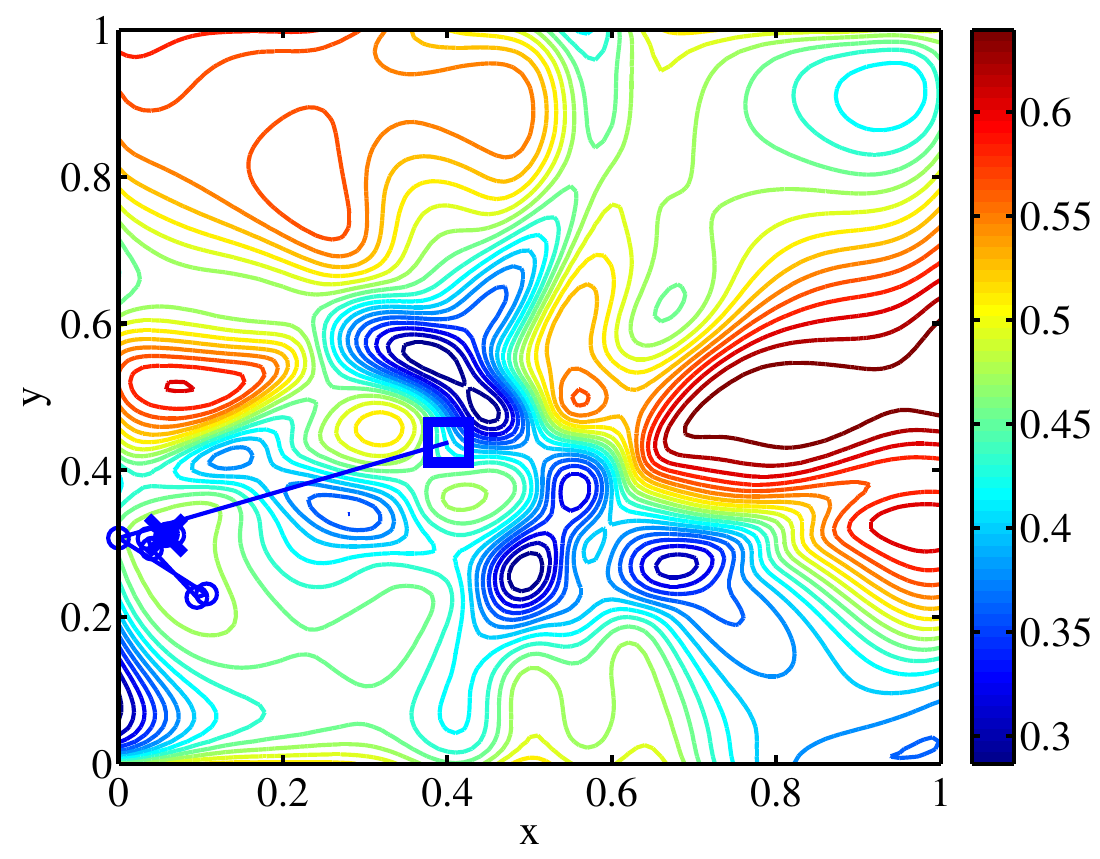}}} 
  \mbox{\subfigure[$N=11$, $M=11$]
    {\includegraphics[width=0.48\textwidth]{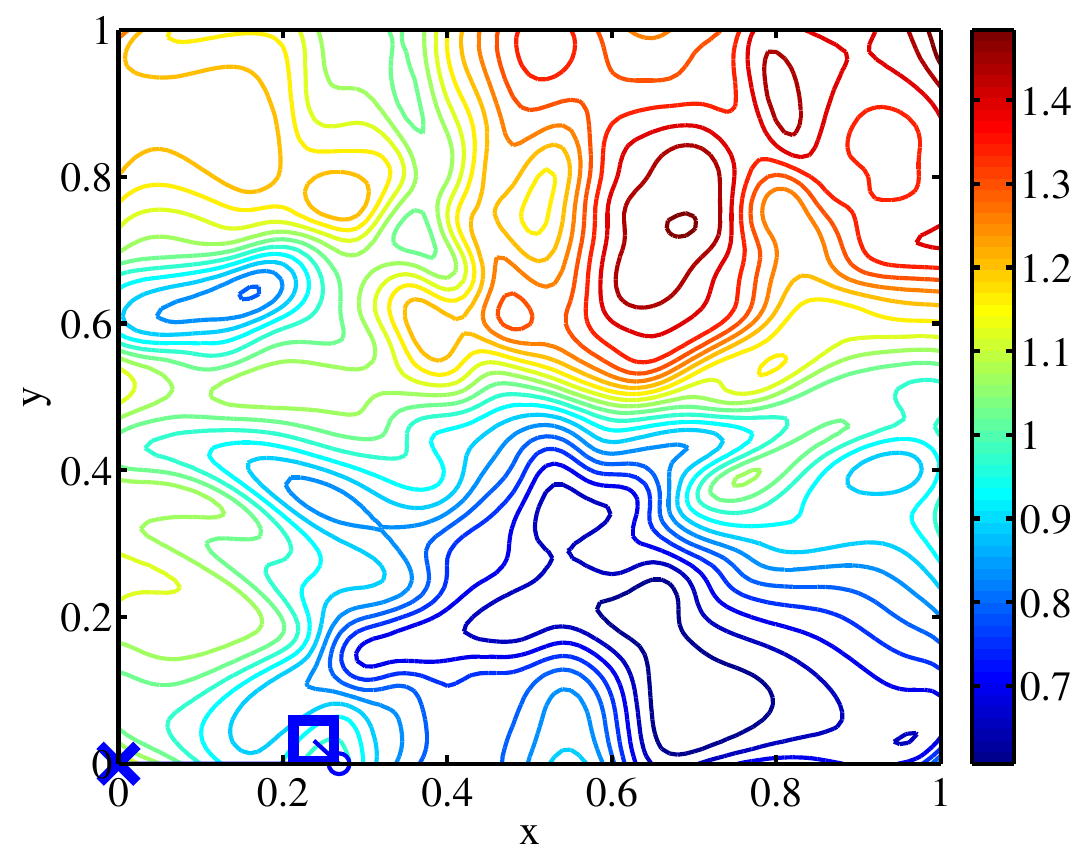}}} 
  \mbox{\subfigure[$N=11$, $M=101$]
    {\includegraphics[width=0.48\textwidth]{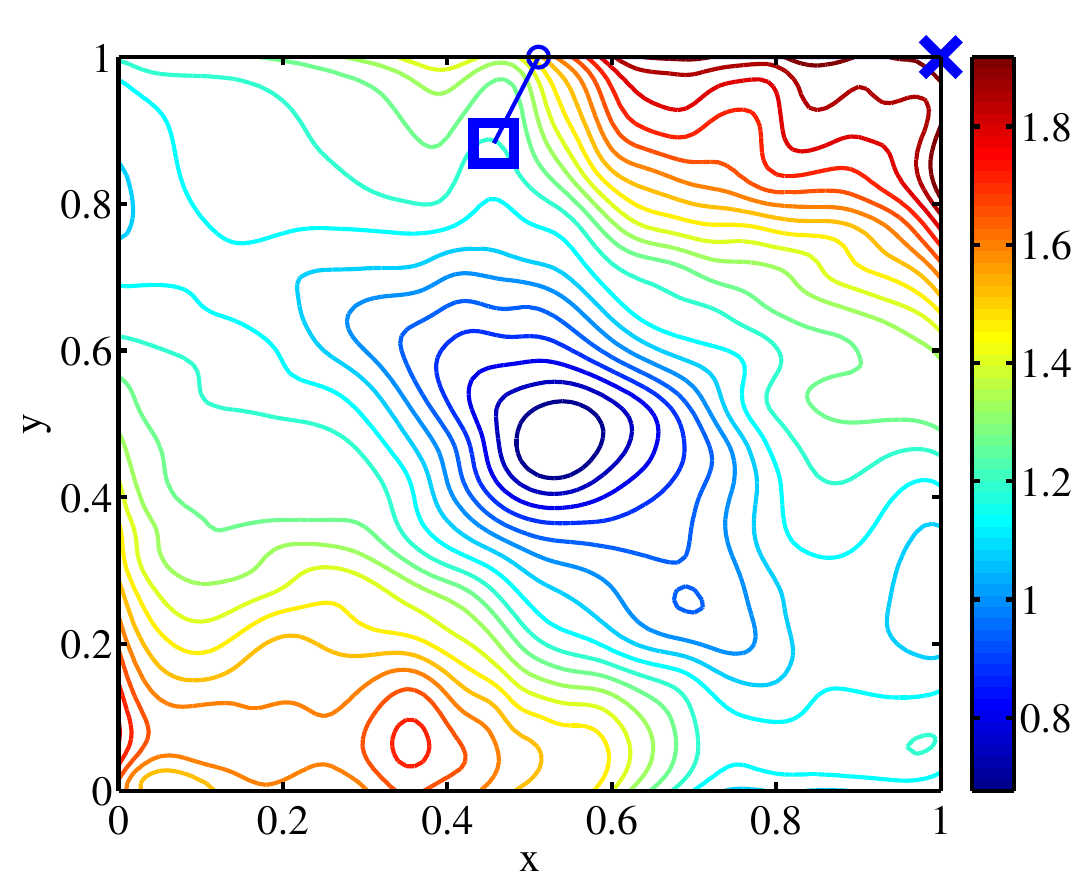}}} 
  \mbox{\subfigure[$N=11$, $M=1001$]
    {\includegraphics[width=0.48\textwidth]{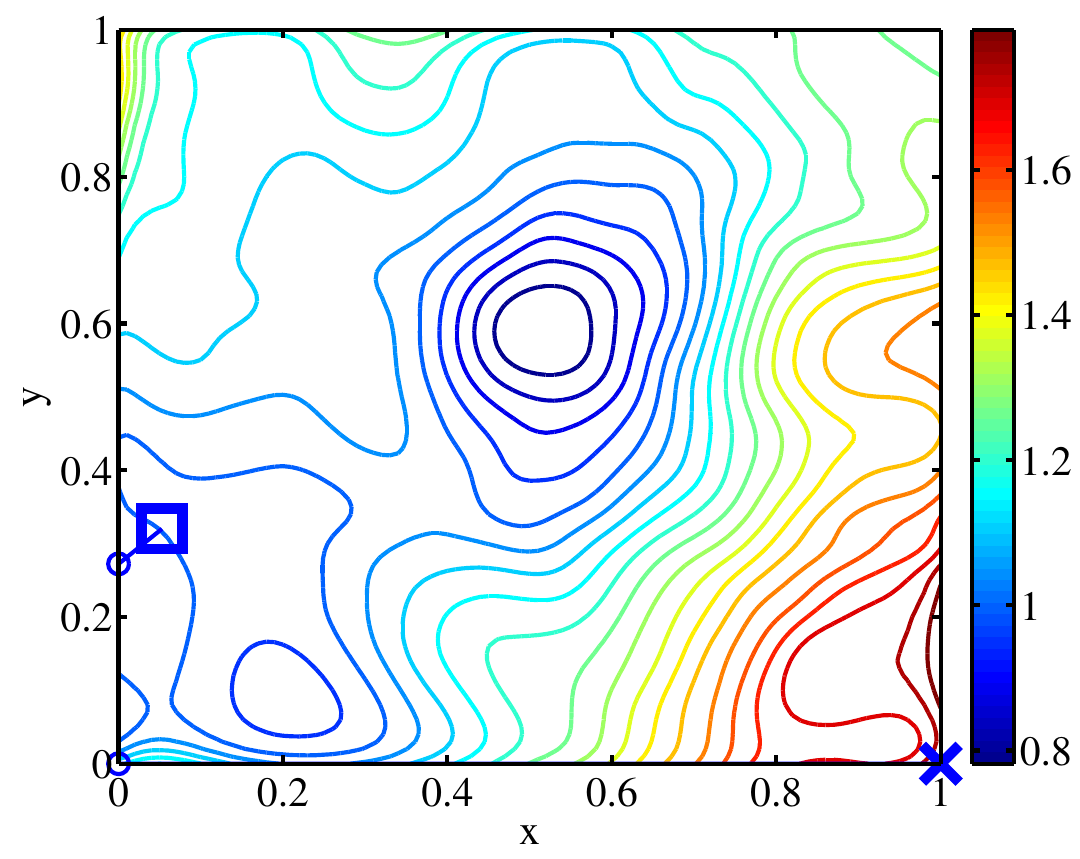}}} 
  \caption{Realizations of the objective function surface using SAA,
    and corresponding steps of BFGS, with $N=11$. The large $\square$
    is the starting position and the large $\times$ is the final
    position.}
  \label{f:SAAInstance12}
\end{figure}

\begin{figure}[htb]
  \centering 
  \mbox{\subfigure[$N=101$, $M=2$]
    {\includegraphics[width=0.48\textwidth]{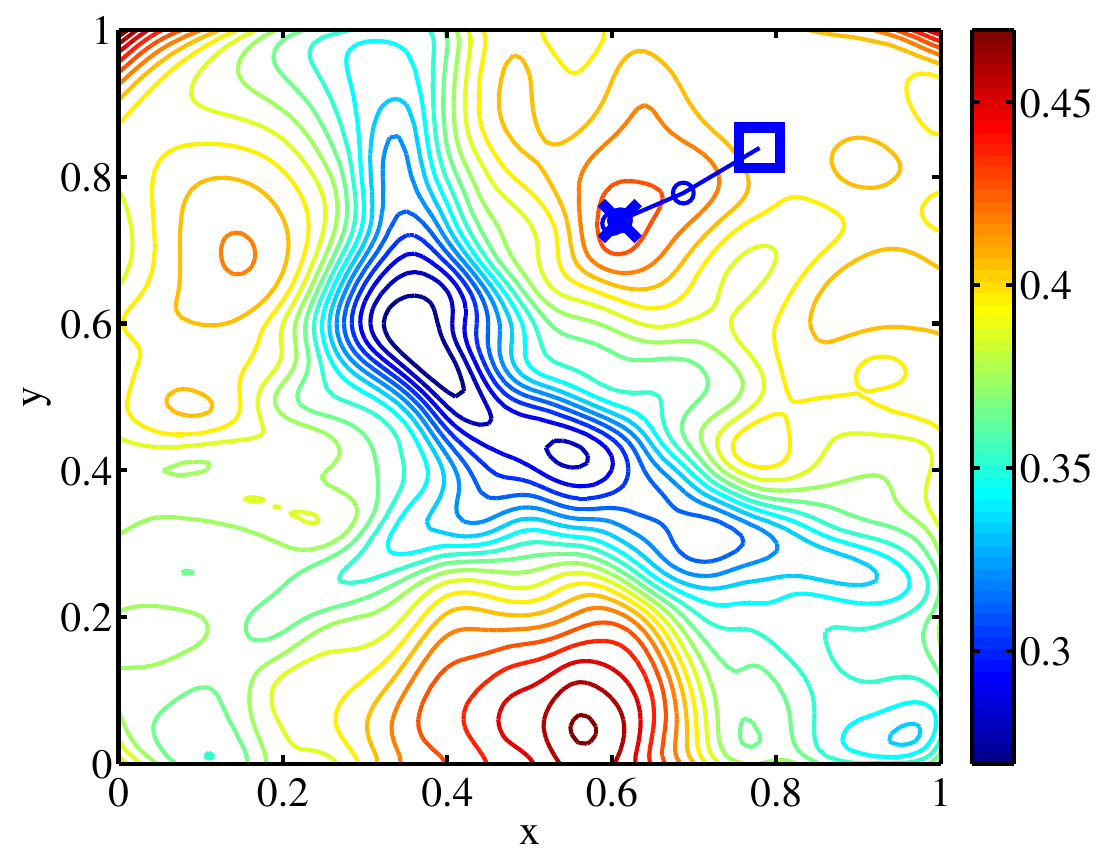}}} 
  \mbox{\subfigure[$N=101$, $M=11$]
    {\includegraphics[width=0.48\textwidth]{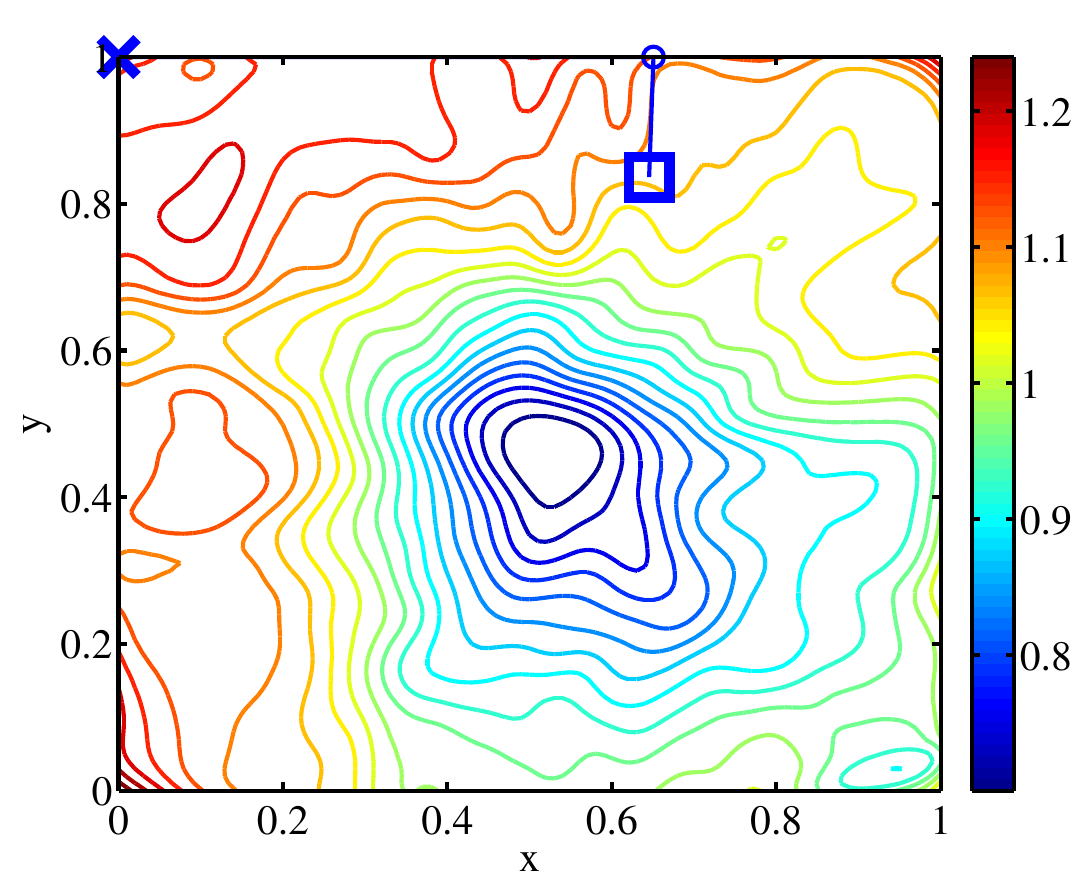}}} 
  \mbox{\subfigure[$N=101$, $M=101$]
    {\includegraphics[width=0.48\textwidth]{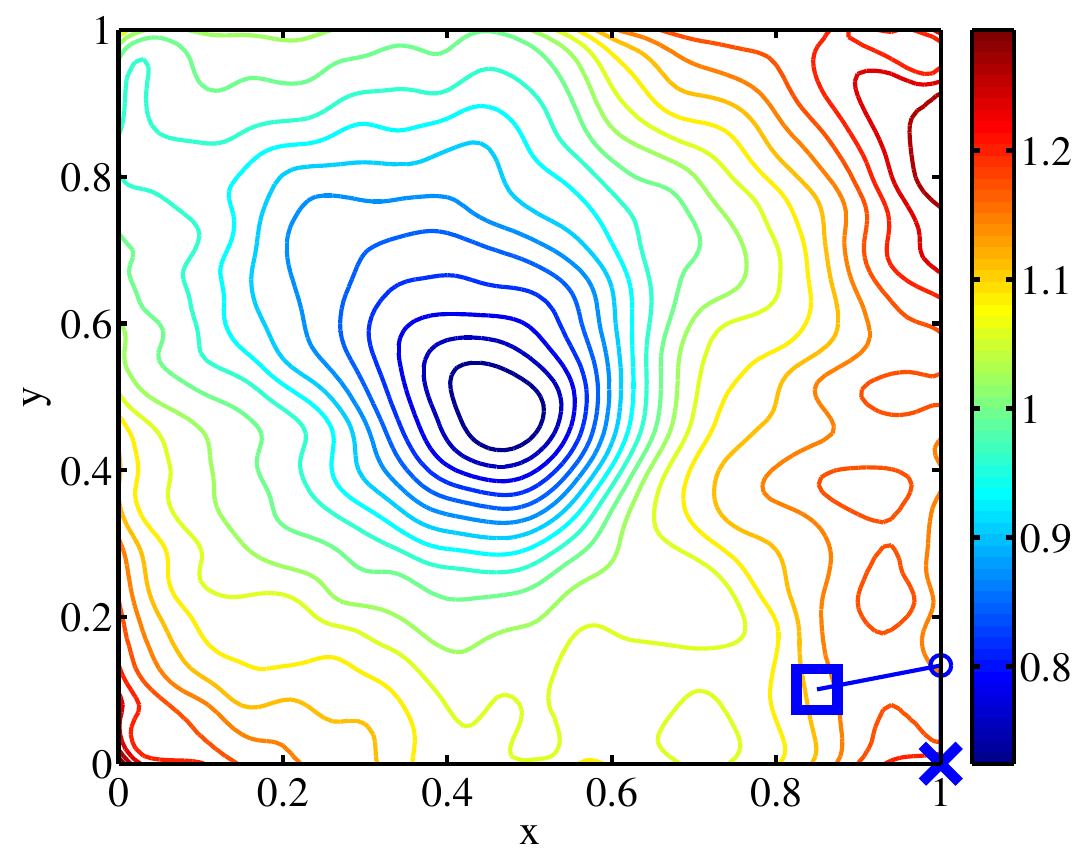}}} 
  \mbox{\subfigure[$N=101$, $M=1001$]
    {\includegraphics[width=0.48\textwidth]{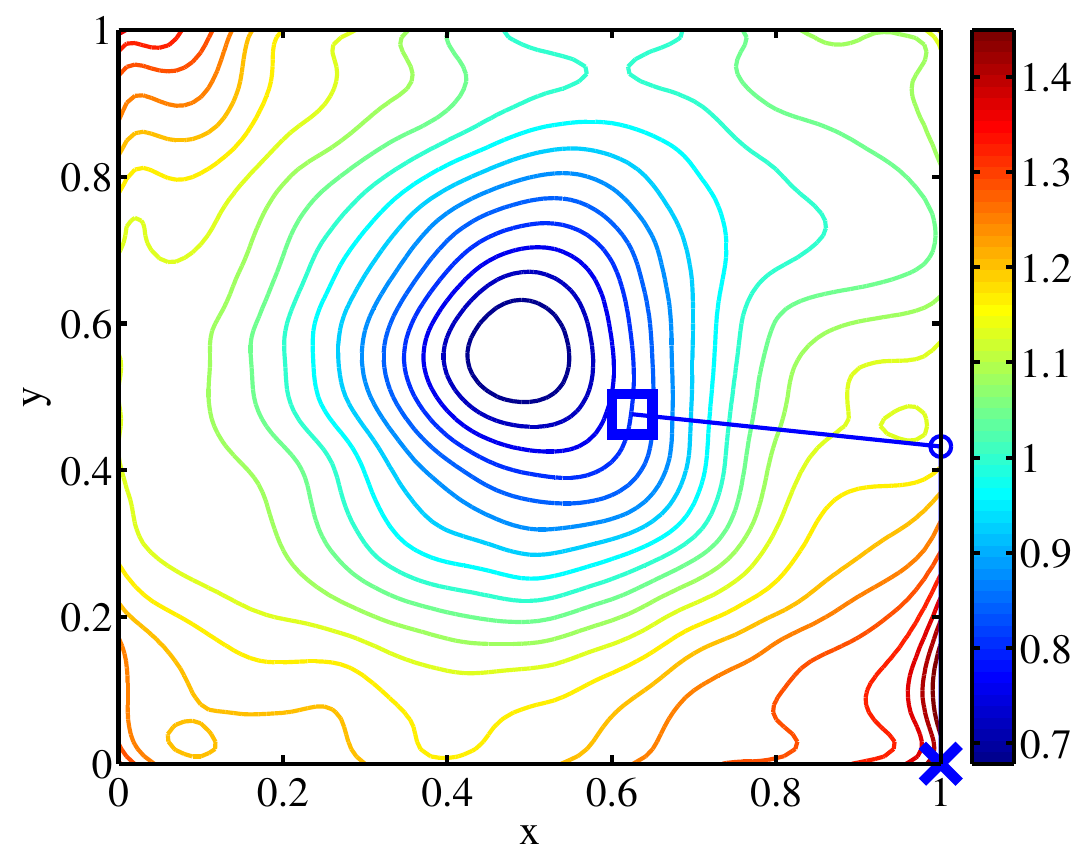}}} 
  \caption{Realizations of the objective function surface using SAA,
    and corresponding steps of BFGS, with $N=101$. The large $\square$
    is the starting position and the large $\times$ is the final
    position.}
  \label{f:SAAInstance13}
\end{figure}

For SAA-BFGS, each choice of the sample set $w_x^t$ yields a different
deterministic objective; example realizations of this objective
surface are shown in
Figures~\ref{f:SAAInstance11}--\ref{f:SAAInstance13}. For each
realization, a local maximum is found efficiently by the BFGS
algorithm, requiring only a few (usually less than 10) iterations. For
each set of results corresponding to a particular $N$ (i.e., each of
Figures \ref{f:SAAInstance11}--\ref{f:SAAInstance13}), the random
numbers used for smaller values of $M$ are proper subsets of those
used for larger $M$. We thus expect some similarity and a sense of
convergence among the subplots in each figure. Note also that when $N$
is low, realizations of the objective can be extremely different from
Figure~\ref{f:EUMC3D} (for example, the plots in
Figure~\ref{f:SAAInstance11} have local maxima near the center of the
domain), although improvement is observed as $N$ is increased. In
general, each deterministic problem in SAA can have very different
features than the underlying objective function. None of the
realizations encountered here has maxima at the corners, or is even
symmetric. Nonetheless, when sampling over many SAA subproblems, even
a low $N$ can provide reasonably good results. This will be shown in
Tables~\ref{t:FinalXHist} and~\ref{t:FinalEUHigh}, and discussed in
detail below.

%%%%%%%%
To compare the performance of RM and SAA-BFGS, 1000 independent runs
are conducted for each algorithm, over a matrix of $N$ and $M$
values. The starting locations of these runs are sampled from a
uniform distribution over the design space. We make reasonable choices
for the numerical parameters in each algorithm (e.g., gain schedule
scaling, termination criteria) leading to similar run times.
Histograms of the final design parameters (sensor positions) resulting
from each set of 1000 optimization runs are shown in
Table~\ref{t:FinalXHist}. The top figures in each major row represent
RM results, while the bottom figures in each major row correspond to
SAA-BFGS results. Columns correspond to different values of $M$. It is
immediately apparent that more designs cluster at the corners of the
domain as $N$ and $M$ are increased. For the case with the largest
number of samples ($N=101$ and $M=1001$), each corner has around 250
designs, suggesting that higher sample sizes cannot further improve
the optimization results.
%
% (turns out it cannot improve run time either because the minimum
% number of iterations is reached, as shown in Table~\ref{t:FinalIter}).
%
An ``overlap'' in quality across the different $N$ cases is also
observed: for example, results of the $N=101$, $M=2$ case are worse
than those of the $N=11$, $M=1001$ case. A balance is thus needed in
choosing samples sizes $N$ and $M$, and it is not ideal to heavily
favor sampling either the inner or outer Monte Carlo loop in
$\hU_{N,M}$.  Overall, comparing the RM and SAA-BFGS plots at
intermediate values of $M$ and $N$, we see that RM has a slight
advantage over SAA-BFGS by placing more designs at the corners.

The distribution of final designs alone does not reflect the
robustness of the optimization results. For example, if $U$ is very
flat near the optimum, then suboptimal designs need not be very close
to the true optimum in the design space to be considered good designs
in practice. To evaluate robustness, a ``high-quality'' objective
estimate $\hU_{1001,1001}$ is computed for each of the 1000 final
designs considered above. The resulting histograms are shown in
Table~\ref{t:FinalEUHigh}, where again the top subrows are for RM and
the bottom subrows are for SAA-BFGS, with the results covering a full
range of $N$ and $M$ values. In keeping with our previous
observations, performance is improved as $N$ and $M$ are
increased---in that the mean (over the optimization runs) expected
information gain increases, while the variance in the expected
information gain decreases. Note, however, that even if all 1000
optimization runs produced identical final designs, this variance will
not reach zero, as there exists a ``floor'' corresponding to the
variance of the estimator $\hU_{1001,1001}$. This minimum variance can
be observed in the histograms of the RM results with $N=101$ and
$M=101$ or $1001$.

One interesting feature of the histograms in Table~\ref{t:FinalEUHigh}
is their bimodality. The higher mode reflects designs near the four
corners, while the lower mode encompasses all other suboptimal
designs. As $N$ or $M$ increase, we observe a transfer of probability
mass from the lower mode to the upper mode. However, the sample sizes
are not large enough for the lower mode to completely disappear for
most cases; it is only absent in the two RM cases with the largest
sample sizes. Overall, the histograms are similar in shape for both
algorithms, but RM appears to produce less variability in the expected
information gain, particularly at high $N$ values.

%%%%%%%%% OPTIMALITY GAP discussions

Table~\ref{t:optGap} shows histograms of optimality gap estimates from
the 1000 SAA-BFGS runs. Since we are dealing with a
\textit{maximization} problem (for the expected information gain), the
estimator from \S\ref{s:SAA} is reversed in sign, such that the upper
bound is now $\barh_{N}$ and the lower bound is
$\hh_{N^\prime}(\hx_s^t, w_{s^\prime}^t)$. The lower bound must be
evaluated with the same inner-loop Monte Carlo sample size $M$ used in
the optimization run in order to represent an identically-biased
underlying objective; hence, the lower bound values will \textit{not}
be the same as the ``high-quality'' objective estimates
$\hU_{1001,1001}$ discussed above. From the table, we observe that
as $N$ increases, values of the optimality gap estimate decrease. This
is a result of the lower bound rising with $N$ (since the optimization
is better able to find designs in regions of large $\bar{U}_M$, e.g.,
corners of the domains in Table~\ref{t:FinalXHist}), and the upper
bound simultaneously falling (since its positive bias monotonically
decreases with $N$~\cite{norkin:1998:aba}). Consequently, both bounds
become tighter and the gap estimates tend toward zero. 
As $M$ increases, the variance of the gap estimates increases. Since
the upper bound ($\barh_{N}$) is fixed for a given set of SAA runs,
the spread is only affected by the variability of the lower
bound. Indeed, from Figure~\ref{f:EUMC3D}, it is apparent that the
objective becomes less flat as $M$ increases, with the highest
gradients (considering the good design regions only) occurring at the
corners. This translates to a higher sensitivity, as a small
``imperfection'' in the design would lead to larger changes in
objective estimate; one then would expect the variation of
$\hh_{N^\prime}(\hx_s^t, w_{s^\prime}^t)$ to become higher as well,
leading to greater variance in the gap estimates.
Finally, as $M$ increases, the histogram values tend to increase, but
they increase more slowly for larger values of $N$. Some intuition for
this result may be obtained by considering the \textit{relative} rates
of change of the upper and lower bounds with respect to $M$, given
different values of $N$. Again referring to Figure~\ref{f:EUMC3D}, the
objective values generally increase with $M$, indicating an increase
of the lower bound. This increase should be more pronounced for larger
$N$, since the optimization converges to designs closer to the
corners, where, as mentioned earlier, the objective has larger
gradient.  The upper bound increases with $M$ as well, as indicated by
the contour levels in
Figures~\ref{f:SAAInstance11}--\ref{f:SAAInstance13}. But this rate of
increase is observed to be slowest at the highest $N$ (i.e., in
Figure~\ref{f:SAAInstance13}). Combining these two effects, it is
reasonable that as $N$ increases, the gap estimate will increase with
$M$ at a slower rate.

Can the optimality gap be used to choose values of $M$ and $N$? For a
fixed $M$, we certainly have convergence as $N$ increases, and the
gap estimate can be a good indicator of solution quality. However,
because different values of $M$ correspond to different objective
surfaces (due to the bias of $\hU_{N,M}$), the optimality gap is
unsuitable for comparisons across different values of $M$; indeed, in
our example, even though solution quality is improved with $M$, the
gap estimates appear looser and noisier.

%%%
Another performance metric we extract from the stochastic optimization
runs is the number of iterations required to reach a solution;
histograms of iteration number for RM and SAA, for the same matrix of
$M$ and $N$ values, are shown in Table~\ref{t:FinalIter}. At low
sample sizes, many of the SAA-BFGS runs take only a few iterations,
while almost all of the RM runs terminate at the maximum
  allowable number of iterations (50 in this case). This difference
again reflects the efficiency of BFGS for deterministic optimization
problems. As $N$ and $M$ are increased, the histograms show a
``transfer of mass'' from higher iteration numbers to lower iteration
numbers, coinciding somewhat with the bimodal behavior described
previously. The reduction in iteration number with increased sample
size implies that an $n-$fold increase in sample size leads to an
increase in computational time that is often \textit{much less} than a
factor of $n$.  Accounting for this sublinear relationship when
allocating computational resources, especially if samples can be drawn
in parallel, can lead to substantial savings.  Although SAA-BFGS
generally requires fewer iterations, each iteration takes longer than
a step of RM. RM thus offers a higher ``resolution'' in run times,
potentially giving more freedom to the user in stopping the algorithm.
RM thus becomes more attractive as the evaluation of the objective
function becomes more expensive.

\begin{figure}[htb] \centering
\mbox{\subfigure[RM]
{\includegraphics[width=0.49\textwidth]{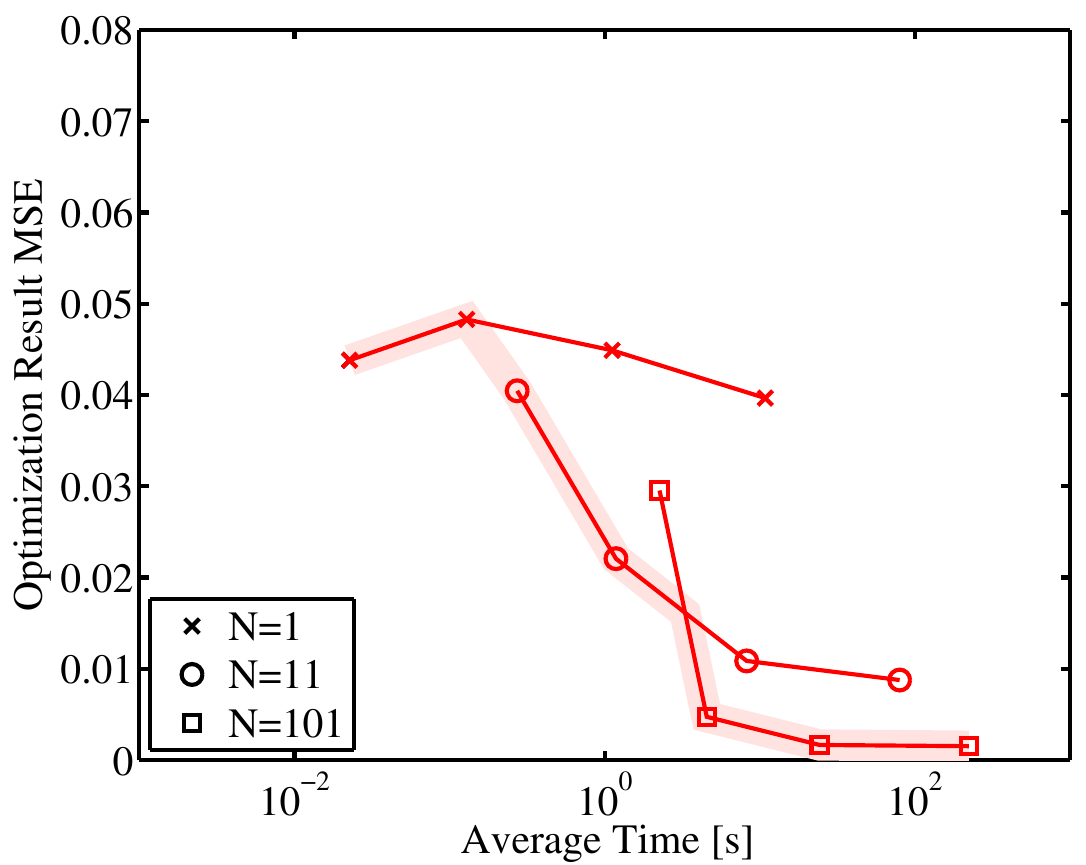}}}
\mbox{\subfigure[SAA-BFGS]
{\includegraphics[width=0.49\textwidth]{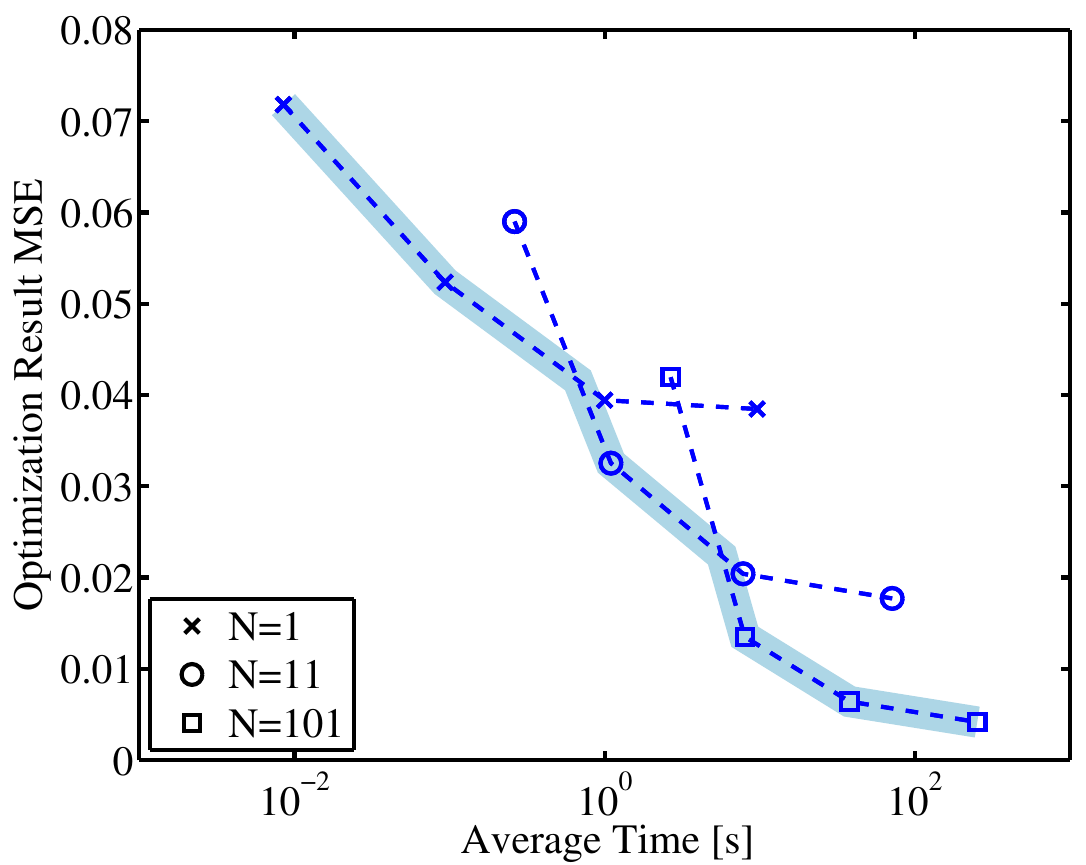}}}
% \mbox{\subfigure[SARM $a = \,$SPSA]
% {\includegraphics[width=0.49\textwidth]{figures/1Exp/p12n1000000Design1ExpSARMSPSAGainSeqMSEVsTime-eps-converted-to.pdf}}}
% \mbox{\subfigure[SPSA]
% {\includegraphics[width=0.49\textwidth]{figures/1Exp/p12n1000000Design1ExpSPSAMSEVsTime-eps-converted-to.pdf}}}
\mbox{\subfigure[RM and SAA-BFGS ``optimal fronts'']
{\includegraphics[width=0.49\textwidth]{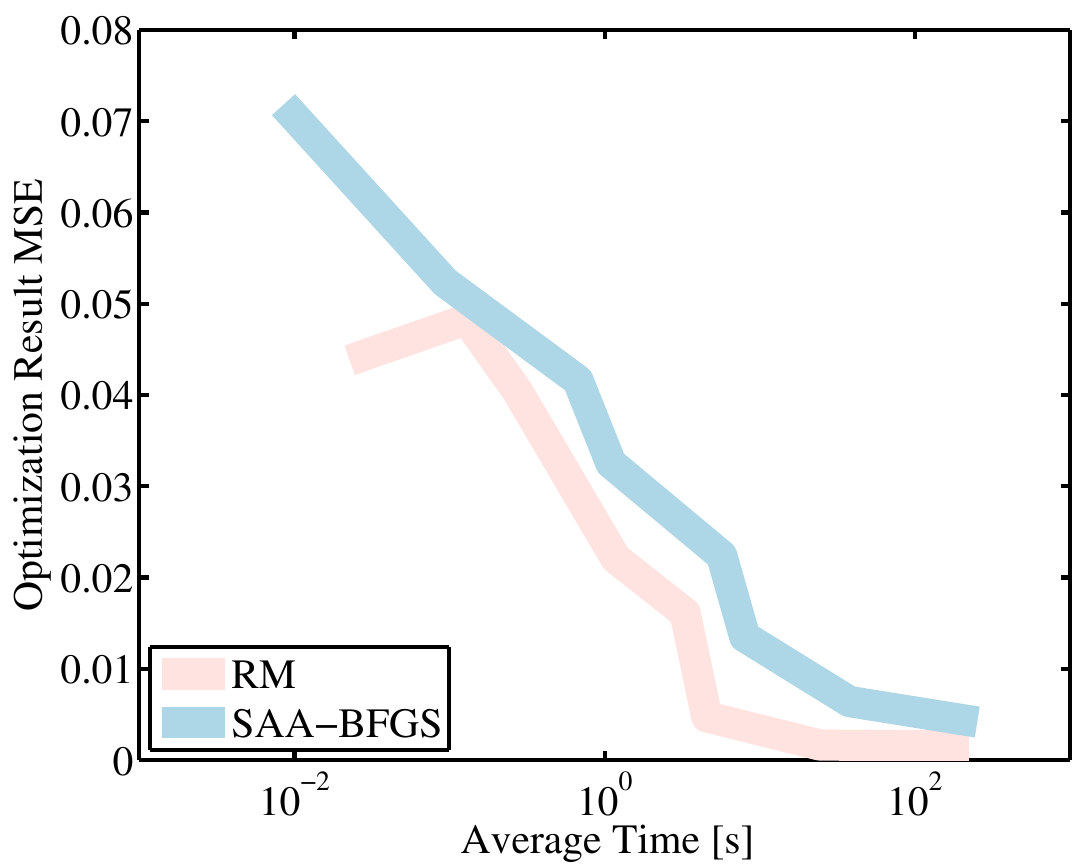}}}
\caption{Mean square error, defined in (\ref{e:mse}), versus average
  run time for each optimization algorithm and various choices of
  inner-loop and outer-loop sample sizes. The highlighted curves are
  ``optimal fronts'' for RM (light red) and SAA-BFGS (light blue).}
  \label{f:MSEVsTime}
\end{figure}

As a single integrated measure of the quality of the stochastic
optimization solutions, we evaluate the following mean square error
(MSE):
\begin{eqnarray}
  \mathrm{MSE} = \frac{1}{T} \sum_{t=1}^{T}
  \(\hU_{1001,1001}(\bd^t, \btheta_{s'}^t, \bz_{s'}^t)-U^{\mathrm{ref}}\)^2,
\label{e:mse}
\end{eqnarray}
where $\bd^t$, $t=1\ldots T$, are the final designs from a given
optimization algorithm, and $U^{\mathrm{ref}}$ is the true optimal
value of the expected information gain. Since the true optimum is
unavailable in this study, $U^{\mathrm{ref}}$ is taken to be the
maximum value of the objective over all runs.  Recall that the MSE
combines the effects of bias and variance; here it reflects the
variance in objective values plus the difference (squared) between the
mean objective value and the true optimum, calculated via $T=1000$
replicated optimization runs. Figure~\ref{f:MSEVsTime} relates
solution quality to computational effort by plotting the MSE against
average computational time (per run). Each symbol represents a
particular value of $N$ ($\times$, $\bigcirc$, and $\square$ represent
$N=1$, 11, and 101, respectively), while the four different $M$ values
are reflected through the average run times. These plots confirm the
behavior we have previously encountered. Solution quality generally
improves (lower MSE) with increasing sample sizes, although a balanced
allocation of samples must be chosen. For instance, a large $N$ with
small $M$ can yield inferior solutions to a smaller $N$ with larger
$M$; while, for any given $N$, continued increases in $M$ beyond some
threshold yield minimal improvements in MSE. The best sample
allocation is described by the minimum of all the curves. We highlight
these ``optimal fronts'' in light red for RM and in light blue for
SAA-BFGS.  Monte Carlo error in the ``high-quality'' estimator
$\hU_{1001,1001}$ may also be reflected in the non-zero MSE asymptote
for the high-$N$ RM cases.

According to Figure~\ref{f:MSEVsTime}, RM outperforms SAA-BFGS by
consistently achieving smaller MSE for a given computational
effort. One should be cautious, however, in generalizing from these
numerical experiments. The advantage of RM is relatively small, and
other factors such as code optimization, choices of algorithm
parameters, and of course the experimental design problem itself can
affect or even reverse this advantage.

\section{Conclusions}
\label{s:conclusions}

This paper has explored the stochastic optimization problem arising
from a general nonlinear formulation of optimal Bayesian experimental
design. In particular, we employed an objective that reflects the
\textit{expected information gain} in model parameters due to an
experiment, and formulated two gradient-based approaches to stochastic
optimization in this context: Robbins-Monro (RM) stochastic
approximation, and sample average approximation (SAA) coupled with
BFGS. Both of these algorithms require gradient information derived
from Monte Carlo approximations of the objective: an unbiased gradient
estimator in the former case, and gradients of a finite-sample Monte
Carlo estimate in the latter case. Methods for extracting this
gradient information must contend with an estimator of expected
information gain that is not a simple Monte Carlo sum, but rather
contains nested Monte Carlo estimates. It is therefore expensive to
evaluate, and biased for finite inner-loop sample sizes. To circumvent
these challenges, we approximate the forward model embedded in the
likelihood function with a polynomial chaos expansion, and maximize
the expected information gain computed via this approximation
instead. Gradient information is readily extracted from the polynomial
chaos expansion, with the help of a simple perturbation analysis.

We analyze the performance of the two stochastic optimization
approaches using the problem of sensor placement for source inversion,
cast as optimal experimental design over a \textit{continuous} design
space. Numerical experiments, performed over a matrix of inner- and
outer-loop sample sizes, examine the impact of bias and variance in
the objective function and gradient estimates on the efficiency of the
optimization algorithms and on the quality of the resulting
solutions. These experiments suggest (unsurprisingly) that solution
quality improves as sample sizes increase, but also that optimization
runs may converge in fewer iterations for larger sample sizes. Also, a
\textit{balanced} allocation of computational resources between the
inner and outer Monte Carlo sums is important for computational
efficiency. Arbitrarily increasing the inner-loop sample size, for
instance, yields little improvement in solution quality when the
outer-loop samples are too few. Our results also suggest that RM has a
consistent performance advantage over SAA-BFGS, but this conclusion is
necessarily problem-dependent. Instead of declaring one algorithm to
be superior, our broader goal is to illustrate the differences between
the two algorithms and provide some selection guidelines based on
their properties.

The SAA approach may provide more flexibility than SA, as it can be
combined with any deterministic optimization algorithm, whereas the SA
approach essentially specifies the form of each optimization
iteration. SAA's flexibility allows one to take advantage of problem
structure: if realizations of the objective surface are known to be
``well-behaved'' and smooth, gradient-based algorithms such as BFGS
can exploit this regularity, as in the present source inversion
example. On the other hand, if the objective is not smooth, or if
gradients are not available, some gradient-free deterministic
algorithm may be more appropriate.
Estimates of optimality gap, obtained from replicate SAA solutions,
can be used to adaptively adjust the outer-loop Monte Carlo sample
size, but are unsuitable for assessing the inner-loop sample size
because of bias effects.  Future work could employ the common random number
stream approach in~\cite{mak:1999:mcb} to obtain a lower-variance
estimate of optimality gap (along with a confidence interval), or the
jackknife technique proposed in~\cite{bayraksan:2009:asq} for bias
reduction.

The RM algorithm and other stochastic approximation methods must use a
stochastic gradient estimator.  This can lead to poor performance if
only high-variance gradient estimates are available. In the current
context, increasing the outer-loop sample size reduces variance and
the RM algorithm performed relatively well. Note that the frequent
(yet cheaper) steps of RM effectively provide a finer resolution in
run time than SAA, giving the user more freedom to terminate the
algorithm without losing much progress between the termination time
and the previous optimization iteration. Therefore, RM may become more
attractive as objective evaluations become more
expensive.\footnote{Even if a polynomial chaos expansion is used as a
  surrogate for the forward model, its evaluation can become expensive
  if the stochastic dimension and polynomial order are high, though it
  remains much cheaper than the original model.}

% With regard to future work, we note that the...
The present approach used a \textit{global} polynomial chaos
surrogate, constructed over the product of the parameter space $\CH$
and the design space $\CD$. In model-based methods for
\textit{deterministic} derivative-free optimization, one might prefer
to construct {local} surrogates valid over increasingly smaller
intervals of $\CD$, particularly as one approaches the
optimum. Pursuing similar ideas in the stochastic context could
possibly offer additional accuracy, but sampling errors in the
stochastic optimization solution will always limit potential gains.

Finally, as we pointed out in Section~\ref{s:oedform}, this paper has
focused on batch or open-loop experimental design, where the
parameters for all experiments are chosen before data are actually
collected. An important target for future work is rigorous sequential
or closed-loop design, where the data from one set of experiments are
used to guide the choice of the next set. Here we expect stochastic
optimization algorithms, for expected information gain and other
objectives, to continue playing a crucial role.

%% The Acknowledgements part is started with the command \acknowledgements;
%% acknowledgements are then done as normal sections before appendix
%% \acknowledgements

%\acknowledgements
\section{Acknowledgements}

This work was partially supported by the Computational Mathematics
Program of the Air Force Office of Scientific Research and by the
National Science Foundation under award number ECCS-1128147.

\clearpage
\section{Figures and Tables}

% \begin{figure}[htb]
%   \centering 
%   \mbox{\subfigure[$\bx_{\mathrm{sensor}}=(0.0,0.0)$]
%     {\includegraphics[width=0.45\textwidth]{figures/1Exp/postMesh1ExpSourcex0_62y0_60Sensorx0y0p12n1000000}}} 
%   \mbox{\subfigure[$\bx_{\mathrm{sensor}}=(0.0,1.0)$]
%     {\includegraphics[width=0.45\textwidth]{figures/1Exp/postMesh1ExpSourcex0_62y0_60Sensorx0y1p12n1000000}}} 
%   \mbox{\subfigure[$\bx_{\mathrm{sensor}}=(1.0,0.0)$]
%     {\includegraphics[width=0.45\textwidth]{figures/1Exp/postMesh1ExpSourcex0_62y0_60Sensorx1y0p12n1000000}}} 
%   \mbox{\subfigure[$\bx_{\mathrm{sensor}}=(1.0,1.0)$]
%     {\includegraphics[width=0.45\textwidth]{figures/1Exp/postMesh1ExpSourcex0_62y0_60Sensorx1y1p12n1000000}}} 
%   \mbox{\subfigure[$\bx_{\mathrm{sensor}}=(0.5,0.5)$]
%     {\includegraphics[width=0.45\textwidth]{figures/1Exp/postMesh1ExpSourcex0_62y0_60Sensorx0_5y0_5p12n1000000}}} 
%   \caption{Parameter posterior density functions for different sensor
%     placements and when the source is actually at
%     $\bx_{\mathrm{src}}=(0.62,0.60)$.}
%   \label{f:posteriorsSourcex0_62y0_60}
% \end{figure}

\begin{table}[htb]
  \centering
  \begin{tabular}{c|cccc}
    \backslashbox{$N$}{$M$} & 2 & 11 & 101 & 1001 \\ \hline
    \multirow{2}{*}{$1$} 
    & \includegraphics[width=0.2\textwidth]{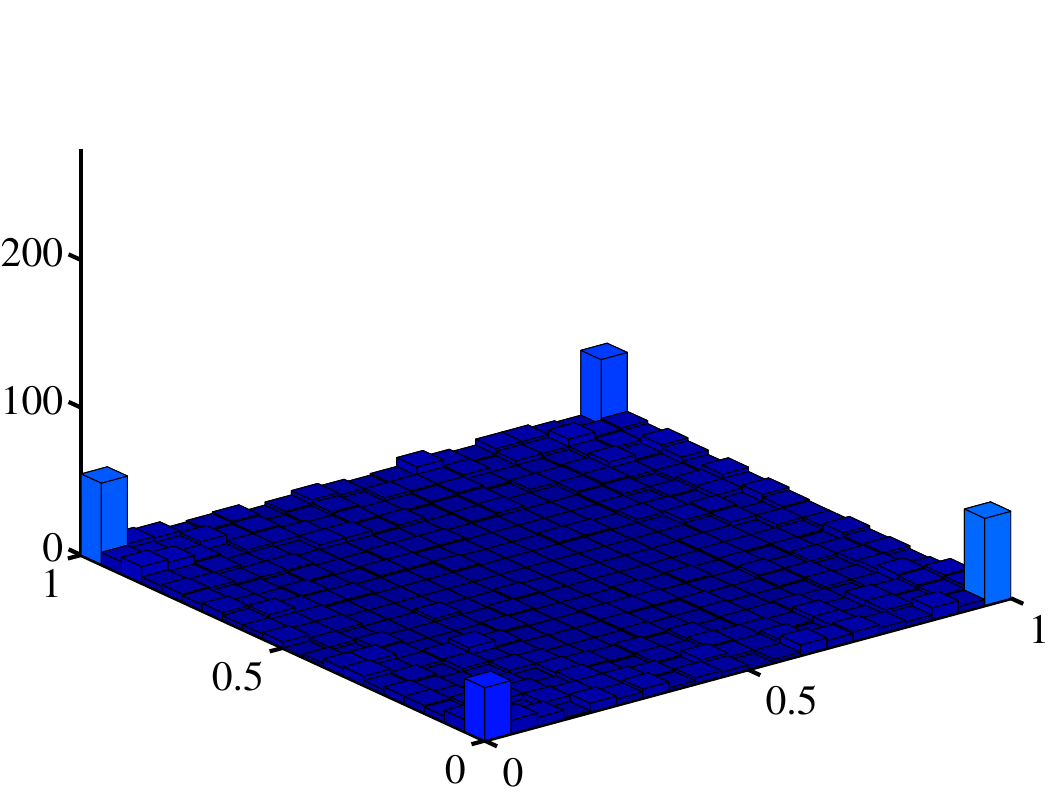}
    & 
    \includegraphics[width=0.2\textwidth]{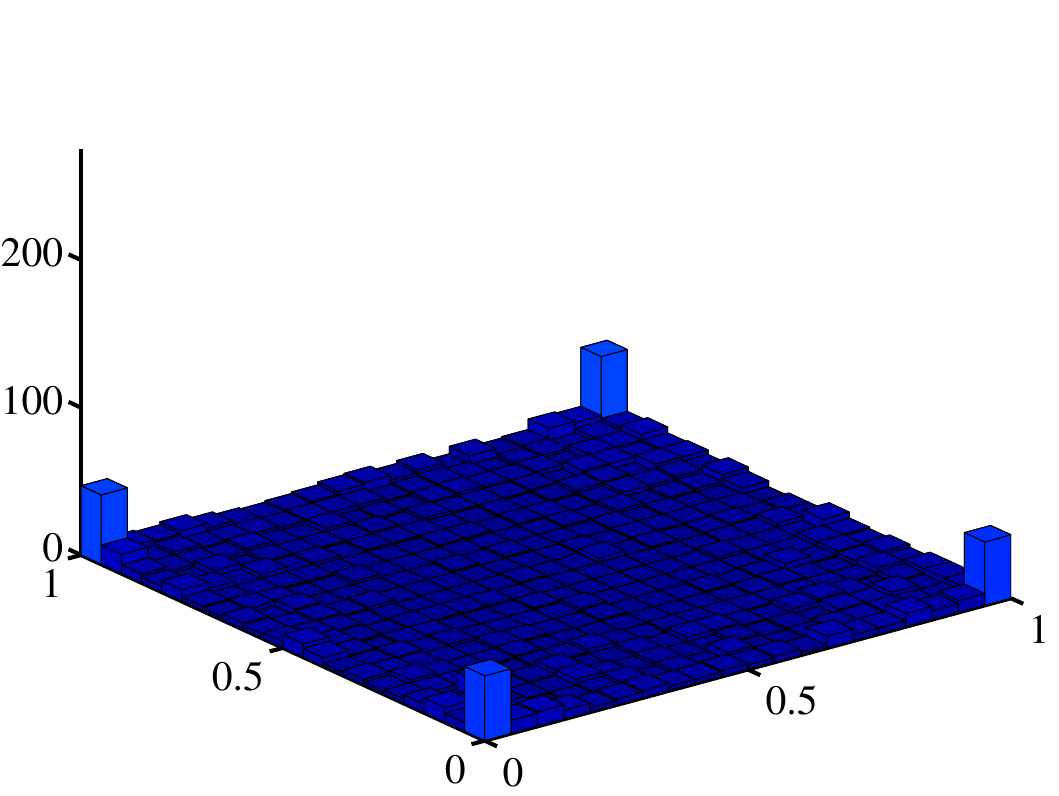}
    & 
    \includegraphics[width=0.2\textwidth]{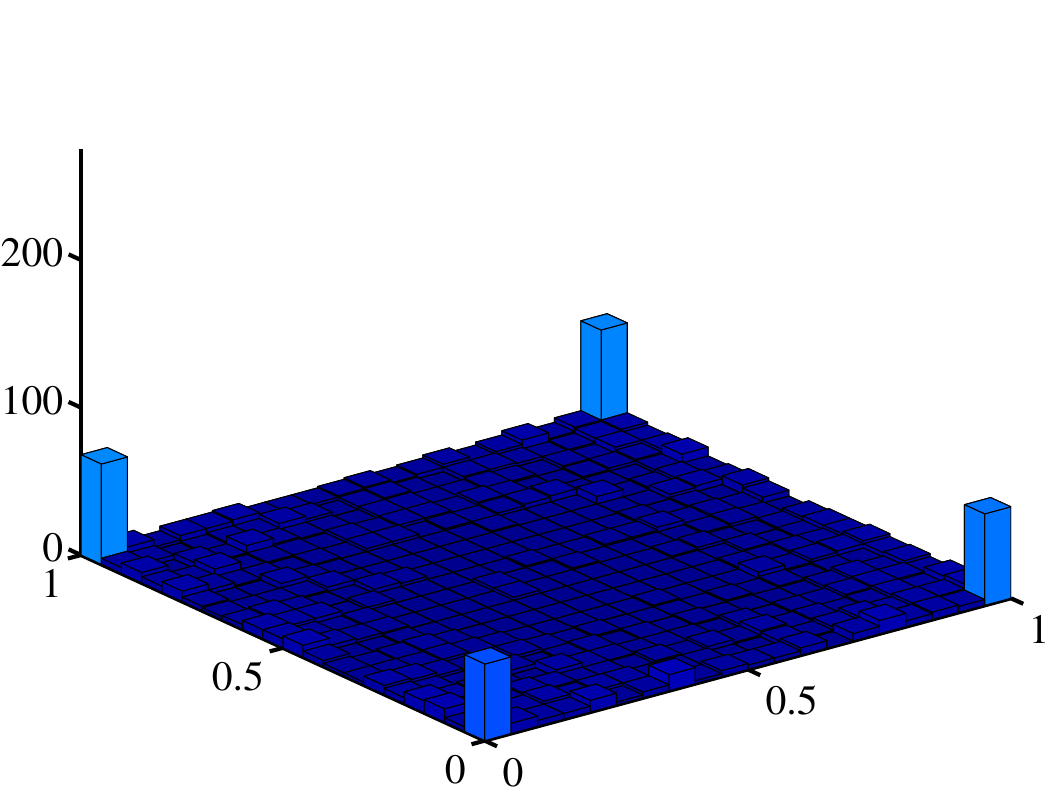}
    & 
    \includegraphics[width=0.2\textwidth]{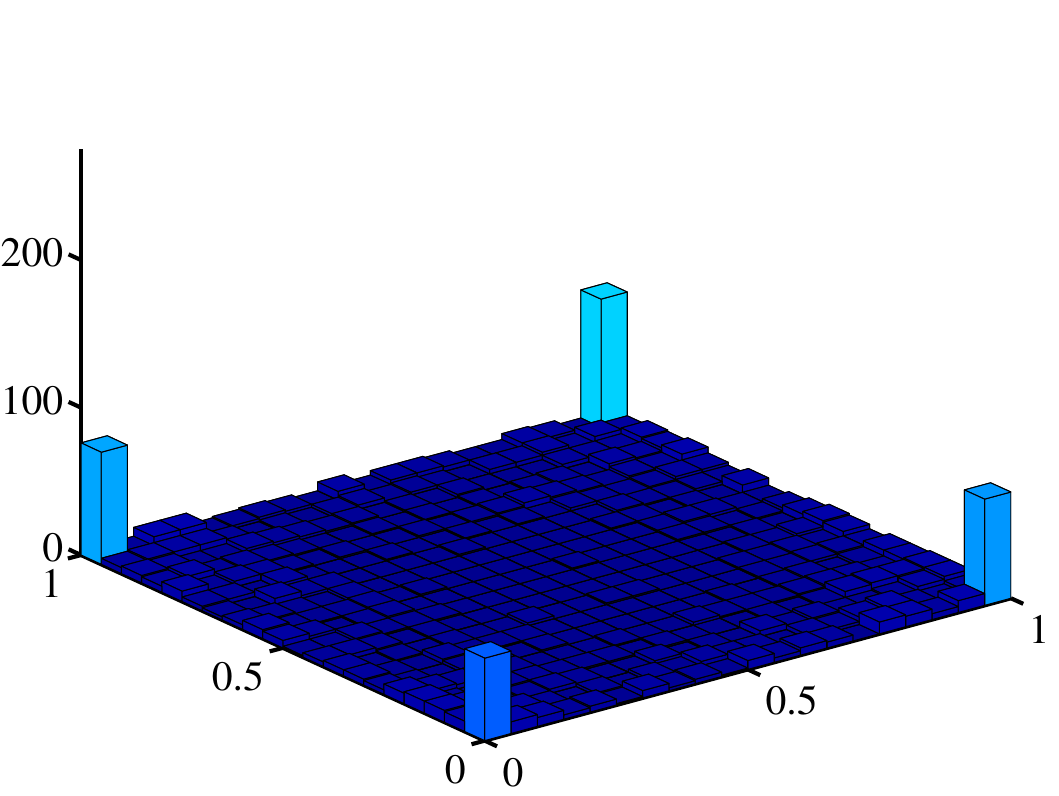}
    \\
    & \includegraphics[width=0.2\textwidth]{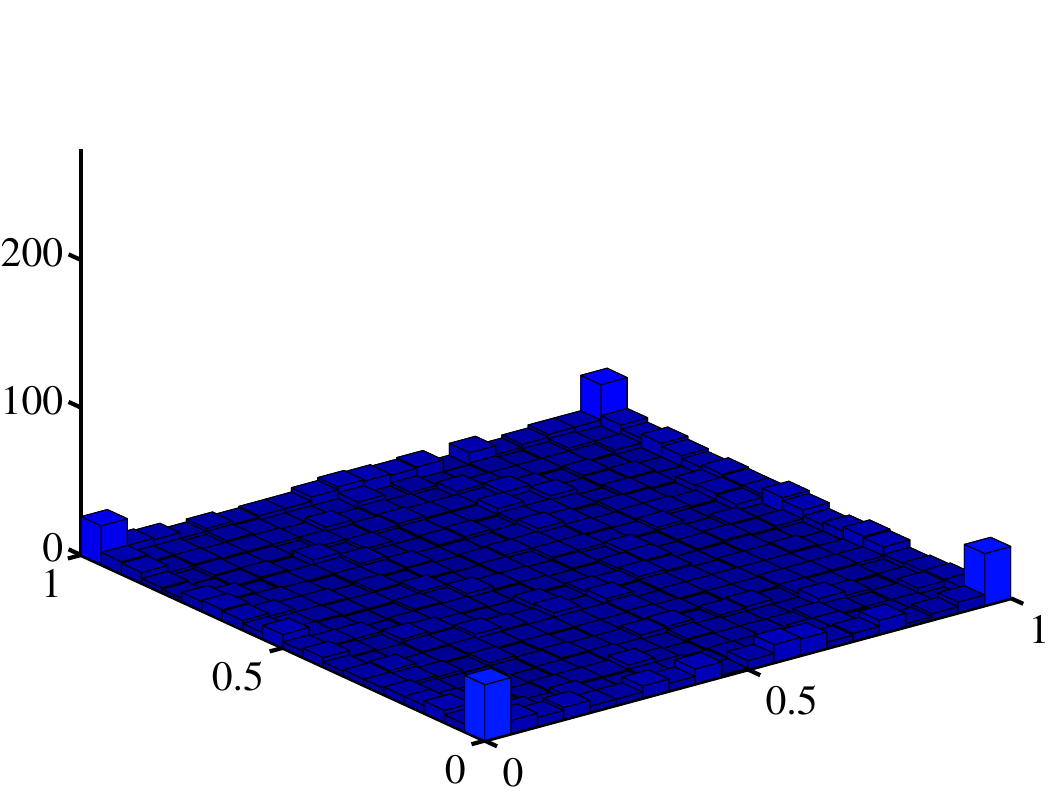}
    & 
    \includegraphics[width=0.2\textwidth]{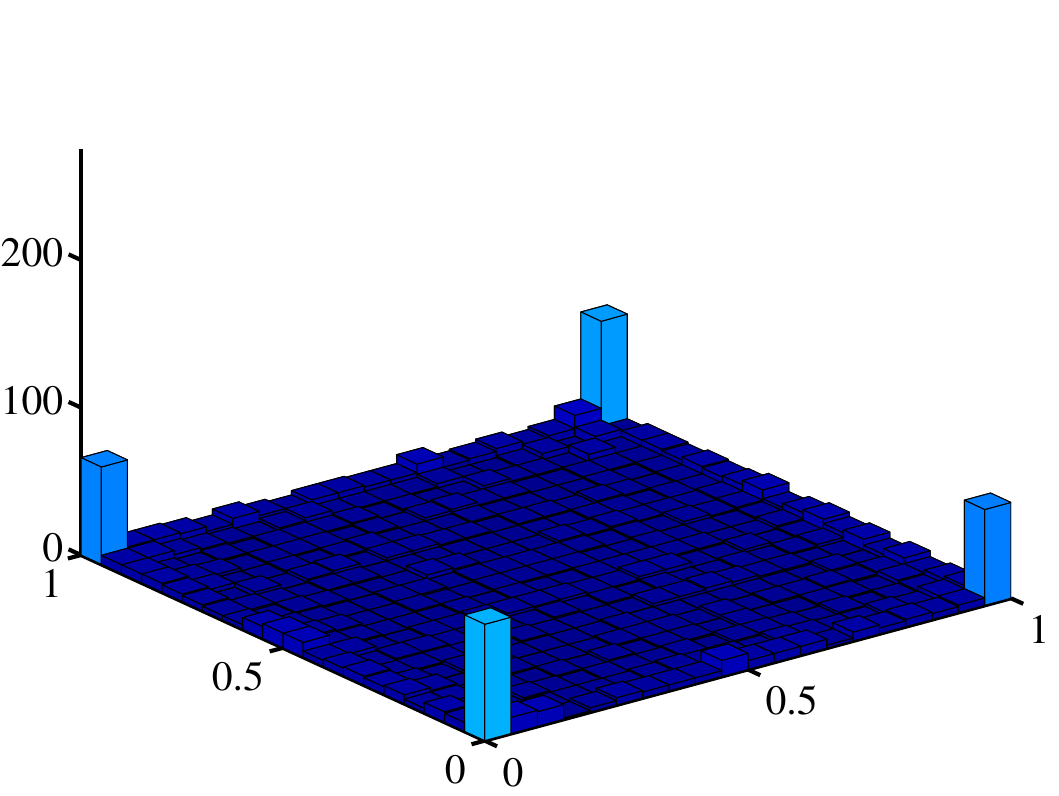}
    & 
    \includegraphics[width=0.2\textwidth]{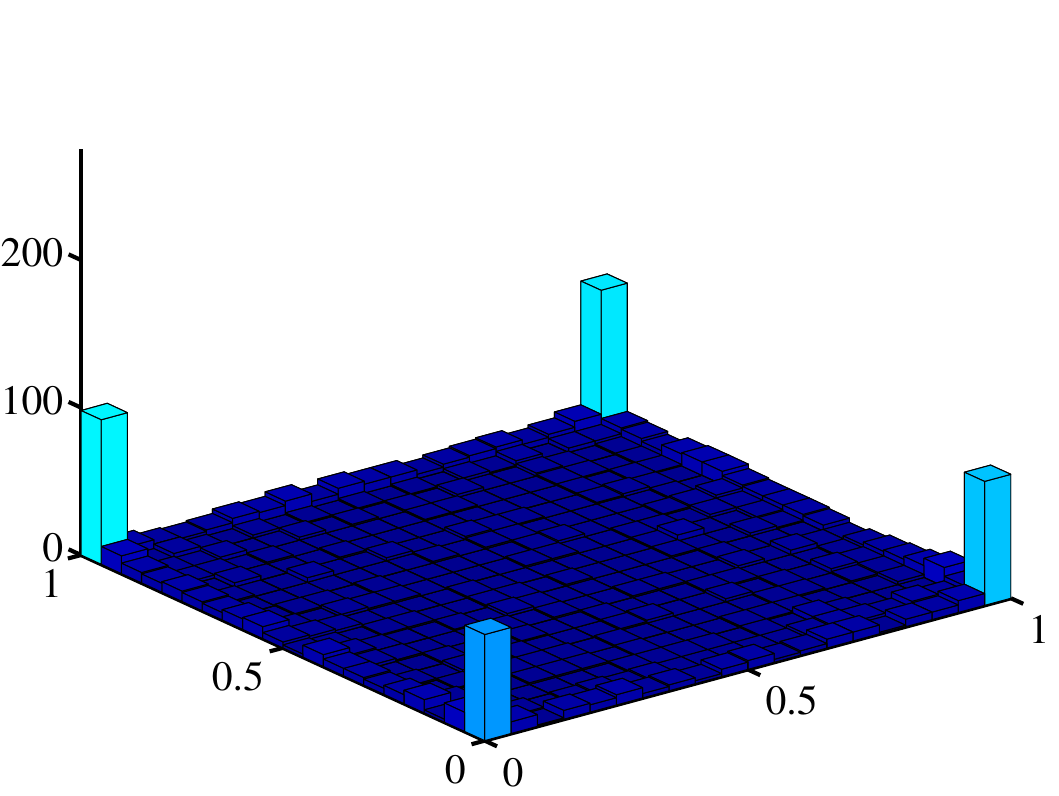}
    & 
    \includegraphics[width=0.2\textwidth]{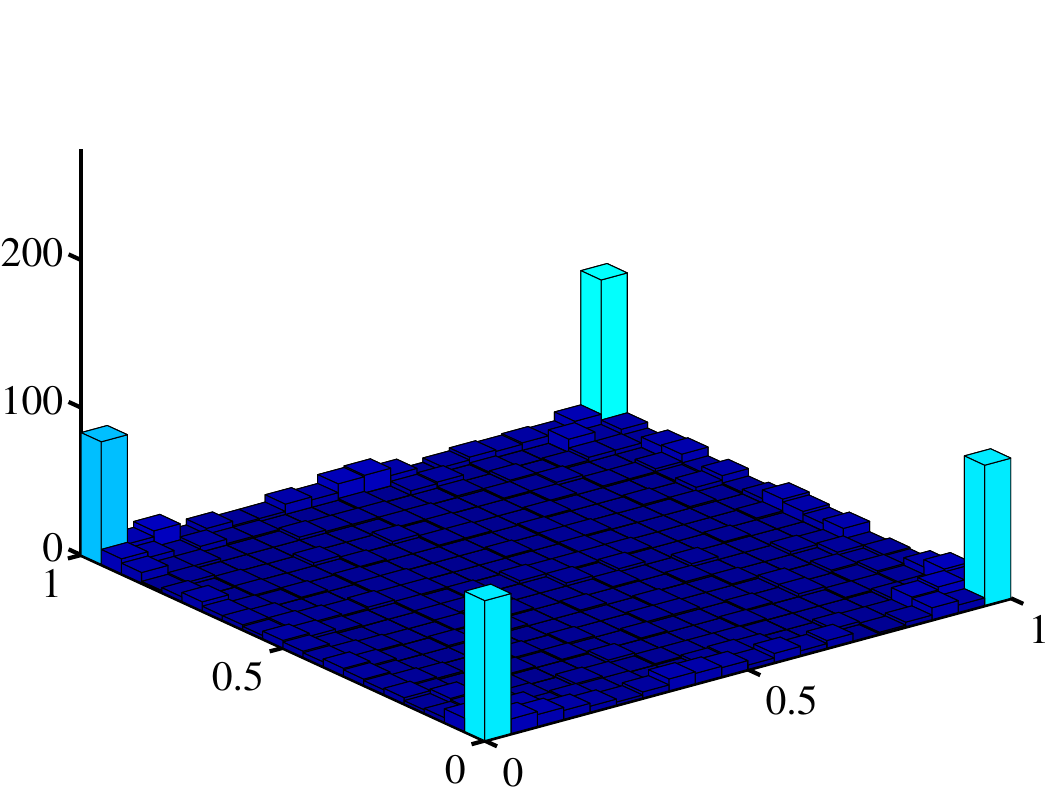}
    \\ \hline
    \multirow{2}{*}{$11$} 
    & \includegraphics[width=0.2\textwidth]{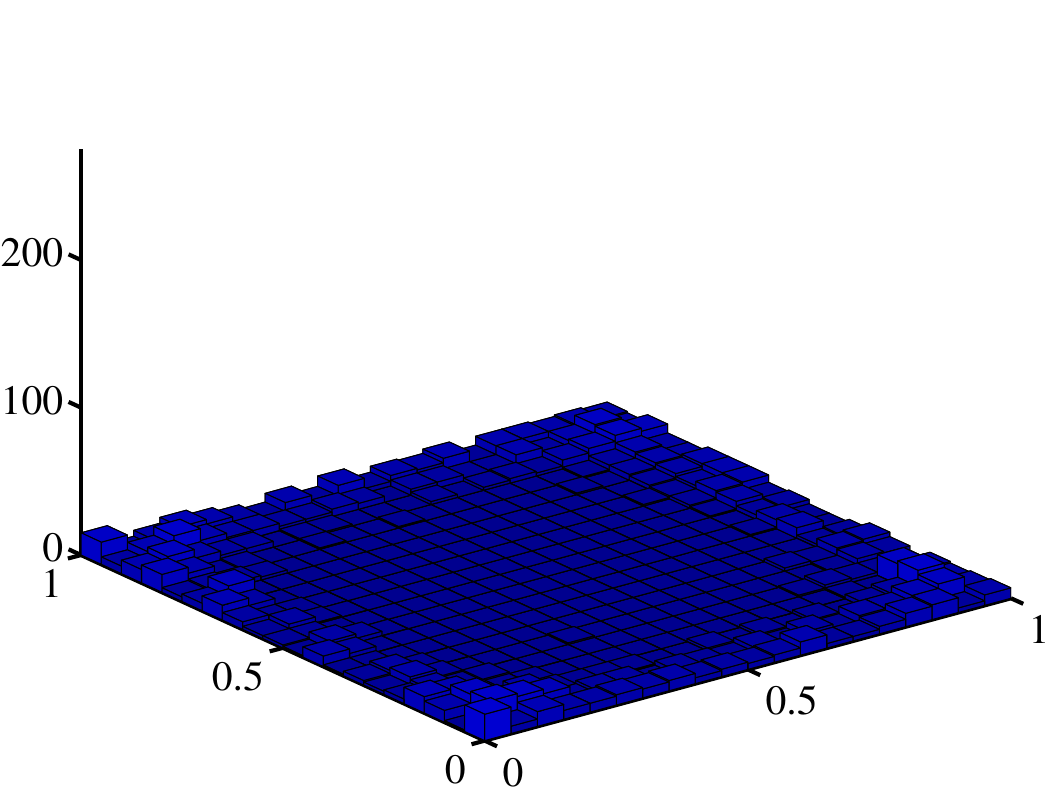}
    & 
    \includegraphics[width=0.2\textwidth]{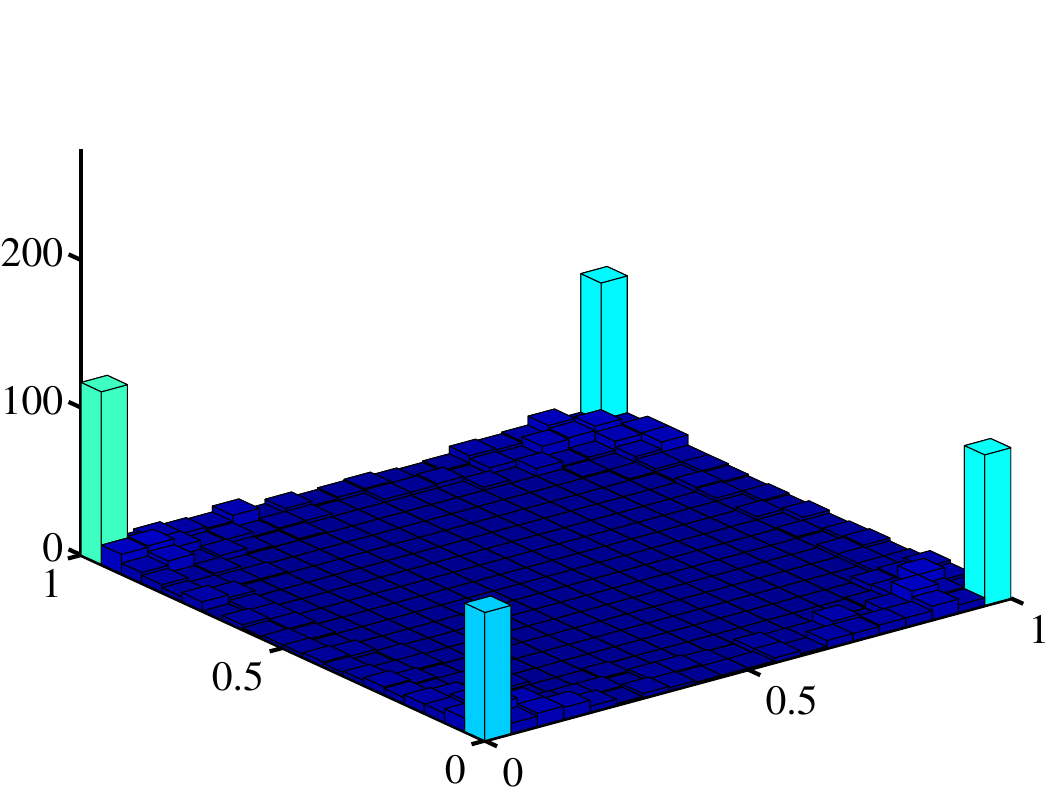}
    & 
    \includegraphics[width=0.2\textwidth]{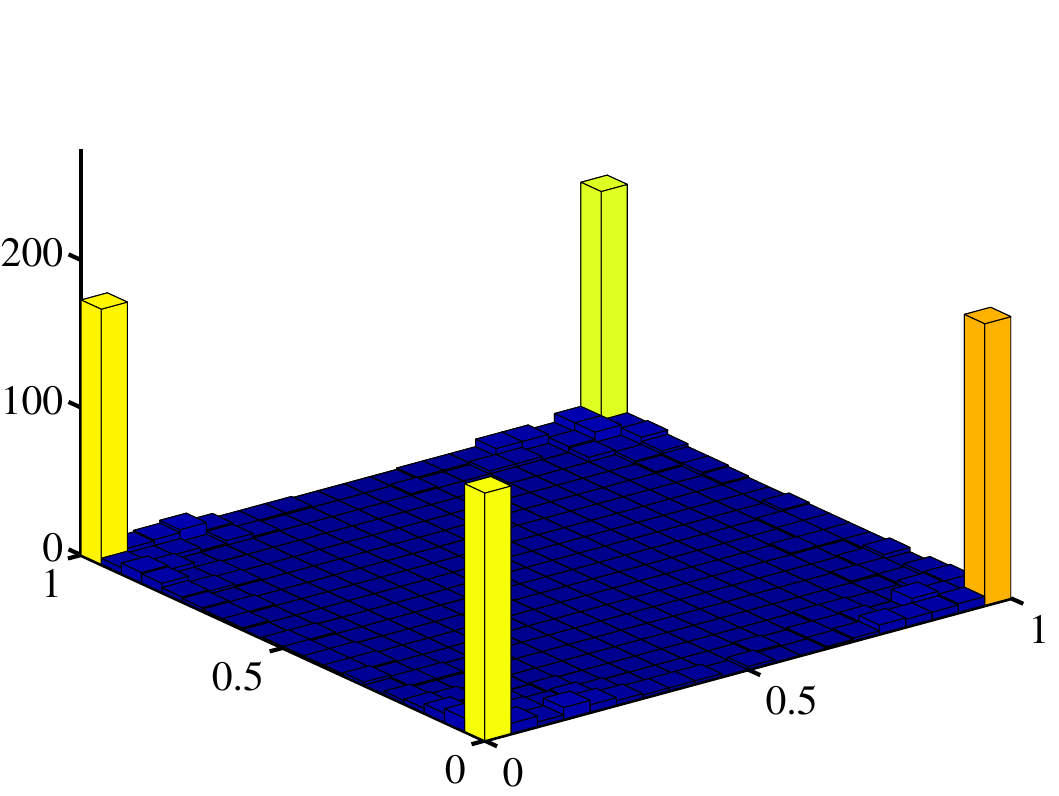}
    & 
    \includegraphics[width=0.2\textwidth]{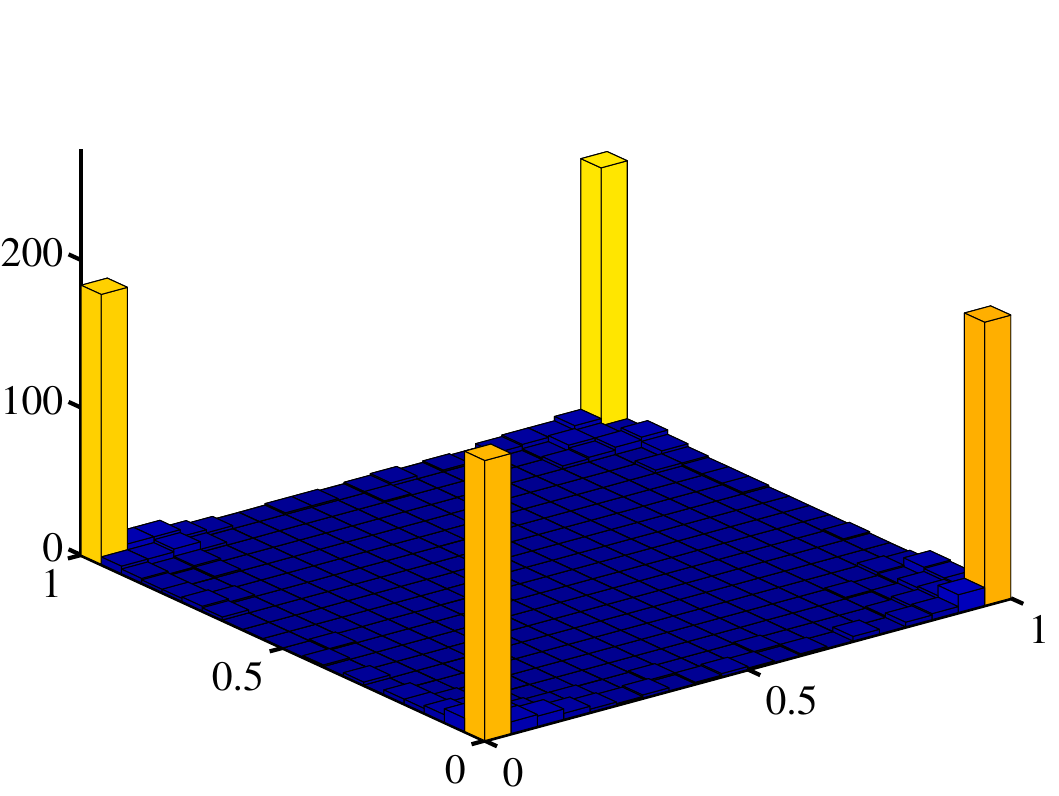}
    \\
    & \includegraphics[width=0.2\textwidth]{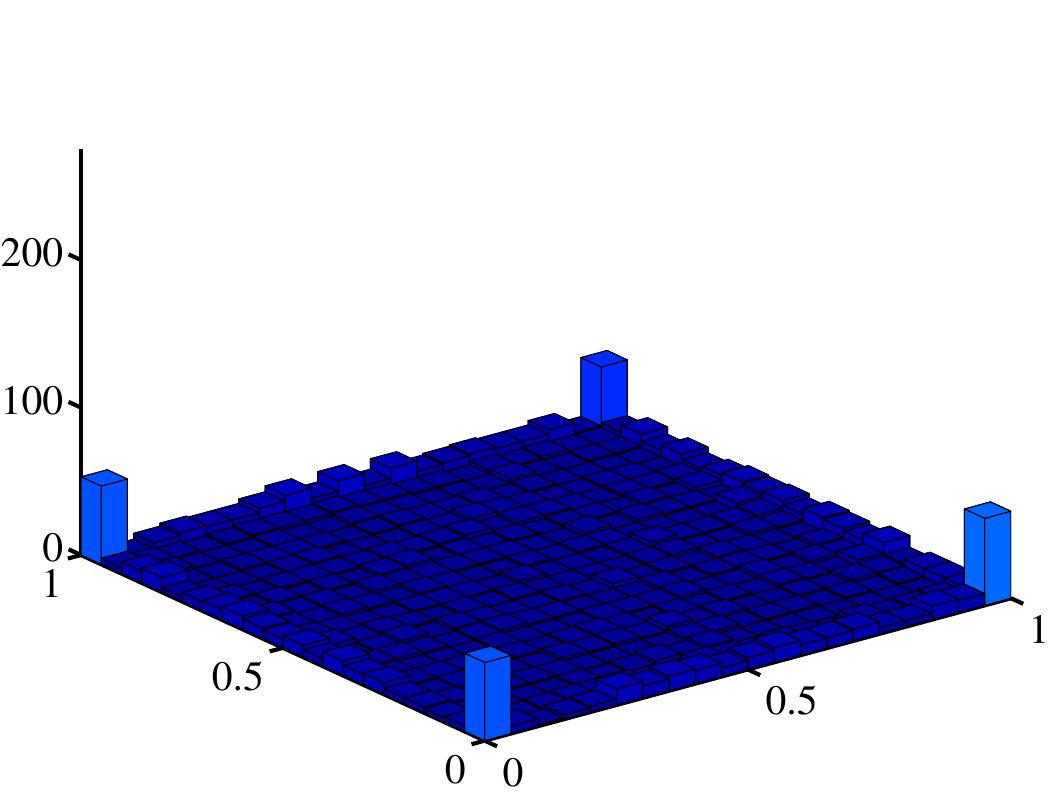}
    & 
    \includegraphics[width=0.2\textwidth]{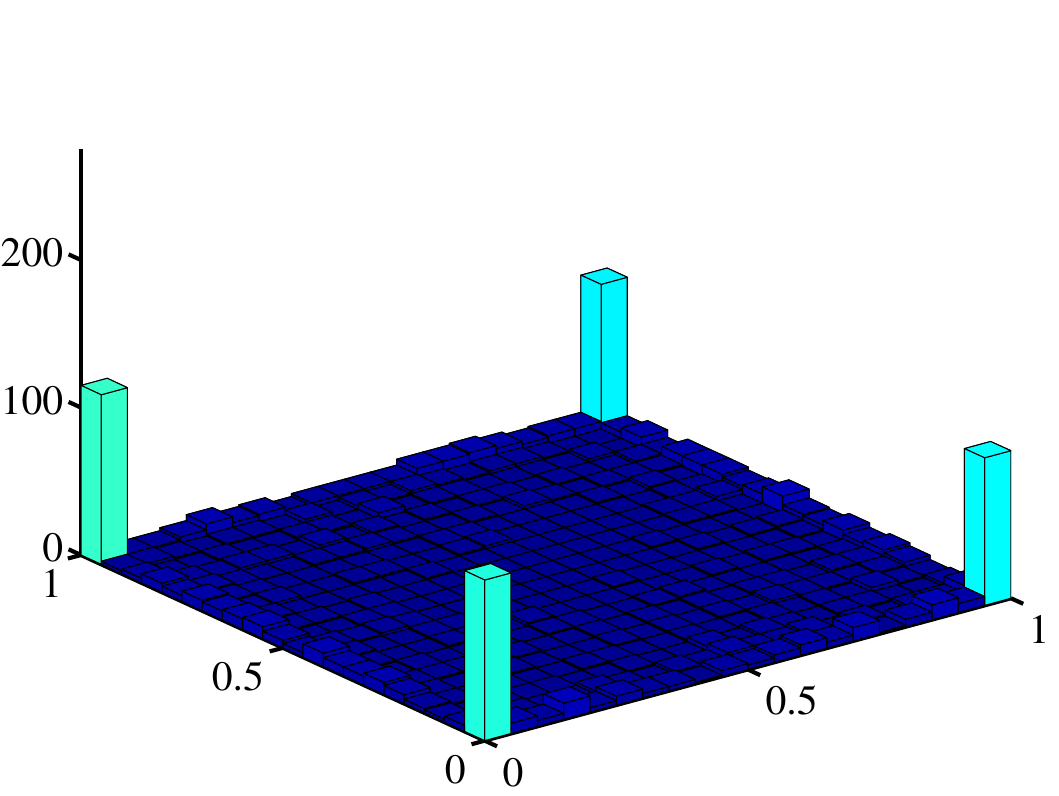}
    & 
    \includegraphics[width=0.2\textwidth]{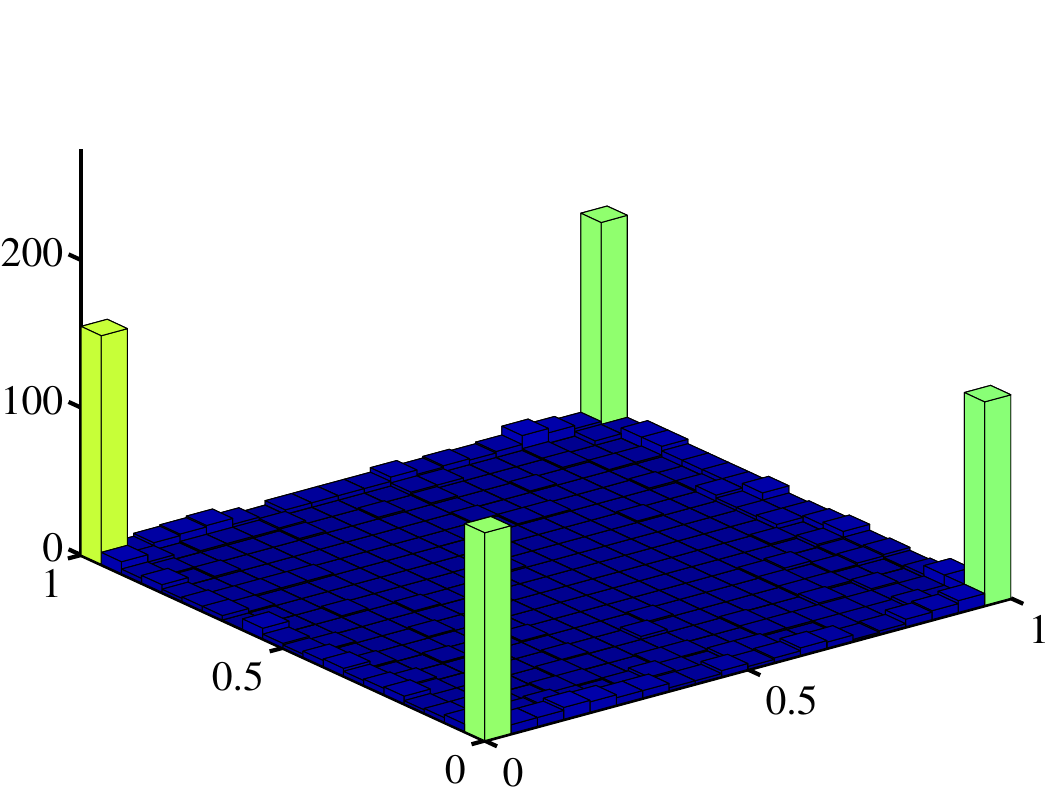}
    & 
    \includegraphics[width=0.2\textwidth]{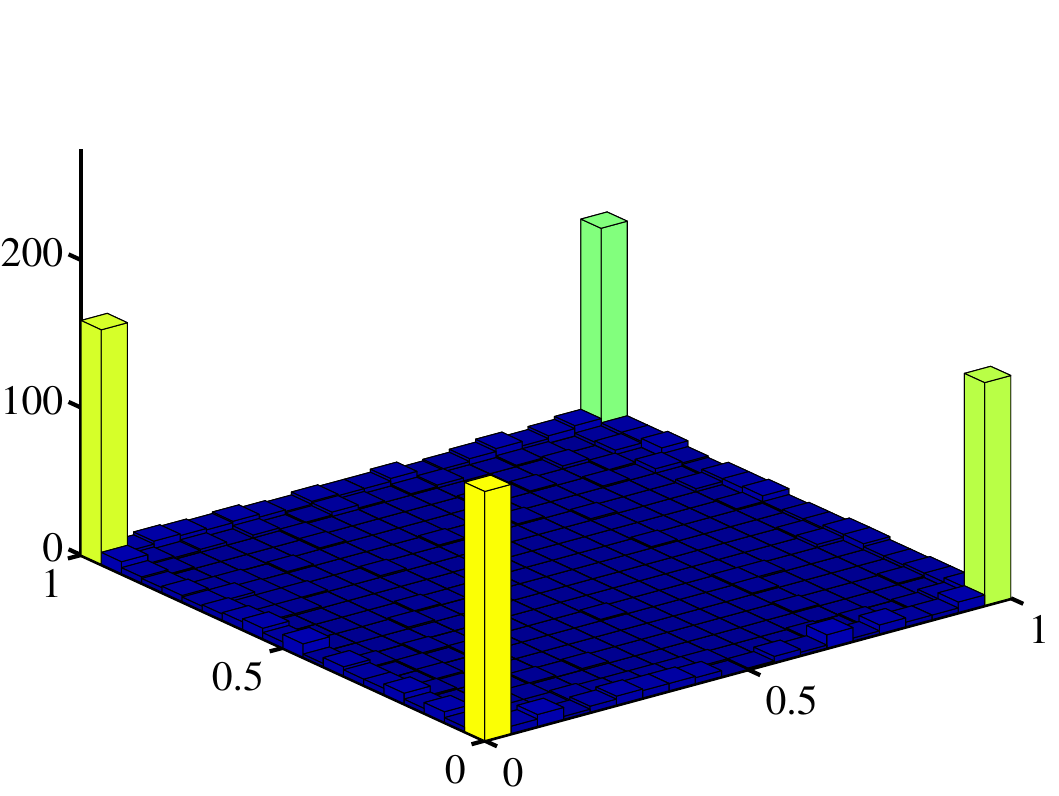}
    \\ \hline
    \multirow{2}{*}{$101$} 
    & \includegraphics[width=0.2\textwidth]{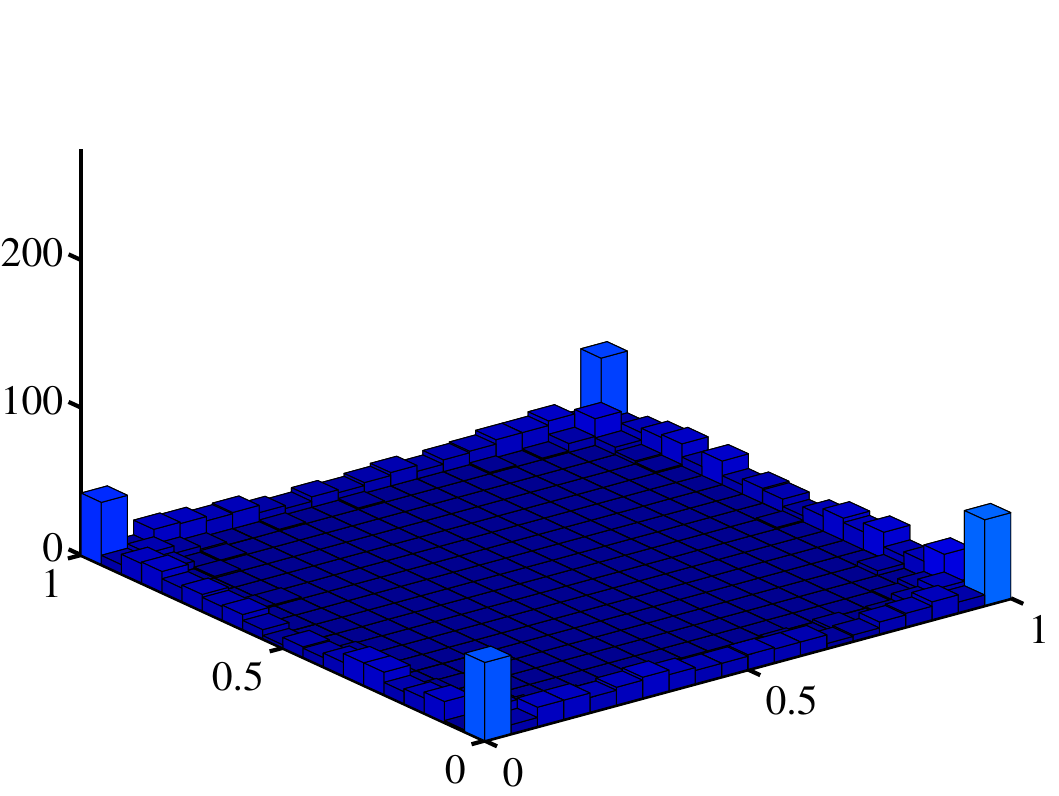}
    & 
    \includegraphics[width=0.2\textwidth]{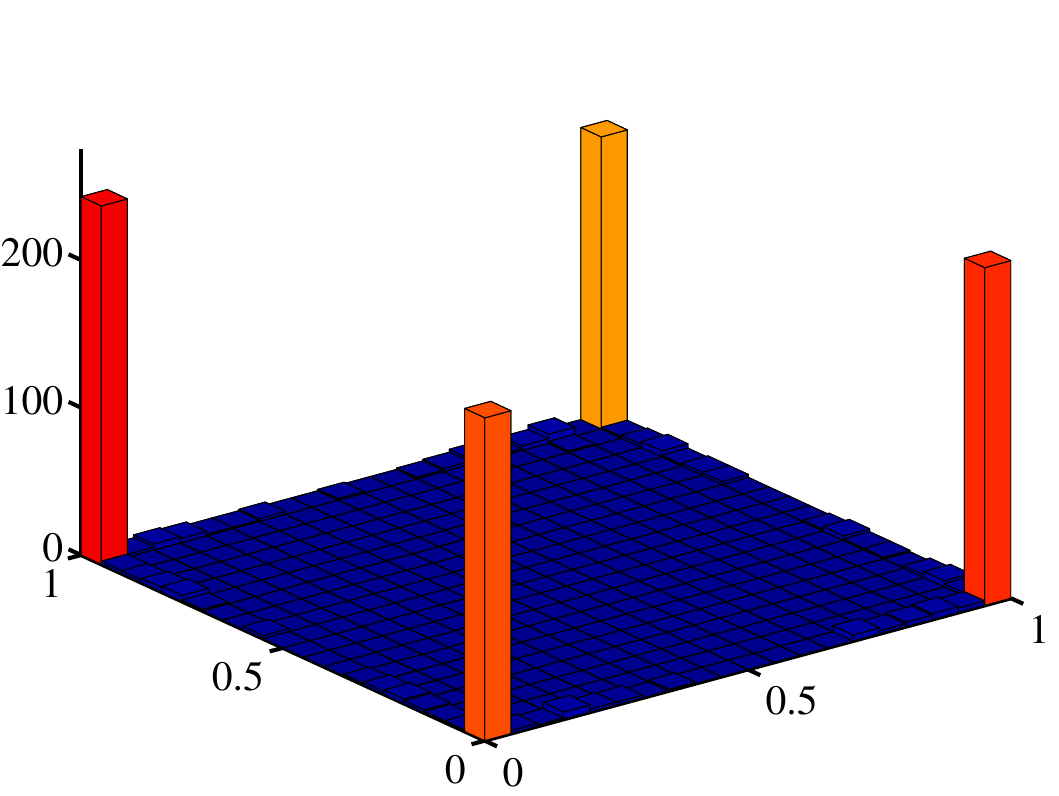}
    & 
    \includegraphics[width=0.2\textwidth]{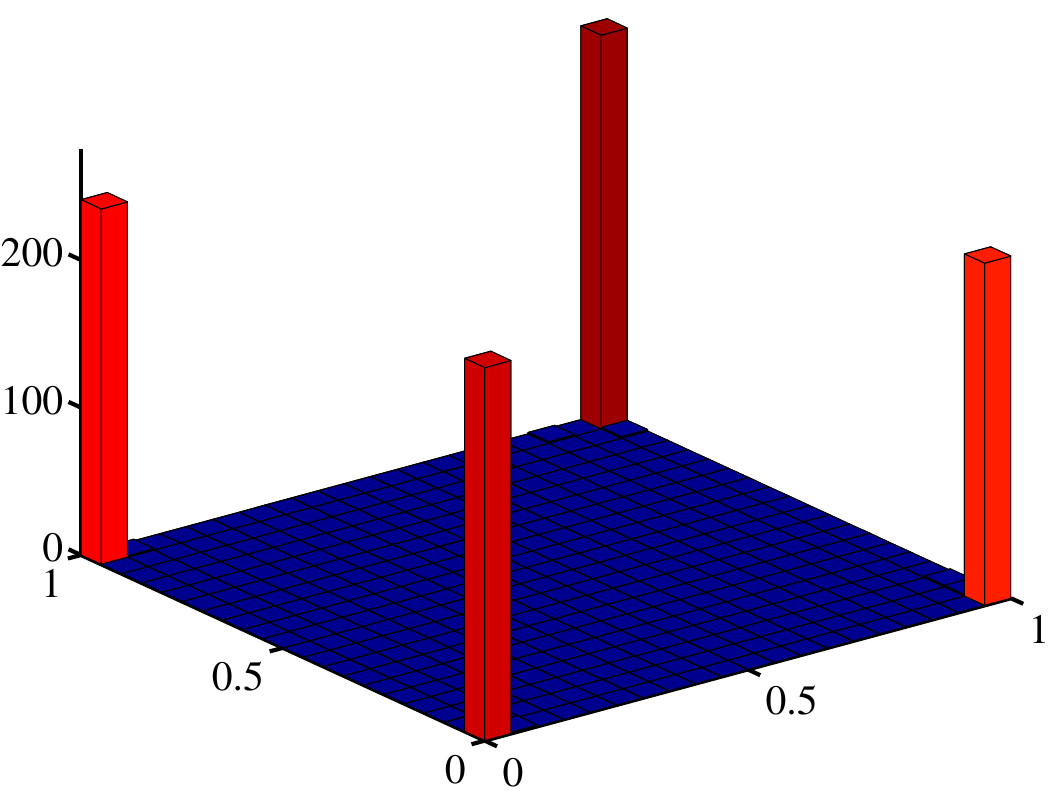}
    & 
    \includegraphics[width=0.2\textwidth]{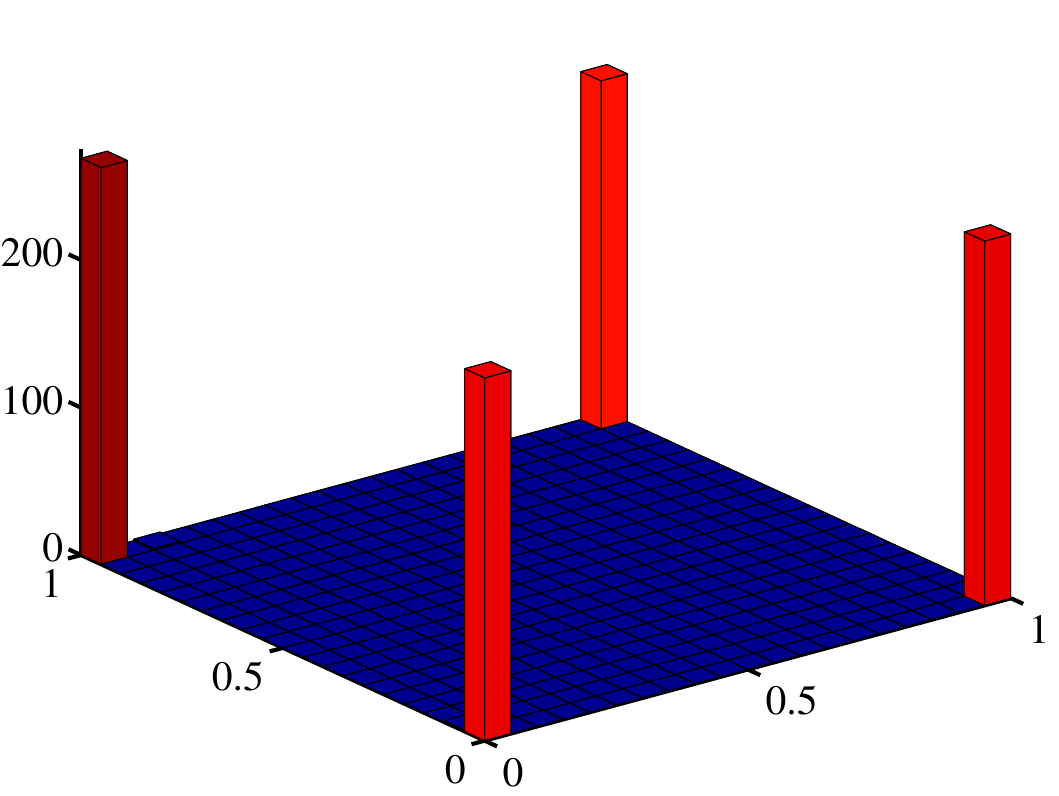}
    \\
    & \includegraphics[width=0.2\textwidth]{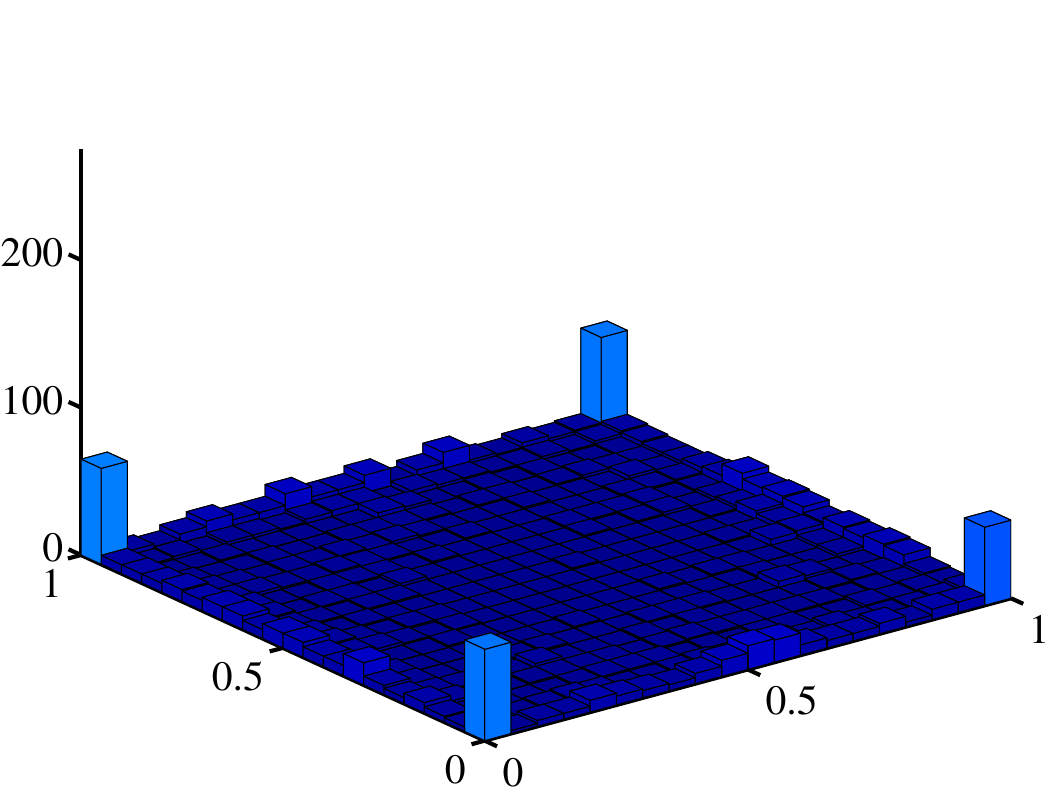}
    & 
    \includegraphics[width=0.2\textwidth]{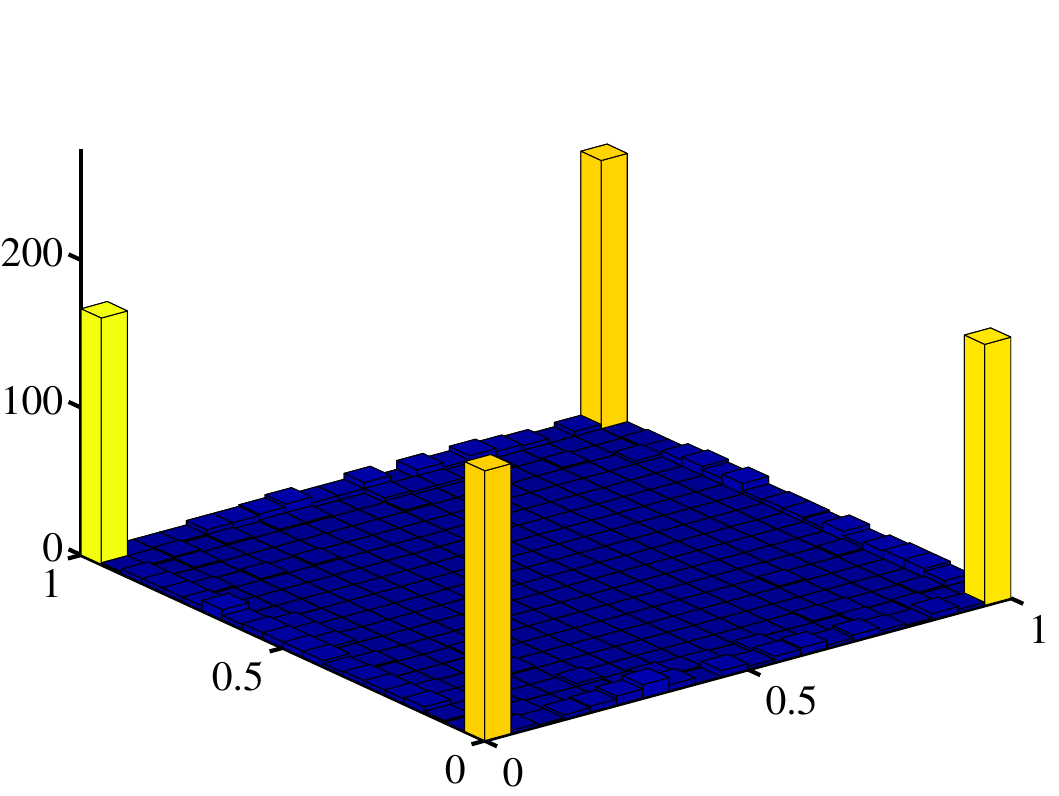}
    & 
    \includegraphics[width=0.2\textwidth]{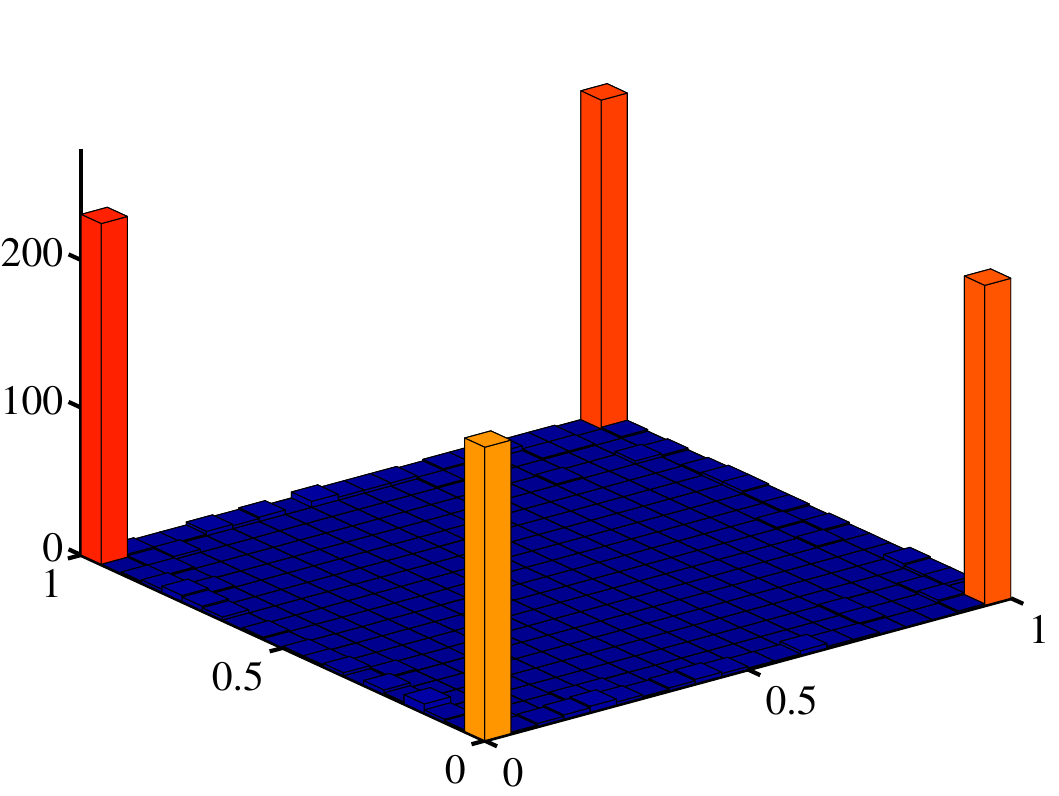}
    & 
    \includegraphics[width=0.2\textwidth]{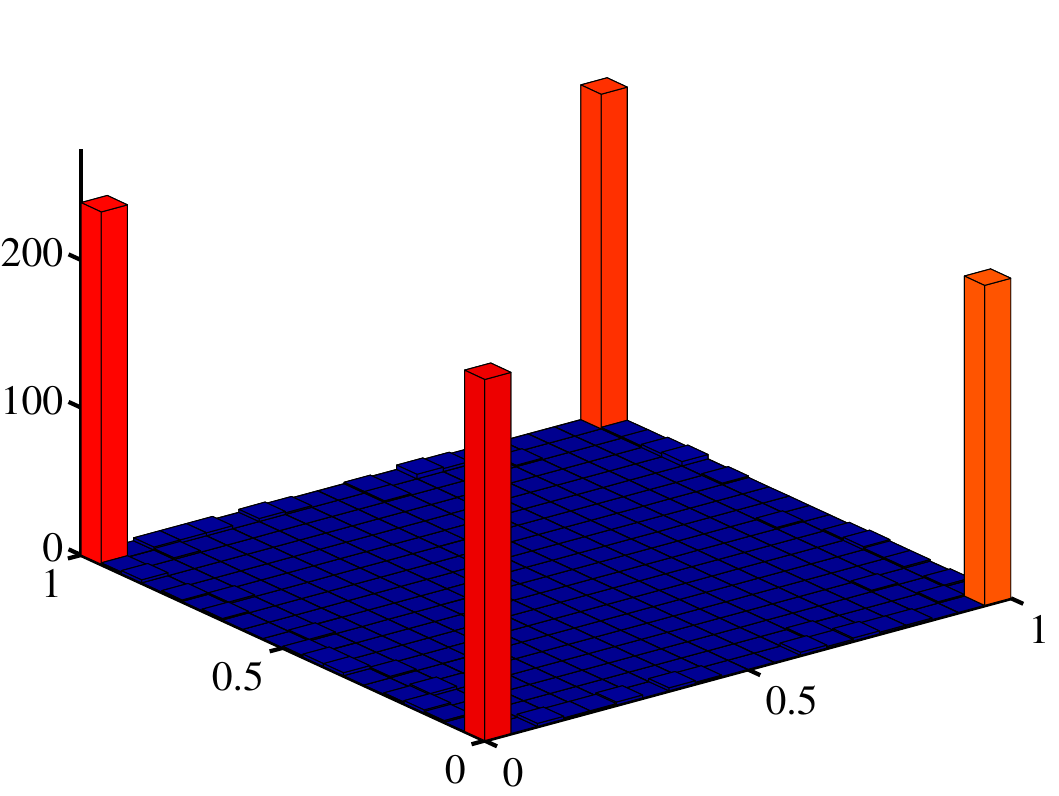}
    \\ \hline
  \end{tabular}
  \caption{Histograms of final search positions resulting from 1000 independent runs
    of RM (top subrows) and SAA (bottom subrows) over a matrix of $N$
    and $M$ sample sizes. For each histogram, the bottom-right and
    bottom-left axes represent the sensor coordinates $x$ and $y$,
    respectively, while the vertical axis represents frequency.}
  \label{t:FinalXHist}
\end{table}

\begin{table}[htb]
  \centering
  \begin{tabular}{c|cccc}
    \backslashbox{$N$}{$M$} & 2 & 11 & 101 & 1001 \\ \hline
    \multirow{2}{*}{1} 
    & \includegraphics[width=0.2\textwidth]{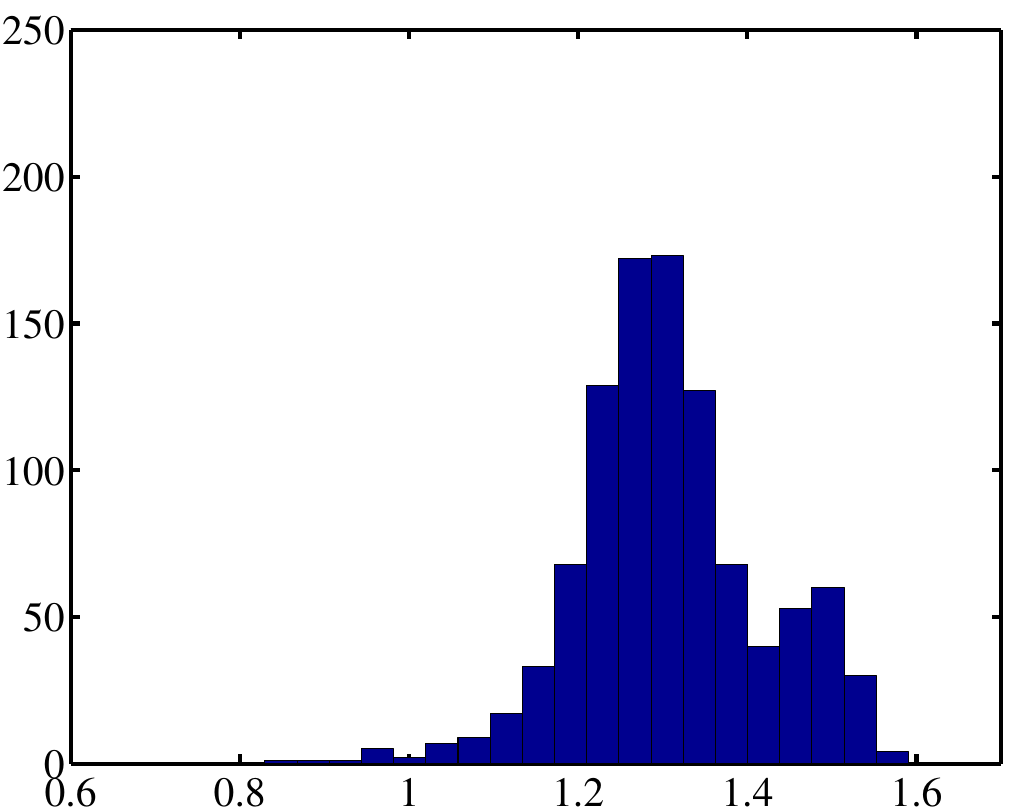}
    & 
    \includegraphics[width=0.2\textwidth]{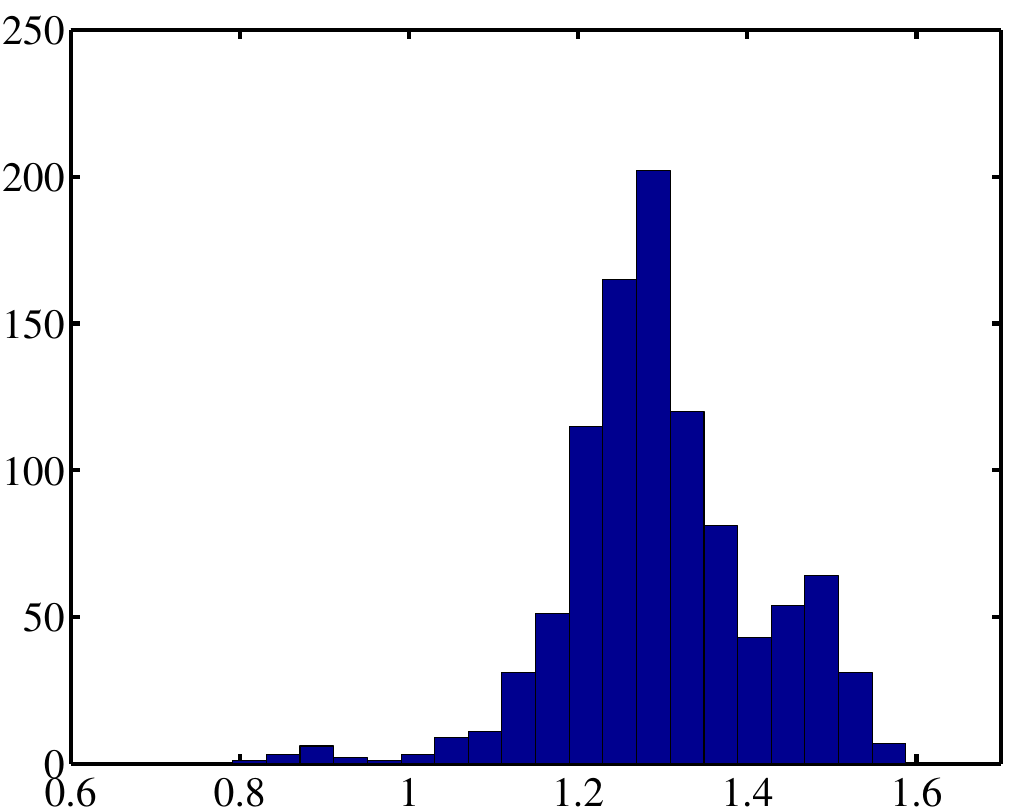}
    & 
    \includegraphics[width=0.2\textwidth]{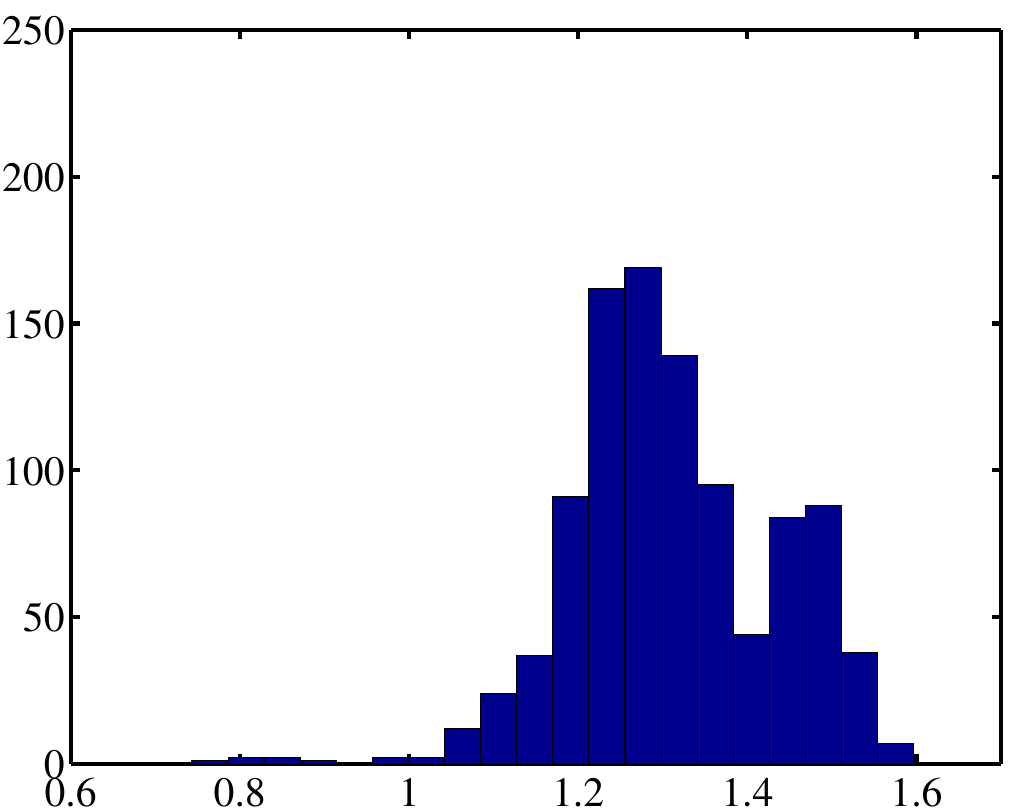}
    & 
    \includegraphics[width=0.2\textwidth]{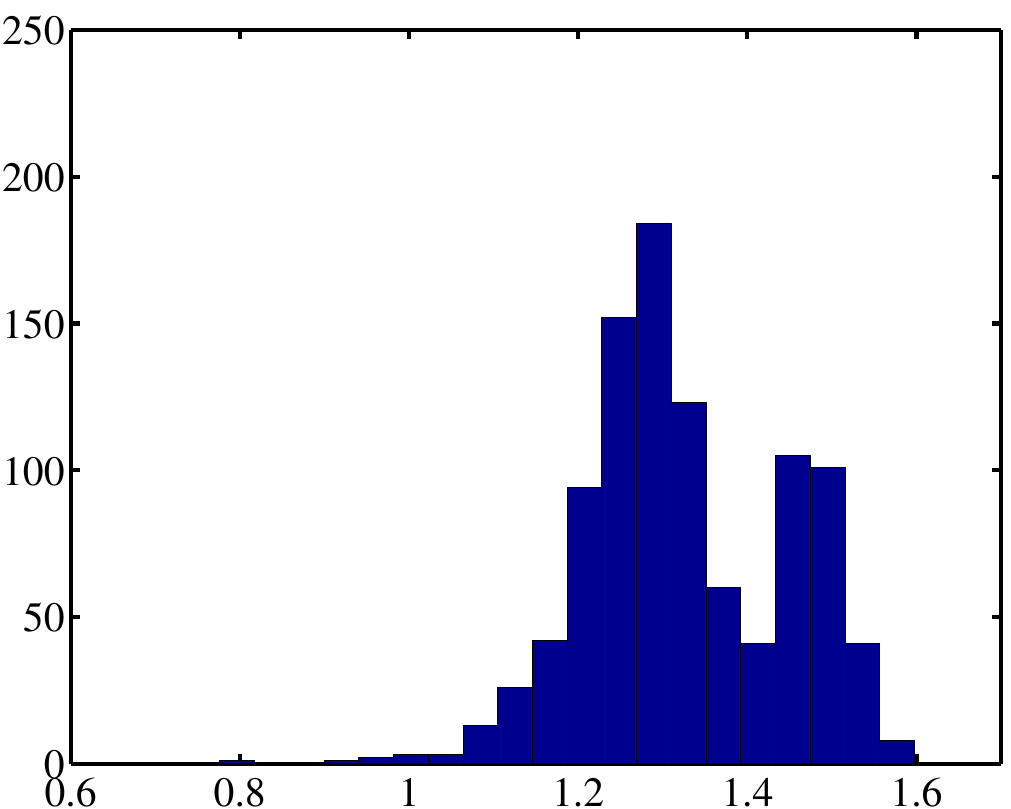}
    \\
    & \includegraphics[width=0.2\textwidth]{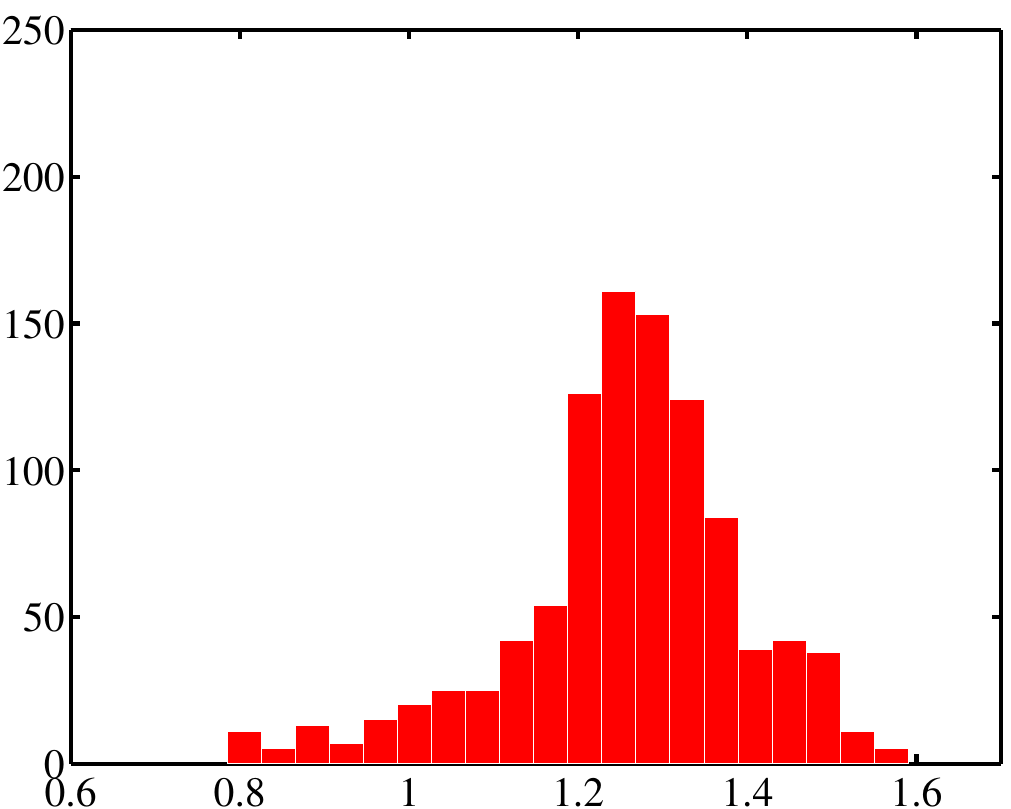}
    & 
    \includegraphics[width=0.2\textwidth]{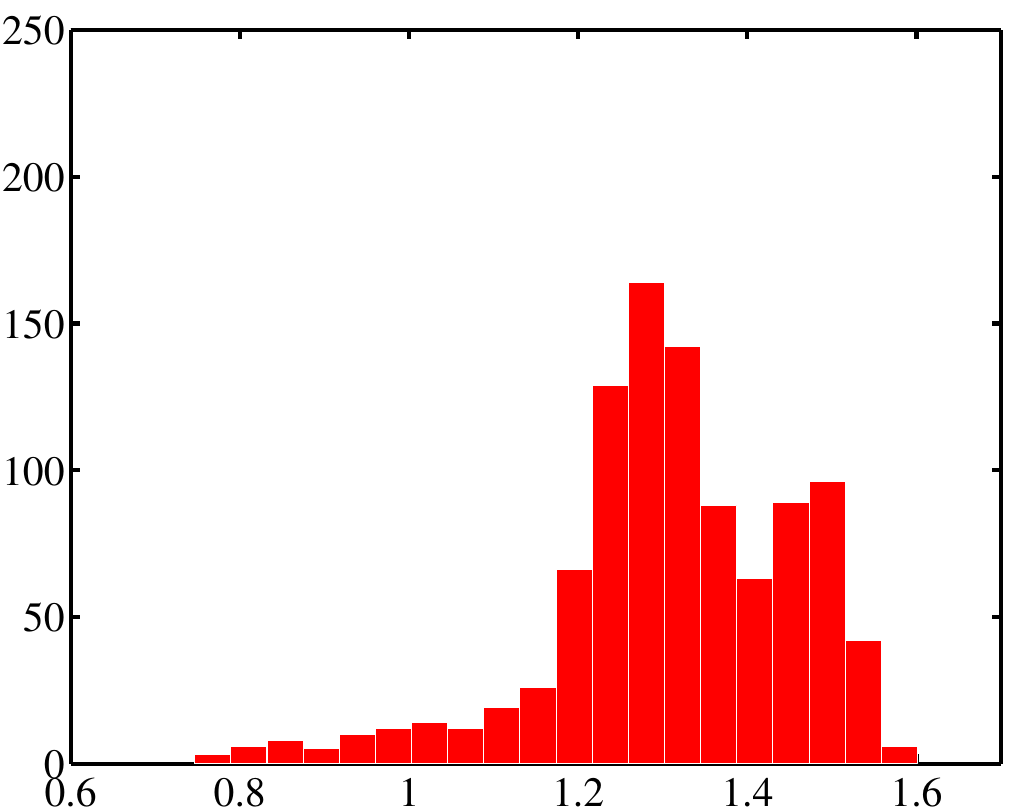}
    & 
    \includegraphics[width=0.2\textwidth]{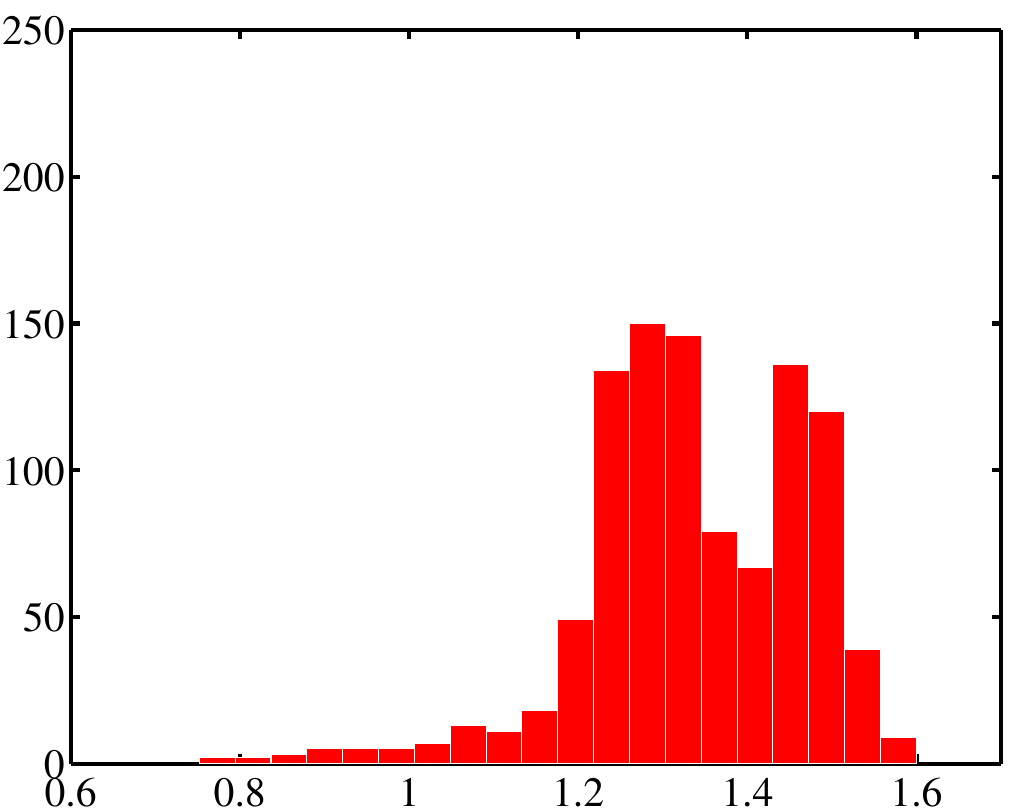}
    & 
    \includegraphics[width=0.2\textwidth]{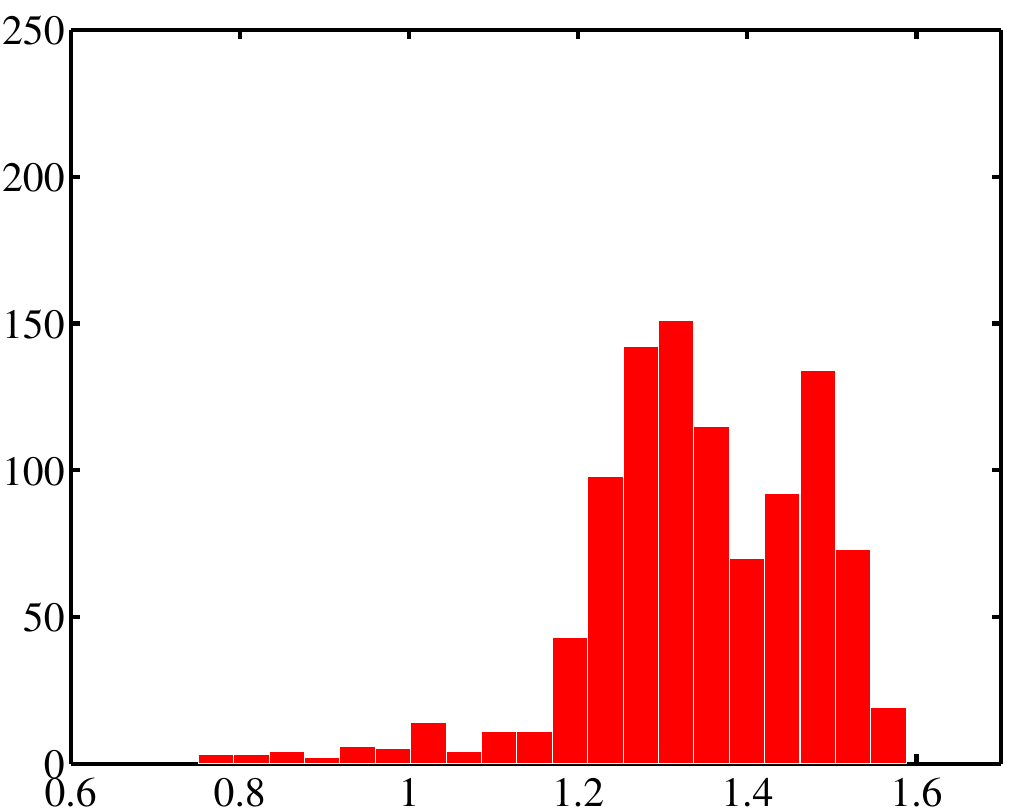}
    \\ \hline
    \multirow{2}{*}{11} 
    & \includegraphics[width=0.2\textwidth]{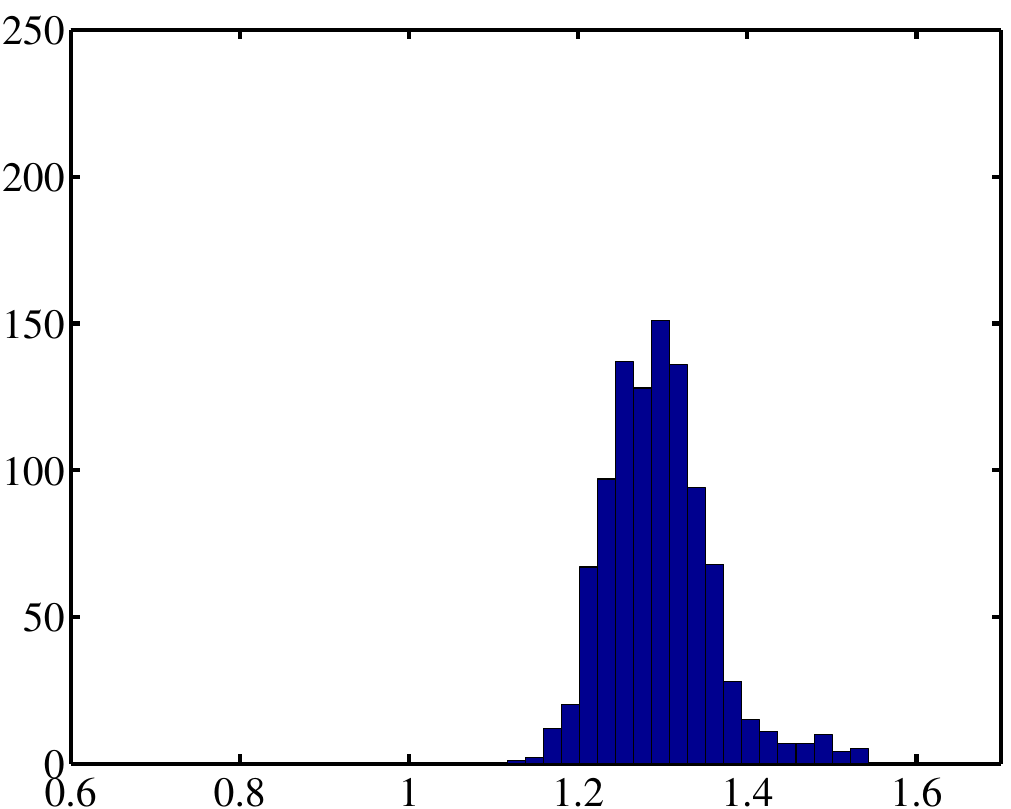}
    & 
    \includegraphics[width=0.2\textwidth]{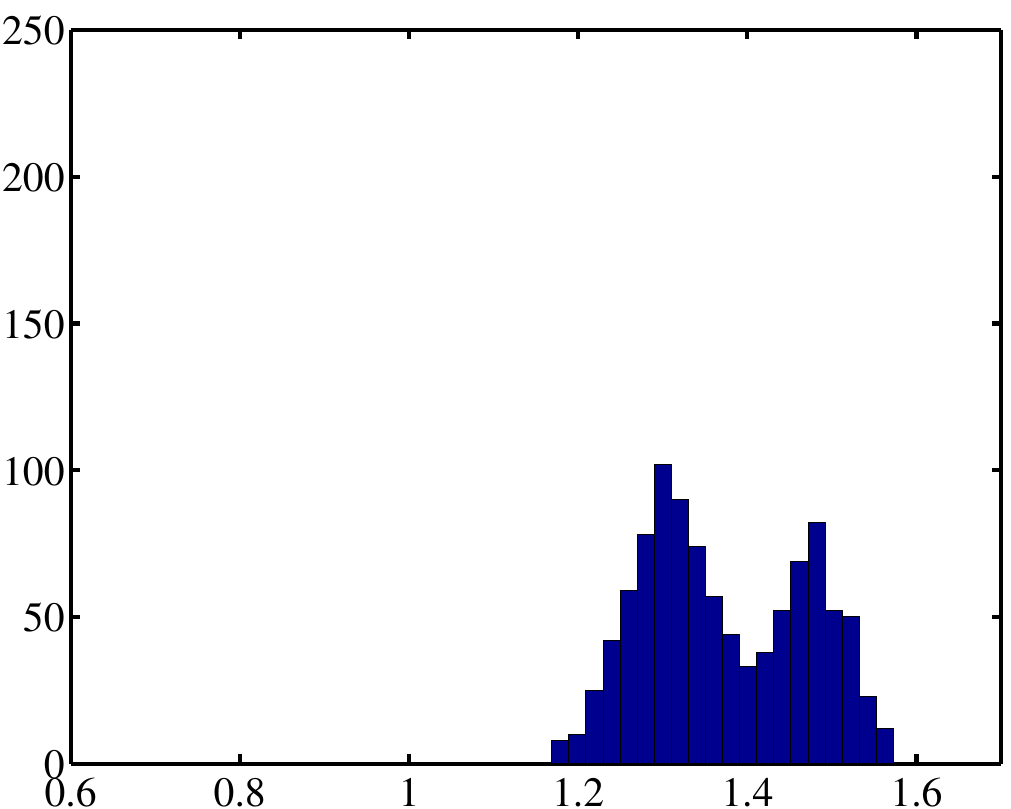}
    & 
    \includegraphics[width=0.2\textwidth]{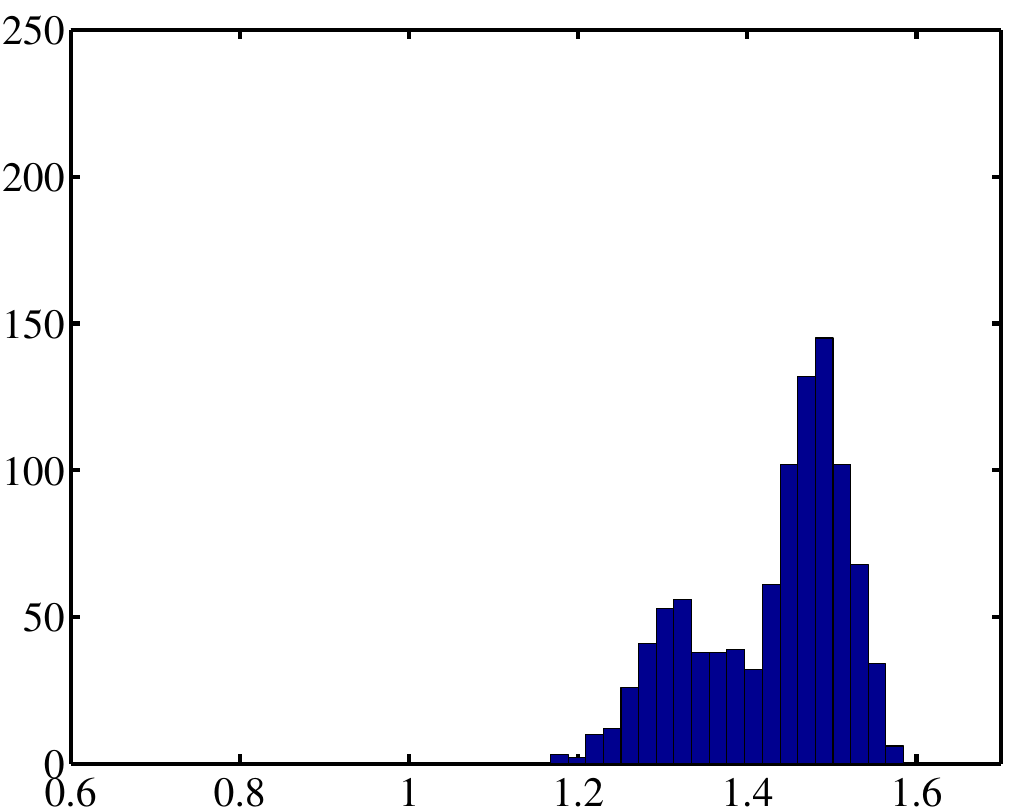}
    & 
    \includegraphics[width=0.2\textwidth]{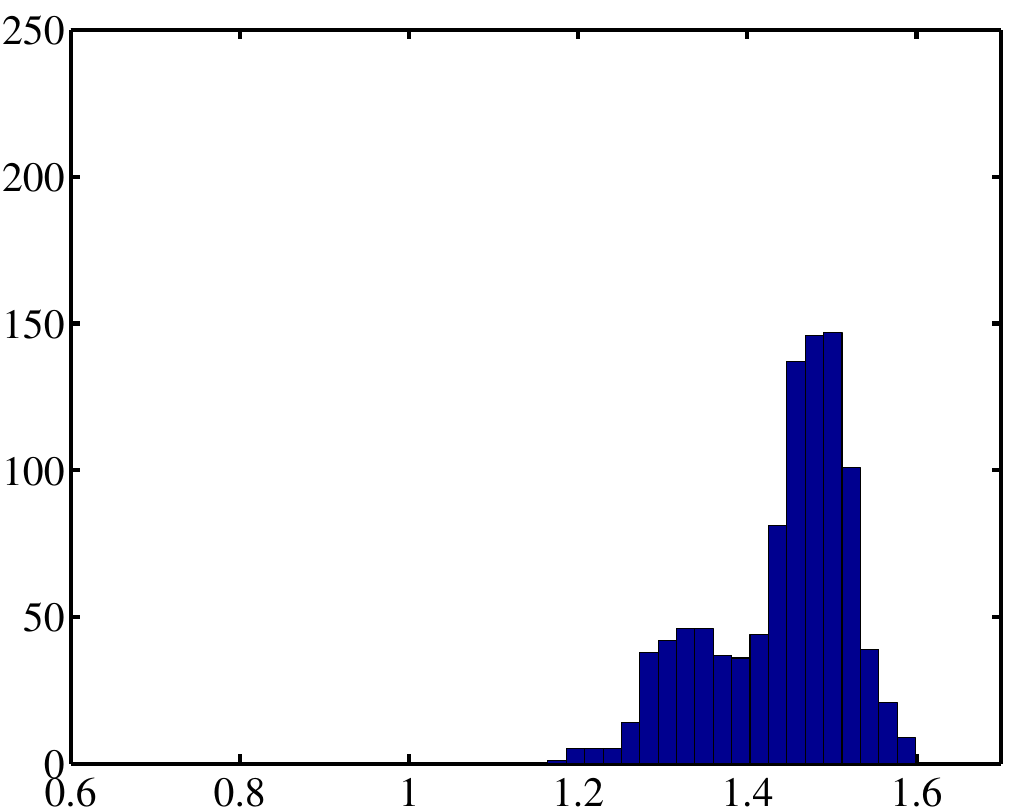}
    \\
    & \includegraphics[width=0.2\textwidth]{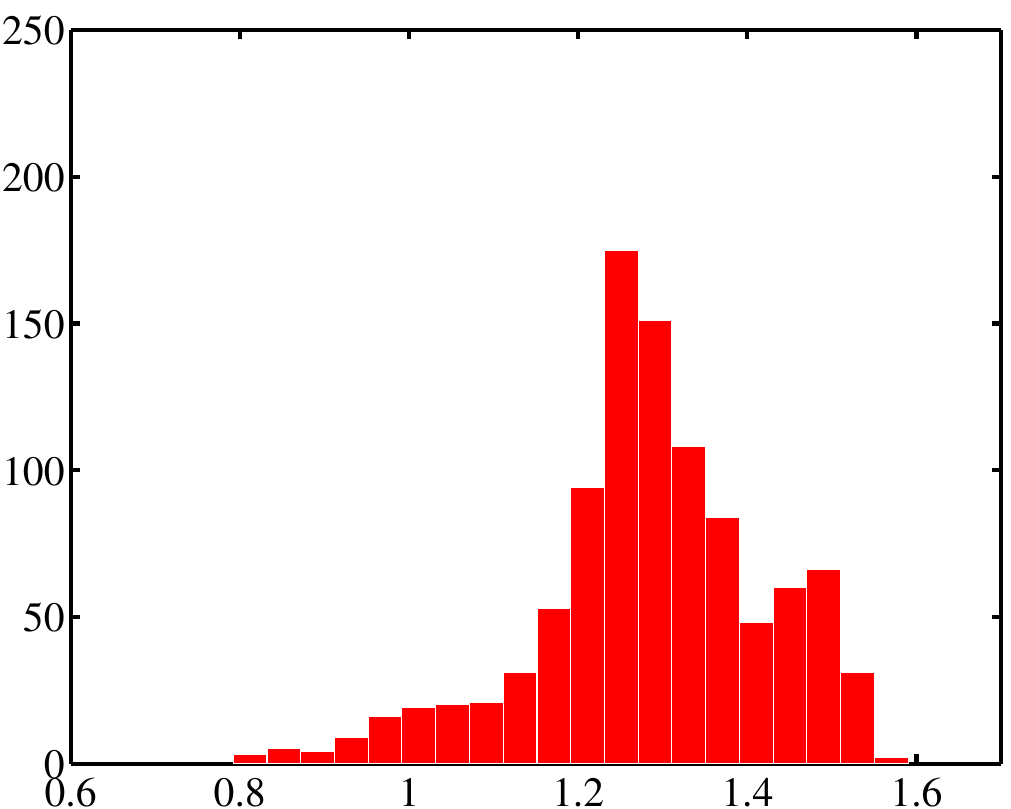}
    & 
    \includegraphics[width=0.2\textwidth]{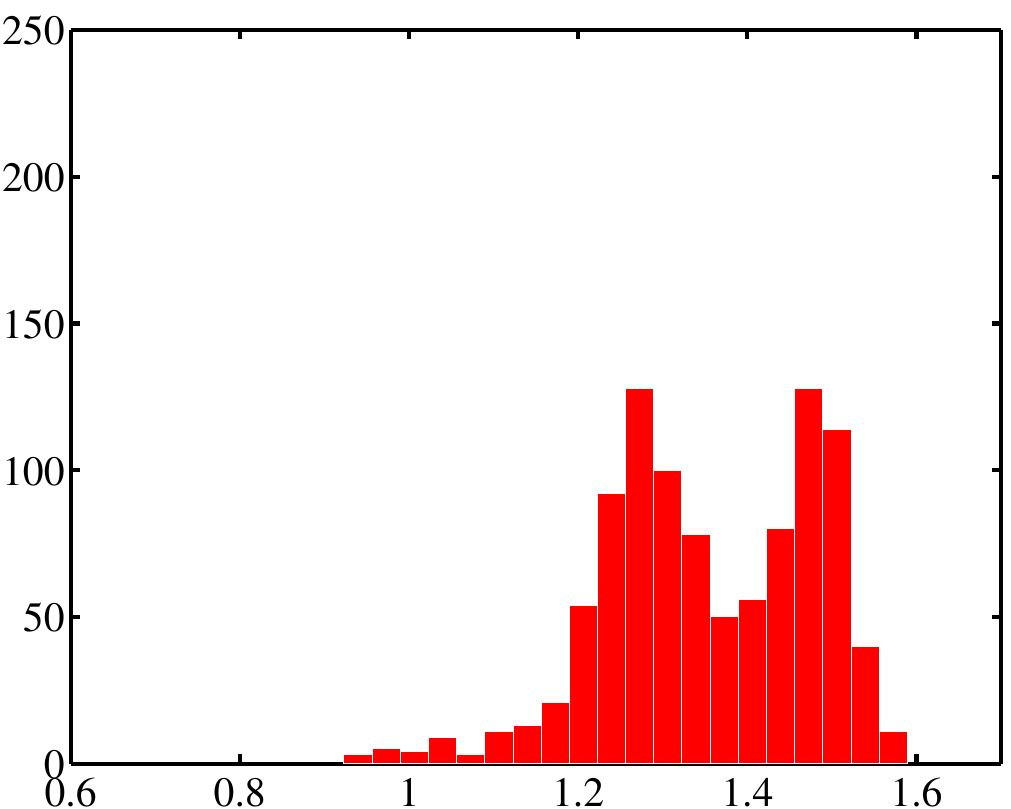}
    & 
    \includegraphics[width=0.2\textwidth]{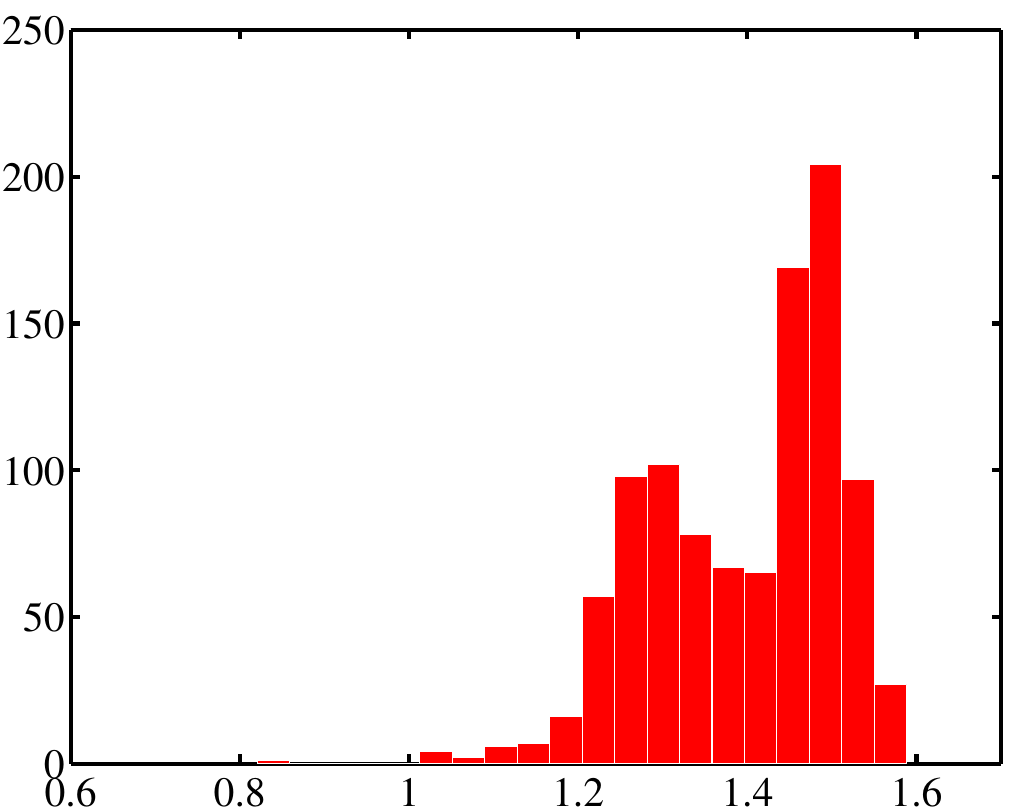}
    & 
    \includegraphics[width=0.2\textwidth]{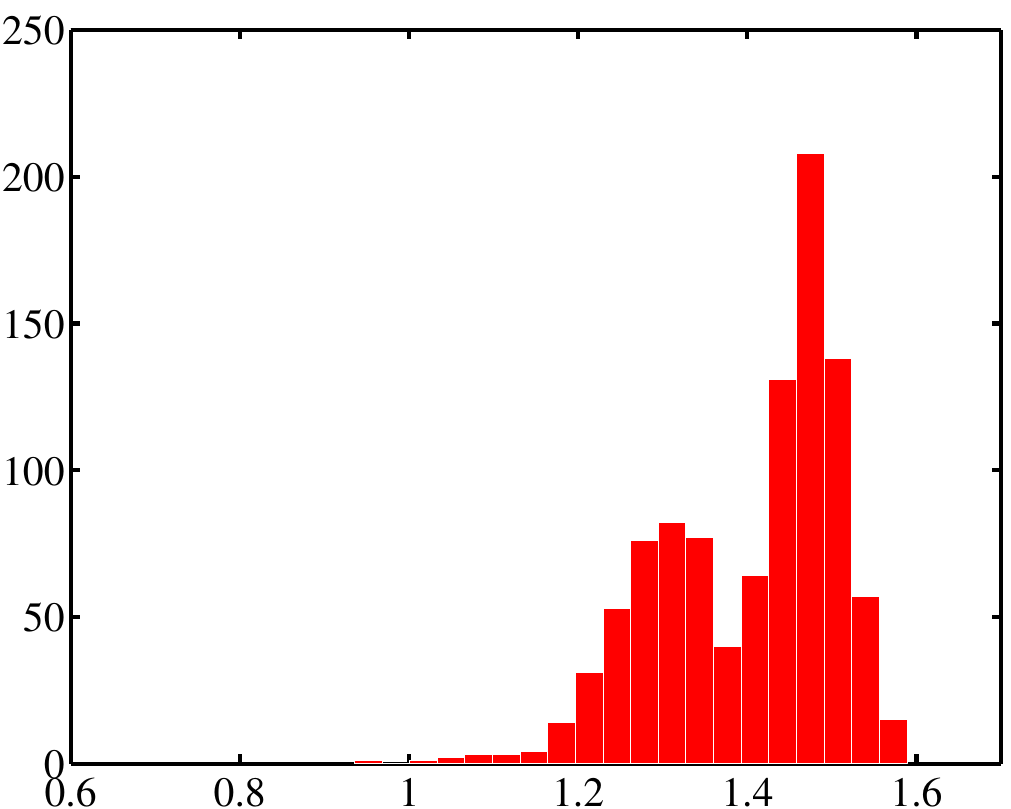}
    \\ \hline
    \multirow{2}{*}{101} 
    & \includegraphics[width=0.2\textwidth]{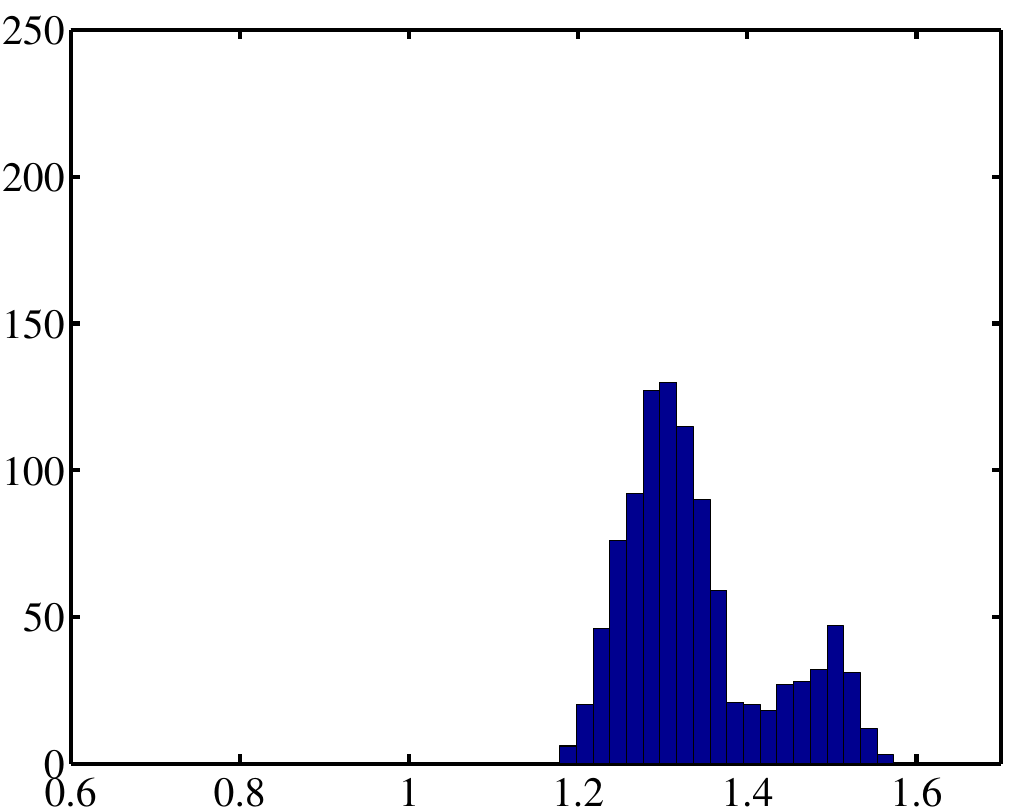}
    & 
    \includegraphics[width=0.2\textwidth]{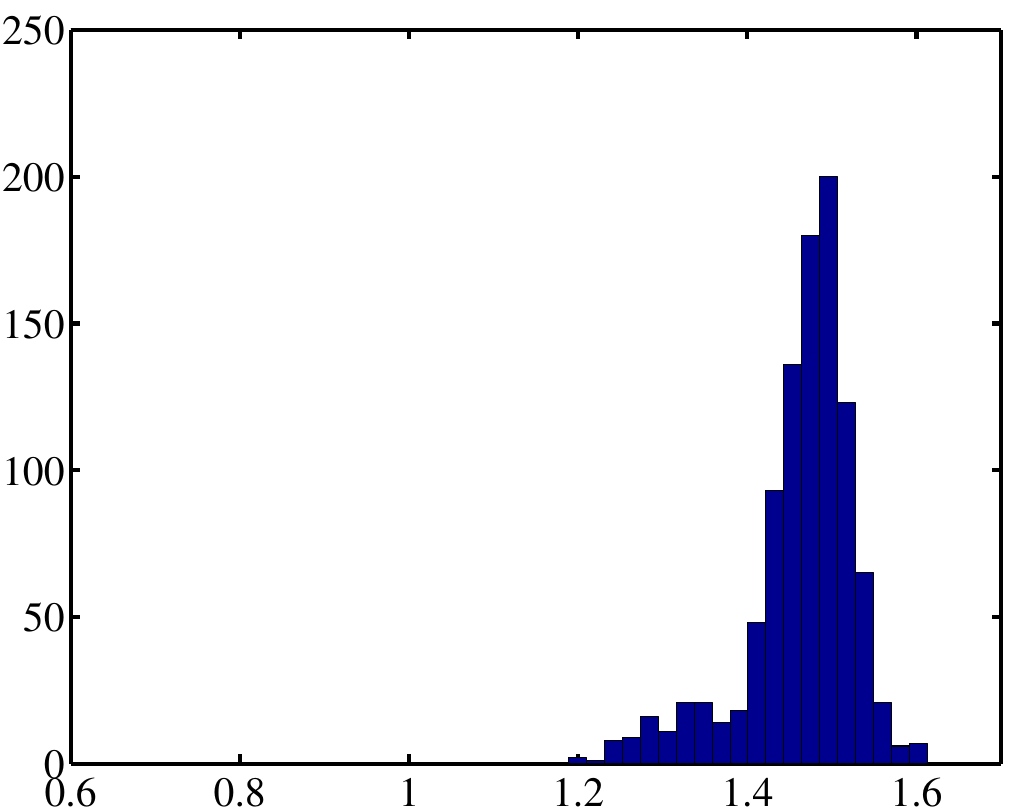}
    & 
    \includegraphics[width=0.2\textwidth]{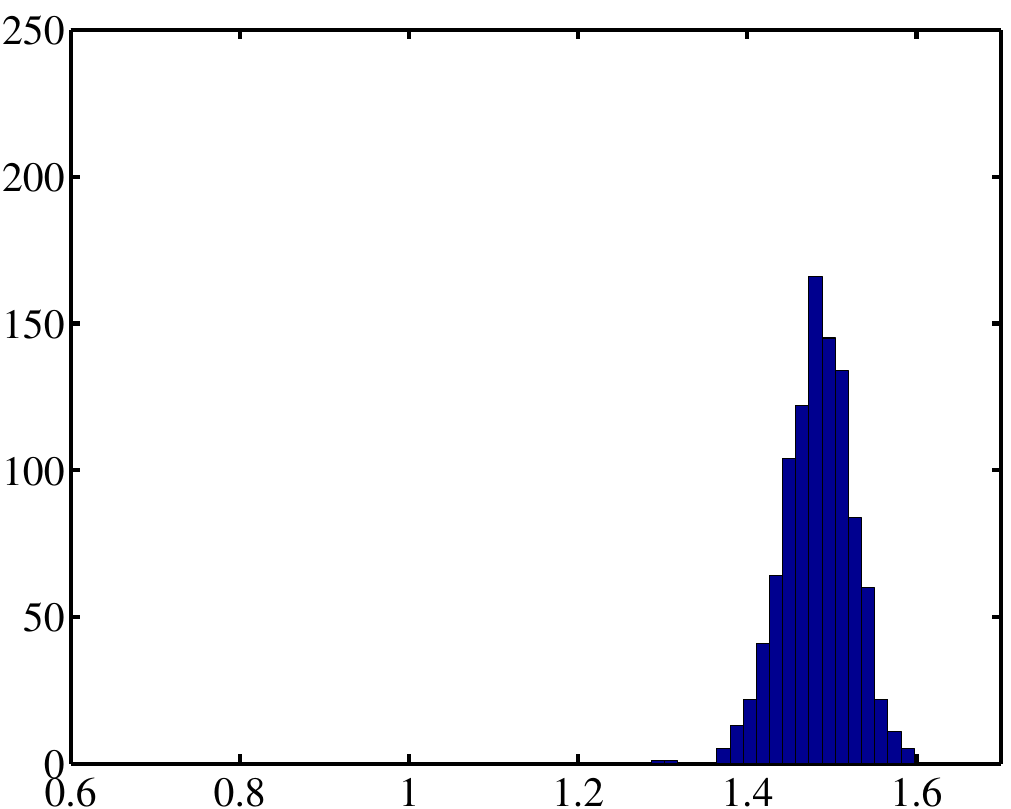}
    & 
    \includegraphics[width=0.2\textwidth]{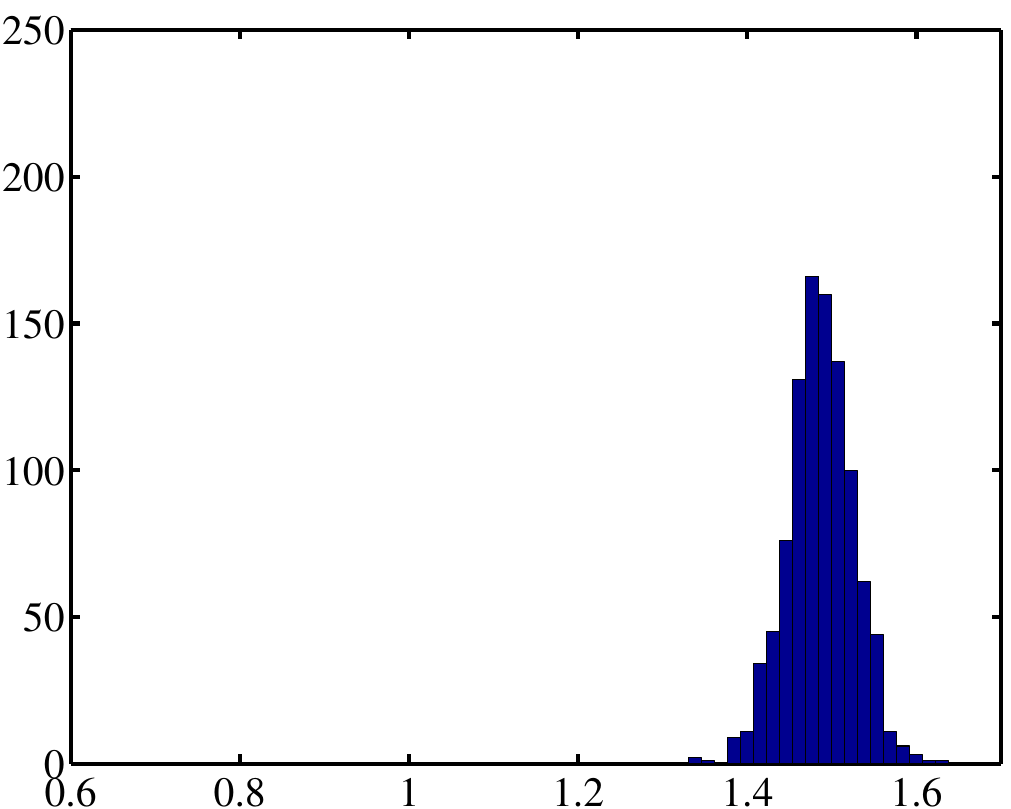}
    \\
    & \includegraphics[width=0.2\textwidth]{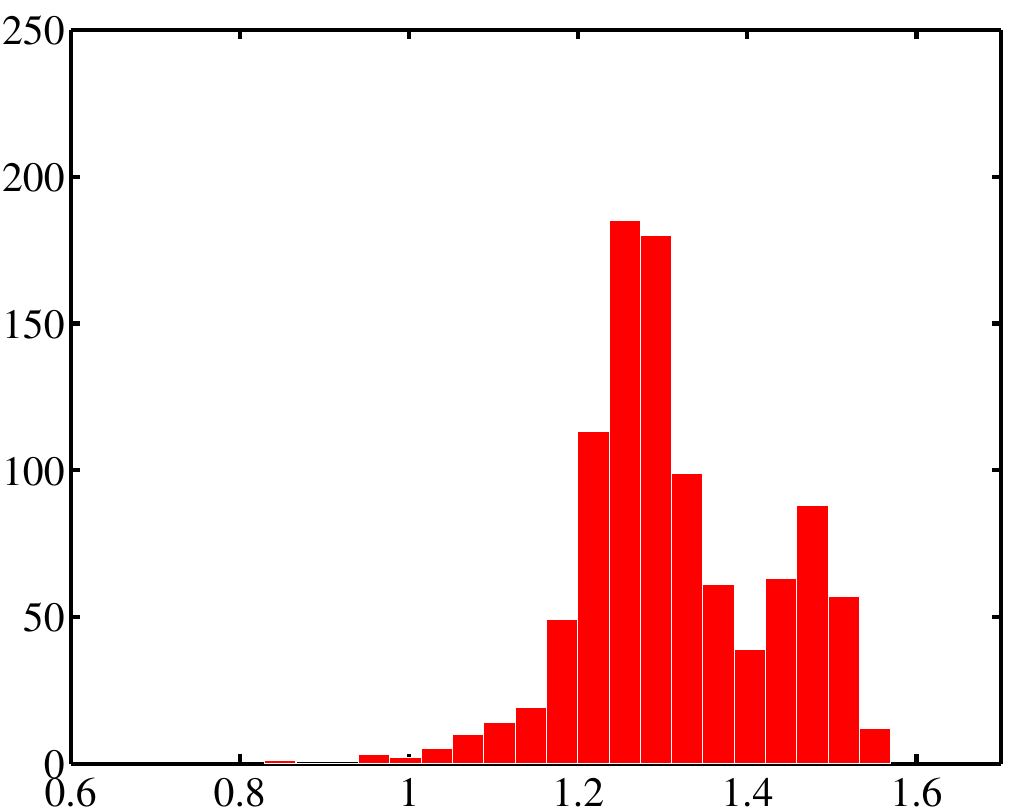}
    & 
    \includegraphics[width=0.2\textwidth]{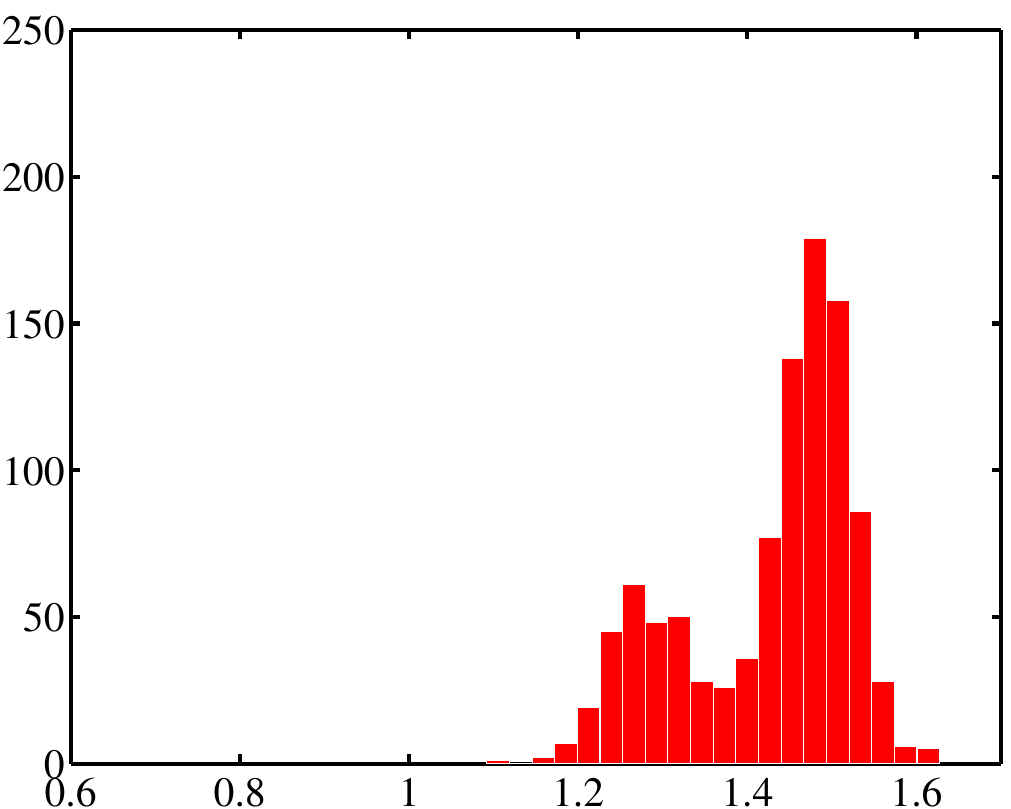}
    & 
    \includegraphics[width=0.2\textwidth]{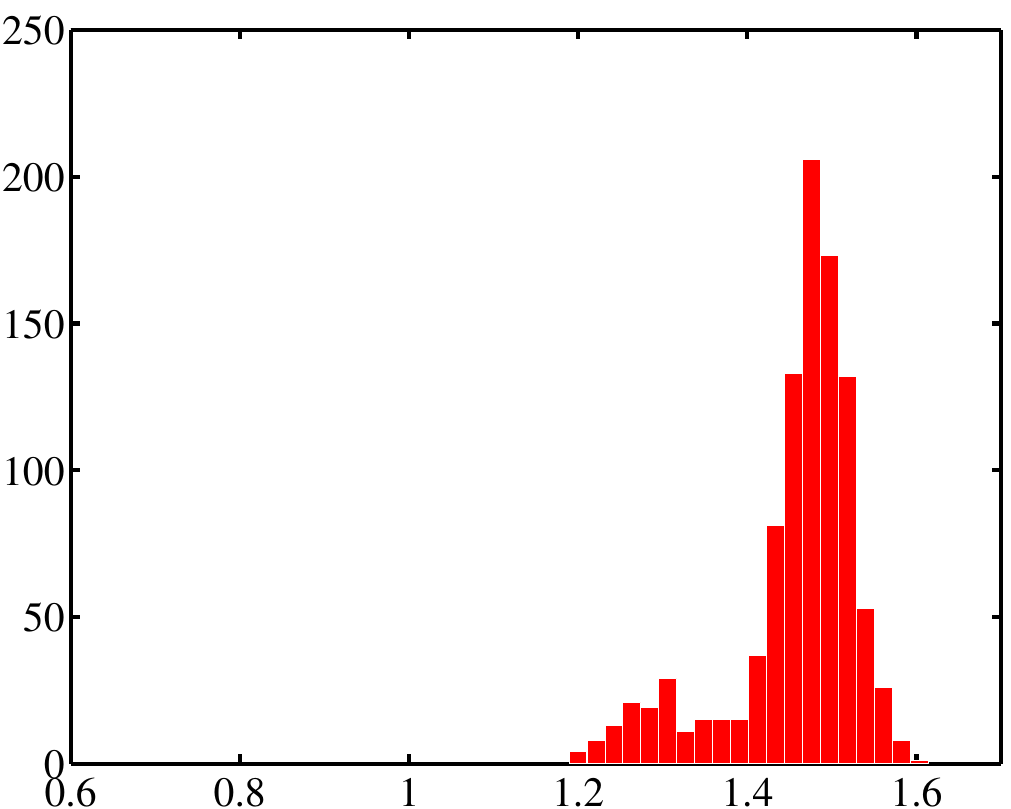}
    & 
    \includegraphics[width=0.2\textwidth]{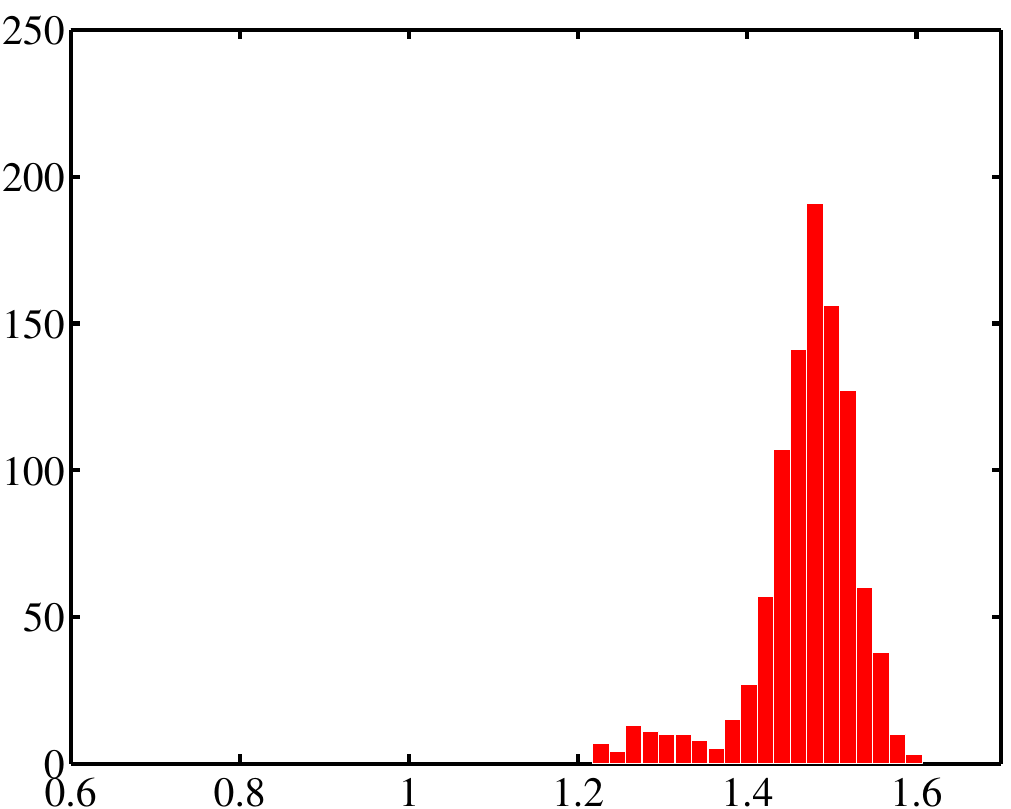}
    \\ \hline
  \end{tabular}
  \caption{High-quality expected information gain estimates at the final sensor positions resulting
    from 1000 independent runs of RM (top subrows, blue) and SAA-BFGS
    (bottom subrows, red). For each histogram, the horizontal axis
    represents values of $\hU_{M=1001, N=1001}$ and the vertical axis represents frequency.}
  \label{t:FinalEUHigh}
\end{table}

% \begin{figure}[htb]
%   \centering 
%   \mbox{\subfigure[$N=101$, $M=11$]
%     {\includegraphics[]{figures/1Exp/p12n1000000Design1ExpSAABFGSo101i11FinalOptGap.eps,angle=0,width=0.48\textwidth}}} 
%   \mbox{\subfigure[$N=101$, $M=1001$]
%     {\includegraphics[]{figures/1Exp/p12n1000000Design1ExpSAABFGSo101i1001FinalOptGap.eps,angle=0,width=0.48\textwidth}}} 
%   \caption{Histograms of optimality gap estimates for selected runs of
%     SAA-BFGS.}
%   \label{f:optGap}
% \end{figure}

\newcolumntype{g}{>{\centering\arraybackslash} m{1.6cm} }  %# New column type
\newcolumntype{h}{>{\centering\arraybackslash} m{3.65cm} }  %# New column type
\begin{table}[htb]
  \centering
  \begin{tabular}{g|hhhh}
    \backslashbox{$N$}{$M$} & 2 & 11 & 101 & 1001 \\ \hline
    1
    & \includegraphics[width=0.2\textwidth]{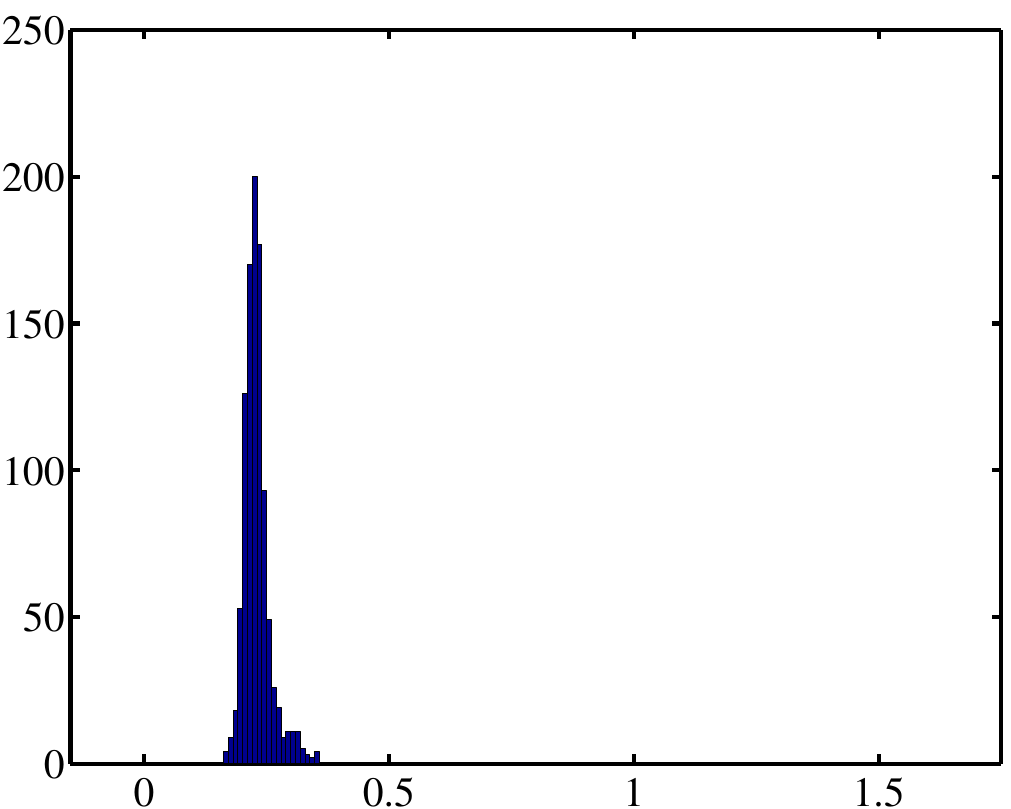}
    & 
    \includegraphics[width=0.2\textwidth]{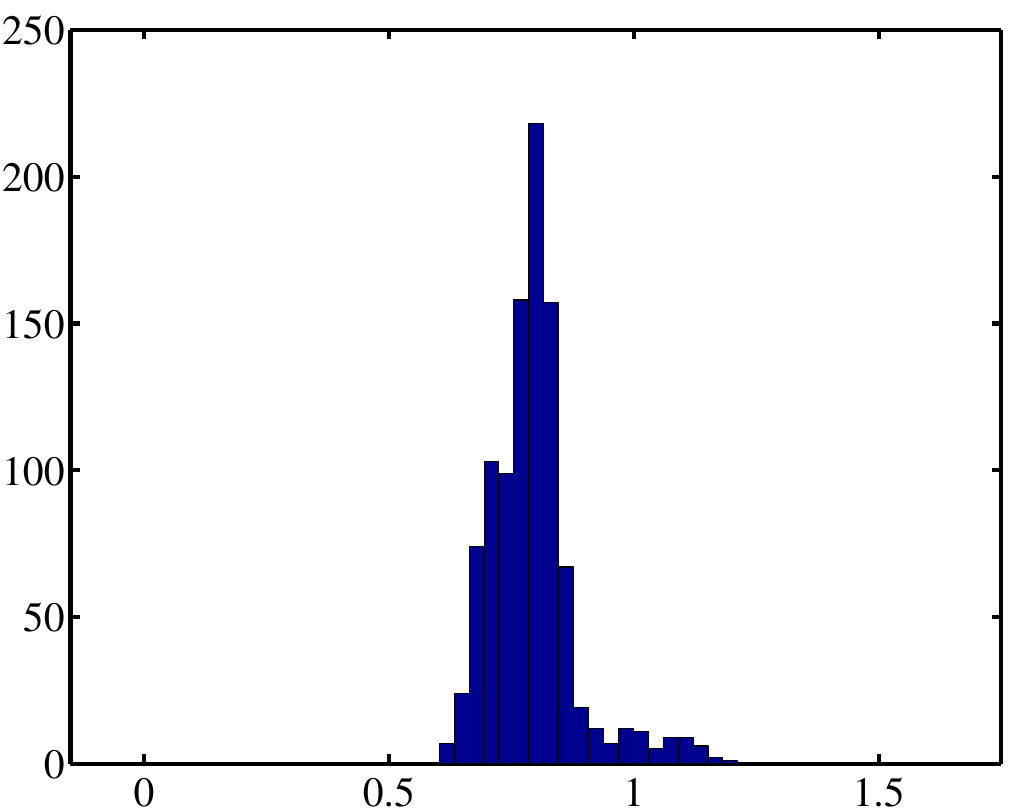}
    & 
    \includegraphics[width=0.2\textwidth]{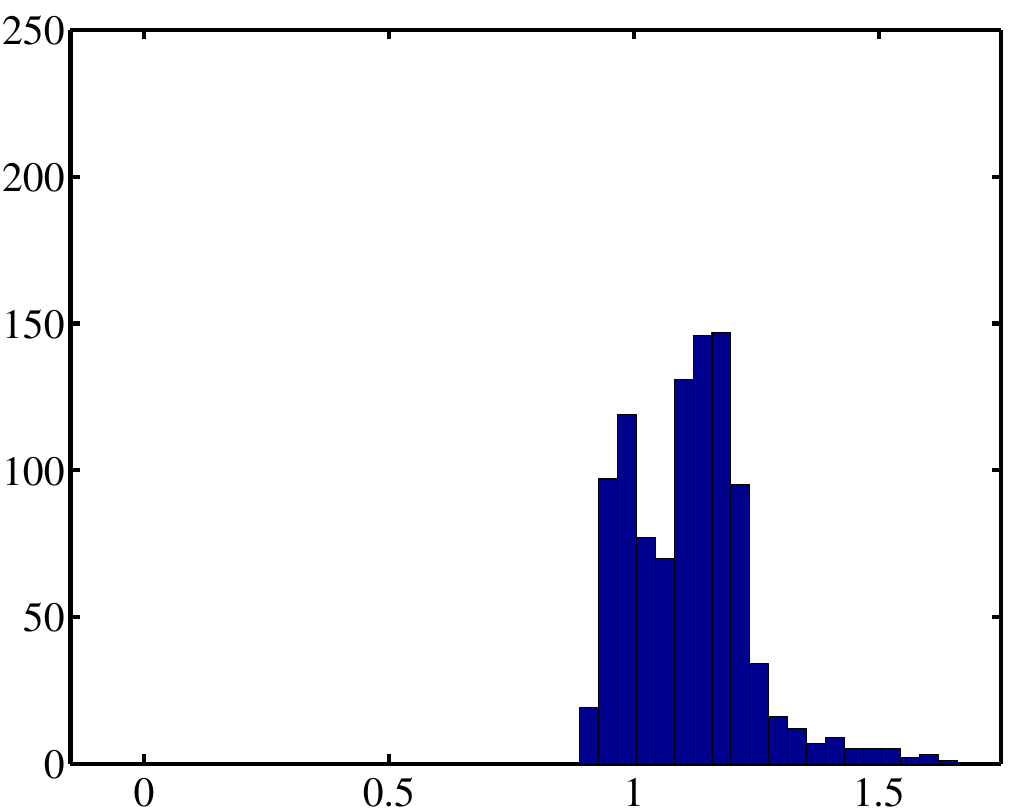}
    & 
    \includegraphics[width=0.2\textwidth]{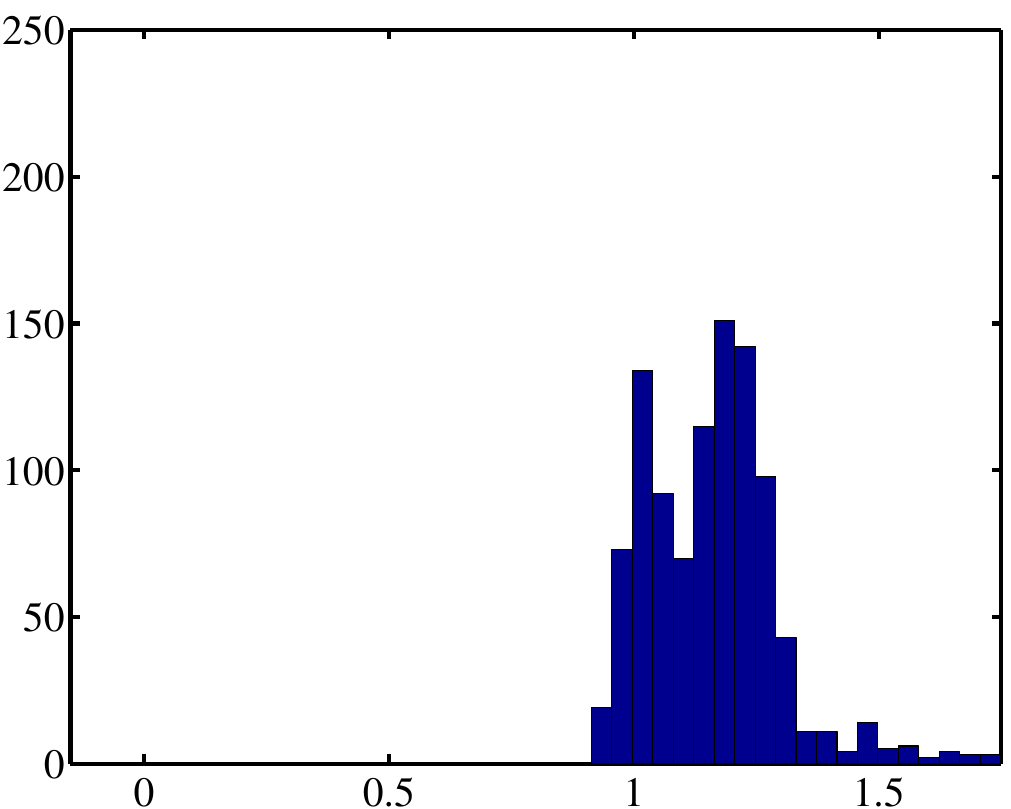}
    \\ \hline
    11
    & \includegraphics[width=0.2\textwidth]{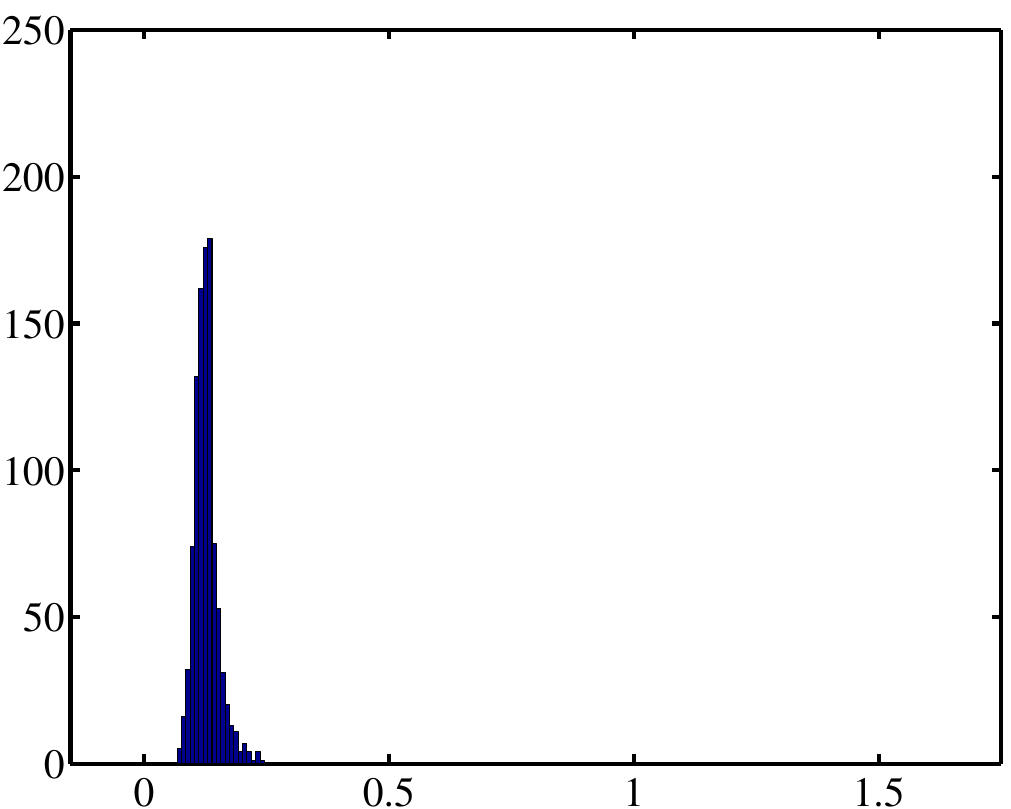}
    & 
    \includegraphics[width=0.2\textwidth]{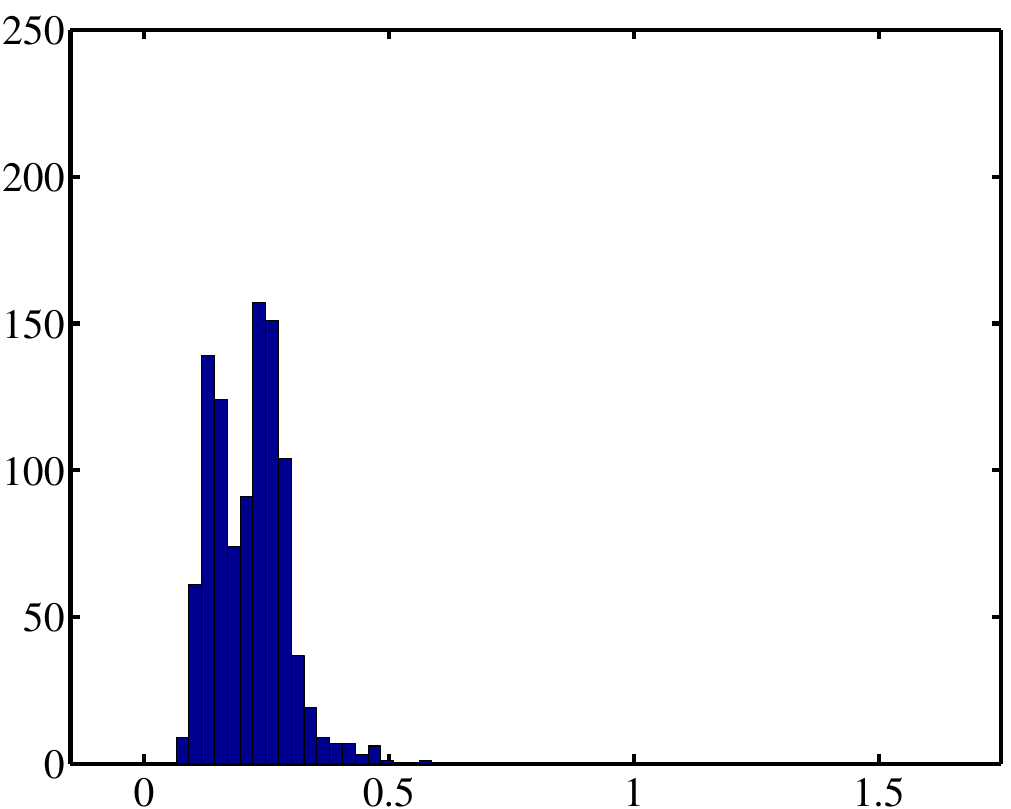}
    & 
    \includegraphics[width=0.2\textwidth]{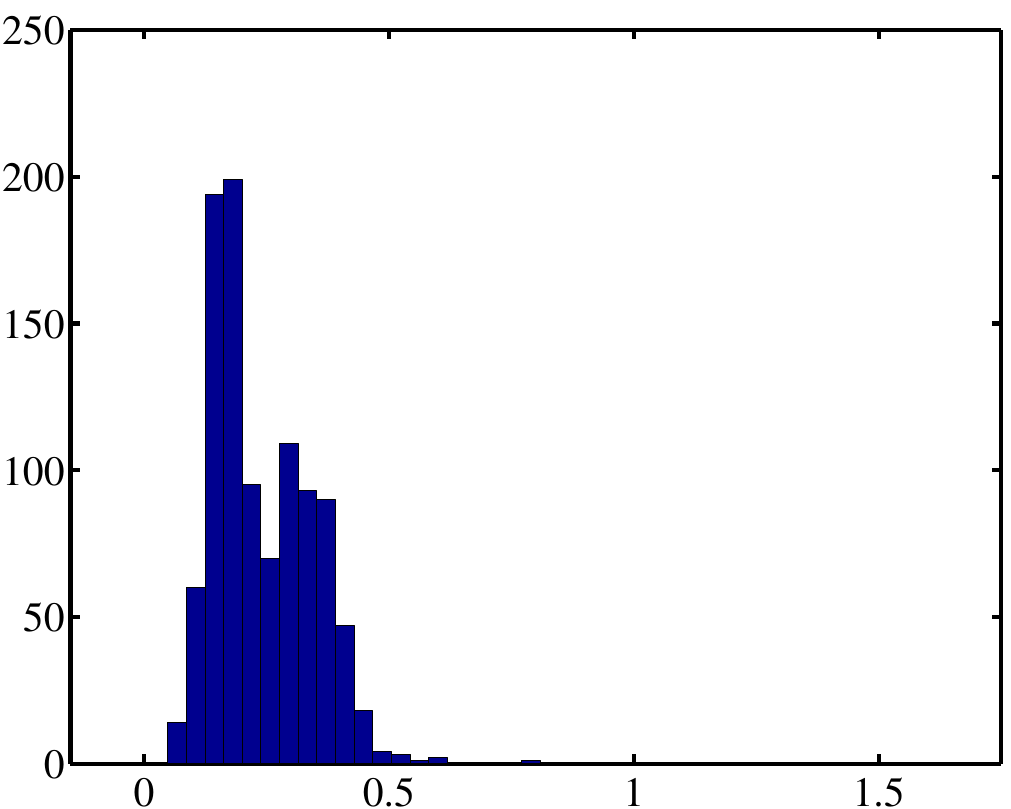}
    & 
    \includegraphics[width=0.2\textwidth]{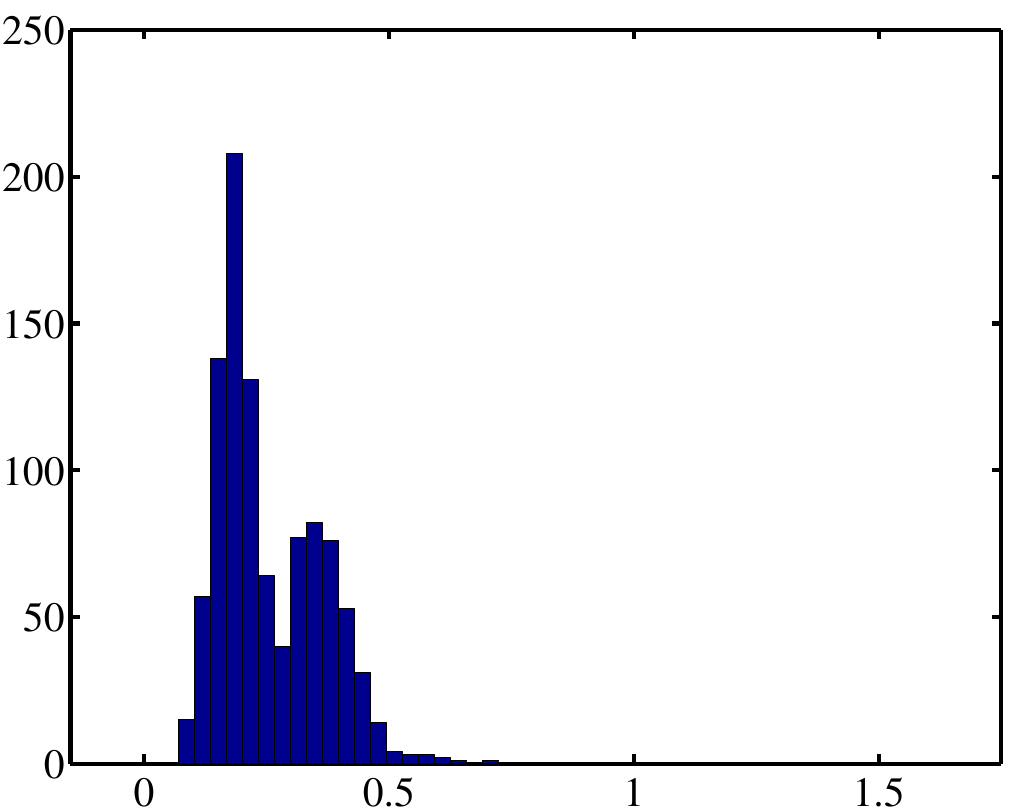}
    \\ \hline
    101
    & \includegraphics[width=0.2\textwidth]{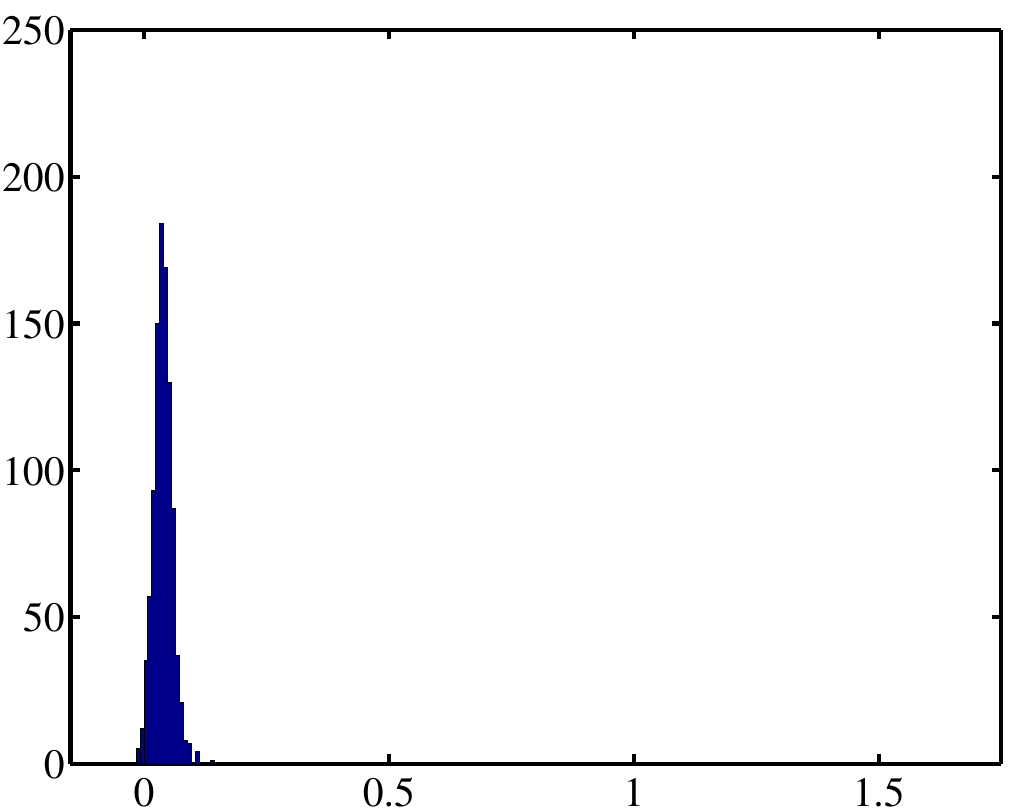}
    & 
    \includegraphics[width=0.2\textwidth]{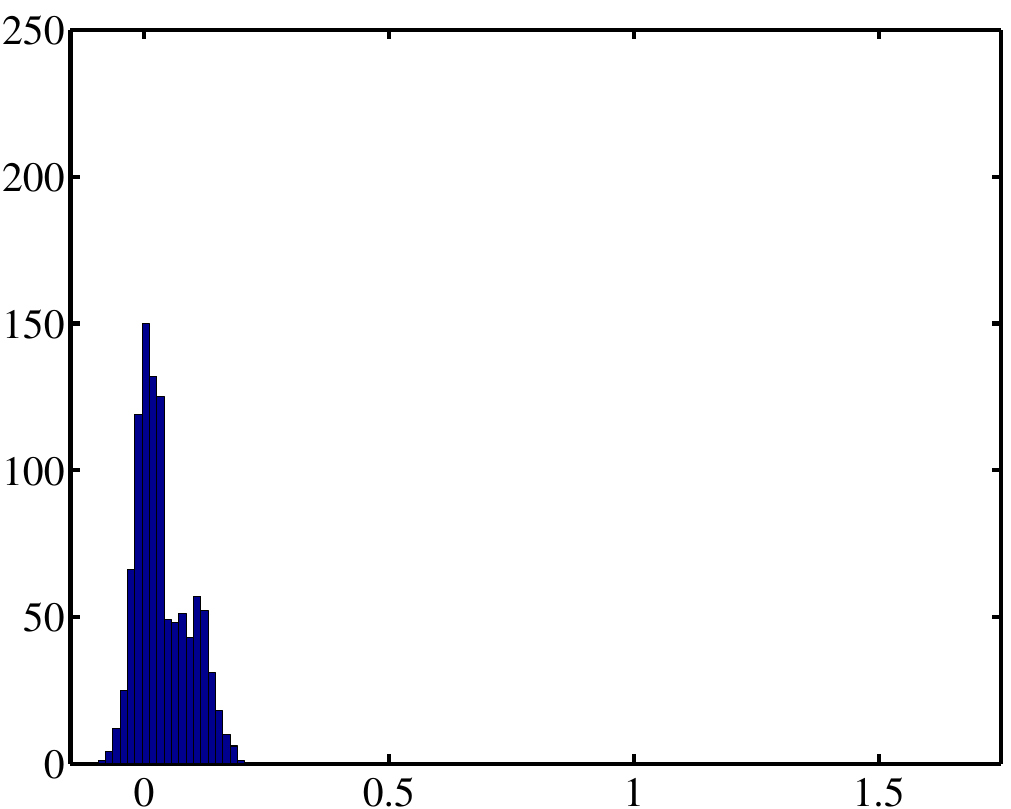}
    & 
    \includegraphics[width=0.2\textwidth]{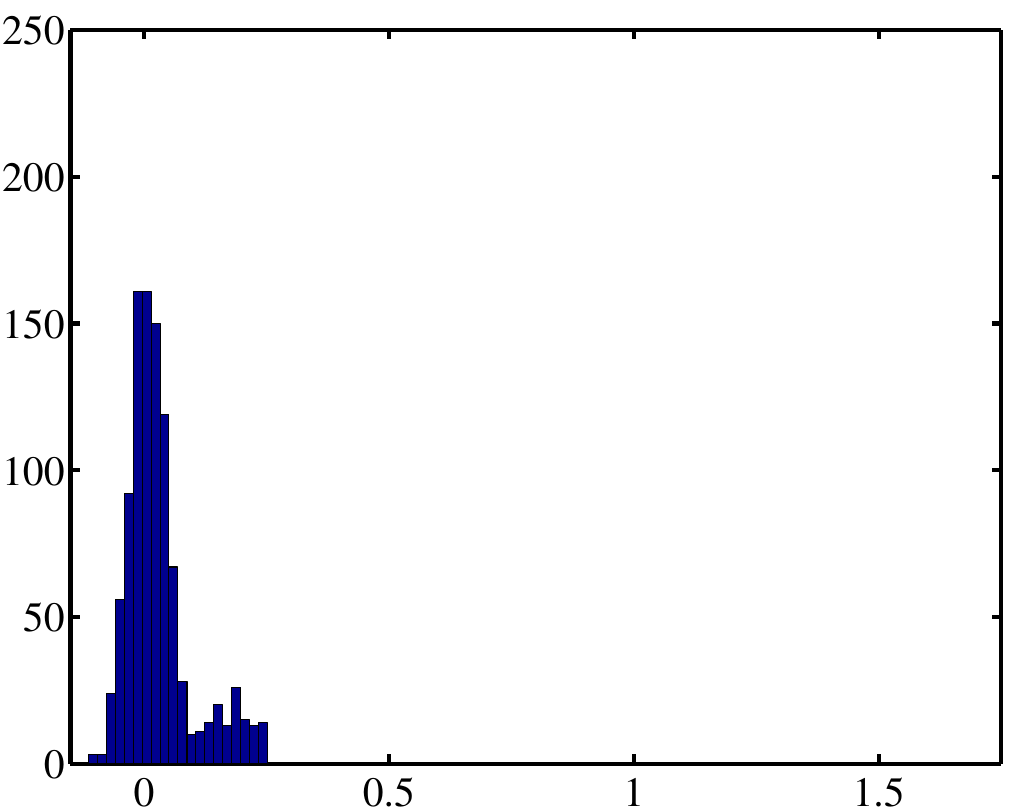}
    & 
    \includegraphics[width=0.2\textwidth]{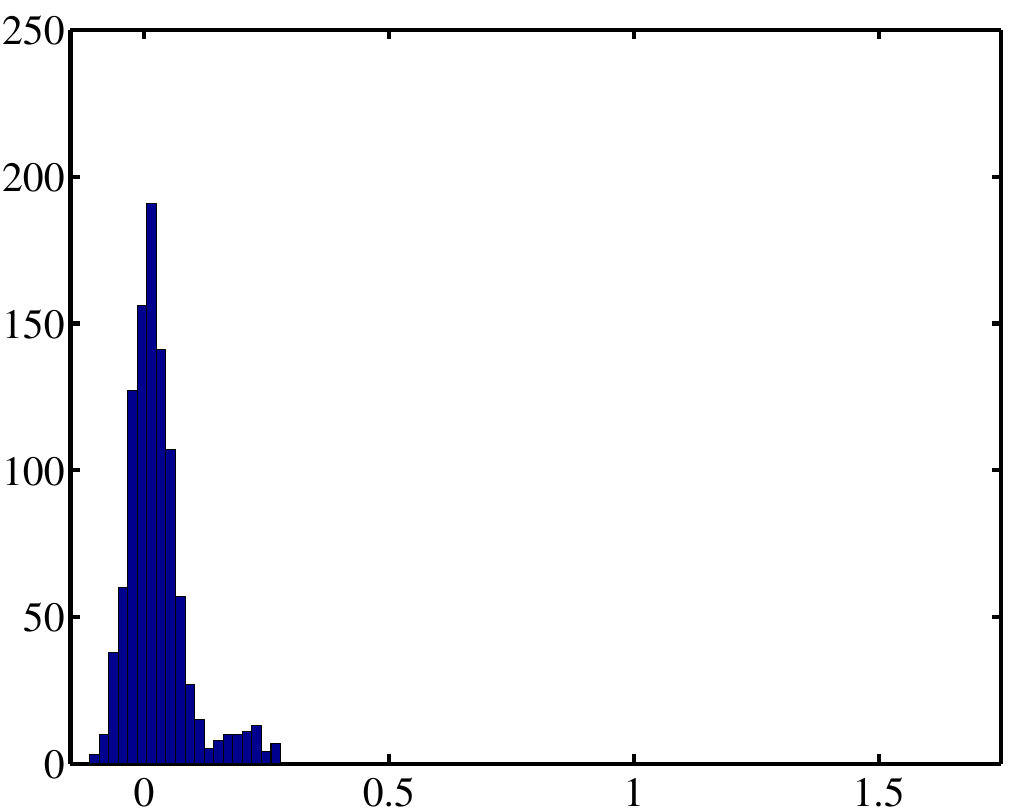}
    \\ \hline
  \end{tabular}
  \caption{Histograms of optimality gap estimates for SAA-BFGS, over a
    matrix of samples sizes $M$ and $N$. For each histogram, the horizontal axis
    represents value of the gap estimate and the vertical axis represents frequency.}
  \label{t:optGap}
\end{table}

\begin{table}[htb]
  \centering
  \begin{tabular}{c|cccc}
    \backslashbox{$N$}{$M$} & 2 & 11 & 101 & 1001 \\ \hline
    \multirow{2}{*}{1} 
    & \includegraphics[width=0.2\textwidth]{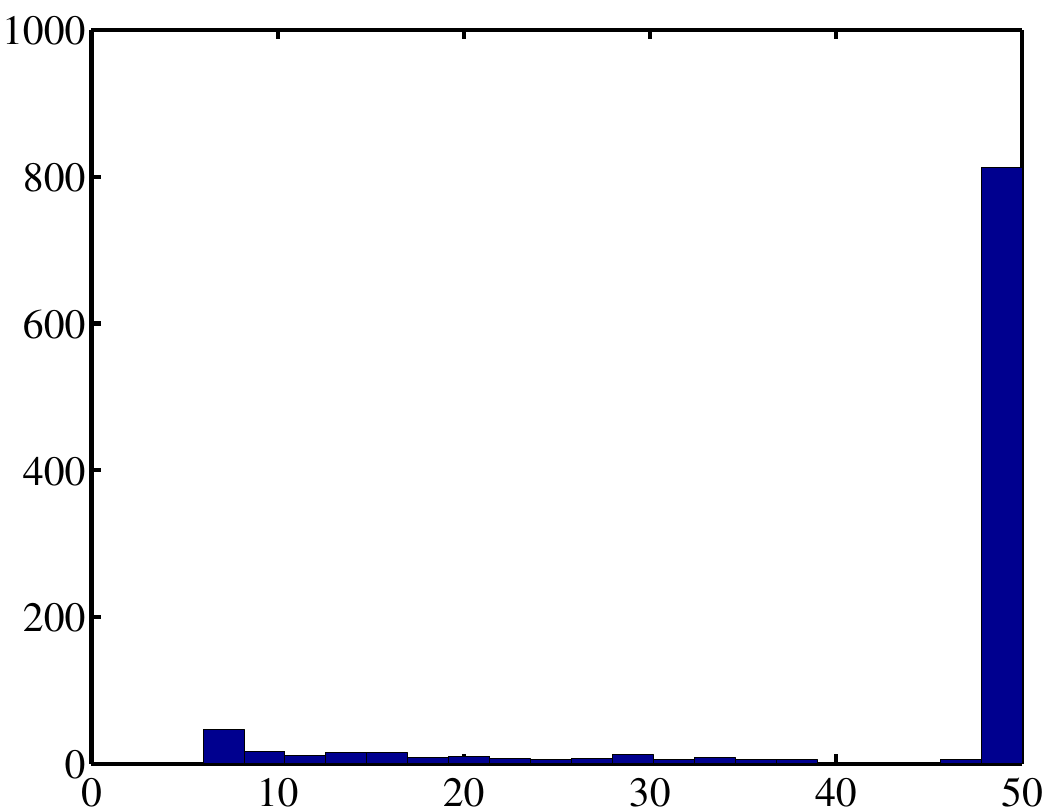}
    & 
    \includegraphics[width=0.2\textwidth]{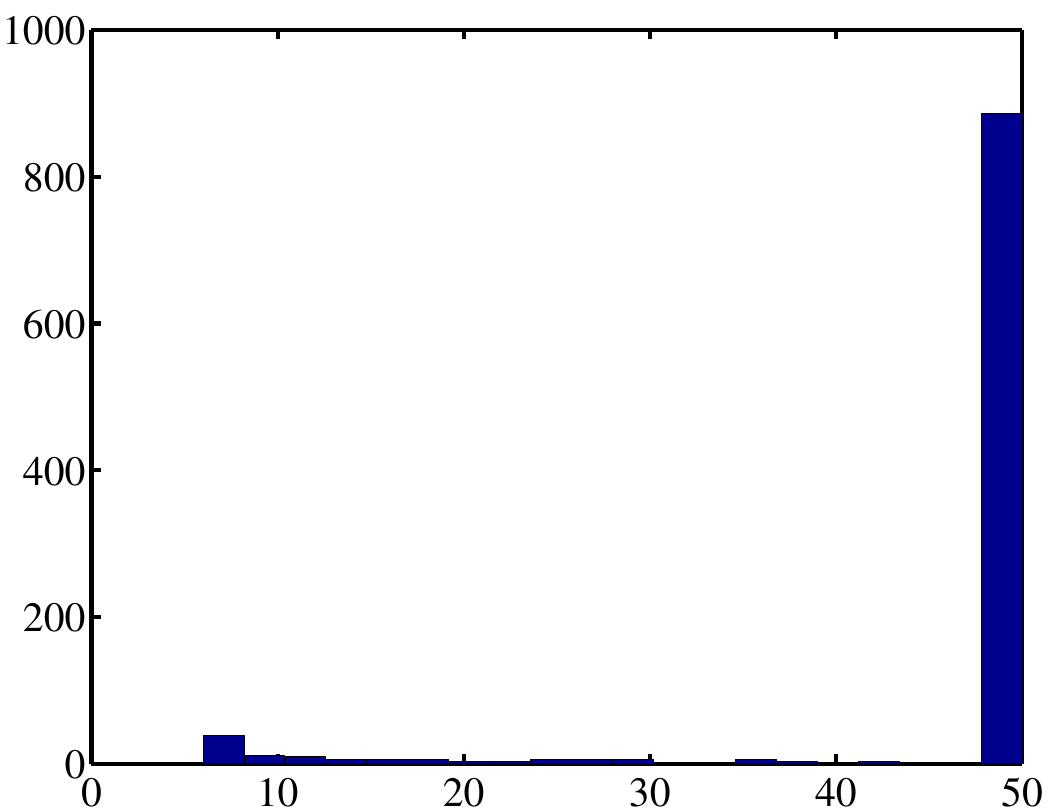}
    & 
    \includegraphics[width=0.2\textwidth]{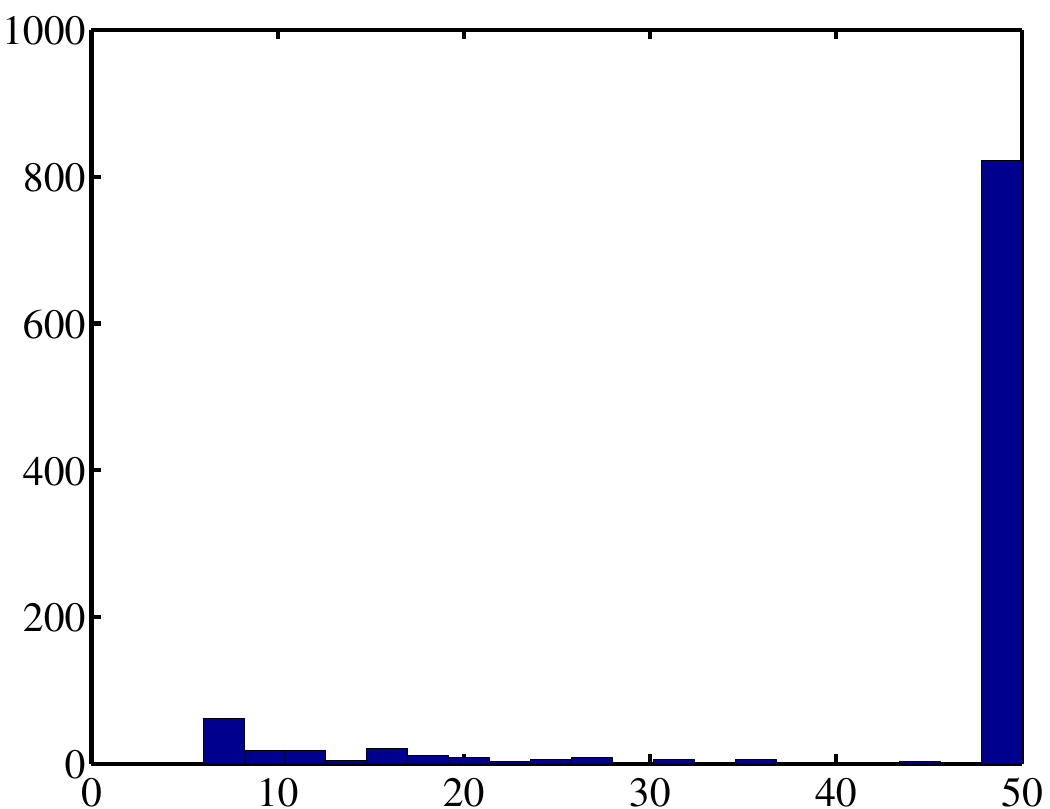}
    & 
    \includegraphics[width=0.2\textwidth]{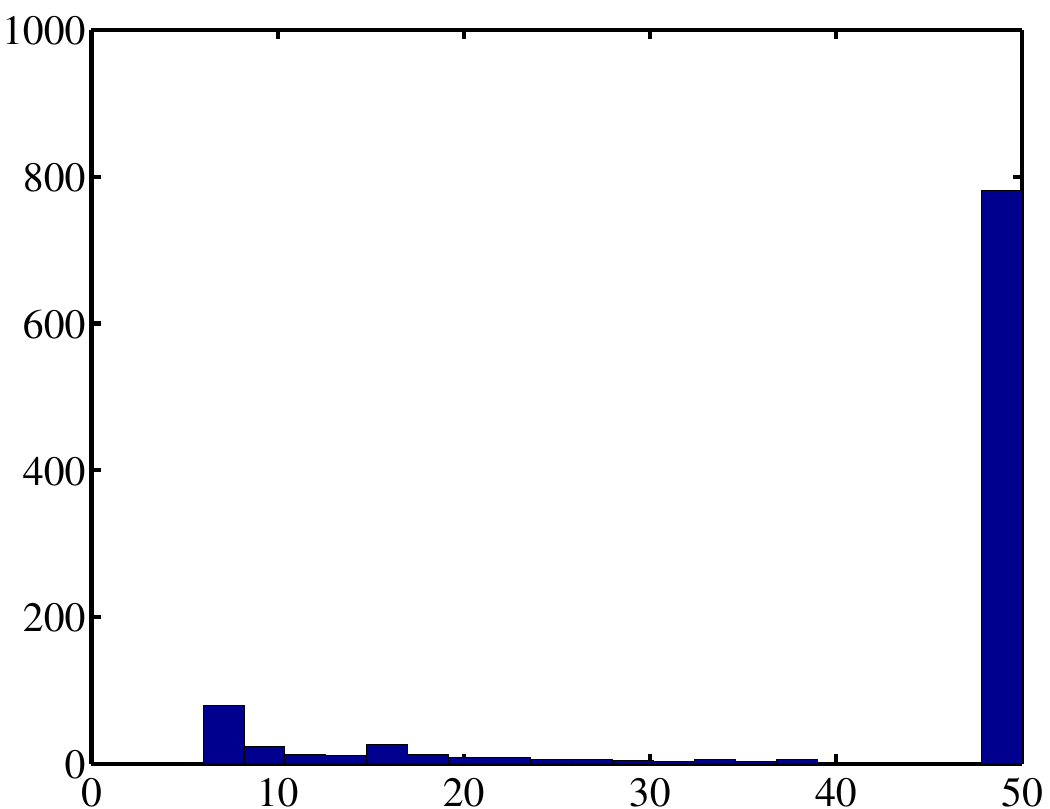}
    \\
    & \includegraphics[width=0.2\textwidth]{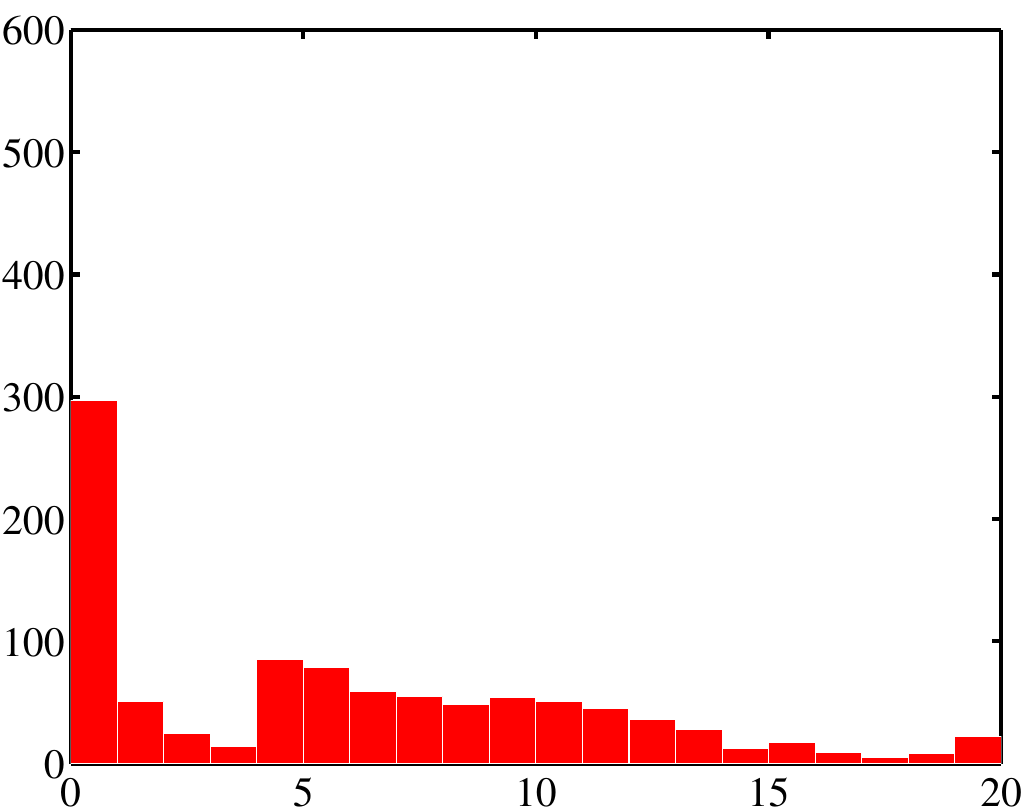}
    & 
    \includegraphics[width=0.2\textwidth]{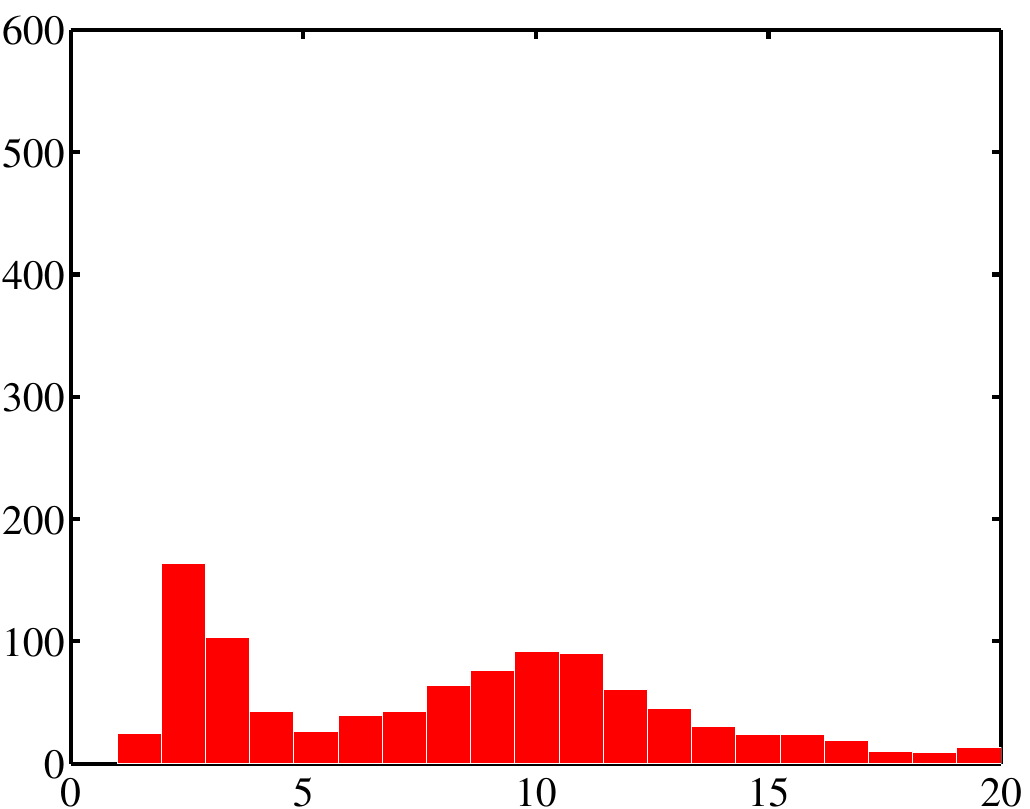}
    & 
    \includegraphics[width=0.2\textwidth]{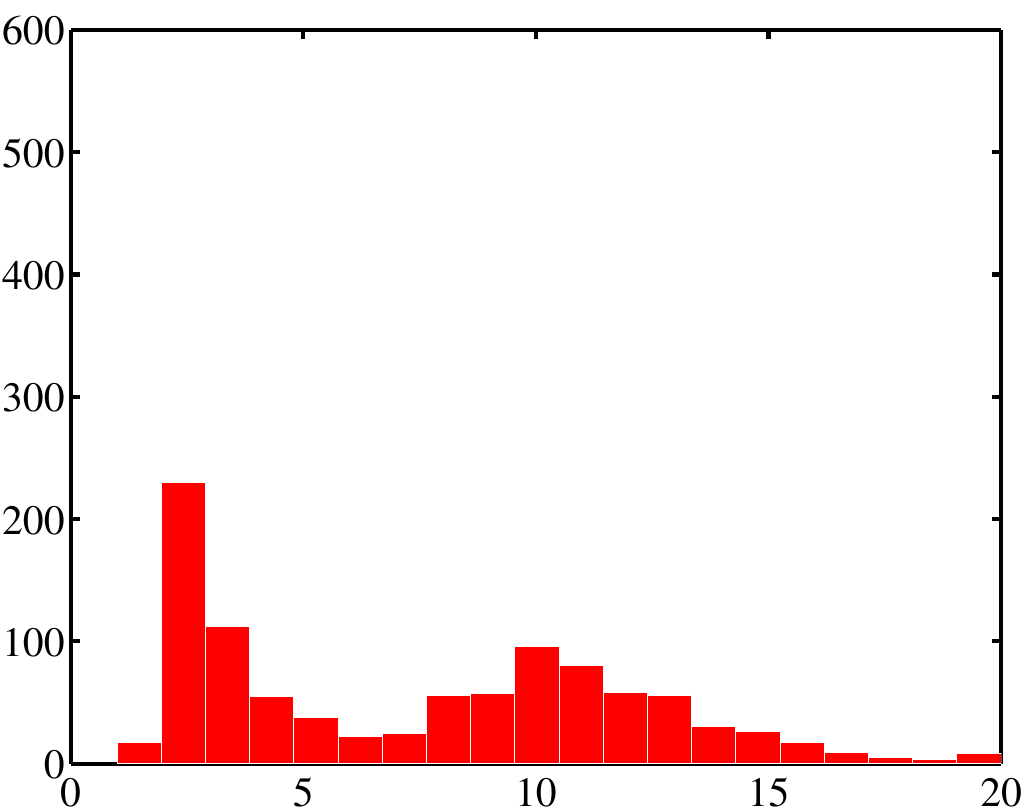}
    & 
    \includegraphics[width=0.2\textwidth]{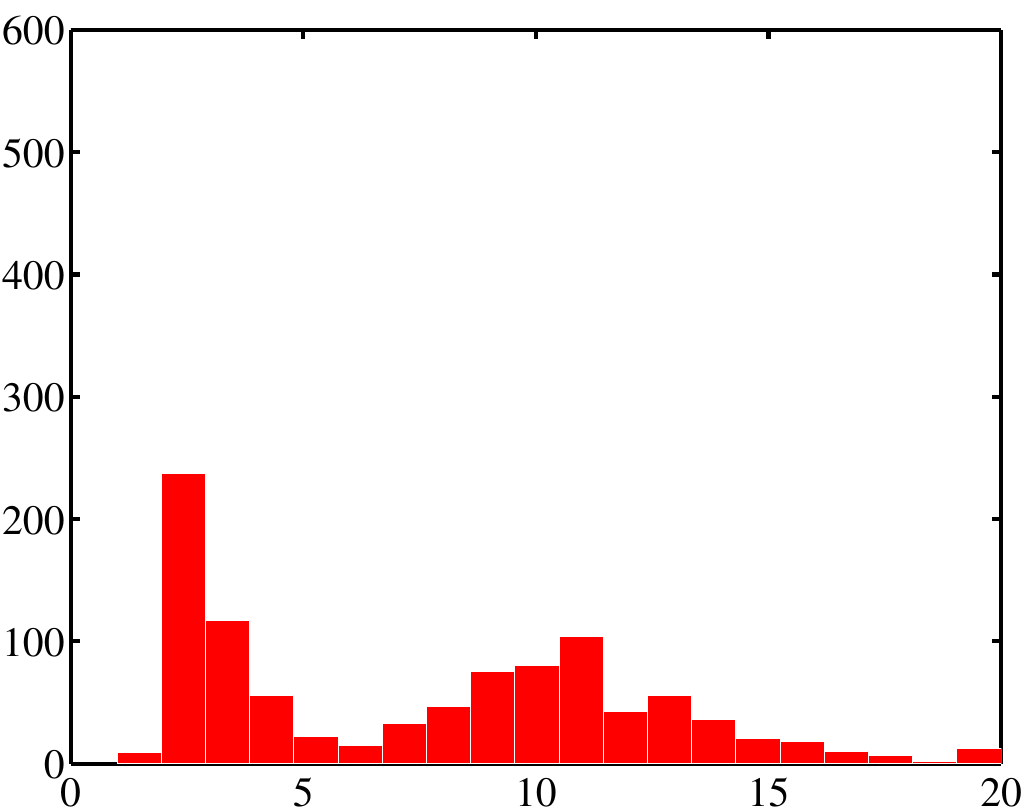}
    \\ \hline
    \multirow{2}{*}{11} 
    & \includegraphics[width=0.2\textwidth]{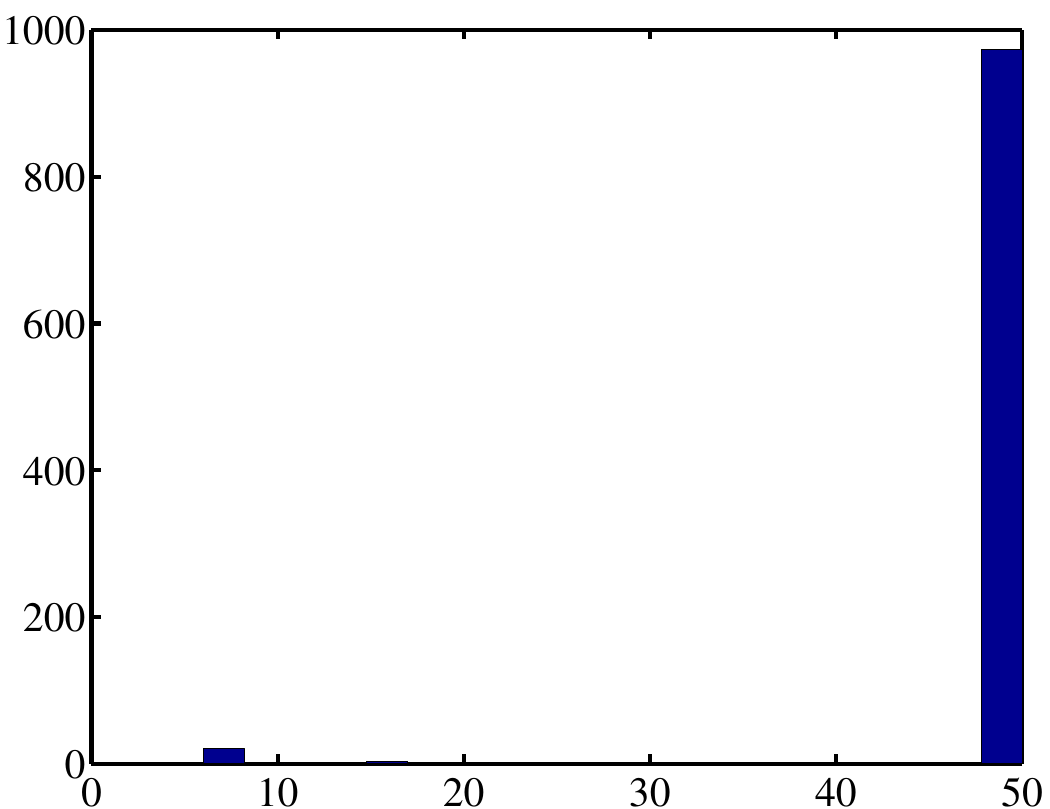}
    & 
    \includegraphics[width=0.2\textwidth]{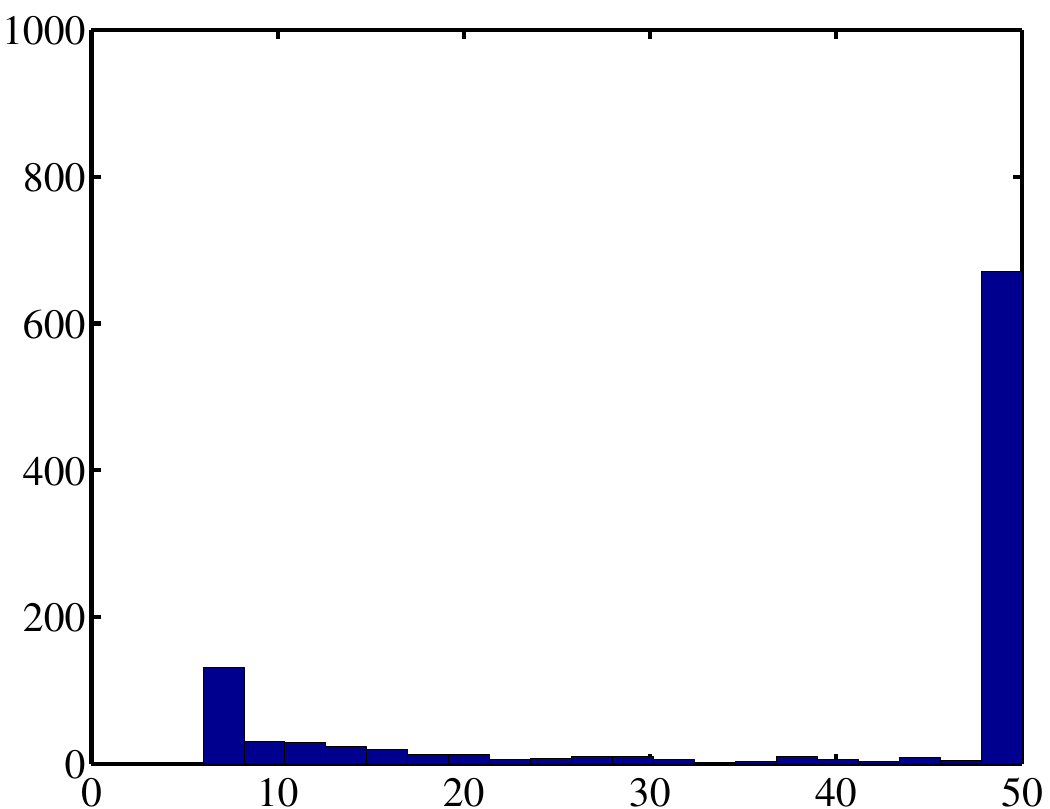}
    & 
    \includegraphics[width=0.2\textwidth]{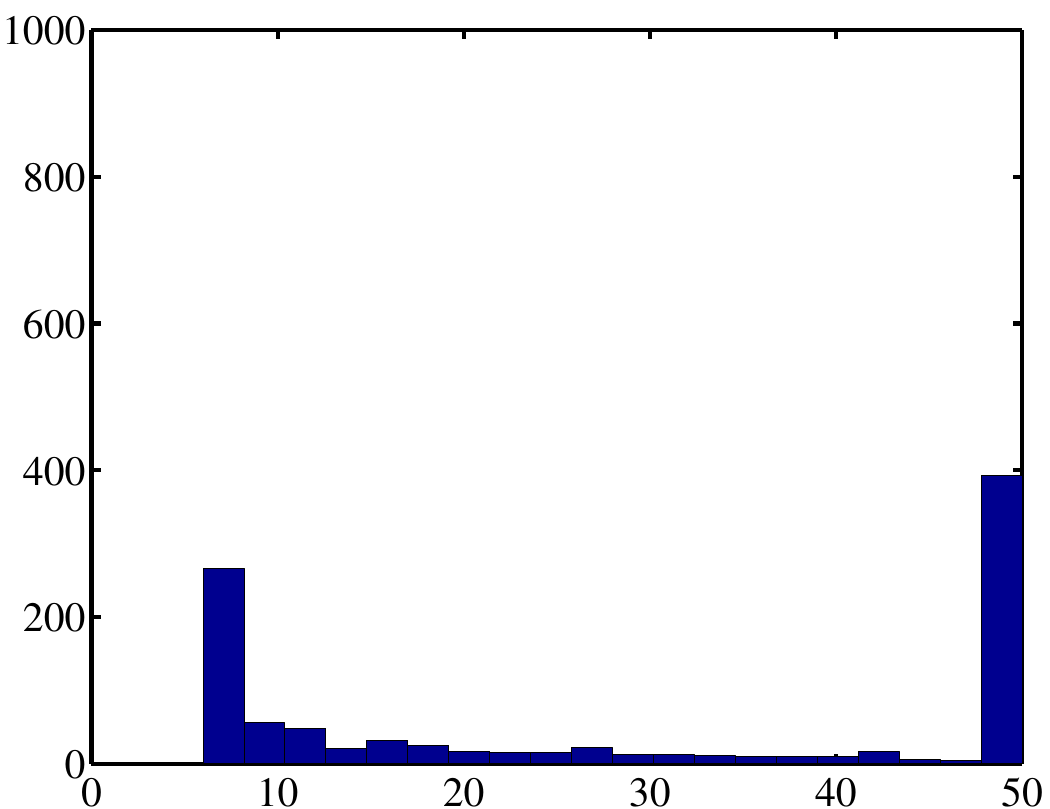}
    & 
    \includegraphics[width=0.2\textwidth]{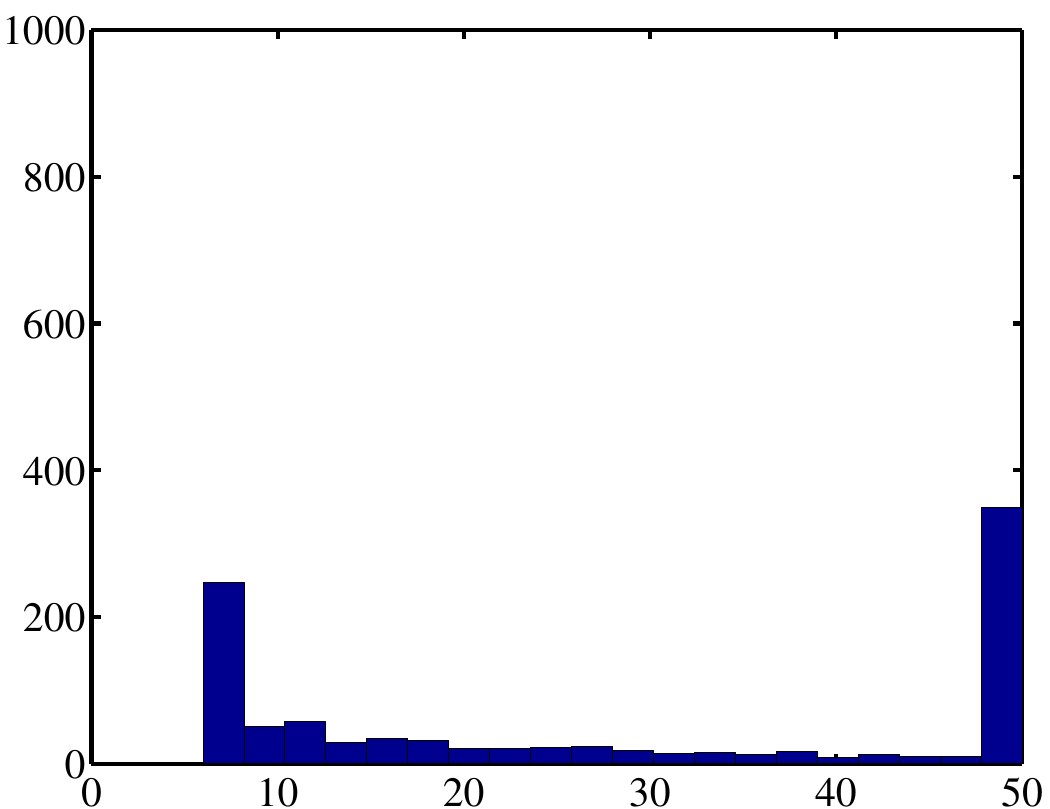}
    \\
    & \includegraphics[width=0.2\textwidth]{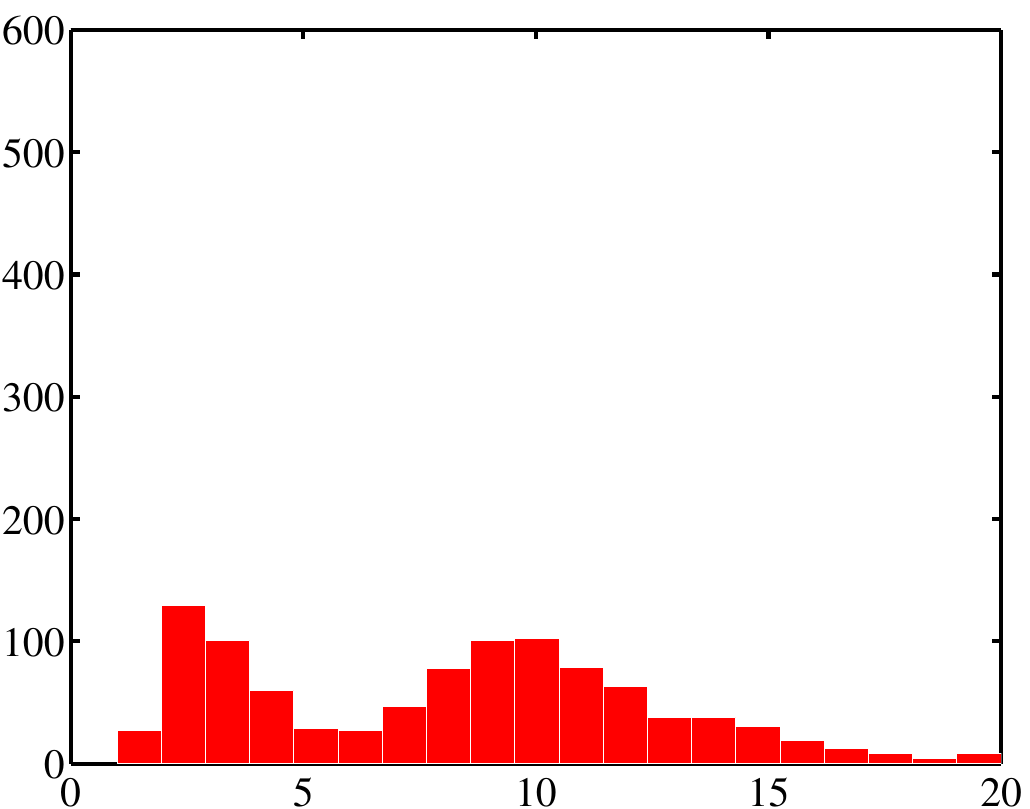}
    & 
    \includegraphics[width=0.2\textwidth]{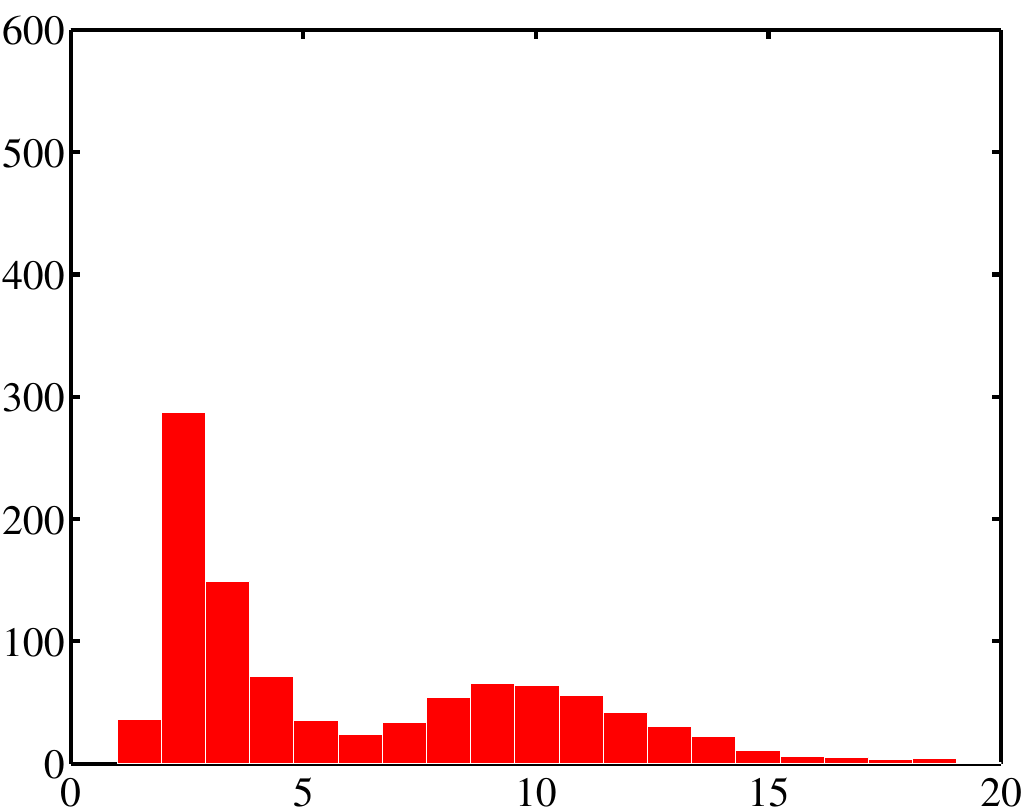}
    & 
    \includegraphics[width=0.2\textwidth]{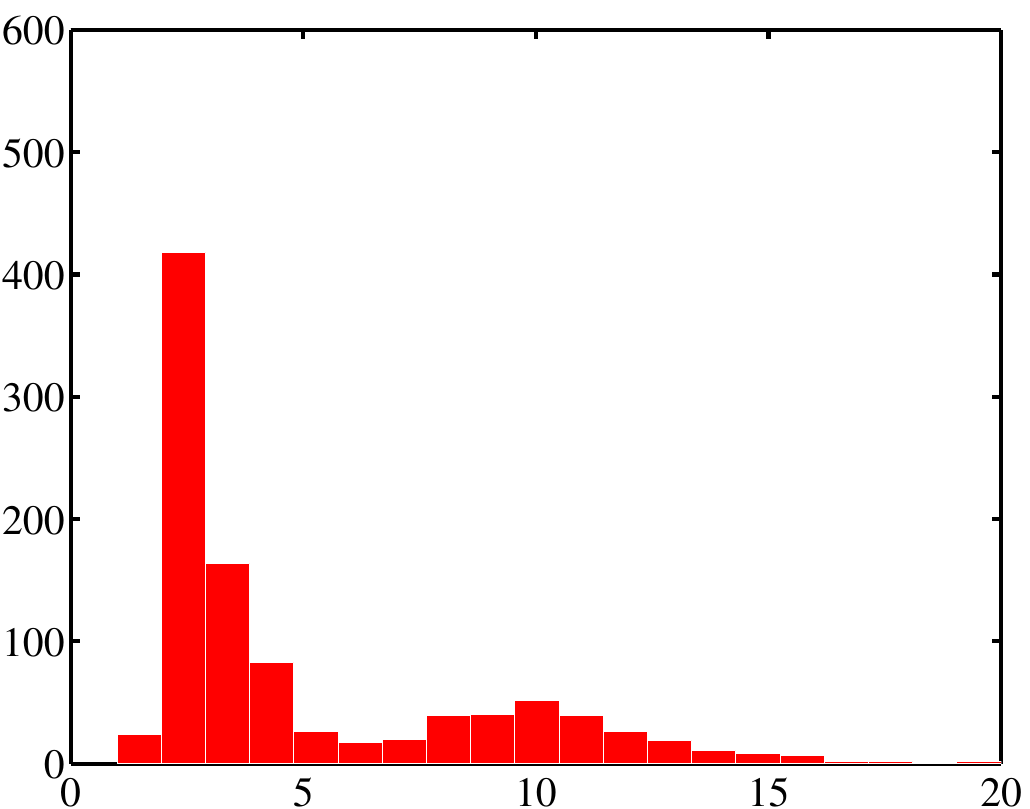}
    & 
    \includegraphics[width=0.2\textwidth]{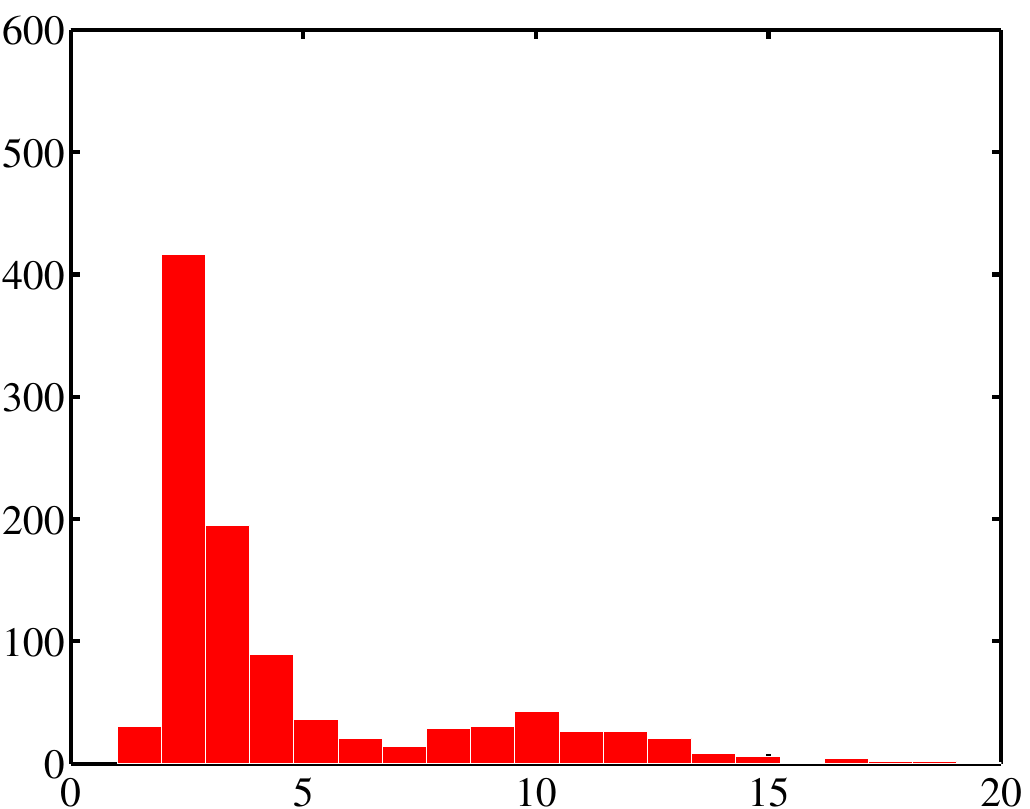}
    \\ \hline
    \multirow{2}{*}{101} 
    & \includegraphics[width=0.2\textwidth]{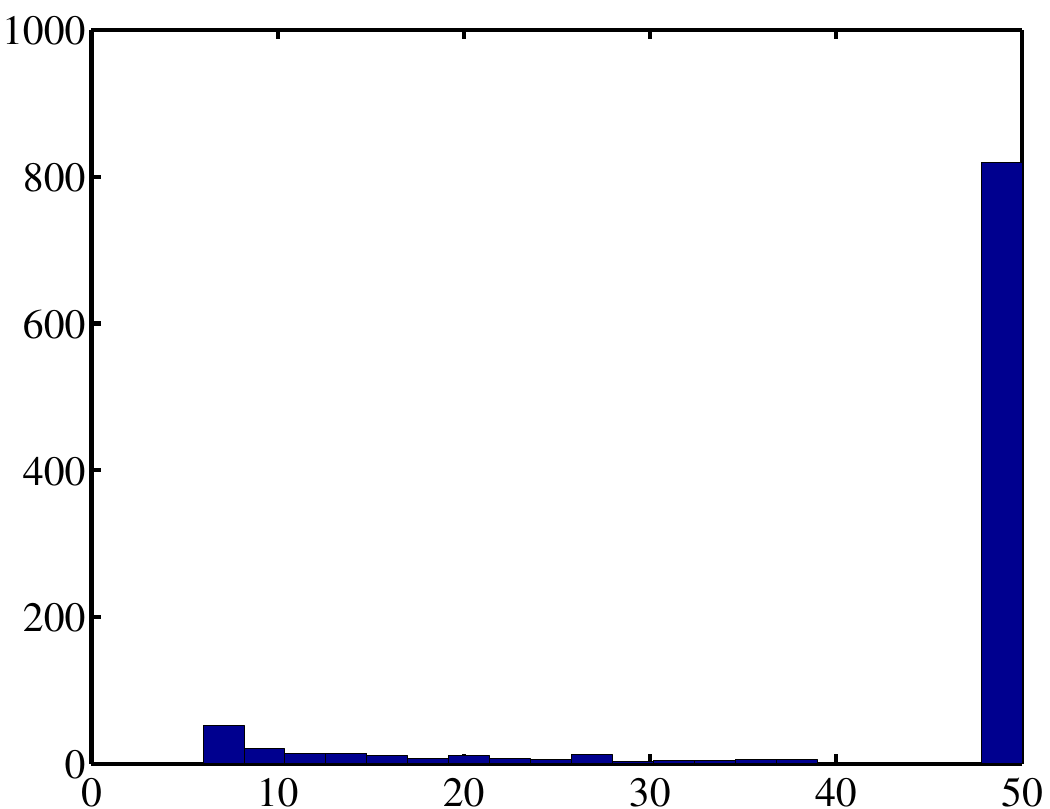}
    & 
    \includegraphics[width=0.2\textwidth]{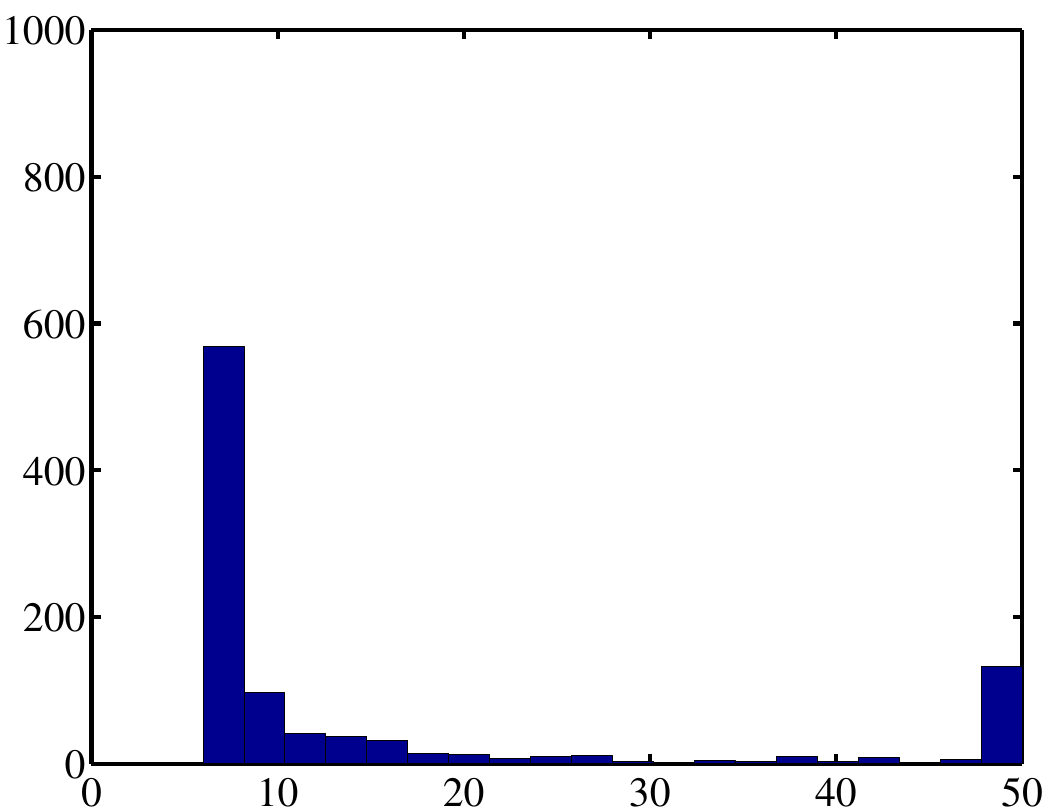}
    & 
    \includegraphics[width=0.2\textwidth]{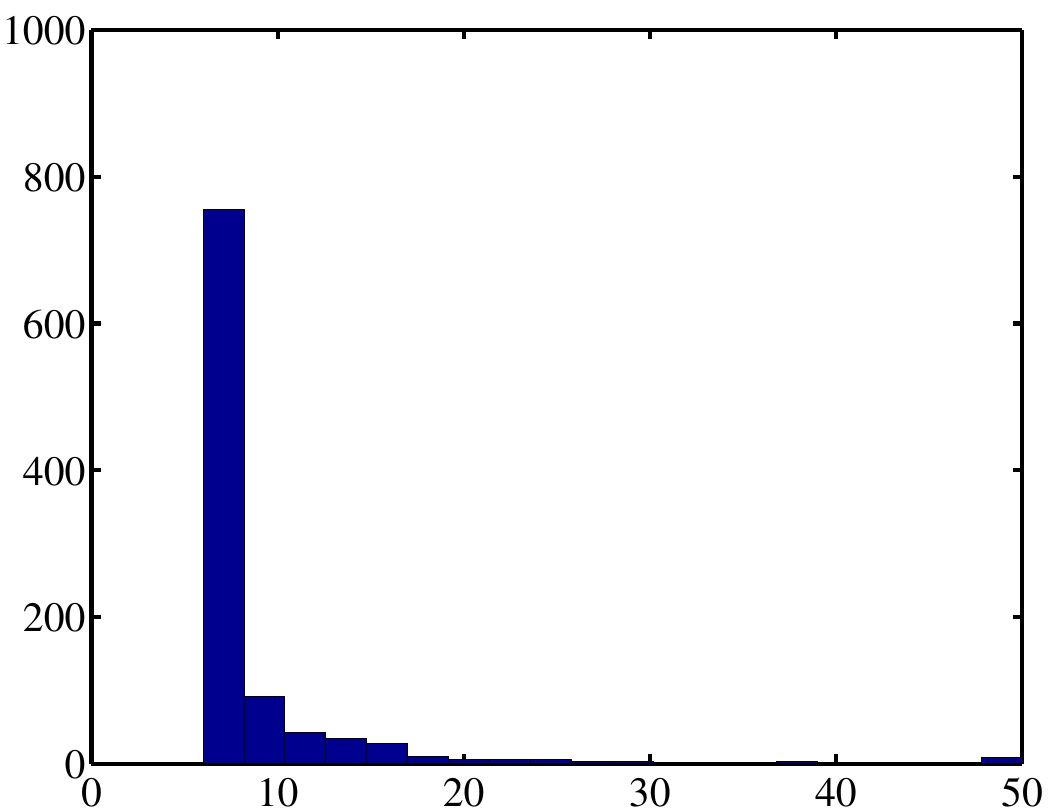}
    & 
    \includegraphics[width=0.2\textwidth]{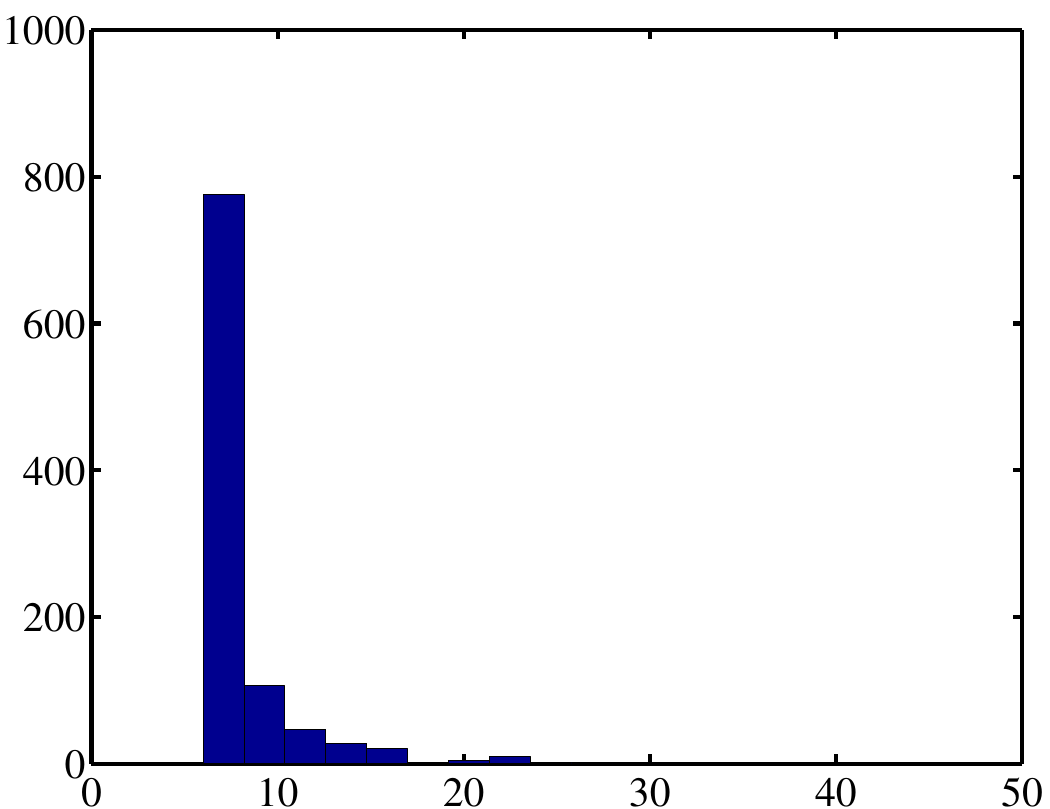}
    \\
    & \includegraphics[width=0.2\textwidth]{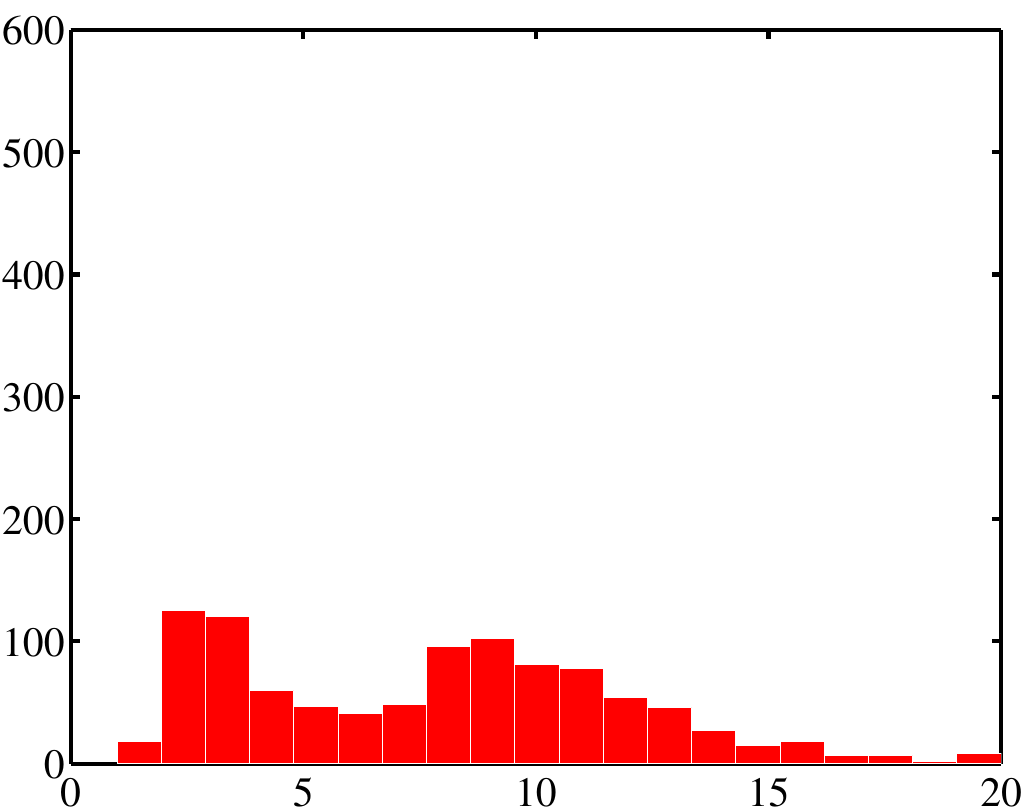}
    & 
    \includegraphics[width=0.2\textwidth]{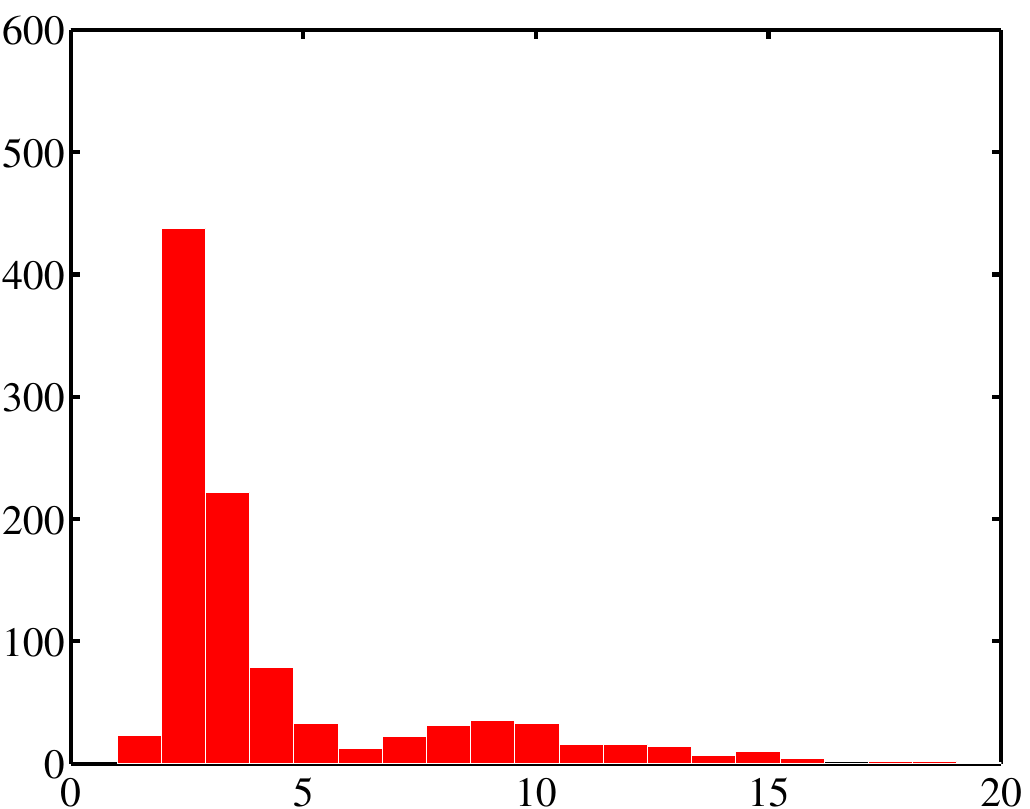}
    & 
    \includegraphics[width=0.2\textwidth]{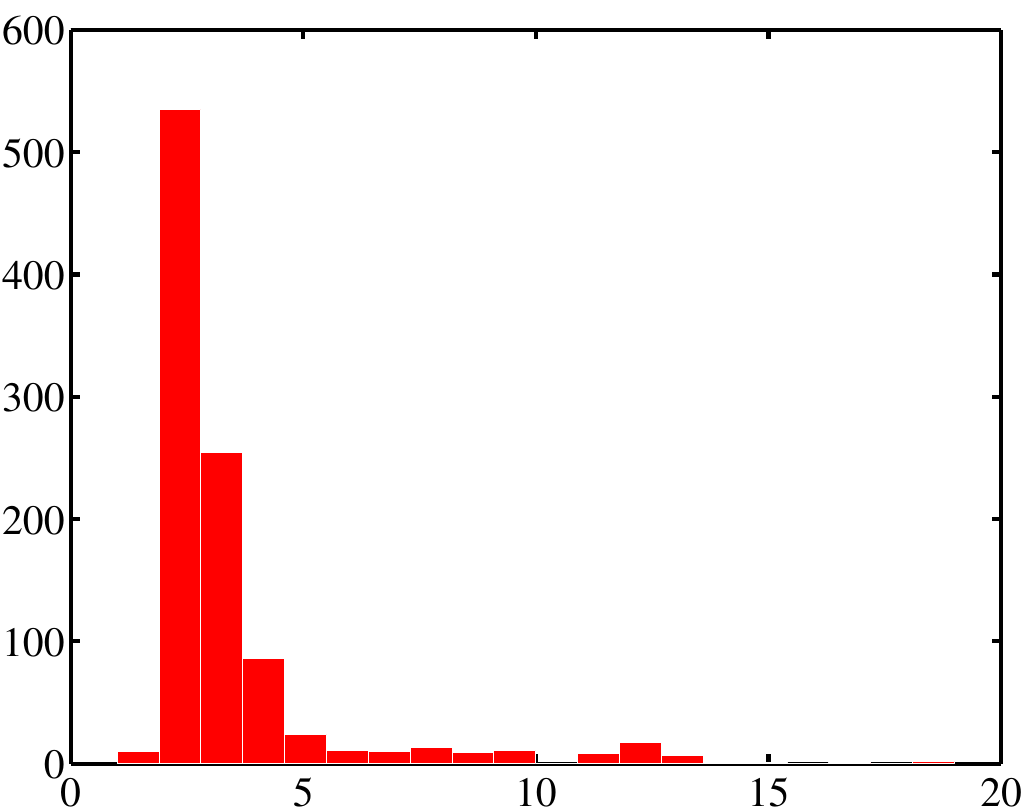}
    & 
    \includegraphics[width=0.2\textwidth]{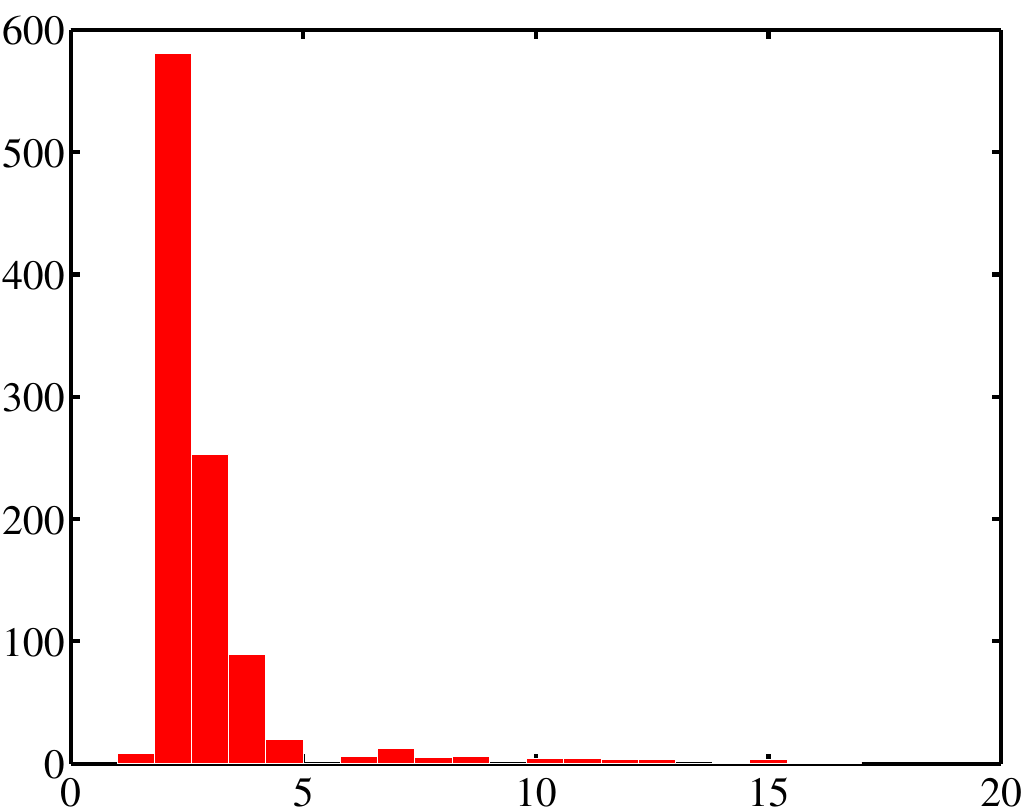}
    \\ \hline
  \end{tabular}
  \caption{Number of iterations in each independent run of RM (top subrows, blue) and SAA-BFGS
    (bottom subrows, red), over a matrix of sample sizes $M$ and $N$. For each histogram, the horizontal axis
    represents iteration number and the vertical axis represents frequency.}
  \label{t:FinalIter}
\end{table}

\cleardoublepage
%% appendix sections are then done as normal sections and after Acknowledgements
\appendix

\section*{Appendix: Analytical Derivation of the Unbiased Gradient
  Estimator}
%\label{app:hUDerivative}

In this section, we derive the analytical form of the unbiased
gradient estimator
$\nabla\hU_{N,M}(\bd,\btheta_s,\bz_s)$,\footnote{Recall that this
  estimator is unbiased with respect to the gradient of $\barU_M$.}
following the method presented in Section~\ref{s:ipa}.

The estimator $\hU_{N,M}(\bd,\btheta_s,\bz_s)$ is defined in
(\ref{e:EUEstimator2}). Its gradient in component form is
\begin{eqnarray}
  \nabla \hU_{N,M}(\bd,\btheta_s,\bz_s) = \[\begin{array}{c}
    \pp{}{d_1}\hU_{N,M}(\bd,\btheta_s,\bz_s) \\
    \pp{}{d_2}\hU_{N,M}(\bd,\btheta_s,\bz_s) \\ \vdots \\
    \pp{}{d_a}\hU_{N,M}(\bd,\btheta_s,\bz_s) \\ \vdots \\ 
    \pp{}{d_{n_d}}\hU_{N,M}(\bd,\btheta_s,\bz_s)  \end{array}\], \label{e:gradVec}
\end{eqnarray}
where $n_d$ is the dimension of the design parameters $\bd$ and $d_a$ denotes the $a$th
component of $\bd$. The $a$th component of the gradient is then
\begin{eqnarray}
  \pp{}{d_a}\hU_{N,M}(\bd,\btheta_s,\bz_s) &=& \frac{1}{N}
  \sum_{i=1}^{N}
  \left\{\frac{\pp{}{d_a}f_{\bY|\bTheta,\bd}
      \(\bG(\btheta^{(i)},\bd)+\bC(\btheta^{(i)},\bd)\bz^{(i)} \left |
        \btheta^{(i)},\bd
      \right. \)}
    {f_{\bY|\bTheta,\bd} \(
      \bG(\btheta^{(i)},\bd)+\bC(\btheta^{(i)},\bd)\bz^{(i)} \left | \btheta^{(i)},\bd
      \right . \)}
  \right. \nonumber\\ && \left. -
    \frac{
      \sum_{j=1}^{M} 
      \pp{}{d_a}f_{\bY|\bTheta,\bd} \(
      \bG(\btheta^{(i)},\bd)+\bC(\btheta^{(i)},\bd)\bz^{(i)} \left | \btheta^{(i,j)},\bd
      \right . \)
    }{\sum_{j^\prime=1}^{M}
      f_{\bY|\bTheta,\bd} \(
      \bG(\btheta^{(i)},\bd)+\bC(\btheta^{(i)},\bd)\bz^{(i)} \left | \btheta^{(i,j^\prime)},\bd
      \right . \)
    }\right\}. \label{e:dUhatdda}
\end{eqnarray}
Partial derivatives of the likelihood function with respect to $\bd$
are required above. We assume that each component of
$\bC(\btheta^{(i)},\bd)$ is of the form $\alpha_c+\beta_c
|G_c(\btheta^{(i)},\bd)|$, $c=1\ldots n_y$, where $n_y$ is the
dimension of the data vector $\bY$, and $\alpha_c, \beta_c$ are
constants. Also, let the random vectors $\bz^{(i)}$ be mutually
independent and composed of i.i.d.\ components, such that the data are
conditionally independent given $\btheta$ and $\bd$. The derivative of
the likelihood function then becomes
\begin{eqnarray}
  &&
  \pp{}{d_a}f_{\bY|\bTheta,\bd} \(
  \bG(\btheta^{(i)},\bd)+\bC(\btheta^{(i)},\bd)\bz^{(i)} \left | \btheta^{(i,j)},\bd
  \right . \)
  \nonumber\\ 
  &=&
  \pp{}{d_a}\[\prod_{c=1}^{n_y}f_{Y_c|\bTheta,\bd} \(G_c(\btheta^{(i)},\bd)+(\alpha_c+\beta_c
  |G_c(\btheta^{(i)},\bd)|)z_c^{(i)} \left | \btheta^{(i,j)},\bd
  \right . \) \] \nonumber \\
  &=& 
  \sum_{k=1}^{n_y}\[\pp{}{d_a}f_{Y_k|\bTheta,\bd} \(G_k(\btheta^{(i)},\bd)+(\alpha_k+\beta_k
  |G_k(\btheta^{(i)},\bd)|)z_k^{(i)} \left | \btheta^{(i,j)},\bd  \right . \) \right. \nonumber\\ 
&& \left. \prod_{\substack{c=1 \\c\neq
      k}}^{n_y}f_{Y_c|\bTheta,\bd} \( G_c(\btheta^{(i)},\bd)+(\alpha_c+\beta_c 
  |G_c(\btheta^{(i)},\bd)|)z_c^{(i)} \left | \btheta^{(i,j)},\bd
    \right . \) \].
\end{eqnarray}
Introducing a standard normal density for each $z_c^{(i)}$, the
likelihood associated with a single component of the data vector is
\begin{footnotesize}
\begin{eqnarray}
  &&f_{Y_c|\bTheta,\bd} \( G_c(\btheta^{(i)},\bd)+(\alpha_c+\beta_c
  |G_c(\btheta^{(i)},\bd)|)z_c^{(i)} \left | \btheta^{(i,j)},\bd
  \right . \) \nonumber\\
  &=& \frac{1}{\sqrt{2\pi}\(\alpha_c +
    \beta_c |G_c(\btheta^{(i,j)},\bd)|\)} \nonumber\\
  && \times
  \exp\[-\frac{\(G_c(\btheta^{(i,j)},\bd)-(G_c(\btheta^{(i)},\bd)+(\alpha_c+\beta_c 
    |G_c(\btheta^{(i)},\bd)|)z_c^{(i)})\)^2}{2\(\alpha_c + \beta_c
    |G_c(\btheta^{(i,j)},\bd)|\)^2}\],
\end{eqnarray}
\end{footnotesize}
and its derivatives are
%\begin{scriptsize}
\begin{eqnarray}
  && \pp{}{d_a}f_{Y_c|\bTheta,\bd} \(G_c(\btheta^{(i)},\bd)+(\alpha_c+\beta_c
  |G_c(\btheta^{(i)},\bd)|)z_c^{(i)} \left | \btheta^{(i,j)},\bd
  \right . \) \nonumber\\
  &=& \frac{-\beta_c \sgn(G_c(\btheta^{(i,j)},\bd))\pp{}{d_a}
    G_c(\btheta^{(i,j)},\bd)}{\sqrt{2\pi}\(\alpha_c+\beta_c 
    |G_c(\btheta^{(i,j)},\bd)|\)^2}\nonumber\\
  &&\times \exp\[-\frac{\(G_c(\btheta^{(i,j)},\bd)-(G_c(\btheta^{(i)},\bd)+(\alpha_c+\beta_c
    |G_c(\btheta^{(i)},\bd)|)z_c^{(i)})\)^2}{2\(\alpha_c + \beta_c
    |G_c(\btheta^{(i,j)},\bd)|\)^2}\] \nonumber\\
  &&+ \frac{1}{\sqrt{2\pi}\(\alpha_c +
    \beta_c |G_c(\btheta^{(i,j)},\bd)|\)} \nonumber\\
  &&\times
  \exp\[-\frac{\(G_c(\btheta^{(i,j)},\bd)-(G_c(\btheta^{(i)},\bd)+(\alpha_c+\beta_c 
    |G_c(\btheta^{(i)},\bd)|)z_c^{(i)})\)^2}{2\(\alpha_c + \beta_c
    |G_c(\btheta^{(i,j)},\bd)|\)^2}\] \nonumber\\
  && \times
  \left\{-\frac{\(G_c(\btheta^{(i,j)},\bd)-(G_c(\btheta^{(i)},\bd)+(\alpha_c+\beta_c 
      |G_c(\btheta^{(i)},\bd)|)z_c^{(i)})\)}{\(\alpha_c + \beta_c
      |G_c(\btheta^{(i,j)},\bd)|\)^2}\right. \nonumber\\
  && \times \(\pp{}{d_a}G_c(\btheta^{(i,j)},\bd)-\(\pp{}{d_a}
      G_c(\btheta^{(i)},\bd)(1+\beta_c
      \sgn(G_c(\btheta^{(i)},\bd))z_c^{(i)})\)\) \nonumber\\ 
  && + \frac{\(G_c(\btheta^{(i,j)},\bd)-(G_c(\btheta^{(i)},\bd)+(\alpha_c+\beta_c
      |G_c(\btheta^{(i)},\bd)|)z_c^{(i)})\)^2}{\(\alpha_c + \beta_c
      |G_c(\btheta^{(i,j)},\bd)|\)^3}\nonumber\\
  && \left. \times \beta_c \sgn(G_c(\btheta^{(i,j)},\bd))\pp{}{d_a}
      G_c(\btheta^{(i,j)},\bd)\right\}.
\end{eqnarray}
%\end{scriptsize}
In cases where conditioning on $\btheta^{(i,j)}$ is replaced by
conditioning on $\btheta^{(i)}$
(i.e., for the first summation term in equation~(\ref{e:dUhatdda})), the
expressions simplify to
\begin{eqnarray}
  && f_{Y_c|\bTheta,\bd}(G_c(\btheta^{(i)},\bd)+(\alpha_c+\beta_c
  |G_c(\btheta^{(i)},\bd)|)z_c^{(i)}|\btheta^{(i)},\bd) \nonumber\\
  &=& \frac{1}{\sqrt{2\pi}\(\alpha_c +
    \beta_c |G_c(\btheta^{(i)},\bd)|\)} \exp\[-\frac{\(z_c^{(i)}\)^2}{2}\]
\end{eqnarray}
and 
\begin{eqnarray}
  && \pp{}{d_a}f_{Y_c|\bTheta,\bd}(G_c(\btheta^{(i)},\bd)+(\alpha_c+\beta_c
  G_c(\btheta^{(i)},\bd))z_c^{(i)}|\btheta^{(i)},\bd) \nonumber\\
  &=& \frac{-\beta_c \sgn(G_c(\btheta^{(i)},\bd)) \pp{}{d_a}
    G_c(\btheta^{(i)},\bd)}{\sqrt{2\pi}\(\alpha_c+\beta_c 
    |G_c(\btheta^{(i)},\bd)|\)^2}
  \exp\[-\frac{\(z_c^{(i)}\)^2}{2}\]. 
\end{eqnarray}

We now require the derivative of each model output $G_c$ with respect
to $\bd$. In most cases, this quantity will not be available
analytically. One could use an adjoint method to evaluate the
derivatives, or instead employ a finite difference approximation, but
embedding these approaches in a Monte Carlo sum may be prohibitive,
particularly if each forward model evaluation is computationally
expensive.
The polynomial chaos surrogate introduced in Section~\ref{s:pce}
addresses this problem by replacing the forward model with polynomial
expansions for either $G_c$
\begin{eqnarray}
  G_c(\btheta^{(i)},\bd) \approx
%  G_c^{p,n_\mathrm{quad}}(\btheta^{(i)},\bd) = 
  \sum_{\bb \in \mathcal{J}} g_{\bb}
  \Psi_{\bb}\(\bxi(\btheta^{(i)},\bd)\)
  \label{e:pcReplace}
\end{eqnarray}
or $\ln G_c$
\begin{eqnarray}
  G_c(\btheta^{(i)},\bd) \approx
%  \exp\[G_{\ln,c}^{p,n_\mathrm{quad}}(\btheta^{(i)},\bd)\] =
  \exp\[\sum_{\bb \in \mathcal{J}} g_{\bb}
  \Psi_{\bb}\(\bxi(\btheta^{(i)},\bd)\)\].
  \label{e:pcReplaceLog}
\end{eqnarray}
Here $g_{\bb}$ are the expansion coefficients and $\mathcal{J}$ is an
admissible multi-index set indicating which polynomial terms are in
the expansion. For instance, if $n_\theta$ is the dimension of
$\btheta$ and $n_d$ is the dimension of $\bd$, such that $n_\theta +
n_d$ is the dimension of $\bxi$, then $\mathcal{J} := \{ \bb \in
\mathbb{N}_0^{n_\theta + n_d} : |\bb|_1 \leq p \}$ is a total-order
expansion of degree $p$. This expansion converges in the $L^2$ sense as $p
\rightarrow \infty$. 

% \footnote{In Equations~\ref{e:pcReplace} and~\ref{e:pcReplaceLog}, we
%   used ``$=$'' rather than ``$\approx$'' in indicating the replacement
%   of $G_c$ by $G_c^{p,n_\mathrm{quad}}$ (or
%   $\exp\[G_{\mathrm{ln},c}^{p,n_\mathrm{quad}}\]$). The PC expansions
%   are indeed approximations to the original forward model, but here we
%   let $G_c$ represent whatever model used in the optimization. The
%   approximation of surrogate is thus grouped into when $U$ is
%   approximated by $\hU_{M,N}$. The surrogate approximation does not
%   change the consistency of $\hU_{M,N}$ since PC expansions converge
%   to the original model as $p$ and $n_{\mathrm{quad}}$ tend to
%   infinity.}
% YMM: I think we mostly covered this issue earlier: that we are
% optimizing the $U$ obtained by direct substitution of the surrogate,
% so we don't in fact worry about the polynomial approximation
% error. Maybe the only thing to emphasize here is that the
% approximation is consistent and can be made as accurate as desired.

Consider the latter (ln-$G_c$) case; here, the derivative of the
polynomial chaos expansion is
\begin{eqnarray}
  \pp{}{d_a}G_c(\btheta^{(i)},\bd) =
  \exp\[\sum_{\bb} g_{\bb}
  \Psi_{\bb}\(\bxi(\btheta^{(i)},\bd)\)\]\sum_{\bb} g_{\bb}
  \pp{}{d_a}\Psi_{\bb}(\bxi(\btheta^{(i)},\bd)). \label{e:dddGc}
\end{eqnarray}
In the former ($G_c$ without the logarithm) case, we obtain the same
expression except without the $\exp\[\cdot\]$ term.

To complete the derivation, we assume that each component of the input
parameters $\bTheta$ and design variables $\bd$ is represented by
an affine transformation of corresponding basis random variable
$\Xi$:
\begin{eqnarray}
  \Theta_l &=& \gamma_l + \delta_l \Xi_l, \\
  d_{l^\prime-n_{\theta}} &=& \gamma_{l^\prime} + \delta_{l^\prime} \Xi_{l^\prime}, \label{e:assumption4b}
\end{eqnarray}
where $\gamma_{(\cdot)}$ and $\delta_{(\cdot)} \neq 0$ are constants,
and $l=1,\ldots,n_{\theta}$ and
$l^\prime=n_{\theta}+1,\ldots,n_{\theta}+n_d$. This is a reasonable
assumption since $\Xi$ can be typically chosen such that their
distributions are of the same family as the prior on $\btheta$ (or the
uniform ``prior'' on $\bd$); this choice avoids any need for
approximate representations of the prior. The derivative of
$\Psi_{\bb}(\bxi(\btheta^{(i)},\bd))$ from equation~(\ref{e:dddGc}) is
thus
\begin{eqnarray}
  \pp{}{d_a}\Psi_{\bb}(\bxi(\btheta^{(i)},\bd)) &=&
  \pp{}{d_a}\prod_{l=1}^{n_{\theta}}
  \psi_{b_l}\(\xi_{l}(\theta_l^{(i)})\) \prod_{l^\prime=n_{\theta}+1}^{n_{\theta}+n_d}
  \psi_{b_{l^\prime}}\(\xi_{l^\prime}(d_{l^\prime-n_{\theta}})\) \nonumber\\ &=& 
  \prod_{l=1}^{n_{\theta}}
  \psi_{b_l}\(\xi_{l}(\theta_l^{(i)})\)
  \[\prod_{\substack{l^\prime=n_{\theta}+1 \\ l^\prime-n_{\theta}\neq a}}^{n_{\theta}+n_d}
  \psi_{b_{l^\prime}}\(\xi_{l^\prime}(d_{l^\prime-n_{\theta}})\)\] \pp{}{d_a}
  \psi_{b_{a+n_{\theta}}}\(\xi_{a+n_{\theta}}(d_{a})\),
\end{eqnarray}
and the derivative of the univariate basis function $\psi$ with respect to
$d_a$ is
\begin{eqnarray}
  \pp{}{d_a}\psi_{b_{a+n_{\theta}}}\(\xi_{a+n_{\theta}}(d_{a})\) &=&
  \pp{}{\xi_{a+n_{\theta}}}\psi_{b_{a+n_{\theta}}}\(\xi_{a+n_{\theta}}\)
  \pp{}{d_a} \xi_{a+n_{\theta}}(d_{a}) \nonumber \\ &=&
  \pp{}{\xi_{a+n_{\theta}}}\psi_{b_{a+n_{\theta}}}\(\xi_{a+n_{\theta}}\)
  \frac{1}{\delta_{a+n_{\theta}}},
\end{eqnarray}
where the second equality is a result of using
equation~(\ref{e:assumption4b}). The derivative of the polynomial basis
function with respect to its argument is available analytically for
many standard orthogonal polynomials, and may be evaluated using
recurrence relationships~\cite{abramowitz:1964:hom}. For example, in
the case of Legendre polynomials, the usual derivative recurrence
relationship is $ \pp{}{\xi} \psi_n(\xi) = \[-b\xi \psi_n(\xi) + b
\psi_{n-1}(\xi)\] / (1-\xi^2)$, where $n$ is the polynomial
degree. However, division by $(1-\xi^2)$ presents numerical
difficulties when evaluated on $\xi$ that fall on or near
the boundaries of the domain. Instead, a more robust alternative that
requires both previous polynomial function and derivative evaluations
can be obtained by directly differentiating the three-term recurrence
relationship for the polynomial, and is preferable in practice:
\begin{eqnarray}
  \pp{}{\xi} \psi_n(\xi) = \frac{2n - 1}{n} \psi_{n-1}(\xi)
  + \frac{2n - 1}{n} \xi \pp{}{\xi} \psi_{n-1}(\xi)
  - \frac{n-1}{n} \pp{}{\xi} \psi_{n-2}(\xi).
\end{eqnarray}

This concludes the derivation of the analytical gradient estimator
$\nabla\hU_{N,M}(\bd,\btheta_s,\bz_s)$.

% References with bibTeX database:

\bibliographystyle{IJ4UQ_Bibliography_Style}

\bibliography{referencesAll}

\end{document}